# PRE-MAIN SEQUENCE EVOLUTIONS OF SOLAR ABUNDANCE LOW MASS STARS

**Youn Kil Jung and Y. -C. Kim**

Department of Astronomy, Yonsei University, Seoul 120-749, Korea
email: nvbrain@galaxy.yonsei.ac.kr, kim@galaxy.yonsei.ac.kr

## ABSTRACT

We present the Pre-Main-Sequence (PMS) evolutionary tracks of stars with $0.065 \sim 5.0 M_\odot$. The models were evolved from the PMS stellar birthline to the onset of hydrogen burning in the core. The convective turnover timescales which enables an observational test of theoretical model, particulary in the stellar dynamic activity, are also calculated. All models have Sun-like metal abundances, typically considered as the stars in the Galactic disk and the star formation region of Population I star. The convection phenomenon is treated by the usual mixing length approximation. All evolutionary tracks are available upon request.

## 1. INTRODUCTION

Young Stellar Objects (YSOs), mainly observed in the star formation regions, are the PMS stars located in the evolutionary stages from protostar to Zero-Age Main Sequence (ZAMS). Recent observations reveal that YSOs are the key to understand star forming regions because of their outrageous levels of X-ray activity (Feigelson et al. 1993, Casanova et al. 1995, Preibisch et al. 1996). These strong X-ray sources not only photoionize surrounding materials, also have an effect on interaction between ionized materials and magnetic fields. Moreover, apparently cool dwarf stars have more intensive X-ray emission.

Theoretically, these important characteristics of YSOs are thought to be connected directly to the stellar dynamic action, since the interaction between rotation and convection is widely considered as the important characteristics of the generation of stellar dynamics and the observed stellar magnetic activities. This is firstly supposed by Durney & Latour (1978) and tested by the number of researchers (Mangeney & Praderie 1984, Hartmann et al. 1984). In this hypothesis, the most significant concept is the Rossby number, defined as the ratio of the rotation period to the local convective turnover time. In the dynamo mechanism, the dynamic action is placed at the base of the convection region, just above the radiative layers. It makes the convective turnover timescale to the most relevant property in the evaluation of Rossby number.

The purpose of this paper is to provide the dynamic and thermal parameters of the PMS evolution tracks of stars with mass range of $0.065$ to $5.0 M_\odot$. For the implication of observational test, the local and global convective overturn timescales, which can be served as an input parameter of Rossby number circulation, are also computed. All models assume to have solar-like metal abundances. The convection is handled by the mixing length approximation in usual way.

## 2. INPUT PHYSICS AND MODELING

The evolutionary tracks of PMS stars have been constructed by using Yale Stellar Evolution Code (Guenther et al. 1992, Guenther & Demarque 1997) with realistic physics. The opacities of Alexander & Ferguson (1994) at low temperatures and the OPAL Rosseland opacities at higher temperatures (Iglesias & Rogers 2001) are used. The equation of state is the OPAL equation of state (Iglesias & Rogers 1996). The energy generation rate is taken from Bahcall & Pinsonneault (1992) and the helium diffusion value is obtained from Thoul et al. (1994). Eddington $T(\tau)$ relation is used in the atmosphere.

In the stellar model, the atmospheric regions are assumed to be grey. In the convectively unstable regions, the thermal structure is derived by the mixing length approximation (MLT). Since the computation contains the helium diffusion, the initial hydrogen abundance is different from the evolved model.

In the MLT, mixing length is assumed to be proportional to the local pressure scale height. As we mainly focused on the stars, typically formed in the Population I formation region, the PMS model is supposed to have the same metal abundance of the Sun. The initial hydrogen and metal abundance as well as mixing length are derived by fitting the values of luminosity, radius, and $Z/X$ (=0.0244; Grevesse et al. 1996) of the stellar models, to the observed solar values. The adopted values are $(X, Z, \alpha)_{initial} = (0.7149, 0.0181, 1.7432)$. These values are then applied to all PMS models.

Operationally, all models in our evolutions are started off from the stellar birthline, where stars initially become visible objects (Palla & Stahler 1993). The stars are assumed to be spherically symmetric, and not affected by any influence of rotation or magnetic fields. The initial models constructed from the polytropic models are located in the invisible region of the H-R diagram. Because they are quite different from the real stellar model in their thermal structures, a relaxation process is required. This process is carried out by YREC. The resulting models are then evolved down along their Hayashi lines until they fulfills the mass-radius relationship, proposed by Palla & Stahler (1991). These PMS evolutionary computations are physically self-consistent (cf. Yi et al. 2001).

Since our evolutionary models, which have different ages and different mass, are assumed to have same mixing ratio, the convective turnover time computed by the stellar model may be not fully correct. However, studies of convective motions with numerical methods verify that the MLT is still adequate in the deep convection zone where the temperature gradient is almost adiabatic (Kim et al. 1996). As the convective turnover timescale is measured near the base of convection zone, where the temperature gradient is fully adiabatic, our assumption and computation are probably reasonable.

## 3. RESULTS

The evolutionary track is computed using an iteration method for the evolution of the star, which yields some initial relaxation period of adjustment during early evolutionary phase. To avoid this uncertain stage, we present only the $2 \times 10^{-2}$ Myr and on. The terminating point of the PMS phase is chosen where the MS stars obtain most of their luminosity from the nuclear burning process. The stars whose mass is greater than $3.0 M_\odot$ are thought to have some non-negligible nuclear burning process just before they settle in the MS, since they have the unstable convective core and some portion of energy is generated by the CN burning in the PMS. By considering these properties, we select the terminating points of these large mass stars as the points that the energy generated by the gravity has the lowest value for the first time.

Figure 1 presents the H-R diagram of all computed PMS stars. For convenience, the reference stars with different line type are marked. The first position of each track is uncertain, since it depends on the choice of the first time step of each evolution. The evolutionary sequences are begun from the stellar birthline and then evolved along the Hayashi track before they reach the ZAMS. As shown in Figure 1, stars with different mass experience quite distinct PMS evolutionary stages. It is originated from the different magnitude of the gravitational contraction. The timescale of the stellar evolution is mainly relied on the mass of the star. The more mass presents, the faster is the evolution of the star through the fuel consuming processes.

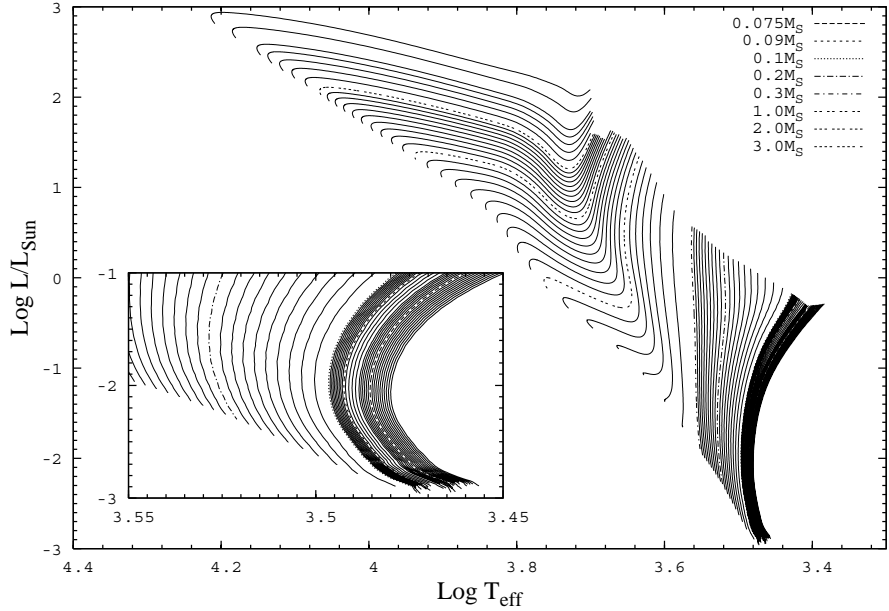

Figure 1. The evolutionary tracks of theoretical PMS stars.

The PMS stage is characterized by the gravitational contraction. During the evolution stage of the protostar, with the hydrostatic equilibrium, the stellar radius is decreased by the gravitational contraction. This accreting protostar heats up in its interior. The increase of the central temperature $T_c$ yields the ionization of materials in the core that the central core becomes opaque. It builds the temperature gradient in the interior, to transport the contraction energy outward. Consequently the central convection is set on. As the protostar evolves, the core convection is expanded to the envelope.

The stars $M \leq 1.3 M_\odot$, originated from the low mass protostars, are fully convective when they arrive at the stellar birth line (the top of the Hayashi line). These fully convective objects contract continuously until $T_c$ is an order of $10^6$K. On the condition of the contraction combined with the convection, the luminosity is rapidly decreased, while the effective temperature $T_{eff}$ is a roughly constant. It is compensated with the transfer downward along the nearly vertical path known as the Hayashi track. At the $T_c$ of about $10^6$K, the full ionization of H in the core decreases the central opacity. Simultaneously, all the $^{12}$C begin to burn into $^{14}$N, and the first reaction of *pp* chain become relevant. Thus, the portion of the energy transport by radiation gradually excesses the contraction energy, which slows down the rate of contraction. This is appeared in Figure 1 as the increasing luminosity and increasing T$_{\text{eff}}$. Subsequently, with the depletion of $^{12}$C in the central region, the convective core is disappeared and the evolution is dominated by radiative transport, a phase which continues until T$_c$ reaches about $10^8$K. At that point, ordinary hydrogen ignites at the centre and gradually outward. For the stars $M \leq 0.5 M_\odot$, because the radiative core grows to include the most of stellar mass (i.e. Virial Theorem), the luminosity does not rise appreciably before they reach the MS. Note that the star, which has lower mass than the critical mass $0.08 M_\odot$, never reaches the central nuclear burning phase, since the central temperature is not sufficient to ignite the hydrogen.

The massive stars evolve faster, because of their strong gravitational contraction that results in the higher central temperature. Therefore, by the time they appear in the stellar birth line, their central regions are already developed to the radiative cores. They also contract continuously until the central hydrogen burns. In contrast to the low mass star, however, the strong contraction makes the central temperature of massive stars approach to about $10^6$K quickly. This is the reason why the more massive star leaves the Hayashi track more rapidly. In addition, as the radiative energy formed in the core exceeds the gravitational energy, the envelope convection is diminished, and finally disappeared. After this detachment, the principal source of opacity during the later contraction is now the electron scattering. Since this type of opacity is independent of temperature and density, the luminosity along the PMS track remains roughly constant as the result of increased T$_{\text{eff}}$. The central core is transformed into the convective one, because the nuclear burning for the equilibrium of $^{12}$C depletion is governed by the CNO cycle instead of the *pp* chain.

Table 1 summarized the detailed characteristics of theoretical PMS models with several physical quantities at fixed ages from the $2 \times 10^{-2}$ Myr to the terminating point of each PMS star. This table also can be used to derive isochrones, but more complete evolutionary data are available for the astronomers upon request. The columns are mostly self-explanatory. Columns (5) in the first row gives the gravitational acceleration. The Y$_c$ is the helium abundance in the central core. The energy generation rates by PP1 cycle and gravity are provided at Columns (7) and (8), respectively. The absences of global and local convective turnover times in Columns (10) and (11) mean that no convection in the envelope was driven.

To describe the characteristics of surface convection, the two parameters, local and global turnover time, have been computed. The local convective turnover timescale, used to explain the convective overturn, was calculated at each time step. The time scale was estimated based on the local convective velocity at a distance of a half the mixing length above the base of the convection region (Gilliland 1986). The nonlocal (or global) convective turnover time was also derived at each time step using the equation

$$\tau_{gc} = \int_{R_b}^{R_*} \frac{dr}{v} \quad (1)$$

where $R_b$ is the location of the bottom of the surface convection region, $R_*$ is the total radius of the stellar model, and $v$ is the local convective velocity (Kim & Demarque 1996). If the MLT describes the deep convection phenomenon with reasonable implementation, these two timescales have identical results except for some scaling factor.

The evolutions of global and local timescale are shown in Figure 2. In the PMS phase, the thermal equilibrium between the contraction and the central pressure modifies the thermal structure within the Kelvin-Helmholtz time. It causes the rapid change of both convective time. Moreover, the timescales of massive stars are less than that of low mass stars, because they have more shallow envelope convection regions. At the T$_{\text{eff}}$ of about 4000K encountered during the early contraction, the absorption by the H$^-$ that releases the free electron increases the opacity of the photospheric layer. It brings about the swell of stellar radius which, in turn, affects on the turnover time. This is the reason that intermediate mass stars experience the increase of the convective turnover time in the early PMS stage. As the amount of the $H^-$ in the envelope decreases, stars start to contract again and the corresponding overturn time is also diminished. For the star which has the convective envelope in near the main sequence, both time scales remain almost constant and may be a function of mass. Since MS stars do not change their thermal structures before the end of their long phase of central hydrogen burning.

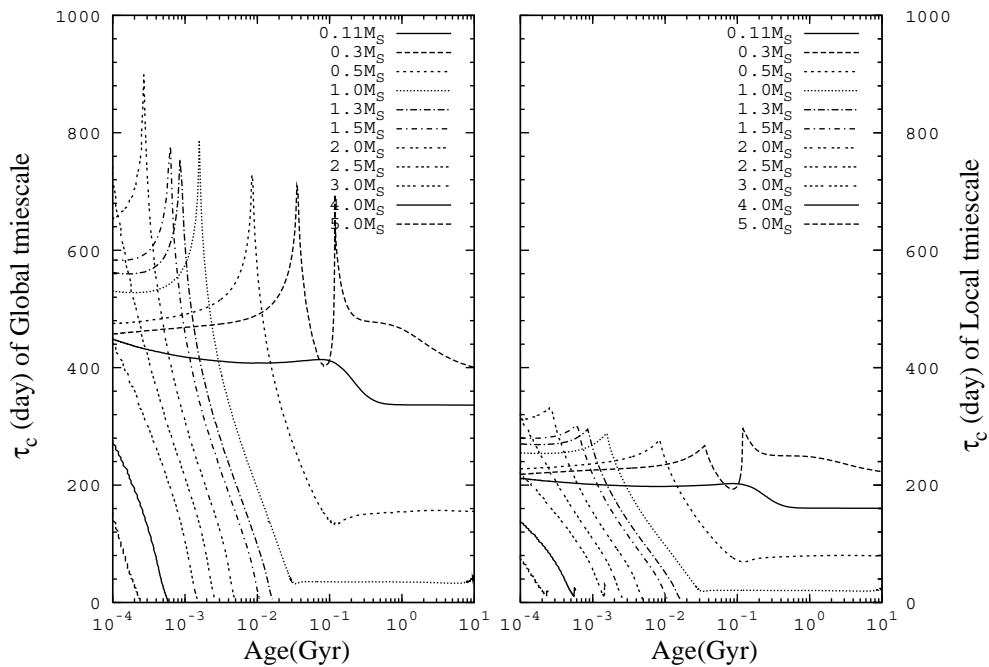

Figure 2. The global (or nonlocal) and local convective time scale vs age for some reference stars. Both time scales are nearly constant in the vicinity of the main sequence for a given mass. The local value is almost same with the one-half of the global value.

The different PMS evolutionary tracks are constructed by D'Antona & Mazzitelli (1994) with different evolution code and input physics. In particular, they examine the effect of the convection treatment between the MLT and the semi-analytical turbulent convection model proposed by Canuto & Mazzitelli (CM) (1991, 1992). Form their approach, it is found that $T_{eff}$ values are dependent on the choice of the opacity and the convection treatment. According to this test, the best reasonable model is the Alexander opacity and the CM set. However, considering the large uncertainties embedded in the theoretical assumptions and the input physics, it is somewhat meaningless to determine what model is more reasonable. Instead, the agreement between the model and the observation may be a criterion to verify the usefulness of the model. The application of updated input physics and physical parameters may also enhance the reliableness of the theoretical data. For our cases, based on the most updated input physics and solar calibration, Yi et al. (2001) already testifies the validity through the comparison with not only the observational globular cluster 47 Tuc also the open cluster IC 2391 and Pleiades.

## 4. SUMMARY

We computed the several evolution tracks of the pre-main sequence stars based on the up-to-date input physics. With fixed age steps, the fundamental dynamic and thermal parameters are presented. The metal abundance is assumed to be that of the Sun. Since the PMS stars are considered to have the convection motion in their envelope, we also provide the two time scales, the global and the local convective turnover timescales. These time scales can be applied to an observational test of theoretical model. The full computed tracks of all PMS stars with more detailed thermal and dynamic variables are also available upon request.

**ACKNOWLEDGEMENTS:** YKJ and YCK are supported by the Astrophysical Center for the Structure and Evolution of the Cosmos (ARCSEC) of Korea Science and Engineering Foundation (KOSEF) through the Science Research Center (SRC) program.

Table 1. Evolutionary properties of PMS stars.

| Age(Gyr) | logT$_{\text{eff}}$ | logL/L$_\odot$ | R(cm) | g(cm/s$^2$) | Y$_c$ | L(PP1) | L$_g$ | X$_{\text{env}}$ | $\tau_{gc}$ (day) | $\tau_{lc}$ (day) |
|---|---|---|---|---|---|---|---|---|---|---|
| | | | | 0.065$M_\odot$ | | | | | | |
| 0.00002 | 3.3837 | −0.2884 | 2.8502 × 10$^{11}$ | 1.0620 × 10$^2$ | 0.26700050 | 0.0000 | 1.0000 | 0.7149 | 4.7221 × 10$^2$ | 2.0956 × 10$^2$ |
| 0.00010 | 3.4360 | −0.8188 | 1.2159 × 10$^{11}$ | 5.8351 × 10$^2$ | 0.26700050 | 0.0000 | 1.0000 | 0.7149 | 4.1012 × 10$^2$ | 1.9251 × 10$^2$ |
| 0.00020 | 3.4507 | −1.0168 | 9.0504 × 10$^{10}$ | 1.0532 × 10$^3$ | 0.26700050 | 0.0000 | 1.0000 | 0.7149 | 3.9465 × 10$^2$ | 1.8749 × 10$^2$ |
| 0.00030 | 3.4578 | −1.1322 | 7.6687 × 10$^{10}$ | 1.4670 × 10$^3$ | 0.26700050 | 0.0000 | 1.0000 | 0.7149 | 3.8719 × 10$^2$ | 1.8493 × 10$^2$ |
| 0.00050 | 3.4649 | −1.2800 | 6.2605 × 10$^{10}$ | 2.2011 × 10$^3$ | 0.26700050 | 0.0000 | 1.0000 | 0.7149 | 3.8007 × 10$^2$ | 1.8241 × 10$^2$ |
| 0.00070 | 3.4687 | −1.3778 | 5.4965 × 10$^{10}$ | 2.8556 × 10$^3$ | 0.26700050 | 0.0000 | 1.0000 | 0.7149 | 3.7641 × 10$^2$ | 1.8112 × 10$^2$ |
| 0.00100 | 3.4720 | −1.4822 | 4.8006 × 10$^{10}$ | 3.7434 × 10$^3$ | 0.26700050 | 0.0000 | 1.0000 | 0.7149 | 3.7332 × 10$^2$ | 1.8005 × 10$^2$ |
| 0.00200 | 3.4767 | −1.6869 | 3.7121 × 10$^{10}$ | 6.2607 × 10$^3$ | 0.26700050 | 0.0000 | 1.0000 | 0.7149 | 3.6885 × 10$^2$ | 1.7866 × 10$^2$ |
| 0.00300 | 3.4785 | −1.8075 | 3.2039 × 10$^{10}$ | 8.4044 × 10$^3$ | 0.26700050 | 0.0000 | 1.0000 | 0.7149 | 3.6656 × 10$^2$ | 1.7768 × 10$^2$ |
| 0.00500 | 3.4797 | −1.9607 | 2.6709 × 10$^{10}$ | 1.2094 × 10$^4$ | 0.26700050 | 0.0000 | 1.0000 | 0.7149 | 3.6521 × 10$^2$ | 1.7750 × 10$^2$ |
| 0.00700 | 3.4800 | −2.0619 | 2.3740 × 10$^{10}$ | 1.5308 × 10$^4$ | 0.26700050 | 0.0000 | 1.0000 | 0.7149 | 3.6531 × 10$^2$ | 1.7766 × 10$^2$ |
| 0.01000 | 3.4797 | −2.1706 | 2.0975 × 10$^{10}$ | 1.9610 × 10$^4$ | 0.26700050 | 0.0000 | 1.0000 | 0.7149 | 3.6715 × 10$^2$ | 1.7863 × 10$^2$ |
| 0.02000 | 3.4768 | −2.3871 | 1.6567 × 10$^{10}$ | 3.1431 × 10$^4$ | 0.26700050 | 0.0000 | 1.0000 | 0.7149 | 3.7004 × 10$^2$ | 1.8106 × 10$^2$ |
| 0.03000 | 3.4740 | −2.5150 | 1.4486 × 10$^{10}$ | 4.1109 × 10$^4$ | 0.26700050 | 0.0000 | 1.0000 | 0.7149 | 3.7284 × 10$^2$ | 1.8306 × 10$^2$ |
| 0.05000 | 3.4695 | −2.6758 | 1.2288 × 10$^{10}$ | 5.7134 × 10$^4$ | 0.26700050 | 0.0000 | 1.0000 | 0.7149 | 3.7870 × 10$^2$ | 1.8605 × 10$^2$ |
| 0.07000 | 3.4636 | −2.7928 | 1.1038 × 10$^{10}$ | 7.0813 × 10$^4$ | 0.26700050 | 0.0000 | 1.0000 | 0.7149 | 3.8632 × 10$^2$ | 1.9047 × 10$^2$ |
| 0.08900 | 3.4564 | −2.8918 | 1.0177 × 10$^{10}$ | 8.3303 × 10$^4$ | 0.26700050 | 0.0000 | 1.0000 | 0.7149 | 3.9506 × 10$^2$ | 1.9474 × 10$^2$ |
| | | | | 0.066$M_\odot$ | | | | | | |
| 0.00002 | 3.3858 | −0.2910 | 2.8138 × 10$^{11}$ | 1.1064 × 10$^2$ | 0.26700050 | 0.0000 | 1.0000 | 0.7149 | 4.7156 × 10$^2$ | 2.0977 × 10$^2$ |
| 0.00010 | 3.4368 | −0.8110 | 1.2229 × 10$^{11}$ | 5.8575 × 10$^2$ | 0.26700050 | 0.0000 | 1.0000 | 0.7149 | 4.1120 × 10$^2$ | 1.9304 × 10$^2$ |
| 0.00020 | 3.4512 | −1.0084 | 9.1146 × 10$^{10}$ | 1.0544 × 10$^3$ | 0.26700050 | 0.0000 | 1.0000 | 0.7149 | 3.9583 × 10$^2$ | 1.8806 × 10$^2$ |
| 0.00030 | 3.4584 | −1.1234 | 7.7268 × 10$^{10}$ | 1.4672 × 10$^3$ | 0.26700050 | 0.0000 | 1.0000 | 0.7149 | 3.8840 × 10$^2$ | 1.8550 × 10$^2$ |
| 0.00050 | 3.4654 | −1.2710 | 6.3102 × 10$^{10}$ | 2.1999 × 10$^3$ | 0.26700050 | 0.0000 | 1.0000 | 0.7149 | 3.8130 × 10$^2$ | 1.8299 × 10$^2$ |
| 0.00070 | 3.4692 | −1.3689 | 5.5403 × 10$^{10}$ | 2.8538 × 10$^3$ | 0.26700050 | 0.0000 | 1.0000 | 0.7149 | 3.7759 × 10$^2$ | 1.8171 × 10$^2$ |
| 0.00100 | 3.4725 | −1.4732 | 4.8393 × 10$^{10}$ | 3.7404 × 10$^3$ | 0.26700050 | 0.0000 | 1.0000 | 0.7149 | 3.7455 × 10$^2$ | 1.8063 × 10$^2$ |
| 0.00200 | 3.4772 | −1.6776 | 3.7433 × 10$^{10}$ | 6.2515 × 10$^3$ | 0.26700050 | 0.0000 | 1.0000 | 0.7149 | 3.7003 × 10$^2$ | 1.7923 × 10$^2$ |
| 0.00300 | 3.4790 | −1.7982 | 3.2308 × 10$^{10}$ | 8.3921 × 10$^3$ | 0.26700050 | 0.0000 | 1.0000 | 0.7149 | 3.6777 × 10$^2$ | 1.7828 × 10$^2$ |
| 0.00500 | 3.4802 | −1.9512 | 2.6934 × 10$^{10}$ | 1.2075 × 10$^4$ | 0.26700050 | 0.0000 | 1.0000 | 0.7149 | 3.6641 × 10$^2$ | 1.7806 × 10$^2$ |
| 0.00700 | 3.4806 | −2.0524 | 2.3933 × 10$^{10}$ | 1.5293 × 10$^4$ | 0.26700050 | 0.0000 | 1.0000 | 0.7149 | 3.6627 × 10$^2$ | 1.7817 × 10$^2$ |
| 0.01000 | 3.4803 | −2.1610 | 2.1147 × 10$^{10}$ | 1.9588 × 10$^4$ | 0.26700050 | 0.0000 | 1.0000 | 0.7149 | 3.6821 × 10$^2$ | 1.7910 × 10$^2$ |
| 0.02000 | 3.4776 | −2.3770 | 1.6700 × 10$^{10}$ | 3.1409 × 10$^4$ | 0.26700050 | 0.0000 | 1.0000 | 0.7149 | 3.7093 × 10$^2$ | 1.8149 × 10$^2$ |
| 0.03000 | 3.4748 | −2.5048 | 1.4600 × 10$^{10}$ | 4.1092 × 10$^4$ | 0.26700050 | 0.0000 | 1.0000 | 0.7149 | 3.7369 × 10$^2$ | 1.8342 × 10$^2$ |
| 0.05000 | 3.4704 | −2.6656 | 1.2380 × 10$^{10}$ | 5.7152 × 10$^4$ | 0.26700050 | 0.0000 | 1.0000 | 0.7149 | 3.7946 × 10$^2$ | 1.8628 × 10$^2$ |
| 0.07000 | 3.4653 | −2.7793 | 1.1123 × 10$^{10}$ | 7.0804 × 10$^4$ | 0.26700050 | 0.0000 | 1.0000 | 0.7149 | 3.8609 × 10$^2$ | 1.9026 × 10$^2$ |
| 0.08861 | 3.4581 | −2.8751 | 1.0297 × 10$^{10}$ | 8.2611 × 10$^4$ | 0.26700050 | 0.0000 | 1.0000 | 0.7149 | 3.9490 × 10$^2$ | 1.9419 × 10$^2$ |
| | | | | 0.067$M_\odot$ | | | | | | |
| 0.00002 | 3.3879 | −0.2938 | 2.7777 × 10$^{11}$ | 1.1525 × 10$^2$ | 0.26700050 | 0.0000 | 1.0000 | 0.7149 | 4.7078 × 10$^2$ | 2.0989 × 10$^2$ |
| 0.00010 | 3.4374 | −0.8035 | 1.2296 × 10$^{11}$ | 5.8816 × 10$^2$ | 0.26700050 | 0.0000 | 1.0000 | 0.7149 | 4.1228 × 10$^2$ | 1.9357 × 10$^2$ |
| 0.00020 | 3.4518 | −1.0001 | 9.1777 × 10$^{10}$ | 1.0557 × 10$^3$ | 0.26700050 | 0.0000 | 1.0000 | 0.7149 | 3.9702 × 10$^2$ | 1.8860 × 10$^2$ |
| 0.00030 | 3.4589 | −1.1148 | 7.7853 × 10$^{10}$ | 1.4671 × 10$^3$ | 0.26700050 | 0.0000 | 1.0000 | 0.7149 | 3.8963 × 10$^2$ | 1.8607 × 10$^2$ |
| 0.00050 | 3.4659 | −1.2621 | 6.3598 × 10$^{10}$ | 2.1985 × 10$^3$ | 0.26700050 | 0.0000 | 1.0000 | 0.7149 | 3.8253 × 10$^2$ | 1.8358 × 10$^2$ |
| 0.00070 | 3.4697 | −1.3599 | 5.5851 × 10$^{10}$ | 2.8507 × 10$^3$ | 0.26700050 | 0.0000 | 1.0000 | 0.7149 | 3.7884 × 10$^2$ | 1.8222 × 10$^2$ |
| 0.00100 | 3.4730 | −1.4642 | 4.8786 × 10$^{10}$ | 3.7362 × 10$^3$ | 0.26700050 | 0.0000 | 1.0000 | 0.7149 | 3.7577 × 10$^2$ | 1.8120 × 10$^2$ |
| 0.00200 | 3.4777 | −1.6684 | 3.7743 × 10$^{10}$ | 6.2425 × 10$^3$ | 0.26700050 | 0.0000 | 1.0000 | 0.7149 | 3.7121 × 10$^2$ | 1.7977 × 10$^2$ |
| 0.00300 | 3.4795 | −1.7889 | 3.2580 × 10$^{10}$ | 8.3776 × 10$^3$ | 0.26700050 | 0.0000 | 1.0000 | 0.7149 | 3.6899 × 10$^2$ | 1.7886 × 10$^2$ |
| 0.00500 | 3.4808 | −1.9420 | 2.7155 × 10$^{10}$ | 1.2059 × 10$^4$ | 0.26700050 | 0.0000 | 1.0000 | 0.7149 | 3.6799 × 10$^2$ | 1.7861 × 10$^2$ |
| 0.00700 | 3.4812 | −2.0430 | 2.4130 × 10$^{10}$ | 1.5272 × 10$^4$ | 0.26700050 | 0.0000 | 1.0000 | 0.7149 | 3.6740 × 10$^2$ | 1.7870 × 10$^2$ |
| 0.01000 | 3.4809 | −2.1513 | 2.1323 × 10$^{10}$ | 1.9558 × 10$^4$ | 0.26700050 | 0.0000 | 1.0000 | 0.7149 | 3.6922 × 10$^2$ | 1.7956 × 10$^2$ |
| 0.02000 | 3.4784 | −2.3671 | 1.6831 × 10$^{10}$ | 3.1392 × 10$^4$ | 0.26700050 | 0.0000 | 1.0000 | 0.7149 | 3.7188 × 10$^2$ | 1.8194 × 10$^2$ |
| 0.03000 | 3.4756 | −2.4947 | 1.4716 × 10$^{10}$ | 4.1065 × 10$^4$ | 0.26700050 | 0.0000 | 1.0000 | 0.7149 | 3.7451 × 10$^2$ | 1.8376 × 10$^2$ |
| 0.05000 | 3.4713 | −2.6555 | 1.2474 × 10$^{10}$ | 5.7150 × 10$^4$ | 0.26700050 | 0.0000 | 1.0000 | 0.7149 | 3.8024 × 10$^2$ | 1.8656 × 10$^2$ |
| 0.07000 | 3.4667 | −2.7671 | 1.1206 × 10$^{10}$ | 7.0808 × 10$^4$ | 0.26700050 | 0.0000 | 1.0000 | 0.7149 | 3.8621 × 10$^2$ | 1.9028 × 10$^2$ |
| 0.09262 | 3.4588 | −2.8791 | 1.0216 × 10$^{10}$ | 8.5205 × 10$^4$ | 0.26700050 | 0.0000 | 1.0000 | 0.7149 | 3.9548 × 10$^2$ | 1.9444 × 10$^2$ |
| | | | | 0.068$M_\odot$ | | | | | | |
| 0.00002 | 3.3899 | −0.2964 | 2.7443 × 10$^{11}$ | 1.1984 × 10$^2$ | 0.26700050 | 0.0000 | 1.0000 | 0.7149 | 4.7042 × 10$^2$ | 2.1019 × 10$^2$ |
| 0.00010 | 3.4381 | −0.7962 | 1.2361 × 10$^{11}$ | 5.9065 × 10$^2$ | 0.26700050 | 0.0000 | 1.0000 | 0.7149 | 4.1337 × 10$^2$ | 1.9410 × 10$^2$ |
| 0.00020 | 3.4524 | −0.9918 | 9.2408 × 10$^{10}$ | 1.0569 × 10$^3$ | 0.26700050 | 0.0000 | 1.0000 | 0.7149 | 3.9819 × 10$^2$ | 1.8915 × 10$^2$ |
| 0.00030 | 3.4594 | −1.1064 | 7.8409 × 10$^{10}$ | 1.4680 × 10$^3$ | 0.26700050 | 0.0000 | 1.0000 | 0.7149 | 3.9085 × 10$^2$ | 1.8662 × 10$^2$ |
| 0.00050 | 3.4665 | −1.2535 | 6.4086 × 10$^{10}$ | 2.1975 × 10$^3$ | 0.26700050 | 0.0000 | 1.0000 | 0.7149 | 3.8373 × 10$^2$ | 1.8416 × 10$^2$ |
| 0.00070 | 3.4702 | −1.3511 | 5.6289 × 10$^{10}$ | 2.8484 × 10$^3$ | 0.26700050 | 0.0000 | 1.0000 | 0.7149 | 3.8003 × 10$^2$ | 1.8278 × 10$^2$ |
| 0.00100 | 3.4735 | −1.4553 | 4.9179 × 10$^{10}$ | 3.7315 × 10$^3$ | 0.26700050 | 0.0000 | 1.0000 | 0.7149 | 3.7703 × 10$^2$ | 1.8173 × 10$^2$ |
| 0.00200 | 3.4782 | −1.6592 | 3.8053 × 10$^{10}$ | 6.2327 × 10$^3$ | 0.26700050 | 0.0000 | 1.0000 | 0.7149 | 3.7231 × 10$^2$ | 1.8029 × 10$^2$ |

Table 1. Continued

| Age(Gyr) | $\log T_{\rm eff}$ | $\log L/L_\odot$ | R(cm) | g(cm/s$^2$) | $Y_c$ | L(PP1) | $L_g$ | $X_{\rm env}$ | $\tau_{gc}$ (day) | $\tau_{lc}$ (day) |
|---|---|---|---|---|---|---|---|---|---|---|
| 0.00300 | 3.4800 | −1.7798 | $3.2848 \times 10^{10}$ | $8.3645 \times 10^3$ | 0.26700050 | 0.0000 | 1.0000 | 0.7149 | $3.7023 \times 10^2$ | $1.7943 \times 10^2$ |
| 0.00500 | 3.4813 | −1.9328 | $2.7377 \times 10^{10}$ | $1.2042 \times 10^4$ | 0.26700050 | 0.0000 | 1.0000 | 0.7149 | $3.6886 \times 10^2$ | $1.7909 \times 10^2$ |
| 0.00700 | 3.4817 | −2.0336 | $2.4329 \times 10^{10}$ | $1.5248 \times 10^4$ | 0.26700050 | 0.0000 | 1.0000 | 0.7149 | $3.6853 \times 10^2$ | $1.7921 \times 10^2$ |
| 0.01000 | 3.4816 | −2.1417 | $2.1500 \times 10^{10}$ | $1.9524 \times 10^4$ | 0.26700050 | 0.0000 | 1.0000 | 0.7149 | $3.7021 \times 10^2$ | $1.8002 \times 10^2$ |
| 0.02000 | 3.4792 | −2.3569 | $1.6967 \times 10^{10}$ | $3.1351 \times 10^4$ | 0.26700050 | 0.0000 | 1.0000 | 0.7149 | $3.7271 \times 10^2$ | $1.8234 \times 10^2$ |
| 0.03000 | 3.4765 | −2.4845 | $1.4832 \times 10^{10}$ | $4.1027 \times 10^4$ | 0.26700050 | 0.0000 | 1.0000 | 0.7149 | $3.7530 \times 10^2$ | $1.8408 \times 10^2$ |
| 0.05000 | 3.4722 | −2.6455 | $1.2568 \times 10^{10}$ | $5.7137 \times 10^4$ | 0.26700050 | 0.0000 | 1.0000 | 0.7149 | $3.8098 \times 10^2$ | $1.8682 \times 10^2$ |
| 0.07000 | 3.4681 | −2.7549 | $1.1291 \times 10^{10}$ | $7.0790 \times 10^4$ | 0.26700050 | 0.0000 | 1.0000 | 0.7149 | $3.8640 \times 10^2$ | $1.9019 \times 10^2$ |
| 0.09552 | 3.4594 | −2.8771 | $1.0209 \times 10^{10}$ | $8.6592 \times 10^4$ | 0.26700050 | 0.0000 | 1.0000 | 0.7149 | $3.9709 \times 10^2$ | $1.9493 \times 10^2$ |
| $0.069 M_\odot$ | | | | | | | | | | |
| 0.00002 | 3.3916 | −0.2996 | $2.7122 \times 10^{11}$ | $1.2449 \times 10^2$ | 0.26700050 | 0.0000 | 1.0000 | 0.7149 | $4.7038 \times 10^2$ | $2.1060 \times 10^2$ |
| 0.00010 | 3.4388 | −0.7891 | $1.2424 \times 10^{11}$ | $5.9328 \times 10^2$ | 0.26700050 | 0.0000 | 1.0000 | 0.7149 | $4.1446 \times 10^2$ | $1.9464 \times 10^2$ |
| 0.00020 | 3.4529 | −0.9839 | $9.3024 \times 10^{10}$ | $1.0583 \times 10^3$ | 0.26700050 | 0.0000 | 1.0000 | 0.7149 | $3.9937 \times 10^2$ | $1.8970 \times 10^2$ |
| 0.00030 | 3.4599 | −1.0981 | $7.8978 \times 10^{10}$ | $1.4682 \times 10^3$ | 0.26700050 | 0.0000 | 1.0000 | 0.7149 | $3.9207 \times 10^2$ | $1.8721 \times 10^2$ |
| 0.00050 | 3.4669 | −1.2450 | $6.4568 \times 10^{10}$ | $2.1966 \times 10^3$ | 0.26700050 | 0.0000 | 1.0000 | 0.7149 | $3.8491 \times 10^2$ | $1.8473 \times 10^2$ |
| 0.00070 | 3.4707 | −1.3425 | $5.6723 \times 10^{10}$ | $2.8463 \times 10^3$ | 0.26700050 | 0.0000 | 1.0000 | 0.7149 | $3.8122 \times 10^2$ | $1.8337 \times 10^2$ |
| 0.00100 | 3.4740 | −1.4466 | $4.9560 \times 10^{10}$ | $3.7285 \times 10^3$ | 0.26700050 | 0.0000 | 1.0000 | 0.7149 | $3.7818 \times 10^2$ | $1.8231 \times 10^2$ |
| 0.00200 | 3.4787 | −1.6502 | $3.8359 \times 10^{10}$ | $6.2238 \times 10^3$ | 0.26700050 | 0.0000 | 1.0000 | 0.7149 | $3.7343 \times 10^2$ | $1.8083 \times 10^2$ |
| 0.00300 | 3.4805 | −1.7709 | $3.3111 \times 10^{10}$ | $8.3530 \times 10^3$ | 0.26700050 | 0.0000 | 1.0000 | 0.7149 | $3.7142 \times 10^2$ | $1.8000 \times 10^2$ |
| 0.00500 | 3.4818 | −1.9236 | $2.7602 \times 10^{10}$ | $1.2021 \times 10^4$ | 0.26700050 | 0.0000 | 1.0000 | 0.7149 | $3.7000 \times 10^2$ | $1.7964 \times 10^2$ |
| 0.00700 | 3.4823 | −2.0244 | $2.4528 \times 10^{10}$ | $1.5222 \times 10^4$ | 0.26700050 | 0.0000 | 1.0000 | 0.7149 | $3.6966 \times 10^2$ | $1.7972 \times 10^2$ |
| 0.01000 | 3.4822 | −2.1322 | $2.1675 \times 10^{10}$ | $1.9493 \times 10^4$ | 0.26700050 | 0.0000 | 1.0000 | 0.7149 | $3.7120 \times 10^2$ | $1.8049 \times 10^2$ |
| 0.02000 | 3.4799 | −2.3472 | $1.7098 \times 10^{10}$ | $3.1327 \times 10^4$ | 0.26700050 | 0.0000 | 1.0000 | 0.7149 | $3.7371 \times 10^2$ | $1.8277 \times 10^2$ |
| 0.03000 | 3.4773 | −2.4745 | $1.4945 \times 10^{10}$ | $4.1004 \times 10^4$ | 0.26700050 | 0.0000 | 1.0000 | 0.7149 | $3.7607 \times 10^2$ | $1.8442 \times 10^2$ |
| 0.05000 | 3.4730 | −2.6357 | $1.2661 \times 10^{10}$ | $5.7130 \times 10^4$ | 0.26700050 | 0.0000 | 1.0000 | 0.7149 | $3.8176 \times 10^2$ | $1.8711 \times 10^2$ |
| 0.07000 | 3.4694 | −2.7433 | $1.1376 \times 10^{10}$ | $7.0769 \times 10^4$ | 0.26700050 | 0.0000 | 1.0000 | 0.7149 | $3.8666 \times 10^2$ | $1.9021 \times 10^2$ |
| 0.10000 | 3.4601 | −2.8825 | $1.0115 \times 10^{10}$ | $8.9516 \times 10^4$ | 0.26700050 | 0.0000 | 1.0000 | 0.7149 | $3.9795 \times 10^2$ | $1.9556 \times 10^2$ |
| 0.10265 | 3.4597 | −2.8931 | $1.0012 \times 10^{10}$ | $9.1361 \times 10^4$ | 0.26700050 | 0.0000 | 1.0000 | 0.7149 | $3.9933 \times 10^2$ | $1.9658 \times 10^2$ |
| $0.070 M_\odot$ | | | | | | | | | | |
| 0.00002 | 3.3937 | −0.3011 | $2.6823 \times 10^{11}$ | $1.2913 \times 10^2$ | 0.26700050 | 0.0000 | 1.0000 | 0.7149 | $4.7017 \times 10^2$ | $2.1094 \times 10^2$ |
| 0.00010 | 3.4394 | −0.7822 | $1.2486 \times 10^{11}$ | $5.9589 \times 10^2$ | 0.26700050 | 0.0000 | 1.0000 | 0.7149 | $4.1556 \times 10^2$ | $1.9519 \times 10^2$ |
| 0.00020 | 3.4535 | −0.9760 | $9.3631 \times 10^{10}$ | $1.0598 \times 10^3$ | 0.26700050 | 0.0000 | 1.0000 | 0.7149 | $4.0052 \times 10^2$ | $1.9025 \times 10^2$ |
| 0.00030 | 3.4604 | −1.0901 | $7.9527 \times 10^{10}$ | $1.4690 \times 10^3$ | 0.26700050 | 0.0000 | 1.0000 | 0.7149 | $3.9325 \times 10^2$ | $1.8776 \times 10^2$ |
| 0.00050 | 3.4674 | −1.2367 | $6.5040 \times 10^{10}$ | $2.1963 \times 10^3$ | 0.26700050 | 0.0000 | 1.0000 | 0.7149 | $3.8607 \times 10^2$ | $1.8530 \times 10^2$ |
| 0.00070 | 3.4712 | −1.3340 | $5.7152 \times 10^{10}$ | $2.8443 \times 10^3$ | 0.26700050 | 0.0000 | 1.0000 | 0.7149 | $3.8239 \times 10^2$ | $1.8393 \times 10^2$ |
| 0.00100 | 3.4745 | −1.4380 | $4.9943 \times 10^{10}$ | $3.7247 \times 10^3$ | 0.26700050 | 0.0000 | 1.0000 | 0.7149 | $3.7940 \times 10^2$ | $1.8281 \times 10^2$ |
| 0.00200 | 3.4792 | −1.6417 | $3.8653 \times 10^{10}$ | $6.2185 \times 10^3$ | 0.26700050 | 0.0000 | 1.0000 | 0.7149 | $3.7464 \times 10^2$ | $1.8137 \times 10^2$ |
| 0.00300 | 3.4810 | −1.7623 | $3.3366 \times 10^{10}$ | $8.3450 \times 10^3$ | 0.26700050 | 0.0000 | 1.0000 | 0.7149 | $3.7261 \times 10^2$ | $1.8056 \times 10^2$ |
| 0.00500 | 3.4824 | −1.9149 | $2.7814 \times 10^{10}$ | $1.2009 \times 10^4$ | 0.26700050 | 0.0000 | 1.0000 | 0.7149 | $3.7112 \times 10^2$ | $1.8015 \times 10^2$ |
| 0.00700 | 3.4828 | −2.0154 | $2.4721 \times 10^{10}$ | $1.5202 \times 10^4$ | 0.26700050 | 0.0000 | 1.0000 | 0.7149 | $3.7077 \times 10^2$ | $1.8023 \times 10^2$ |
| 0.01000 | 3.4827 | −2.1232 | $2.1844 \times 10^{10}$ | $1.9471 \times 10^4$ | 0.26700050 | 0.0000 | 1.0000 | 0.7149 | $3.7227 \times 10^2$ | $1.8097 \times 10^2$ |
| 0.02000 | 3.4806 | −2.3377 | $1.7230 \times 10^{10}$ | $3.1294 \times 10^4$ | 0.26700050 | 0.0000 | 1.0000 | 0.7149 | $3.7469 \times 10^2$ | $1.8325 \times 10^2$ |
| 0.03000 | 3.4780 | −2.4652 | $1.5056 \times 10^{10}$ | $4.0986 \times 10^4$ | 0.26700050 | 0.0000 | 1.0000 | 0.7149 | $3.7700 \times 10^2$ | $1.8478 \times 10^2$ |
| 0.05000 | 3.4739 | −2.6259 | $1.2753 \times 10^{10}$ | $5.7122 \times 10^4$ | 0.26700050 | 0.0000 | 1.0000 | 0.7149 | $3.8251 \times 10^2$ | $1.8739 \times 10^2$ |
| 0.07000 | 3.4705 | −2.7323 | $1.1460 \times 10^{10}$ | $7.0745 \times 10^4$ | 0.26700050 | 0.0000 | 1.0000 | 0.7149 | $3.8704 \times 10^2$ | $1.9028 \times 10^2$ |
| 0.09670 | 3.4618 | −2.8567 | $1.0340 \times 10^{10}$ | $8.6890 \times 10^4$ | 0.26700050 | 0.0000 | 1.0000 | 0.7149 | $3.9893 \times 10^2$ | $1.9532 \times 10^2$ |
| $0.071 M_\odot$ | | | | | | | | | | |
| 0.00002 | 3.3956 | −0.3026 | $2.6537 \times 10^{11}$ | $1.3381 \times 10^2$ | 0.26700050 | 0.0000 | 1.0000 | 0.7149 | $4.6943 \times 10^2$ | $2.1095 \times 10^2$ |
| 0.00010 | 3.4401 | −0.7755 | $1.2546 \times 10^{11}$ | $5.9865 \times 10^2$ | 0.26700050 | 0.0000 | 1.0000 | 0.7149 | $4.1665 \times 10^2$ | $1.9574 \times 10^2$ |
| 0.00020 | 3.4541 | −0.9682 | $9.4230 \times 10^{10}$ | $1.0613 \times 10^3$ | 0.26700050 | 0.0000 | 1.0000 | 0.7149 | $4.0162 \times 10^2$ | $1.9077 \times 10^2$ |
| 0.00030 | 3.4609 | −1.0819 | $8.0082 \times 10^{10}$ | $1.4694 \times 10^3$ | 0.26700050 | 0.0000 | 1.0000 | 0.7149 | $3.9439 \times 10^2$ | $1.8830 \times 10^2$ |
| 0.00050 | 3.4679 | −1.2284 | $6.5521 \times 10^{10}$ | $2.1951 \times 10^3$ | 0.26700050 | 0.0000 | 1.0000 | 0.7149 | $3.8725 \times 10^2$ | $1.8586 \times 10^2$ |
| 0.00070 | 3.4716 | −1.3257 | $5.7581 \times 10^{10}$ | $2.8421 \times 10^3$ | 0.26700050 | 0.0000 | 1.0000 | 0.7149 | $3.8356 \times 10^2$ | $1.8449 \times 10^2$ |
| 0.00100 | 3.4749 | −1.4295 | $5.0328 \times 10^{10}$ | $3.7204 \times 10^3$ | 0.26700050 | 0.0000 | 1.0000 | 0.7149 | $3.8055 \times 10^2$ | $1.8336 \times 10^2$ |
| 0.00200 | 3.4796 | −1.6330 | $3.8960 \times 10^{10}$ | $6.2082 \times 10^3$ | 0.26700050 | 0.0000 | 1.0000 | 0.7149 | $3.7573 \times 10^2$ | $1.8195 \times 10^2$ |
| 0.00300 | 3.4814 | −1.7535 | $3.3631 \times 10^{10}$ | $8.3313 \times 10^3$ | 0.26700050 | 0.0000 | 1.0000 | 0.7149 | $3.7391 \times 10^2$ | $1.8109 \times 10^2$ |
| 0.00500 | 3.4829 | −1.9059 | $2.8035 \times 10^{10}$ | $1.1990 \times 10^4$ | 0.26700050 | 0.0000 | 1.0000 | 0.7149 | $3.7221 \times 10^2$ | $1.8064 \times 10^2$ |
| 0.00700 | 3.4833 | −2.0064 | $2.4917 \times 10^{10}$ | $1.5178 \times 10^4$ | 0.26700050 | 0.0000 | 1.0000 | 0.7149 | $3.7188 \times 10^2$ | $1.8074 \times 10^2$ |
| 0.01000 | 3.4833 | −2.1142 | $2.2013 \times 10^{10}$ | $1.9446 \times 10^4$ | 0.26700050 | 0.0000 | 1.0000 | 0.7149 | $3.7330 \times 10^2$ | $1.8143 \times 10^2$ |
| 0.02000 | 3.4813 | −2.3284 | $1.7363 \times 10^{10}$ | $3.1258 \times 10^4$ | 0.26700050 | 0.0000 | 1.0000 | 0.7149 | $3.7566 \times 10^2$ | $1.8371 \times 10^2$ |
| 0.03000 | 3.4788 | −2.4560 | $1.5166 \times 10^{10}$ | $4.0969 \times 10^4$ | 0.26700050 | 0.0000 | 1.0000 | 0.7149 | $3.7789 \times 10^2$ | $1.8516 \times 10^2$ |
| 0.05000 | 3.4748 | −2.6164 | $1.2844 \times 10^{10}$ | $5.7124 \times 10^4$ | 0.26700050 | 0.0000 | 1.0000 | 0.7149 | $3.8323 \times 10^2$ | $1.8766 \times 10^2$ |
| 0.07000 | 3.4716 | −2.7218 | $1.1542 \times 10^{10}$ | $7.0741 \times 10^4$ | 0.26700050 | 0.0000 | 1.0000 | 0.7149 | $3.8746 \times 10^2$ | $1.9039 \times 10^2$ |
| 0.10000 | 3.4624 | −2.8573 | $1.0305 \times 10^{10}$ | $8.8742 \times 10^4$ | 0.26700050 | 0.0000 | 1.0000 | 0.7149 | $4.0015 \times 10^2$ | $1.9587 \times 10^2$ |

Table 1. Continued

| Age(Gyr) | $\log T_{\rm eff}$ | $\log L/L_\odot$ | R(cm) | g(cm/s$^2$) | $Y_c$ | L(PP1) | $L_g$ | $X_{\rm env}$ | $\tau_{gc}$ (day) | $\tau_{lc}$ (day) |
|---|---|---|---|---|---|---|---|---|---|---|
| \multicolumn{11}{c}{$0.072 M_\odot$} |
| 0.10578 | 3.4616 | $-2.8788$ | $1.0086 \times 10^{10}$ | $9.2629 \times 10^4$ | 0.26700050 | 0.0000 | 1.0000 | 0.7149 | $4.0042 \times 10^2$ | $1.9660 \times 10^2$ |
| 0.00002 | 3.3975 | $-0.3040$ | $2.6269 \times 10^{11}$ | $1.3848 \times 10^2$ | 0.26700050 | 0.0000 | 1.0000 | 0.7149 | $4.6845 \times 10^2$ | $2.1083 \times 10^2$ |
| 0.00010 | 3.4407 | $-0.7689$ | $1.2606 \times 10^{11}$ | $6.0135 \times 10^2$ | 0.26700050 | 0.0000 | 1.0000 | 0.7149 | $4.1774 \times 10^2$ | $1.9626 \times 10^2$ |
| 0.00020 | 3.4546 | $-0.9603$ | $9.4837 \times 10^{10}$ | $1.0625 \times 10^3$ | 0.26700050 | 0.0000 | 1.0000 | 0.7149 | $4.0266 \times 10^2$ | $1.9129 \times 10^2$ |
| 0.00030 | 3.4615 | $-1.0739$ | $8.0627 \times 10^{10}$ | $1.4700 \times 10^3$ | 0.26700050 | 0.0000 | 1.0000 | 0.7149 | $3.9546 \times 10^2$ | $1.8880 \times 10^2$ |
| 0.00050 | 3.4684 | $-1.2202$ | $6.6003 \times 10^{10}$ | $2.1936 \times 10^3$ | 0.26700050 | 0.0000 | 1.0000 | 0.7149 | $3.8848 \times 10^2$ | $1.8634 \times 10^2$ |
| 0.00070 | 3.4721 | $-1.3175$ | $5.8007 \times 10^{10}$ | $2.8400 \times 10^3$ | 0.26700050 | 0.0000 | 1.0000 | 0.7149 | $3.8469 \times 10^2$ | $1.8507 \times 10^2$ |
| 0.00100 | 3.4754 | $-1.4212$ | $5.0709 \times 10^{10}$ | $3.7163 \times 10^3$ | 0.26700050 | 0.0000 | 1.0000 | 0.7149 | $3.8170 \times 10^2$ | $1.8391 \times 10^2$ |
| 0.00200 | 3.4801 | $-1.6246$ | $3.9262 \times 10^{10}$ | $6.1992 \times 10^3$ | 0.26700050 | 0.0000 | 1.0000 | 0.7149 | $3.7690 \times 10^2$ | $1.8249 \times 10^2$ |
| 0.00300 | 3.4819 | $-1.7449$ | $3.3896 \times 10^{10}$ | $8.3174 \times 10^3$ | 0.26700050 | 0.0000 | 1.0000 | 0.7149 | $3.7501 \times 10^2$ | $1.8163 \times 10^2$ |
| 0.00500 | 3.4834 | $-1.8969$ | $2.8257 \times 10^{10}$ | $1.1968 \times 10^4$ | 0.26700050 | 0.0000 | 1.0000 | 0.7149 | $3.7324 \times 10^2$ | $1.8113 \times 10^2$ |
| 0.00700 | 3.4839 | $-1.9976$ | $2.5112 \times 10^{10}$ | $1.5154 \times 10^4$ | 0.26700050 | 0.0000 | 1.0000 | 0.7149 | $3.7297 \times 10^2$ | $1.8123 \times 10^2$ |
| 0.01000 | 3.4839 | $-2.1053$ | $2.2184 \times 10^{10}$ | $1.9418 \times 10^4$ | 0.26700050 | 0.0000 | 1.0000 | 0.7149 | $3.7427 \times 10^2$ | $1.8189 \times 10^2$ |
| 0.02000 | 3.4820 | $-2.3192$ | $1.7494 \times 10^{10}$ | $3.1223 \times 10^4$ | 0.26700050 | 0.0000 | 1.0000 | 0.7149 | $3.7662 \times 10^2$ | $1.8417 \times 10^2$ |
| 0.03000 | 3.4795 | $-2.4467$ | $1.5279 \times 10^{10}$ | $4.0936 \times 10^4$ | 0.26700050 | 0.0000 | 1.0000 | 0.7149 | $3.7878 \times 10^2$ | $1.8552 \times 10^2$ |
| 0.05000 | 3.4756 | $-2.6067$ | $1.2936 \times 10^{10}$ | $5.7102 \times 10^4$ | 0.26700050 | 0.0000 | 1.0000 | 0.7149 | $3.8395 \times 10^2$ | $1.8792 \times 10^2$ |
| 0.07000 | 3.4726 | $-2.7120$ | $1.1620 \times 10^{10}$ | $7.0769 \times 10^4$ | 0.26700050 | 0.0000 | 1.0000 | 0.7149 | $3.8821 \times 10^2$ | $1.9065 \times 10^2$ |
| 0.10000 | 3.4639 | $-2.8450$ | $1.0377 \times 10^{10}$ | $8.8739 \times 10^4$ | 0.26700050 | 0.0000 | 1.0000 | 0.7149 | $4.0010 \times 10^2$ | $1.9575 \times 10^2$ |
| 0.10417 | 3.4627 | $-2.8609$ | $1.0244 \times 10^{10}$ | $9.1056 \times 10^4$ | 0.26700050 | 0.0000 | 1.0000 | 0.7149 | $4.0213 \times 10^2$ | $1.9682 \times 10^2$ |
| \multicolumn{11}{c}{$0.073 M_\odot$} |
| 0.00002 | 3.3992 | $-0.3050$ | $2.6032 \times 10^{11}$ | $1.4297 \times 10^2$ | 0.26700050 | 0.0000 | 1.0000 | 0.7149 | $4.6889 \times 10^2$ | $2.1136 \times 10^2$ |
| 0.00010 | 3.4414 | $-0.7620$ | $1.2667 \times 10^{11}$ | $6.0385 \times 10^2$ | 0.26700050 | 0.0000 | 1.0000 | 0.7149 | $4.1867 \times 10^2$ | $1.9672 \times 10^2$ |
| 0.00020 | 3.4552 | $-0.9528$ | $9.5414 \times 10^{10}$ | $1.0643 \times 10^3$ | 0.26700050 | 0.0000 | 1.0000 | 0.7149 | $4.0374 \times 10^2$ | $1.9178 \times 10^2$ |
| 0.00030 | 3.4620 | $-1.0659$ | $8.1166 \times 10^{10}$ | $1.4707 \times 10^3$ | 0.26700050 | 0.0000 | 1.0000 | 0.7149 | $3.9651 \times 10^2$ | $1.8932 \times 10^2$ |
| 0.00050 | 3.4688 | $-1.2123$ | $6.6459 \times 10^{10}$ | $2.1936 \times 10^3$ | 0.26700050 | 0.0000 | 1.0000 | 0.7149 | $3.8963 \times 10^2$ | $1.8688 \times 10^2$ |
| 0.00070 | 3.4725 | $-1.3095$ | $5.8423 \times 10^{10}$ | $2.8386 \times 10^3$ | 0.26700050 | 0.0000 | 1.0000 | 0.7149 | $3.8585 \times 10^2$ | $1.8561 \times 10^2$ |
| 0.00100 | 3.4758 | $-1.4131$ | $5.1080 \times 10^{10}$ | $3.7133 \times 10^3$ | 0.26700050 | 0.0000 | 1.0000 | 0.7149 | $3.8284 \times 10^2$ | $1.8447 \times 10^2$ |
| 0.00200 | 3.4805 | $-1.6163$ | $3.9558 \times 10^{10}$ | $6.1915 \times 10^3$ | 0.26700050 | 0.0000 | 1.0000 | 0.7149 | $3.7807 \times 10^2$ | $1.8302 \times 10^2$ |
| 0.00300 | 3.4824 | $-1.7363$ | $3.4153 \times 10^{10}$ | $8.3063 \times 10^3$ | 0.26700050 | 0.0000 | 1.0000 | 0.7149 | $3.7610 \times 10^2$ | $1.8213 \times 10^2$ |
| 0.00500 | 3.4839 | $-1.8883$ | $2.8472 \times 10^{10}$ | $1.1952 \times 10^4$ | 0.26700050 | 0.0000 | 1.0000 | 0.7149 | $3.7430 \times 10^2$ | $1.8165 \times 10^2$ |
| 0.00700 | 3.4844 | $-1.9889$ | $2.5303 \times 10^{10}$ | $1.5133 \times 10^4$ | 0.26700050 | 0.0000 | 1.0000 | 0.7149 | $3.7403 \times 10^2$ | $1.8173 \times 10^2$ |
| 0.01000 | 3.4844 | $-2.0966$ | $2.2350 \times 10^{10}$ | $1.9396 \times 10^4$ | 0.26700050 | 0.0000 | 1.0000 | 0.7149 | $3.7533 \times 10^2$ | $1.8236 \times 10^2$ |
| 0.02000 | 3.4826 | $-2.3101$ | $1.7626 \times 10^{10}$ | $3.1187 \times 10^4$ | 0.26700050 | 0.0000 | 1.0000 | 0.7149 | $3.7754 \times 10^2$ | $1.8462 \times 10^2$ |
| 0.03000 | 3.4802 | $-2.4375$ | $1.5391 \times 10^{10}$ | $4.0901 \times 10^4$ | 0.26700050 | 0.0000 | 1.0000 | 0.7149 | $3.7966 \times 10^2$ | $1.8587 \times 10^2$ |
| 0.05000 | 3.4764 | $-2.5974$ | $1.3029 \times 10^{10}$ | $5.7077 \times 10^4$ | 0.26700050 | 0.0000 | 1.0000 | 0.7149 | $3.8477 \times 10^2$ | $1.8822 \times 10^2$ |
| 0.07000 | 3.4733 | $-2.7029$ | $1.1702 \times 10^{10}$ | $7.0756 \times 10^4$ | 0.26700050 | 0.0000 | 1.0000 | 0.7149 | $3.8910 \times 10^2$ | $1.9098 \times 10^2$ |
| 0.10000 | 3.4658 | $-2.8312$ | $1.0451 \times 10^{10}$ | $8.8708 \times 10^4$ | 0.26700050 | 0.0000 | 1.0000 | 0.7149 | $3.9963 \times 10^2$ | $1.9542 \times 10^2$ |
| 0.12458 | 3.4611 | $-2.9180$ | $9.6650 \times 10^{9}$ | $1.0372 \times 10^5$ | 0.26700050 | 0.0000 | 1.0000 | 0.7149 | $4.0633 \times 10^2$ | $2.0023 \times 10^2$ |
| \multicolumn{11}{c}{$0.074 M_\odot$} |
| 0.00002 | 3.4008 | $-0.3074$ | $2.5772 \times 10^{11}$ | $1.4787 \times 10^2$ | 0.26700050 | 0.0000 | 1.0000 | 0.7149 | $4.6886 \times 10^2$ | $2.1165 \times 10^2$ |
| 0.00010 | 3.4421 | $-0.7553$ | $1.2723 \times 10^{11}$ | $6.0671 \times 10^2$ | 0.26700050 | 0.0000 | 1.0000 | 0.7149 | $4.1954 \times 10^2$ | $1.9717 \times 10^2$ |
| 0.00020 | 3.4558 | $-0.9454$ | $9.5984 \times 10^{10}$ | $1.0661 \times 10^3$ | 0.26700050 | 0.0000 | 1.0000 | 0.7149 | $4.0479 \times 10^2$ | $1.9230 \times 10^2$ |
| 0.00030 | 3.4625 | $-1.0583$ | $8.1687 \times 10^{10}$ | $1.4719 \times 10^3$ | 0.26700050 | 0.0000 | 1.0000 | 0.7149 | $3.9760 \times 10^2$ | $1.8980 \times 10^2$ |
| 0.00050 | 3.4693 | $-1.2044$ | $6.6919 \times 10^{10}$ | $2.1932 \times 10^3$ | 0.26700050 | 0.0000 | 1.0000 | 0.7149 | $3.9073 \times 10^2$ | $1.8740 \times 10^2$ |
| 0.00070 | 3.4730 | $-1.3015$ | $5.8842 \times 10^{10}$ | $2.8367 \times 10^3$ | 0.26700050 | 0.0000 | 1.0000 | 0.7149 | $3.8703 \times 10^2$ | $1.8612 \times 10^2$ |
| 0.00100 | 3.4762 | $-1.4051$ | $5.1448 \times 10^{10}$ | $3.7105 \times 10^3$ | 0.26700050 | 0.0000 | 1.0000 | 0.7149 | $3.8395 \times 10^2$ | $1.8503 \times 10^2$ |
| 0.00200 | 3.4809 | $-1.6082$ | $3.9848 \times 10^{10}$ | $6.1853 \times 10^3$ | 0.26700050 | 0.0000 | 1.0000 | 0.7149 | $3.7922 \times 10^2$ | $1.8355 \times 10^2$ |
| 0.00300 | 3.4829 | $-1.7279$ | $3.4406 \times 10^{10}$ | $8.2967 \times 10^3$ | 0.26700050 | 0.0000 | 1.0000 | 0.7149 | $3.7721 \times 10^2$ | $1.8263 \times 10^2$ |
| 0.00500 | 3.4844 | $-1.8799$ | $2.8683 \times 10^{10}$ | $1.1938 \times 10^4$ | 0.26700050 | 0.0000 | 1.0000 | 0.7149 | $3.7556 \times 10^2$ | $1.8218 \times 10^2$ |
| 0.00700 | 3.4849 | $-1.9804$ | $2.5492 \times 10^{10}$ | $1.5114 \times 10^4$ | 0.26700050 | 0.0000 | 1.0000 | 0.7149 | $3.7509 \times 10^2$ | $1.8222 \times 10^2$ |
| 0.01000 | 3.4849 | $-2.0880$ | $2.2518 \times 10^{10}$ | $1.9370 \times 10^4$ | 0.26700050 | 0.0000 | 1.0000 | 0.7149 | $3.7631 \times 10^2$ | $1.8281 \times 10^2$ |
| 0.02000 | 3.4833 | $-2.3011$ | $1.7755 \times 10^{10}$ | $3.1155 \times 10^4$ | 0.26700050 | 0.0000 | 1.0000 | 0.7149 | $3.7844 \times 10^2$ | $1.8506 \times 10^2$ |
| 0.03000 | 3.4809 | $-2.4284$ | $1.5501 \times 10^{10}$ | $4.0873 \times 10^4$ | 0.26700050 | 0.0000 | 1.0000 | 0.7149 | $3.8050 \times 10^2$ | $1.8623 \times 10^2$ |
| 0.05000 | 3.4772 | $-2.5882$ | $1.3122 \times 10^{10}$ | $5.7043 \times 10^4$ | 0.26700050 | 0.0000 | 1.0000 | 0.7149 | $3.8552 \times 10^2$ | $1.8855 \times 10^2$ |
| 0.07000 | 3.4741 | $-2.6940$ | $1.1784 \times 10^{10}$ | $7.0728 \times 10^4$ | 0.26700050 | 0.0000 | 1.0000 | 0.7149 | $3.8994 \times 10^2$ | $1.9129 \times 10^2$ |
| 0.10000 | 3.4677 | $-2.8176$ | $1.0522 \times 10^{10}$ | $8.8703 \times 10^4$ | 0.26700050 | 0.0000 | 1.0000 | 0.7149 | $3.9915 \times 10^2$ | $1.9511 \times 10^2$ |
| 0.11731 | 3.4630 | $-2.8836$ | $9.9699 \times 10^{9}$ | $9.8810 \times 10^4$ | 0.26700050 | 0.0000 | 1.0000 | 0.7149 | $4.0515 \times 10^2$ | $1.9870 \times 10^2$ |
| \multicolumn{11}{c}{$0.075 M_\odot$} |
| 0.00002 | 3.4023 | $-0.3095$ | $2.5534 \times 10^{11}$ | $1.5268 \times 10^2$ | 0.26700050 | 0.0000 | 1.0000 | 0.7149 | $4.6864 \times 10^2$ | $2.1183 \times 10^2$ |
| 0.00010 | 3.4428 | $-0.7487$ | $1.2779 \times 10^{11}$ | $6.0957 \times 10^2$ | 0.26700050 | 0.0000 | 1.0000 | 0.7149 | $4.2046 \times 10^2$ | $1.9762 \times 10^2$ |
| 0.00020 | 3.4563 | $-0.9382$ | $9.6548 \times 10^{10}$ | $1.0679 \times 10^3$ | 0.26700050 | 0.0000 | 1.0000 | 0.7149 | $4.0587 \times 10^2$ | $1.9283 \times 10^2$ |
| 0.00030 | 3.4630 | $-1.0507$ | $8.2229 \times 10^{10}$ | $1.4722 \times 10^3$ | 0.26700050 | 0.0000 | 1.0000 | 0.7149 | $3.9869 \times 10^2$ | $1.9034 \times 10^2$ |

Table 1. Continued

| Age(Gyr) | log$T_{\text{eff}}$ | log$L/L_\odot$ | R(cm) | g(cm/s$^2$) | $Y_c$ | L(PP1) | $L_g$ | $X_{\text{env}}$ | $\tau_{gc}$ (day) | $\tau_{lc}$ (day) |
|---|---|---|---|---|---|---|---|---|---|---|
| 0.00050 | 3.4698 | −1.1966 | 6.7379 × 10$^{10}$ | 2.1926 × 10$^3$ | 0.26700050 | 0.0000 | 1.0000 | 0.7149 | 3.9183 × 10$^2$ | 1.8792 × 10$^2$ |
| 0.00070 | 3.4734 | −1.2937 | 5.9253 × 10$^{10}$ | 2.8352 × 10$^3$ | 0.26700050 | 0.0000 | 1.0000 | 0.7149 | 3.8821 × 10$^2$ | 1.8665 × 10$^2$ |
| 0.00100 | 3.4767 | −1.3971 | 5.1819 × 10$^{10}$ | 3.7071 × 10$^3$ | 0.26700050 | 0.0000 | 1.0000 | 0.7149 | 3.8507 × 10$^2$ | 1.8558 × 10$^2$ |
| 0.00200 | 3.4814 | −1.6003 | 4.0133 × 10$^{10}$ | 6.1802 × 10$^3$ | 0.26700050 | 0.0000 | 1.0000 | 0.7149 | 3.8043 × 10$^2$ | 1.8405 × 10$^2$ |
| 0.00300 | 3.4834 | −1.7198 | 3.4656 × 10$^{10}$ | 8.2879 × 10$^3$ | 0.26700050 | 0.0000 | 1.0000 | 0.7149 | 3.7827 × 10$^2$ | 1.8318 × 10$^2$ |
| 0.00500 | 3.4849 | −1.8717 | 2.8894 × 10$^{10}$ | 1.1923 × 10$^4$ | 0.26700050 | 0.0000 | 1.0000 | 0.7149 | 3.7647 × 10$^2$ | 1.8268 × 10$^2$ |
| 0.00700 | 3.4854 | −1.9721 | 2.5677 × 10$^{10}$ | 1.5098 × 10$^4$ | 0.26700050 | 0.0000 | 1.0000 | 0.7149 | 3.7613 × 10$^2$ | 1.8272 × 10$^2$ |
| 0.01000 | 3.4855 | −2.0796 | 2.2682 × 10$^{10}$ | 1.9348 × 10$^4$ | 0.26700050 | 0.0000 | 1.0000 | 0.7149 | 3.7727 × 10$^2$ | 1.8326 × 10$^2$ |
| 0.02000 | 3.4839 | −2.2924 | 1.7881 × 10$^{10}$ | 3.1133 × 10$^4$ | 0.26700050 | 0.0000 | 1.0000 | 0.7149 | 3.7935 × 10$^2$ | 1.8550 × 10$^2$ |
| 0.03000 | 3.4816 | −2.4194 | 1.5611 × 10$^{10}$ | 4.0844 × 10$^4$ | 0.26700050 | 0.0000 | 1.0000 | 0.7149 | 3.8130 × 10$^2$ | 1.8659 × 10$^2$ |
| 0.05000 | 3.4779 | −2.5792 | 1.3213 × 10$^{10}$ | 5.7018 × 10$^4$ | 0.26700050 | 0.0000 | 1.0000 | 0.7149 | 3.8632 × 10$^2$ | 1.8887 × 10$^2$ |
| 0.07000 | 3.4748 | −2.6852 | 1.1864 × 10$^{10}$ | 7.0716 × 10$^4$ | 0.26700050 | 0.0000 | 1.0000 | 0.7149 | 3.9073 × 10$^2$ | 1.9157 × 10$^2$ |
| 0.10000 | 3.4693 | −2.8057 | 1.0591 × 10$^{10}$ | 8.8742 × 10$^4$ | 0.26700050 | 0.0000 | 1.0000 | 0.7149 | 3.9917 × 10$^2$ | 1.9501 × 10$^2$ |
| 0.12329 | 3.4632 | −2.8913 | 9.8703 × 10$^9$ | 1.0218 × 10$^5$ | 0.26700050 | 0.0000 | 1.0000 | 0.7149 | 4.0708 × 10$^2$ | 1.9972 × 10$^2$ |
| | | | | 0.076$M_\odot$ | | | | | | |
| 0.00002 | 3.4050 | −0.3082 | 2.5252 × 10$^{11}$ | 1.5819 × 10$^2$ | 0.26700050 | 0.0000 | 1.0000 | 0.7149 | 4.6752 × 10$^2$ | 2.1166 × 10$^2$ |
| 0.00010 | 3.4434 | −0.7425 | 1.2832 × 10$^{11}$ | 6.1262 × 10$^2$ | 0.26700050 | 0.0000 | 1.0000 | 0.7149 | 4.2136 × 10$^2$ | 1.9808 × 10$^2$ |
| 0.00020 | 3.4568 | −0.9311 | 9.7099 × 10$^{10}$ | 1.0699 × 10$^3$ | 0.26700050 | 0.0000 | 1.0000 | 0.7149 | 4.0692 × 10$^2$ | 1.9334 × 10$^2$ |
| 0.00030 | 3.4635 | −1.0432 | 8.2744 × 10$^{10}$ | 1.4733 × 10$^3$ | 0.26700050 | 0.0000 | 1.0000 | 0.7149 | 3.9976 × 10$^2$ | 1.9084 × 10$^2$ |
| 0.00050 | 3.4702 | −1.1889 | 6.7837 × 10$^{10}$ | 2.1919 × 10$^3$ | 0.26700050 | 0.0000 | 1.0000 | 0.7149 | 3.9294 × 10$^2$ | 1.8844 × 10$^2$ |
| 0.00070 | 3.4739 | −1.2860 | 5.9661 × 10$^{10}$ | 2.8339 × 10$^3$ | 0.26700050 | 0.0000 | 1.0000 | 0.7149 | 3.8931 × 10$^2$ | 1.8714 × 10$^2$ |
| 0.00100 | 3.4771 | −1.3894 | 5.2180 × 10$^{10}$ | 3.7047 × 10$^3$ | 0.26700050 | 0.0000 | 1.0000 | 0.7149 | 3.8619 × 10$^2$ | 1.8611 × 10$^2$ |
| 0.00200 | 3.4818 | −1.5923 | 4.0427 × 10$^{10}$ | 6.1719 × 10$^3$ | 0.26700050 | 0.0000 | 1.0000 | 0.7149 | 3.8154 × 10$^2$ | 1.8460 × 10$^2$ |
| 0.00300 | 3.4838 | −1.7116 | 3.4915 × 10$^{10}$ | 8.2746 × 10$^3$ | 0.26700050 | 0.0000 | 1.0000 | 0.7149 | 3.7940 × 10$^2$ | 1.8367 × 10$^2$ |
| 0.00500 | 3.4853 | −1.8635 | 2.9108 × 10$^{10}$ | 1.1905 × 10$^4$ | 0.26700050 | 0.0000 | 1.0000 | 0.7149 | 3.7754 × 10$^2$ | 1.8319 × 10$^2$ |
| 0.00700 | 3.4859 | −1.9640 | 2.5860 × 10$^{10}$ | 1.5084 × 10$^4$ | 0.26700050 | 0.0000 | 1.0000 | 0.7149 | 3.7718 × 10$^2$ | 1.8323 × 10$^2$ |
| 0.01000 | 3.4860 | −2.0712 | 2.2847 × 10$^{10}$ | 1.9324 × 10$^4$ | 0.26700050 | 0.0000 | 1.0000 | 0.7149 | 3.7822 × 10$^2$ | 1.8370 × 10$^2$ |
| 0.02000 | 3.4845 | −2.2839 | 1.8007 × 10$^{10}$ | 3.1109 × 10$^4$ | 0.26700050 | 0.0000 | 1.0000 | 0.7149 | 3.8024 × 10$^2$ | 1.8592 × 10$^2$ |
| 0.03000 | 3.4823 | −2.4106 | 1.5720 × 10$^{10}$ | 4.0818 × 10$^4$ | 0.26700050 | 0.0000 | 1.0000 | 0.7149 | 3.8206 × 10$^2$ | 1.8693 × 10$^2$ |
| 0.05000 | 3.4786 | −2.5703 | 1.3303 × 10$^{10}$ | 5.6996 × 10$^4$ | 0.26700050 | 0.0000 | 1.0000 | 0.7149 | 3.8712 × 10$^2$ | 1.8919 × 10$^2$ |
| 0.07000 | 3.4756 | −2.6762 | 1.1943 × 10$^{10}$ | 7.0716 × 10$^4$ | 0.26700050 | 0.0000 | 1.0000 | 0.7149 | 3.9146 × 10$^2$ | 1.9180 × 10$^2$ |
| 0.10000 | 3.4706 | −2.7948 | 1.0661 × 10$^{10}$ | 8.8745 × 10$^4$ | 0.26700050 | 0.0000 | 1.0000 | 0.7149 | 3.9951 × 10$^2$ | 1.9511 × 10$^2$ |
| 0.12819 | 3.4639 | −2.8931 | 9.8176 × 10$^9$ | 1.0465 × 10$^5$ | 0.26700050 | 0.0000 | 1.0000 | 0.7149 | 4.0821 × 10$^2$ | 2.0018 × 10$^2$ |
| | | | | 0.077$M_\odot$ | | | | | | |
| 0.00002 | 3.4069 | −0.3083 | 2.5026 × 10$^{11}$ | 1.6317 × 10$^2$ | 0.26700050 | 0.0000 | 1.0000 | 0.7149 | 4.6695 × 10$^2$ | 2.1167 × 10$^2$ |
| 0.00010 | 3.4441 | −0.7363 | 1.2884 × 10$^{11}$ | 6.1563 × 10$^2$ | 0.26700050 | 0.0000 | 1.0000 | 0.7149 | 4.2230 × 10$^2$ | 1.9854 × 10$^2$ |
| 0.00020 | 3.4573 | −0.9242 | 9.7643 × 10$^{10}$ | 1.0719 × 10$^3$ | 0.26700050 | 0.0000 | 1.0000 | 0.7149 | 4.0787 × 10$^2$ | 1.9381 × 10$^2$ |
| 0.00030 | 3.4640 | −1.0359 | 8.3262 × 10$^{10}$ | 1.4742 × 10$^3$ | 0.26700050 | 0.0000 | 1.0000 | 0.7149 | 4.0084 × 10$^2$ | 1.9138 × 10$^2$ |
| 0.00050 | 3.4707 | −1.1814 | 6.8289 × 10$^{10}$ | 2.1915 × 10$^3$ | 0.26700050 | 0.0000 | 1.0000 | 0.7149 | 3.9403 × 10$^2$ | 1.8898 × 10$^2$ |
| 0.00070 | 3.4743 | −1.2783 | 6.0070 × 10$^{10}$ | 2.8322 × 10$^3$ | 0.26700050 | 0.0000 | 1.0000 | 0.7149 | 3.9039 × 10$^2$ | 1.8765 × 10$^2$ |
| 0.00100 | 3.4775 | −1.3818 | 5.2544 × 10$^{10}$ | 3.7016 × 10$^3$ | 0.26700050 | 0.0000 | 1.0000 | 0.7149 | 3.8771 × 10$^2$ | 1.8663 × 10$^2$ |
| 0.00200 | 3.4822 | −1.5844 | 4.0716 × 10$^{10}$ | 6.1645 × 10$^3$ | 0.26700050 | 0.0000 | 1.0000 | 0.7149 | 3.8262 × 10$^2$ | 1.8509 × 10$^2$ |
| 0.00300 | 3.4842 | −1.7035 | 3.5170 × 10$^{10}$ | 8.2619 × 10$^3$ | 0.26700050 | 0.0000 | 1.0000 | 0.7149 | 3.8047 × 10$^2$ | 1.8419 × 10$^2$ |
| 0.00500 | 3.4858 | −1.8554 | 2.9319 × 10$^{10}$ | 1.1889 × 10$^4$ | 0.26700050 | 0.0000 | 1.0000 | 0.7149 | 3.7864 × 10$^2$ | 1.8368 × 10$^2$ |
| 0.00700 | 3.4864 | −1.9557 | 2.6050 × 10$^{10}$ | 1.5060 × 10$^4$ | 0.26700050 | 0.0000 | 1.0000 | 0.7149 | 3.7823 × 10$^2$ | 1.8370 × 10$^2$ |
| 0.01000 | 3.4865 | −2.0627 | 2.3017 × 10$^{10}$ | 1.9290 × 10$^4$ | 0.26700050 | 0.0000 | 1.0000 | 0.7149 | 3.7915 × 10$^2$ | 1.8414 × 10$^2$ |
| 0.02000 | 3.4851 | −2.2752 | 1.8137 × 10$^{10}$ | 3.1066 × 10$^4$ | 0.26700050 | 0.0000 | 1.0000 | 0.7149 | 3.8115 × 10$^2$ | 1.8636 × 10$^2$ |
| 0.03000 | 3.4830 | −2.4018 | 1.5829 × 10$^{10}$ | 4.0790 × 10$^4$ | 0.26700050 | 0.0000 | 1.0000 | 0.7149 | 3.8292 × 10$^2$ | 1.8729 × 10$^2$ |
| 0.05000 | 3.4794 | −2.5615 | 1.3394 × 10$^{10}$ | 5.6969 × 10$^4$ | 0.26700050 | 0.0000 | 1.0000 | 0.7149 | 3.8793 × 10$^2$ | 1.8952 × 10$^2$ |
| 0.07000 | 3.4764 | −2.6671 | 1.2024 × 10$^{10}$ | 7.0691 × 10$^4$ | 0.26700050 | 0.0000 | 1.0000 | 0.7149 | 3.9213 × 10$^2$ | 1.9202 × 10$^2$ |
| 0.10000 | 3.4718 | −2.7844 | 1.0730 × 10$^{10}$ | 8.8759 × 10$^4$ | 0.26700050 | 0.0000 | 1.0000 | 0.7149 | 3.9998 × 10$^2$ | 1.9529 × 10$^2$ |
| 0.15190 | 3.4627 | −2.9436 | 9.3142 × 10$^9$ | 1.1780 × 10$^5$ | 0.26700049 | 0.0046 | 0.9954 | 0.7149 | 4.0617 × 10$^2$ | 2.0321 × 10$^2$ |
| | | | | 0.078$M_\odot$ | | | | | | |
| 0.00002 | 3.4084 | −0.3092 | 2.4832 × 10$^{11}$ | 1.6788 × 10$^2$ | 0.26700050 | 0.0000 | 1.0000 | 0.7149 | 4.6677 × 10$^2$ | 2.1183 × 10$^2$ |
| 0.00010 | 3.4447 | −0.7304 | 1.2936 × 10$^{11}$ | 6.1868 × 10$^2$ | 0.26700050 | 0.0000 | 1.0000 | 0.7149 | 4.2323 × 10$^2$ | 1.9901 × 10$^2$ |
| 0.00020 | 3.4579 | −0.9172 | 9.8192 × 10$^{10}$ | 1.0737 × 10$^3$ | 0.26700050 | 0.0000 | 1.0000 | 0.7149 | 4.0898 × 10$^2$ | 1.9430 × 10$^2$ |
| 0.00030 | 3.4645 | −1.0288 | 8.3762 × 10$^{10}$ | 1.4755 × 10$^3$ | 0.26700050 | 0.0000 | 1.0000 | 0.7149 | 4.0193 × 10$^2$ | 1.9187 × 10$^2$ |
| 0.00050 | 3.4711 | −1.1739 | 6.8738 × 10$^{10}$ | 2.1910 × 10$^3$ | 0.26700050 | 0.0000 | 1.0000 | 0.7149 | 3.9511 × 10$^2$ | 1.8950 × 10$^2$ |
| 0.00070 | 3.4747 | −1.2706 | 6.0479 × 10$^{10}$ | 2.8303 × 10$^3$ | 0.26700050 | 0.0000 | 1.0000 | 0.7149 | 3.9147 × 10$^2$ | 1.8816 × 10$^2$ |
| 0.00100 | 3.4779 | −1.3741 | 5.2910 × 10$^{10}$ | 3.6980 × 10$^3$ | 0.26700050 | 0.0000 | 1.0000 | 0.7149 | 3.8849 × 10$^2$ | 1.8710 × 10$^2$ |
| 0.00200 | 3.4827 | −1.5763 | 4.1009 × 10$^{10}$ | 6.1557 × 10$^3$ | 0.26700050 | 0.0000 | 1.0000 | 0.7149 | 3.8364 × 10$^2$ | 1.8558 × 10$^2$ |
| 0.00300 | 3.4847 | −1.6956 | 3.5424 × 10$^{10}$ | 8.2498 × 10$^3$ | 0.26700050 | 0.0000 | 1.0000 | 0.7149 | 3.8148 × 10$^2$ | 1.8466 × 10$^2$ |
| 0.00500 | 3.4862 | −1.8475 | 2.9526 × 10$^{10}$ | 1.1875 × 10$^4$ | 0.26700050 | 0.0000 | 1.0000 | 0.7149 | 3.7976 × 10$^2$ | 1.8416 × 10$^2$ |
| 0.00700 | 3.4868 | −1.9477 | 2.6232 × 10$^{10}$ | 1.5044 × 10$^4$ | 0.26700050 | 0.0000 | 1.0000 | 0.7149 | 3.7925 × 10$^2$ | 1.8420 × 10$^2$ |
| 0.01000 | 3.4870 | −2.0545 | 2.3180 × 10$^{10}$ | 1.9267 × 10$^4$ | 0.26700050 | 0.0000 | 1.0000 | 0.7149 | 3.8008 × 10$^2$ | 1.8460 × 10$^2$ |

Table 1. Continued

| Age(Gyr) | $\log T_{\text{eff}}$ | $\log L/L_\odot$ | R(cm) | g(cm/s$^2$) | $Y_c$ | L(PP1) | $L_g$ | $X_{\text{env}}$ | $\tau_{gc}$ (day) | $\tau_{lc}$ (day) |
|---|---|---|---|---|---|---|---|---|---|---|
| 0.02000 | 3.4857 | −2.2668 | 1.8264 × 10$^{10}$ | 3.1036 × 10$^4$ | 0.26700050 | 0.0000 | 1.0000 | 0.7149 | 3.8206 × 10$^2$ | 1.8681 × 10$^2$ |
| 0.03000 | 3.4837 | −2.3933 | 1.5938 × 10$^{10}$ | 4.0756 × 10$^4$ | 0.26700050 | 0.0000 | 1.0000 | 0.7149 | 3.8378 × 10$^2$ | 1.8767 × 10$^2$ |
| 0.05000 | 3.4801 | −2.5530 | 1.3481 × 10$^{10}$ | 5.6959 × 10$^4$ | 0.26700050 | 0.0000 | 1.0000 | 0.7149 | 3.8873 × 10$^2$ | 1.8985 × 10$^2$ |
| 0.07000 | 3.4772 | −2.6583 | 1.2102 × 10$^{10}$ | 7.0689 × 10$^4$ | 0.26700050 | 0.0000 | 1.0000 | 0.7149 | 3.9281 × 10$^2$ | 1.9227 × 10$^2$ |
| 0.10000 | 3.4729 | −2.7745 | 1.0800 × 10$^{10}$ | 8.8757 × 10$^4$ | 0.26700050 | 0.0000 | 1.0000 | 0.7149 | 4.0041 × 10$^2$ | 1.9547 × 10$^2$ |
| 0.14455 | 3.4637 | −2.9179 | 9.5523 × 10$^9$ | 1.1346 × 10$^5$ | 0.26700050 | 0.0049 | 0.9951 | 0.7149 | 4.0627 × 10$^2$ | 2.0237 × 10$^2$ |
| | | | | | 0.079$M_\odot$ | | | | | |
| 0.00002 | 3.4080 | −0.3002 | 2.5135 × 10$^{11}$ | 1.6597 × 10$^2$ | 0.26700050 | 0.0000 | 1.0000 | 0.7149 | 4.6862 × 10$^2$ | 2.1257 × 10$^2$ |
| 0.00010 | 3.4451 | −0.7224 | 1.3029 × 10$^{11}$ | 6.1764 × 10$^2$ | 0.26700050 | 0.0000 | 1.0000 | 0.7149 | 4.2443 × 10$^2$ | 1.9956 × 10$^2$ |
| 0.00020 | 3.4583 | −0.9091 | 9.8898 × 10$^{10}$ | 1.0720 × 10$^3$ | 0.26700050 | 0.0000 | 1.0000 | 0.7149 | 4.1001 × 10$^2$ | 1.9481 × 10$^2$ |
| 0.00030 | 3.4648 | −1.0212 | 8.4362 × 10$^{10}$ | 1.4733 × 10$^3$ | 0.26700050 | 0.0000 | 1.0000 | 0.7149 | 4.0307 × 10$^2$ | 1.9240 × 10$^2$ |
| 0.00050 | 3.4716 | −1.1659 | 6.9227 × 10$^{10}$ | 2.1879 × 10$^3$ | 0.26700050 | 0.0000 | 1.0000 | 0.7149 | 3.9615 × 10$^2$ | 1.8996 × 10$^2$ |
| 0.00070 | 3.4752 | −1.2628 | 6.0907 × 10$^{10}$ | 2.8265 × 10$^3$ | 0.26700050 | 0.0000 | 1.0000 | 0.7149 | 3.9252 × 10$^2$ | 1.8865 × 10$^2$ |
| 0.00100 | 3.4783 | −1.3664 | 5.3284 × 10$^{10}$ | 3.6930 × 10$^3$ | 0.26700050 | 0.0000 | 1.0000 | 0.7149 | 3.8957 × 10$^2$ | 1.8762 × 10$^2$ |
| 0.00200 | 3.4831 | −1.5687 | 4.1299 × 10$^{10}$ | 6.1474 × 10$^3$ | 0.26700050 | 0.0000 | 1.0000 | 0.7149 | 3.8480 × 10$^2$ | 1.8609 × 10$^2$ |
| 0.00300 | 3.4851 | −1.6878 | 3.5673 × 10$^{10}$ | 8.2393 × 10$^3$ | 0.26700050 | 0.0000 | 1.0000 | 0.7149 | 3.8259 × 10$^2$ | 1.8520 × 10$^2$ |
| 0.00500 | 3.4867 | −1.8395 | 2.9738 × 10$^{10}$ | 1.1856 × 10$^4$ | 0.26700050 | 0.0000 | 1.0000 | 0.7149 | 3.8080 × 10$^2$ | 1.8467 × 10$^2$ |
| 0.00700 | 3.4873 | −1.9395 | 2.6425 × 10$^{10}$ | 1.5016 × 10$^4$ | 0.26700050 | 0.0000 | 1.0000 | 0.7149 | 3.8031 × 10$^2$ | 1.8461 × 10$^2$ |
| 0.01000 | 3.4875 | −2.0464 | 2.3345 × 10$^{10}$ | 1.9239 × 10$^4$ | 0.26700050 | 0.0000 | 1.0000 | 0.7149 | 3.8109 × 10$^2$ | 1.8505 × 10$^2$ |
| 0.02000 | 3.4863 | −2.2583 | 1.8395 × 10$^{10}$ | 3.0987 × 10$^4$ | 0.26700050 | 0.0000 | 1.0000 | 0.7149 | 3.8305 × 10$^2$ | 1.8720 × 10$^2$ |
| 0.03000 | 3.4843 | −2.3848 | 1.6048 × 10$^{10}$ | 4.0712 × 10$^4$ | 0.26700050 | 0.0000 | 1.0000 | 0.7149 | 3.8463 × 10$^2$ | 1.8804 × 10$^2$ |
| 0.05000 | 3.4808 | −2.5441 | 1.3576 × 10$^{10}$ | 5.6890 × 10$^4$ | 0.26700050 | 0.0000 | 1.0000 | 0.7149 | 3.8946 × 10$^2$ | 1.9016 × 10$^2$ |
| 0.07000 | 3.4779 | −2.6496 | 1.2183 × 10$^{10}$ | 7.0644 × 10$^4$ | 0.26700050 | 0.0000 | 1.0000 | 0.7149 | 3.9354 × 10$^2$ | 1.9254 × 10$^2$ |
| 0.10000 | 3.4742 | −2.7632 | 1.0873 × 10$^{10}$ | 8.8692 × 10$^4$ | 0.26700050 | 0.0000 | 1.0000 | 0.7149 | 4.0051 × 10$^2$ | 1.9545 × 10$^2$ |
| 0.16418 | 3.4639 | −2.9491 | 9.2051 × 10$^9$ | 1.2374 × 10$^5$ | 0.26700050 | 0.0140 | 0.9860 | 0.7149 | 4.0396 × 10$^2$ | 2.0357 × 10$^2$ |
| | | | | | 0.080$M_\odot$ | | | | | |
| 0.00002 | 3.4088 | −0.2941 | 2.5223 × 10$^{11}$ | 1.6689 × 10$^2$ | 0.26700050 | 0.0000 | 1.0000 | 0.7149 | 4.6940 × 10$^2$ | 2.1301 × 10$^2$ |
| 0.00010 | 3.4456 | −0.7151 | 1.3109 × 10$^{11}$ | 6.1789 × 10$^2$ | 0.26700050 | 0.0000 | 1.0000 | 0.7149 | 4.2547 × 10$^2$ | 2.0006 × 10$^2$ |
| 0.00020 | 3.4588 | −0.9016 | 9.9547 × 10$^{10}$ | 1.0715 × 10$^3$ | 0.26700050 | 0.0000 | 1.0000 | 0.7149 | 4.1108 × 10$^2$ | 1.9531 × 10$^2$ |
| 0.00030 | 3.4653 | −1.0137 | 8.4922 × 10$^{10}$ | 1.4723 × 10$^3$ | 0.26700050 | 0.0000 | 1.0000 | 0.7149 | 4.0417 × 10$^2$ | 1.9291 × 10$^2$ |
| 0.00050 | 3.4720 | −1.1584 | 6.9696 × 10$^{10}$ | 2.1858 × 10$^3$ | 0.26700050 | 0.0000 | 1.0000 | 0.7149 | 3.9725 × 10$^2$ | 1.9049 × 10$^2$ |
| 0.00070 | 3.4756 | −1.2552 | 6.1327 × 10$^{10}$ | 2.8231 × 10$^3$ | 0.26700050 | 0.0000 | 1.0000 | 0.7149 | 3.9359 × 10$^2$ | 1.8916 × 10$^2$ |
| 0.00100 | 3.4787 | −1.3589 | 5.3644 × 10$^{10}$ | 3.6897 × 10$^3$ | 0.26700050 | 0.0000 | 1.0000 | 0.7149 | 3.9057 × 10$^2$ | 1.8817 × 10$^2$ |
| 0.00200 | 3.4835 | −1.5609 | 4.1586 × 10$^{10}$ | 6.1396 × 10$^3$ | 0.26700050 | 0.0000 | 1.0000 | 0.7149 | 3.8582 × 10$^2$ | 1.8657 × 10$^2$ |
| 0.00300 | 3.4855 | −1.6803 | 3.5915 × 10$^{10}$ | 8.2316 × 10$^3$ | 0.26700050 | 0.0000 | 1.0000 | 0.7149 | 3.8365 × 10$^2$ | 1.8569 × 10$^2$ |
| 0.00500 | 3.4871 | −1.8319 | 2.9941 × 10$^{10}$ | 1.1844 × 10$^4$ | 0.26700050 | 0.0000 | 1.0000 | 0.7149 | 3.8190 × 10$^2$ | 1.8515 × 10$^2$ |
| 0.00700 | 3.4878 | −1.9317 | 2.6607 × 10$^{10}$ | 1.4998 × 10$^4$ | 0.26700050 | 0.0000 | 1.0000 | 0.7149 | 3.8136 × 10$^2$ | 1.8508 × 10$^2$ |
| 0.01000 | 3.4880 | −2.0385 | 2.3507 × 10$^{10}$ | 1.9216 × 10$^4$ | 0.26700050 | 0.0000 | 1.0000 | 0.7149 | 3.8204 × 10$^2$ | 1.8549 × 10$^2$ |
| 0.02000 | 3.4868 | −2.2502 | 1.8521 × 10$^{10}$ | 3.0952 × 10$^4$ | 0.26700050 | 0.0000 | 1.0000 | 0.7149 | 3.8395 × 10$^2$ | 1.8762 × 10$^2$ |
| 0.03000 | 3.4849 | −2.3765 | 1.6157 × 10$^{10}$ | 4.0676 × 10$^4$ | 0.26700050 | 0.0000 | 1.0000 | 0.7149 | 3.8548 × 10$^2$ | 1.8843 × 10$^2$ |
| 0.05000 | 3.4815 | −2.5356 | 1.3664 × 10$^{10}$ | 5.6868 × 10$^4$ | 0.26700050 | 0.0000 | 1.0000 | 0.7149 | 3.9020 × 10$^2$ | 1.9051 × 10$^2$ |
| 0.07000 | 3.4787 | −2.6411 | 1.2260 × 10$^{10}$ | 7.0641 × 10$^4$ | 0.26700050 | 0.0000 | 1.0000 | 0.7149 | 3.9430 × 10$^2$ | 1.9283 × 10$^2$ |
| 0.10000 | 3.4751 | −2.7539 | 1.0944 × 10$^{10}$ | 8.8647 × 10$^4$ | 0.26700050 | 0.0000 | 1.0000 | 0.7149 | 4.0106 × 10$^2$ | 1.9571 × 10$^2$ |
| 0.13580 | 3.4668 | −2.8707 | 9.9393 × 10$^9$ | 1.0748 × 10$^5$ | 0.26700050 | 0.0060 | 0.9940 | 0.7149 | 4.0575 × 10$^2$ | 2.0074 × 10$^2$ |
| | | | | | 0.081$M_\odot$ | | | | | |
| 0.00002 | 3.4095 | −0.2882 | 2.5311 × 10$^{11}$ | 1.6780 × 10$^2$ | 0.26700050 | 0.0000 | 1.0000 | 0.7149 | 4.7026 × 10$^2$ | 2.1348 × 10$^2$ |
| 0.00010 | 3.4461 | −0.7080 | 1.3187 × 10$^{11}$ | 6.1823 × 10$^2$ | 0.26700050 | 0.0000 | 1.0000 | 0.7149 | 4.2653 × 10$^2$ | 2.0056 × 10$^2$ |
| 0.00020 | 3.4592 | −0.8943 | 1.0018 × 10$^{11}$ | 1.0711 × 10$^3$ | 0.26700050 | 0.0000 | 1.0000 | 0.7149 | 4.1216 × 10$^2$ | 1.9581 × 10$^2$ |
| 0.00030 | 3.4657 | −1.0063 | 8.5484 × 10$^{10}$ | 1.4712 × 10$^3$ | 0.26700050 | 0.0000 | 1.0000 | 0.7149 | 4.0526 × 10$^2$ | 1.9343 × 10$^2$ |
| 0.00050 | 3.4724 | −1.1510 | 7.0166 × 10$^{10}$ | 2.1837 × 10$^3$ | 0.26700050 | 0.0000 | 1.0000 | 0.7149 | 3.9832 × 10$^2$ | 1.9101 × 10$^2$ |
| 0.00070 | 3.4760 | −1.2477 | 6.1744 × 10$^{10}$ | 2.8199 × 10$^3$ | 0.26700050 | 0.0000 | 1.0000 | 0.7149 | 3.9467 × 10$^2$ | 1.8968 × 10$^2$ |
| 0.00100 | 3.4791 | −1.3512 | 5.4014 × 10$^{10}$ | 3.6849 × 10$^3$ | 0.26700050 | 0.0000 | 1.0000 | 0.7149 | 3.9158 × 10$^2$ | 1.8865 × 10$^2$ |
| 0.00200 | 3.4839 | −1.5532 | 4.1872 × 10$^{10}$ | 6.1317 × 10$^3$ | 0.26700050 | 0.0000 | 1.0000 | 0.7149 | 3.8683 × 10$^2$ | 1.8706 × 10$^2$ |
| 0.00300 | 3.4859 | −1.6727 | 3.6163 × 10$^{10}$ | 8.2204 × 10$^3$ | 0.26700050 | 0.0000 | 1.0000 | 0.7149 | 3.8479 × 10$^2$ | 1.8616 × 10$^2$ |
| 0.00500 | 3.4875 | −1.8242 | 3.0146 × 10$^{10}$ | 1.1830 × 10$^4$ | 0.26700050 | 0.0000 | 1.0000 | 0.7149 | 3.8287 × 10$^2$ | 1.8567 × 10$^2$ |
| 0.00700 | 3.4882 | −1.9239 | 2.6791 × 10$^{10}$ | 1.4978 × 10$^4$ | 0.26700050 | 0.0000 | 1.0000 | 0.7149 | 3.8236 × 10$^2$ | 1.8555 × 10$^2$ |
| 0.01000 | 3.4885 | −2.0306 | 2.3667 × 10$^{10}$ | 1.9193 × 10$^4$ | 0.26700050 | 0.0000 | 1.0000 | 0.7149 | 3.8299 × 10$^2$ | 1.8596 × 10$^2$ |
| 0.02000 | 3.4874 | −2.2421 | 1.8648 × 10$^{10}$ | 3.0915 × 10$^4$ | 0.26700050 | 0.0000 | 1.0000 | 0.7149 | 3.8485 × 10$^2$ | 1.8803 × 10$^2$ |
| 0.03000 | 3.4855 | −2.3683 | 1.6265 × 10$^{10}$ | 4.0637 × 10$^4$ | 0.26700050 | 0.0000 | 1.0000 | 0.7149 | 3.8631 × 10$^2$ | 1.8881 × 10$^2$ |
| 0.05000 | 3.4821 | −2.5276 | 1.3751 × 10$^{10}$ | 5.6852 × 10$^4$ | 0.26700050 | 0.0000 | 1.0000 | 0.7149 | 3.9111 × 10$^2$ | 1.9088 × 10$^2$ |
| 0.07000 | 3.4794 | −2.6328 | 1.2338 × 10$^{10}$ | 7.0619 × 10$^4$ | 0.26700050 | 0.0000 | 1.0000 | 0.7149 | 3.9509 × 10$^2$ | 1.9313 × 10$^2$ |
| 0.10000 | 3.4759 | −2.7454 | 1.1013 × 10$^{10}$ | 8.8642 × 10$^4$ | 0.26700050 | 0.0003 | 0.9997 | 0.7149 | 4.0086 × 10$^2$ | 1.9600 × 10$^2$ |
| 0.15457 | 3.4652 | −2.9120 | 9.5479 × 10$^9$ | 1.1793 × 10$^5$ | 0.26700050 | 0.0160 | 0.9840 | 0.7149 | 4.0471 × 10$^2$ | 2.0263 × 10$^2$ |

Table 1. Continued

| Age(Gyr) | $\log T_{\rm eff}$ | $\log L/L_\odot$ | R(cm) | g(cm/s$^2$) | $Y_c$ | L(PP1) | $L_g$ | $X_{\rm env}$ | $\tau_{gc}$ (day) | $\tau_{lc}$ (day) |
|---|---|---|---|---|---|---|---|---|---|---|
| | | | | $0.083 M_\odot$ | | | | | | |
| 0.00002 | 3.4108 | $-0.2767$ | $2.5489 \times 10^{11}$ | $1.6956 \times 10^2$ | 0.26700050 | 0.0000 | 1.0000 | 0.7149 | $4.7211 \times 10^2$ | $2.1446 \times 10^2$ |
| 0.00010 | 3.4471 | $-0.6940$ | $1.3341 \times 10^{11}$ | $6.1890 \times 10^2$ | 0.26700050 | 0.0000 | 1.0000 | 0.7149 | $4.2857 \times 10^2$ | $2.0154 \times 10^2$ |
| 0.00020 | 3.4602 | $-0.8794$ | $1.0145 \times 10^{11}$ | $1.0704 \times 10^3$ | 0.26700050 | 0.0000 | 1.0000 | 0.7149 | $4.1402 \times 10^2$ | $1.9671 \times 10^2$ |
| 0.00030 | 3.4666 | $-0.9918$ | $8.6571 \times 10^{10}$ | $1.4699 \times 10^3$ | 0.26700050 | 0.0000 | 1.0000 | 0.7149 | $4.0733 \times 10^2$ | $1.9444 \times 10^2$ |
| 0.00050 | 3.4732 | $-1.1366$ | $7.1086 \times 10^{10}$ | $2.1800 \times 10^3$ | 0.26700050 | 0.0000 | 1.0000 | 0.7149 | $4.0042 \times 10^2$ | $1.9206 \times 10^2$ |
| 0.00070 | 3.4767 | $-1.2331$ | $6.2569 \times 10^{10}$ | $2.8139 \times 10^3$ | 0.26700050 | 0.0000 | 1.0000 | 0.7149 | $3.9681 \times 10^2$ | $1.9071 \times 10^2$ |
| 0.00100 | 3.4800 | $-1.3363$ | $5.4745 \times 10^{10}$ | $3.6756 \times 10^3$ | 0.26700050 | 0.0000 | 1.0000 | 0.7149 | $3.9367 \times 10^2$ | $1.8956 \times 10^2$ |
| 0.00200 | 3.4847 | $-1.5383$ | $4.2439 \times 10^{10}$ | $6.1162 \times 10^3$ | 0.26700050 | 0.0000 | 1.0000 | 0.7149 | $3.8891 \times 10^2$ | $1.8809 \times 10^2$ |
| 0.00300 | 3.4867 | $-1.6577$ | $3.6662 \times 10^{10}$ | $8.1960 \times 10^3$ | 0.26700050 | 0.0000 | 1.0000 | 0.7149 | $3.8687 \times 10^2$ | $1.8714 \times 10^2$ |
| 0.00500 | 3.4884 | $-1.8087$ | $3.0563 \times 10^{10}$ | $1.1793 \times 10^4$ | 0.26700050 | 0.0000 | 1.0000 | 0.7149 | $3.8489 \times 10^2$ | $1.8655 \times 10^2$ |
| 0.00700 | 3.4891 | $-1.9086$ | $2.7158 \times 10^{10}$ | $1.4936 \times 10^4$ | 0.26700050 | 0.0000 | 1.0000 | 0.7149 | $3.8436 \times 10^2$ | $1.8647 \times 10^2$ |
| 0.01000 | 3.4894 | $-2.0155$ | $2.3982 \times 10^{10}$ | $1.9154 \times 10^4$ | 0.26700050 | 0.0000 | 1.0000 | 0.7149 | $3.8502 \times 10^2$ | $1.8686 \times 10^2$ |
| 0.02000 | 3.4884 | $-2.2262$ | $1.8899 \times 10^{10}$ | $3.0843 \times 10^4$ | 0.26700050 | 0.0000 | 1.0000 | 0.7149 | $3.8662 \times 10^2$ | $1.8884 \times 10^2$ |
| 0.03000 | 3.4867 | $-2.3521$ | $1.6479 \times 10^{10}$ | $4.0565 \times 10^4$ | 0.26700050 | 0.0000 | 1.0000 | 0.7149 | $3.8792 \times 10^2$ | $1.8957 \times 10^2$ |
| 0.05000 | 3.4833 | $-2.5118$ | $1.3927 \times 10^{10}$ | $5.6796 \times 10^4$ | 0.26700050 | 0.0000 | 1.0000 | 0.7149 | $3.9275 \times 10^2$ | $1.9165 \times 10^2$ |
| 0.07000 | 3.4807 | $-2.6166$ | $1.2494 \times 10^{10}$ | $7.0576 \times 10^4$ | 0.26700050 | 0.0000 | 1.0000 | 0.7149 | $3.9663 \times 10^2$ | $1.9377 \times 10^2$ |
| 0.10000 | 3.4777 | $-2.7274$ | $1.1151 \times 10^{10}$ | $8.8585 \times 10^4$ | 0.26700050 | 0.0014 | 0.9986 | 0.7149 | $4.0047 \times 10^2$ | $1.9638 \times 10^2$ |
| 0.16486 | 3.4659 | $-2.9152$ | $9.4836 \times 10^9$ | $1.2248 \times 10^5$ | 0.26700050 | 0.0288 | 0.9712 | 0.7149 | $4.0327 \times 10^2$ | $2.0260 \times 10^2$ |
| | | | | $0.085 M_\odot$ | | | | | | |
| 0.00002 | 3.4128 | $-0.2636$ | $2.5643 \times 10^{11}$ | $1.7156 \times 10^2$ | 0.26700050 | 0.0000 | 1.0000 | 0.7149 | $4.7285 \times 10^2$ | $2.1494 \times 10^2$ |
| 0.00010 | 3.4482 | $-0.6799$ | $1.3492 \times 10^{11}$ | $6.1978 \times 10^2$ | 0.26700050 | 0.0000 | 1.0000 | 0.7149 | $4.3039 \times 10^2$ | $2.0242 \times 10^2$ |
| 0.00020 | 3.4612 | $-0.8652$ | $1.0267 \times 10^{11}$ | $1.0703 \times 10^3$ | 0.26700050 | 0.0000 | 1.0000 | 0.7149 | $4.1592 \times 10^2$ | $1.9761 \times 10^2$ |
| 0.00030 | 3.4675 | $-0.9775$ | $8.7647 \times 10^{10}$ | $1.4686 \times 10^3$ | 0.26700050 | 0.0000 | 1.0000 | 0.7149 | $4.0925 \times 10^2$ | $1.9533 \times 10^2$ |
| 0.00050 | 3.4740 | $-1.1225$ | $7.1991 \times 10^{10}$ | $2.1768 \times 10^3$ | 0.26700050 | 0.0000 | 1.0000 | 0.7149 | $4.0249 \times 10^2$ | $1.9304 \times 10^2$ |
| 0.00070 | 3.4775 | $-1.2192$ | $6.3359 \times 10^{10}$ | $2.8102 \times 10^3$ | 0.26700050 | 0.0000 | 1.0000 | 0.7149 | $3.9887 \times 10^2$ | $1.9175 \times 10^2$ |
| 0.00100 | 3.4807 | $-1.3219$ | $5.5458 \times 10^{10}$ | $3.6680 \times 10^3$ | 0.26700050 | 0.0000 | 1.0000 | 0.7149 | $3.9569 \times 10^2$ | $1.9051 \times 10^2$ |
| 0.00200 | 3.4855 | $-1.5239$ | $4.2999 \times 10^{10}$ | $6.1017 \times 10^3$ | 0.26700050 | 0.0000 | 1.0000 | 0.7149 | $3.9102 \times 10^2$ | $1.8907 \times 10^2$ |
| 0.00300 | 3.4875 | $-1.6429$ | $3.7145 \times 10^{10}$ | $8.1762 \times 10^3$ | 0.26700050 | 0.0000 | 1.0000 | 0.7149 | $3.8880 \times 10^2$ | $1.8801 \times 10^2$ |
| 0.00500 | 3.4893 | $-1.7936$ | $3.0970 \times 10^{10}$ | $1.1762 \times 10^4$ | 0.26700050 | 0.0000 | 1.0000 | 0.7149 | $3.8680 \times 10^2$ | $1.8746 \times 10^2$ |
| 0.00700 | 3.4900 | $-1.8939$ | $2.7513 \times 10^{10}$ | $1.4904 \times 10^4$ | 0.26700050 | 0.0000 | 1.0000 | 0.7149 | $3.8631 \times 10^2$ | $1.8738 \times 10^2$ |
| 0.01000 | 3.4903 | $-2.0006$ | $2.4297 \times 10^{10}$ | $1.9110 \times 10^4$ | 0.26700050 | 0.0000 | 1.0000 | 0.7149 | $3.8692 \times 10^2$ | $1.8774 \times 10^2$ |
| 0.02000 | 3.4895 | $-2.2108$ | $1.9143 \times 10^{10}$ | $3.0787 \times 10^4$ | 0.26700050 | 0.0000 | 1.0000 | 0.7149 | $3.8840 \times 10^2$ | $1.8965 \times 10^2$ |
| 0.03000 | 3.4879 | $-2.3366$ | $1.6687 \times 10^{10}$ | $4.0515 \times 10^4$ | 0.26700050 | 0.0000 | 1.0000 | 0.7149 | $3.8961 \times 10^2$ | $1.9031 \times 10^2$ |
| 0.05000 | 3.4845 | $-2.4961$ | $1.4100 \times 10^{10}$ | $5.6741 \times 10^4$ | 0.26700050 | 0.0000 | 1.0000 | 0.7149 | $3.9440 \times 10^2$ | $1.9237 \times 10^2$ |
| 0.07000 | 3.4820 | $-2.6008$ | $1.2648 \times 10^{10}$ | $7.0526 \times 10^4$ | 0.26700050 | 0.0000 | 1.0000 | 0.7149 | $3.9816 \times 10^2$ | $1.9445 \times 10^2$ |
| 0.10000 | 3.4792 | $-2.7107$ | $1.1289 \times 10^{10}$ | $8.8528 \times 10^4$ | 0.26700050 | 0.0027 | 0.9973 | 0.7149 | $4.0102 \times 10^2$ | $1.9687 \times 10^2$ |
| 0.15514 | 3.4697 | $-2.8683$ | $9.8386 \times 10^9$ | $1.1655 \times 10^5$ | 0.26700050 | 0.0285 | 0.9715 | 0.7149 | $4.0278 \times 10^2$ | $2.0091 \times 10^2$ |
| | | | | $0.087 M_\odot$ | | | | | | |
| 0.00002 | 3.4149 | $-0.2509$ | $2.5778 \times 10^{11}$ | $1.7377 \times 10^2$ | 0.26700050 | 0.0000 | 1.0000 | 0.7149 | $4.7343 \times 10^2$ | $2.1535 \times 10^2$ |
| 0.00010 | 3.4493 | $-0.6665$ | $1.3635 \times 10^{11}$ | $6.2113 \times 10^2$ | 0.26700050 | 0.0000 | 1.0000 | 0.7149 | $4.3216 \times 10^2$ | $2.0329 \times 10^2$ |
| 0.00020 | 3.4621 | $-0.8516$ | $1.0385 \times 10^{11}$ | $1.0706 \times 10^3$ | 0.26700050 | 0.0000 | 1.0000 | 0.7149 | $4.1784 \times 10^2$ | $1.9851 \times 10^2$ |
| 0.00030 | 3.4683 | $-0.9637$ | $8.8696 \times 10^{10}$ | $1.4678 \times 10^3$ | 0.26700050 | 0.0000 | 1.0000 | 0.7149 | $4.1117 \times 10^2$ | $1.9626 \times 10^2$ |
| 0.00050 | 3.4748 | $-1.1084$ | $7.2885 \times 10^{10}$ | $2.1736 \times 10^3$ | 0.26700050 | 0.0000 | 1.0000 | 0.7149 | $4.0451 \times 10^2$ | $1.9388 \times 10^2$ |
| 0.00070 | 3.4782 | $-1.2053$ | $6.4170 \times 10^{10}$ | $2.8042 \times 10^3$ | 0.26700050 | 0.0000 | 1.0000 | 0.7149 | $4.0104 \times 10^2$ | $1.9269 \times 10^2$ |
| 0.00100 | 3.4815 | $-1.3078$ | $5.6170 \times 10^{10}$ | $3.6598 \times 10^3$ | 0.26700050 | 0.0000 | 1.0000 | 0.7149 | $3.9769 \times 10^2$ | $1.9149 \times 10^2$ |
| 0.00200 | 3.4862 | $-1.5099$ | $4.3552 \times 10^{10}$ | $6.0876 \times 10^3$ | 0.26700050 | 0.0000 | 1.0000 | 0.7149 | $3.9322 \times 10^2$ | $1.9001 \times 10^2$ |
| 0.00300 | 3.4884 | $-1.6284$ | $3.7625 \times 10^{10}$ | $8.1565 \times 10^3$ | 0.26700050 | 0.0000 | 1.0000 | 0.7149 | $3.9067 \times 10^2$ | $1.8893 \times 10^2$ |
| 0.00500 | 3.4902 | $-1.7794$ | $3.1362 \times 10^{10}$ | $1.1740 \times 10^4$ | 0.26700050 | 0.0000 | 1.0000 | 0.7149 | $3.8873 \times 10^2$ | $1.8844 \times 10^2$ |
| 0.00700 | 3.4908 | $-1.8794$ | $2.7868 \times 10^{10}$ | $1.4868 \times 10^4$ | 0.26700050 | 0.0000 | 1.0000 | 0.7149 | $3.8871 \times 10^2$ | $1.8828 \times 10^2$ |
| 0.01000 | 3.4912 | $-1.9858$ | $2.4613 \times 10^{10}$ | $1.9060 \times 10^4$ | 0.26700050 | 0.0000 | 1.0000 | 0.7149 | $3.8877 \times 10^2$ | $1.8860 \times 10^2$ |
| 0.02000 | 3.4905 | $-2.1956$ | $1.9390 \times 10^{10}$ | $3.0711 \times 10^4$ | 0.26700050 | 0.0000 | 1.0000 | 0.7149 | $3.9000 \times 10^2$ | $1.9038 \times 10^2$ |
| 0.03000 | 3.4890 | $-2.3212$ | $1.6898 \times 10^{10}$ | $4.0441 \times 10^4$ | 0.26700050 | 0.0000 | 1.0000 | 0.7149 | $3.9124 \times 10^2$ | $1.9104 \times 10^2$ |
| 0.05000 | 3.4858 | $-2.4805$ | $1.4276 \times 10^{10}$ | $5.6659 \times 10^4$ | 0.26700050 | 0.0000 | 1.0000 | 0.7149 | $3.9590 \times 10^2$ | $1.9307 \times 10^2$ |
| 0.07000 | 3.4833 | $-2.5850$ | $1.2802 \times 10^{10}$ | $7.0453 \times 10^4$ | 0.26700050 | 0.0000 | 1.0000 | 0.7149 | $3.9955 \times 10^2$ | $1.9507 \times 10^2$ |
| 0.10000 | 3.4806 | $-2.6947$ | $1.1427 \times 10^{10}$ | $8.8434 \times 10^4$ | 0.26700050 | 0.0044 | 0.9956 | 0.7149 | $4.0168 \times 10^2$ | $1.9737 \times 10^2$ |
| 0.17558 | 3.4693 | $-2.8939$ | $9.5693 \times 10^9$ | $1.2610 \times 10^5$ | 0.26700050 | 0.0558 | 0.9442 | 0.7149 | $3.9957 \times 10^2$ | $2.0050 \times 10^2$ |
| | | | | $0.089 M_\odot$ | | | | | | |
| 0.00002 | 3.4166 | $-0.2390$ | $2.5921 \times 10^{11}$ | $1.7580 \times 10^2$ | 0.26700050 | 0.0000 | 1.0000 | 0.7149 | $4.7432 \times 10^2$ | $2.1590 \times 10^2$ |
| 0.00010 | 3.4503 | $-0.6533$ | $1.3776 \times 10^{11}$ | $6.2240 \times 10^2$ | 0.26700050 | 0.0000 | 1.0000 | 0.7149 | $4.3387 \times 10^2$ | $2.0410 \times 10^2$ |
| 0.00020 | 3.4630 | $-0.8383$ | $1.0503 \times 10^{11}$ | $1.0708 \times 10^3$ | 0.26700050 | 0.0000 | 1.0000 | 0.7149 | $4.1980 \times 10^2$ | $1.9945 \times 10^2$ |
| 0.00030 | 3.4692 | $-0.9502$ | $8.9747 \times 10^{10}$ | $1.4665 \times 10^3$ | 0.26700050 | 0.0000 | 1.0000 | 0.7149 | $4.1312 \times 10^2$ | $1.9717 \times 10^2$ |
| 0.00050 | 3.4755 | $-1.0948$ | $7.3775 \times 10^{10}$ | $2.1703 \times 10^3$ | 0.26700050 | 0.0000 | 1.0000 | 0.7149 | $4.0647 \times 10^2$ | $1.9481 \times 10^2$ |

Table 1. Continued

| Age(Gyr) | log$T_{\rm eff}$ | log$L/L_\odot$ | R(cm) | g(cm/s$^2$) | $Y_c$ | L(PP1) | $L_g$ | $X_{\rm env}$ | $\tau_{gc}$(day) | $\tau_{lc}$(day) |
|---|---|---|---|---|---|---|---|---|---|---|
| 0.00070 | 3.4790 | −1.1917 | 6.4952 × 10$^{10}$ | 2.7999 × 10$^3$ | 0.26700050 | 0.0000 | 1.0000 | 0.7149 | 4.0290 × 10$^2$ | 1.9365 × 10$^2$ |
| 0.00100 | 3.4822 | −1.2945 | 5.6855 × 10$^{10}$ | 3.6543 × 10$^3$ | 0.26700050 | 0.0000 | 1.0000 | 0.7149 | 3.9970 × 10$^2$ | 1.9248 × 10$^2$ |
| 0.00200 | 3.4869 | −1.4963 | 4.4093 × 10$^{10}$ | 6.0757 × 10$^3$ | 0.26700050 | 0.0000 | 1.0000 | 0.7149 | 3.9523 × 10$^2$ | 1.9096 × 10$^2$ |
| 0.00300 | 3.4891 | −1.6145 | 3.8096 × 10$^{10}$ | 8.1391 × 10$^3$ | 0.26700050 | 0.0000 | 1.0000 | 0.7149 | 3.9259 × 10$^2$ | 1.8988 × 10$^2$ |
| 0.00500 | 3.4909 | −1.7654 | 3.1758 × 10$^{10}$ | 1.1712 × 10$^4$ | 0.26700050 | 0.0000 | 1.0000 | 0.7149 | 3.9070 × 10$^2$ | 1.8937 × 10$^2$ |
| 0.00700 | 3.4917 | −1.8650 | 2.8218 × 10$^{10}$ | 1.4835 × 10$^4$ | 0.26700050 | 0.0000 | 1.0000 | 0.7149 | 3.9000 × 10$^2$ | 1.8912 × 10$^2$ |
| 0.01000 | 3.4920 | −1.9714 | 2.4925 × 10$^{10}$ | 1.9014 × 10$^4$ | 0.26700050 | 0.0000 | 1.0000 | 0.7149 | 3.9056 × 10$^2$ | 1.8945 × 10$^2$ |
| 0.02000 | 3.4915 | −2.1809 | 1.9634 × 10$^{10}$ | 3.0643 × 10$^4$ | 0.26700050 | 0.0000 | 1.0000 | 0.7149 | 3.9177 × 10$^2$ | 1.9113 × 10$^2$ |
| 0.03000 | 3.4900 | −2.3062 | 1.7110 × 10$^{10}$ | 4.0350 × 10$^4$ | 0.26700050 | 0.0000 | 1.0000 | 0.7149 | 3.9294 × 10$^2$ | 1.9181 × 10$^2$ |
| 0.05000 | 3.4870 | −2.4651 | 1.4449 × 10$^{10}$ | 5.6581 × 10$^4$ | 0.26700050 | 0.0000 | 1.0000 | 0.7149 | 3.9733 × 10$^2$ | 1.9375 × 10$^2$ |
| 0.07000 | 3.4846 | −2.5696 | 1.2953 × 10$^{10}$ | 7.0401 × 10$^4$ | 0.26700050 | 0.0000 | 1.0000 | 0.7149 | 4.0091 × 10$^2$ | 1.9569 × 10$^2$ |
| 0.10000 | 3.4817 | −2.6803 | 1.1557 × 10$^{10}$ | 8.8437 × 10$^4$ | 0.26700050 | 0.0061 | 0.9939 | 0.7149 | 4.0281 × 10$^2$ | 1.9803 × 10$^2$ |
| 0.19019 | 3.4700 | −2.9014 | 9.4563 × 10$^9$ | 1.3210 × 10$^5$ | 0.26700050 | 0.0847 | 0.9153 | 0.7149 | 3.9635 × 10$^2$ | 1.9909 × 10$^2$ |
| | | | | 0.090$M_\odot$ | | | | | | |
| 0.00002 | 3.4174 | −0.2329 | 2.6011 × 10$^{11}$ | 1.7656 × 10$^2$ | 0.26700050 | 0.0000 | 1.0000 | 0.7149 | 4.7490 × 10$^2$ | 2.1622 × 10$^2$ |
| 0.00010 | 3.4508 | −0.6467 | 1.3848 × 10$^{11}$ | 6.2286 × 10$^2$ | 0.26700050 | 0.0000 | 1.0000 | 0.7149 | 4.3475 × 10$^2$ | 2.0453 × 10$^2$ |
| 0.00020 | 3.4634 | −0.8318 | 1.0562 × 10$^{11}$ | 1.0707 × 10$^3$ | 0.26700050 | 0.0000 | 1.0000 | 0.7149 | 4.2075 × 10$^2$ | 1.9992 × 10$^2$ |
| 0.00030 | 3.4695 | −0.9435 | 9.0278 × 10$^{10}$ | 1.4656 × 10$^3$ | 0.26700050 | 0.0000 | 1.0000 | 0.7149 | 4.1409 × 10$^2$ | 1.9762 × 10$^2$ |
| 0.00050 | 3.4759 | −1.0880 | 7.4224 × 10$^{10}$ | 2.1682 × 10$^3$ | 0.26700050 | 0.0000 | 1.0000 | 0.7149 | 4.0745 × 10$^2$ | 1.9528 × 10$^2$ |
| 0.00070 | 3.4794 | −1.1848 | 6.5351 × 10$^{10}$ | 2.7969 × 10$^3$ | 0.26700050 | 0.0000 | 1.0000 | 0.7149 | 4.0388 × 10$^2$ | 1.9405 × 10$^2$ |
| 0.00100 | 3.4825 | −1.2877 | 5.7210 × 10$^{10}$ | 3.6496 × 10$^3$ | 0.26700050 | 0.0000 | 1.0000 | 0.7149 | 4.0080 × 10$^2$ | 1.9289 × 10$^2$ |
| 0.00200 | 3.4873 | −1.4892 | 4.4370 × 10$^{10}$ | 6.0676 × 10$^3$ | 0.26700050 | 0.0000 | 1.0000 | 0.7149 | 3.9615 × 10$^2$ | 1.9138 × 10$^2$ |
| 0.00300 | 3.4895 | −1.6076 | 3.8335 × 10$^{10}$ | 8.1284 × 10$^3$ | 0.26700050 | 0.0000 | 1.0000 | 0.7149 | 3.9357 × 10$^2$ | 1.9034 × 10$^2$ |
| 0.00500 | 3.4913 | −1.7584 | 3.1960 × 10$^{10}$ | 1.1694 × 10$^4$ | 0.26700050 | 0.0000 | 1.0000 | 0.7149 | 3.9169 × 10$^2$ | 1.8983 × 10$^2$ |
| 0.00700 | 3.4921 | −1.8580 | 2.8394 × 10$^{10}$ | 1.4816 × 10$^4$ | 0.26700050 | 0.0000 | 1.0000 | 0.7149 | 3.9096 × 10$^2$ | 1.8953 × 10$^2$ |
| 0.01000 | 3.4924 | −1.9643 | 2.5082 × 10$^{10}$ | 1.8988 × 10$^4$ | 0.26700050 | 0.0000 | 1.0000 | 0.7149 | 3.9143 × 10$^2$ | 1.8986 × 10$^2$ |
| 0.02000 | 3.4919 | −2.1737 | 1.9756 × 10$^{10}$ | 3.0605 × 10$^4$ | 0.26700050 | 0.0000 | 1.0000 | 0.7149 | 3.9263 × 10$^2$ | 1.9150 × 10$^2$ |
| 0.03000 | 3.4906 | −2.2988 | 1.7215 × 10$^{10}$ | 4.0306 × 10$^4$ | 0.26700050 | 0.0000 | 1.0000 | 0.7149 | 3.9377 × 10$^2$ | 1.9219 × 10$^2$ |
| 0.05000 | 3.4876 | −2.4580 | 1.4530 × 10$^{10}$ | 5.6576 × 10$^4$ | 0.26700050 | 0.0000 | 1.0000 | 0.7149 | 3.9817 × 10$^2$ | 1.9411 × 10$^2$ |
| 0.07000 | 3.4852 | −2.5623 | 1.3028 × 10$^{10}$ | 7.0377 × 10$^4$ | 0.26700050 | 0.0002 | 0.9998 | 0.7149 | 4.0137 × 10$^2$ | 1.9603 × 10$^2$ |
| 0.10000 | 3.4822 | −2.6735 | 1.1623 × 10$^{10}$ | 8.8420 × 10$^4$ | 0.26700050 | 0.0071 | 0.9929 | 0.7149 | 4.0336 × 10$^2$ | 1.9836 × 10$^2$ |
| 0.20339 | 3.4699 | −2.9136 | 9.3283 × 10$^9$ | 1.3727 × 10$^5$ | 0.26700050 | 0.1107 | 0.8893 | 0.7149 | 3.9348 × 10$^2$ | 1.9778 × 10$^2$ |
| | | | | 0.091$M_\odot$ | | | | | | |
| 0.00002 | 3.4185 | −0.2268 | 2.6065 × 10$^{11}$ | 1.7777 × 10$^2$ | 0.26700050 | 0.0000 | 1.0000 | 0.7149 | 4.7504 × 10$^2$ | 2.1636 × 10$^2$ |
| 0.00010 | 3.4513 | −0.6406 | 1.3916 × 10$^{11}$ | 6.2371 × 10$^2$ | 0.26700050 | 0.0000 | 1.0000 | 0.7149 | 4.3562 × 10$^2$ | 2.0495 × 10$^2$ |
| 0.00020 | 3.4638 | −0.8255 | 1.0619 × 10$^{11}$ | 1.0711 × 10$^3$ | 0.26700050 | 0.0000 | 1.0000 | 0.7149 | 4.2170 × 10$^2$ | 2.0036 × 10$^2$ |
| 0.00030 | 3.4699 | −0.9370 | 9.0791 × 10$^{10}$ | 1.4652 × 10$^3$ | 0.26700050 | 0.0000 | 1.0000 | 0.7149 | 4.1509 × 10$^2$ | 1.9809 × 10$^2$ |
| 0.00050 | 3.4763 | −1.0814 | 7.4665 × 10$^{10}$ | 2.1665 × 10$^3$ | 0.26700050 | 0.0000 | 1.0000 | 0.7149 | 4.0843 × 10$^2$ | 1.9577 × 10$^2$ |
| 0.00070 | 3.4798 | −1.1780 | 6.5744 × 10$^{10}$ | 2.7943 × 10$^3$ | 0.26700050 | 0.0000 | 1.0000 | 0.7149 | 4.0485 × 10$^2$ | 1.9447 × 10$^2$ |
| 0.00100 | 3.4829 | −1.2812 | 5.7558 × 10$^{10}$ | 3.6456 × 10$^3$ | 0.26700050 | 0.0000 | 1.0000 | 0.7149 | 4.0181 × 10$^2$ | 1.9339 × 10$^2$ |
| 0.00200 | 3.4878 | −1.4823 | 4.4643 × 10$^{10}$ | 6.0600 × 10$^3$ | 0.26700050 | 0.0000 | 1.0000 | 0.7149 | 3.9703 × 10$^2$ | 1.9180 × 10$^2$ |
| 0.00300 | 3.4899 | −1.6008 | 3.8571 × 10$^{10}$ | 8.1182 × 10$^3$ | 0.26700050 | 0.0000 | 1.0000 | 0.7149 | 3.9460 × 10$^2$ | 1.9077 × 10$^2$ |
| 0.00500 | 3.4917 | −1.7515 | 3.2157 × 10$^{10}$ | 1.1679 × 10$^4$ | 0.26700050 | 0.0000 | 1.0000 | 0.7149 | 3.9275 × 10$^2$ | 1.9024 × 10$^2$ |
| 0.00700 | 3.4925 | −1.8511 | 2.8568 × 10$^{10}$ | 1.4799 × 10$^4$ | 0.26700050 | 0.0000 | 1.0000 | 0.7149 | 3.9189 × 10$^2$ | 1.8996 × 10$^2$ |
| 0.01000 | 3.4929 | −1.9573 | 2.5235 × 10$^{10}$ | 1.8966 × 10$^4$ | 0.26700050 | 0.0000 | 1.0000 | 0.7149 | 3.9229 × 10$^2$ | 1.9027 × 10$^2$ |
| 0.02000 | 3.4924 | −2.1665 | 1.9878 × 10$^{10}$ | 3.0567 × 10$^4$ | 0.26700050 | 0.0000 | 1.0000 | 0.7149 | 3.9344 × 10$^2$ | 1.9186 × 10$^2$ |
| 0.03000 | 3.4911 | −2.2915 | 1.7320 × 10$^{10}$ | 4.0263 × 10$^4$ | 0.26700050 | 0.0000 | 1.0000 | 0.7149 | 3.9459 × 10$^2$ | 1.9256 × 10$^2$ |
| 0.05000 | 3.4881 | −2.4506 | 1.4617 × 10$^{10}$ | 5.6526 × 10$^4$ | 0.26700050 | 0.0000 | 1.0000 | 0.7149 | 3.9895 × 10$^2$ | 1.9447 × 10$^2$ |
| 0.07000 | 3.4858 | −2.5549 | 1.3105 × 10$^{10}$ | 7.0329 × 10$^4$ | 0.26700050 | 0.0004 | 0.9996 | 0.7149 | 4.0173 × 10$^2$ | 1.9636 × 10$^2$ |
| 0.10000 | 3.4827 | −2.6665 | 1.1690 × 10$^{10}$ | 8.8384 × 10$^4$ | 0.26700050 | 0.0080 | 0.9920 | 0.7149 | 4.0386 × 10$^2$ | 1.9858 × 10$^2$ |
| 0.20000 | 3.4710 | −2.8981 | 9.4499 × 10$^9$ | 1.3525 × 10$^5$ | 0.26700050 | 0.1117 | 0.8883 | 0.7149 | 3.9359 × 10$^2$ | 1.9739 × 10$^2$ |
| 0.22219 | 3.4694 | −2.9335 | 9.1378 × 10$^9$ | 1.4464 × 10$^5$ | 0.26700050 | 0.1506 | 0.8494 | 0.7149 | 3.8925 × 10$^2$ | 1.9591 × 10$^2$ |
| | | | | 0.092$M_\odot$ | | | | | | |
| 0.00002 | 3.4197 | −0.2201 | 2.6119 × 10$^{11}$ | 1.7898 × 10$^2$ | 0.26700050 | 0.0000 | 1.0000 | 0.7149 | 4.7495 × 10$^2$ | 2.1640 × 10$^2$ |
| 0.00010 | 3.4518 | −0.6345 | 1.3983 × 10$^{11}$ | 6.2454 × 10$^2$ | 0.26700050 | 0.0000 | 1.0000 | 0.7149 | 4.3651 × 10$^2$ | 2.0538 × 10$^2$ |
| 0.00020 | 3.4642 | −0.8192 | 1.0676 × 10$^{11}$ | 1.0712 × 10$^3$ | 0.26700050 | 0.0000 | 1.0000 | 0.7149 | 4.2262 × 10$^2$ | 2.0080 × 10$^2$ |
| 0.00030 | 3.4703 | −0.9307 | 9.1301 × 10$^{10}$ | 1.4648 × 10$^3$ | 0.26700050 | 0.0000 | 1.0000 | 0.7149 | 4.1603 × 10$^2$ | 1.9854 × 10$^2$ |
| 0.00050 | 3.4766 | −1.0749 | 7.5101 × 10$^{10}$ | 2.1649 × 10$^3$ | 0.26700050 | 0.0000 | 1.0000 | 0.7149 | 4.0939 × 10$^2$ | 1.9624 × 10$^2$ |
| 0.00070 | 3.4801 | −1.1713 | 6.6136 × 10$^{10}$ | 2.7916 × 10$^3$ | 0.26700050 | 0.0000 | 1.0000 | 0.7149 | 4.0578 × 10$^2$ | 1.9491 × 10$^2$ |
| 0.00100 | 3.4832 | −1.2748 | 5.7890 × 10$^{10}$ | 3.6436 × 10$^3$ | 0.26700050 | 0.0000 | 1.0000 | 0.7149 | 4.0272 × 10$^2$ | 1.9390 × 10$^2$ |
| 0.00200 | 3.4882 | −1.4755 | 4.4912 × 10$^{10}$ | 6.0535 × 10$^3$ | 0.26700050 | 0.0000 | 1.0000 | 0.7149 | 3.9790 × 10$^2$ | 1.9223 × 10$^2$ |
| 0.00300 | 3.4902 | −1.5942 | 3.8801 × 10$^{10}$ | 8.1106 × 10$^3$ | 0.26700050 | 0.0000 | 1.0000 | 0.7149 | 3.9550 × 10$^2$ | 1.9125 × 10$^2$ |
| 0.00500 | 3.4920 | −1.7449 | 3.2349 × 10$^{10}$ | 1.1668 × 10$^4$ | 0.26700050 | 0.0000 | 1.0000 | 0.7149 | 3.9366 × 10$^2$ | 1.9070 × 10$^2$ |
| 0.00700 | 3.4929 | −1.8442 | 2.8742 × 10$^{10}$ | 1.4781 × 10$^4$ | 0.26700050 | 0.0000 | 1.0000 | 0.7149 | 3.9281 × 10$^2$ | 1.9039 × 10$^2$ |
| 0.01000 | 3.4933 | −1.9505 | 2.5385 × 10$^{10}$ | 1.8949 × 10$^4$ | 0.26700050 | 0.0000 | 1.0000 | 0.7149 | 3.9321 × 10$^2$ | 1.9070 × 10$^2$ |

Table 1. Continued

| Age(Gyr) | $\log T_{\rm eff}$ | $\log L/L_\odot$ | R(cm) | g(cm/s$^2$) | $Y_c$ | L(PP1) | $L_g$ | $X_{\rm env}$ | $\tau_{gc}$ (day) | $\tau_{lc}$ (day) |
|---|---|---|---|---|---|---|---|---|---|---|
| 0.02000 | 3.4928 | $-2.1595$ | $1.9996 \times 10^{10}$ | $3.0537 \times 10^4$ | 0.26700050 | 0.0000 | 1.0000 | 0.7149 | $3.9432 \times 10^2$ | $1.9224 \times 10^2$ |
| 0.03000 | 3.4916 | $-2.2843$ | $1.7423 \times 10^{10}$ | $4.0225 \times 10^4$ | 0.26700050 | 0.0000 | 1.0000 | 0.7149 | $3.9537 \times 10^2$ | $1.9292 \times 10^2$ |
| 0.05000 | 3.4887 | $-2.4433$ | $1.4703 \times 10^{10}$ | $5.6483 \times 10^4$ | 0.26700050 | 0.0000 | 1.0000 | 0.7149 | $3.9969 \times 10^2$ | $1.9482 \times 10^2$ |
| 0.07000 | 3.4863 | $-2.5476$ | $1.3180 \times 10^{10}$ | $7.0290 \times 10^4$ | 0.26700050 | 0.0005 | 0.9995 | 0.7149 | $4.0220 \times 10^2$ | $1.9669 \times 10^2$ |
| 0.10000 | 3.4833 | $-2.6589$ | $1.1759 \times 10^{10}$ | $8.8314 \times 10^4$ | 0.26700050 | 0.0090 | 0.9910 | 0.7149 | $4.0420 \times 10^2$ | $1.9875 \times 10^2$ |
| 0.20000 | 3.4718 | $-2.8887$ | $9.5153 \times 10^9$ | $1.3486 \times 10^5$ | 0.26700050 | 0.1192 | 0.8808 | 0.7149 | $3.9308 \times 10^2$ | $1.9684 \times 10^2$ |
| 0.22162 | 3.4705 | $-2.9218$ | $9.2171 \times 10^9$ | $1.4373 \times 10^5$ | 0.26700050 | 0.1582 | 0.8418 | 0.7149 | $3.8839 \times 10^2$ | $1.9502 \times 10^2$ |
| | | | | $0.093 M_\odot$ | | | | | | |
| 0.00002 | 3.4208 | $-0.2138$ | $2.6172 \times 10^{11}$ | $1.8020 \times 10^2$ | 0.26700050 | 0.0000 | 1.0000 | 0.7149 | $4.7495 \times 10^2$ | $2.1647 \times 10^2$ |
| 0.00010 | 3.4523 | $-0.6285$ | $1.4049 \times 10^{11}$ | $6.2538 \times 10^2$ | 0.26700050 | 0.0000 | 1.0000 | 0.7149 | $4.3739 \times 10^2$ | $2.0582 \times 10^2$ |
| 0.00020 | 3.4647 | $-0.8127$ | $1.0733 \times 10^{11}$ | $1.0714 \times 10^3$ | 0.26700050 | 0.0000 | 1.0000 | 0.7149 | $4.2347 \times 10^2$ | $2.0121 \times 10^2$ |
| 0.00030 | 3.4707 | $-0.9245$ | $9.1797 \times 10^{10}$ | $1.4648 \times 10^3$ | 0.26700050 | 0.0000 | 1.0000 | 0.7149 | $4.1701 \times 10^2$ | $1.9901 \times 10^2$ |
| 0.00050 | 3.4770 | $-1.0685$ | $7.5532 \times 10^{10}$ | $2.1635 \times 10^3$ | 0.26700050 | 0.0000 | 1.0000 | 0.7149 | $4.1035 \times 10^2$ | $1.9671 \times 10^2$ |
| 0.00070 | 3.4805 | $-1.1648$ | $6.6523 \times 10^{10}$ | $2.7893 \times 10^3$ | 0.26700050 | 0.0000 | 1.0000 | 0.7149 | $4.0671 \times 10^2$ | $1.9537 \times 10^2$ |
| 0.00100 | 3.4836 | $-1.2680$ | $5.8234 \times 10^{10}$ | $3.6398 \times 10^3$ | 0.26700050 | 0.0000 | 1.0000 | 0.7149 | $4.0361 \times 10^2$ | $1.9431 \times 10^2$ |
| 0.00200 | 3.4886 | $-1.4688$ | $4.5174 \times 10^{10}$ | $6.0486 \times 10^3$ | 0.26700050 | 0.0000 | 1.0000 | 0.7149 | $3.9880 \times 10^2$ | $1.9266 \times 10^2$ |
| 0.00300 | 3.4906 | $-1.5879$ | $3.9020 \times 10^{10}$ | $8.1068 \times 10^3$ | 0.26700050 | 0.0000 | 1.0000 | 0.7149 | $3.9646 \times 10^2$ | $1.9171 \times 10^2$ |
| 0.00500 | 3.4924 | $-1.7384$ | $3.2533 \times 10^{10}$ | $1.1662 \times 10^4$ | 0.26700050 | 0.0000 | 1.0000 | 0.7149 | $3.9460 \times 10^2$ | $1.9110 \times 10^2$ |
| 0.00700 | 3.4933 | $-1.8376$ | $2.8911 \times 10^{10}$ | $1.4768 \times 10^4$ | 0.26700050 | 0.0000 | 1.0000 | 0.7149 | $3.9371 \times 10^2$ | $1.9082 \times 10^2$ |
| 0.01000 | 3.4937 | $-1.9439$ | $2.5532 \times 10^{10}$ | $1.8934 \times 10^4$ | 0.26700050 | 0.0000 | 1.0000 | 0.7149 | $3.9409 \times 10^2$ | $1.9112 \times 10^2$ |
| 0.02000 | 3.4933 | $-2.1526$ | $2.0115 \times 10^{10}$ | $3.0505 \times 10^4$ | 0.26700050 | 0.0000 | 1.0000 | 0.7149 | $3.9513 \times 10^2$ | $1.9260 \times 10^2$ |
| 0.03000 | 3.4921 | $-2.2772$ | $1.7526 \times 10^{10}$ | $4.0187 \times 10^4$ | 0.26700050 | 0.0000 | 1.0000 | 0.7149 | $3.9618 \times 10^2$ | $1.9329 \times 10^2$ |
| 0.05000 | 3.4892 | $-2.4362$ | $1.4788 \times 10^{10}$ | $5.6440 \times 10^4$ | 0.26700050 | 0.0000 | 1.0000 | 0.7149 | $4.0046 \times 10^2$ | $1.9517 \times 10^2$ |
| 0.07000 | 3.4869 | $-2.5405$ | $1.3256 \times 10^{10}$ | $7.0248 \times 10^4$ | 0.26700050 | 0.0008 | 0.9992 | 0.7149 | $4.0279 \times 10^2$ | $1.9703 \times 10^2$ |
| 0.10000 | 3.4840 | $-2.6511$ | $1.1826 \times 10^{10}$ | $8.8251 \times 10^4$ | 0.26700050 | 0.0099 | 0.9901 | 0.7149 | $4.0450 \times 10^2$ | $1.9889 \times 10^2$ |
| 0.20519 | 3.4722 | $-2.8879$ | $9.5091 \times 10^9$ | $1.3651 \times 10^5$ | 0.26700050 | 0.1360 | 0.8640 | 0.7149 | $3.9151 \times 10^2$ | $1.9580 \times 10^2$ |
| | | | | $0.094 M_\odot$ | | | | | | |
| 0.00002 | 3.4219 | $-0.2078$ | $2.6226 \times 10^{11}$ | $1.8139 \times 10^2$ | 0.26700050 | 0.0000 | 1.0000 | 0.7149 | $4.7506 \times 10^2$ | $2.1659 \times 10^2$ |
| 0.00010 | 3.4528 | $-0.6223$ | $1.4116 \times 10^{11}$ | $6.2609 \times 10^2$ | 0.26700050 | 0.0000 | 1.0000 | 0.7149 | $4.3817 \times 10^2$ | $2.0620 \times 10^2$ |
| 0.00020 | 3.4652 | $-0.8062$ | $1.0789 \times 10^{11}$ | $1.0718 \times 10^3$ | 0.26700050 | 0.0000 | 1.0000 | 0.7149 | $4.2425 \times 10^2$ | $2.0158 \times 10^2$ |
| 0.00030 | 3.4711 | $-0.9183$ | $9.2283 \times 10^{10}$ | $1.4650 \times 10^3$ | 0.26700050 | 0.0000 | 1.0000 | 0.7149 | $4.1793 \times 10^2$ | $1.9944 \times 10^2$ |
| 0.00050 | 3.4774 | $-1.0623$ | $7.5952 \times 10^{10}$ | $2.1627 \times 10^3$ | 0.26700050 | 0.0000 | 1.0000 | 0.7149 | $4.1133 \times 10^2$ | $1.9714 \times 10^2$ |
| 0.00070 | 3.4809 | $-1.1583$ | $6.6905 \times 10^{10}$ | $2.7871 \times 10^3$ | 0.26700050 | 0.0000 | 1.0000 | 0.7149 | $4.0763 \times 10^2$ | $1.9581 \times 10^2$ |
| 0.00100 | 3.4840 | $-1.2614$ | $5.8572 \times 10^{10}$ | $3.6365 \times 10^3$ | 0.26700050 | 0.0000 | 1.0000 | 0.7149 | $4.0450 \times 10^2$ | $1.9473 \times 10^2$ |
| 0.00200 | 3.4889 | $-1.4623$ | $4.5433 \times 10^{10}$ | $6.0440 \times 10^3$ | 0.26700050 | 0.0000 | 1.0000 | 0.7149 | $3.9969 \times 10^2$ | $1.9313 \times 10^2$ |
| 0.00300 | 3.4909 | $-1.5812$ | $3.9258 \times 10^{10}$ | $8.0949 \times 10^3$ | 0.26700050 | 0.0000 | 1.0000 | 0.7149 | $3.9754 \times 10^2$ | $1.9214 \times 10^2$ |
| 0.00500 | 3.4929 | $-1.7314$ | $3.2733 \times 10^{10}$ | $1.1644 \times 10^4$ | 0.26700050 | 0.0000 | 1.0000 | 0.7149 | $3.9545 \times 10^2$ | $1.9150 \times 10^2$ |
| 0.00700 | 3.4937 | $-1.8308$ | $2.9085 \times 10^{10}$ | $1.4748 \times 10^4$ | 0.26700050 | 0.0000 | 1.0000 | 0.7149 | $3.9462 \times 10^2$ | $1.9125 \times 10^2$ |
| 0.01000 | 3.4941 | $-1.9372$ | $2.5683 \times 10^{10}$ | $1.8914 \times 10^4$ | 0.26700050 | 0.0000 | 1.0000 | 0.7149 | $3.9498 \times 10^2$ | $1.9155 \times 10^2$ |
| 0.02000 | 3.4937 | $-2.1458$ | $2.0234 \times 10^{10}$ | $3.0471 \times 10^4$ | 0.26700050 | 0.0000 | 1.0000 | 0.7149 | $3.9596 \times 10^2$ | $1.9297 \times 10^2$ |
| 0.03000 | 3.4925 | $-2.2702$ | $1.7628 \times 10^{10}$ | $4.0151 \times 10^4$ | 0.26700050 | 0.0000 | 1.0000 | 0.7149 | $3.9701 \times 10^2$ | $1.9366 \times 10^2$ |
| 0.05000 | 3.4897 | $-2.4291$ | $1.4872 \times 10^{10}$ | $5.6404 \times 10^4$ | 0.26700050 | 0.0000 | 1.0000 | 0.7149 | $4.0122 \times 10^2$ | $1.9551 \times 10^2$ |
| 0.07000 | 3.4874 | $-2.5334$ | $1.3330 \times 10^{10}$ | $7.0211 \times 10^4$ | 0.26700050 | 0.0010 | 0.9990 | 0.7149 | $4.0344 \times 10^2$ | $1.9737 \times 10^2$ |
| 0.10000 | 3.4847 | $-2.6436$ | $1.1893 \times 10^{10}$ | $8.8203 \times 10^4$ | 0.26700050 | 0.0108 | 0.9892 | 0.7149 | $4.0482 \times 10^2$ | $1.9904 \times 10^2$ |
| 0.20000 | 3.4735 | $-2.8706$ | $9.6408 \times 10^9$ | $1.3423 \times 10^5$ | 0.26700050 | 0.1315 | 0.8685 | 0.7149 | $3.9200 \times 10^2$ | $1.9572 \times 10^2$ |
| 0.20738 | 3.4727 | $-2.8824$ | $9.5452 \times 10^9$ | $1.3693 \times 10^5$ | 0.26700050 | 0.1477 | 0.8523 | 0.7149 | $3.9075 \times 10^2$ | $1.9516 \times 10^2$ |
| | | | | $0.095 M_\odot$ | | | | | | |
| 0.00002 | 3.4229 | $-0.2020$ | $2.6279 \times 10^{11}$ | $1.8258 \times 10^2$ | 0.26700050 | 0.0000 | 1.0000 | 0.7149 | $4.7524 \times 10^2$ | $2.1675 \times 10^2$ |
| 0.00010 | 3.4534 | $-0.6159$ | $1.4181 \times 10^{11}$ | $6.2694 \times 10^2$ | 0.26700050 | 0.0000 | 1.0000 | 0.7149 | $4.3887 \times 10^2$ | $2.0654 \times 10^2$ |
| 0.00020 | 3.4656 | $-0.7999$ | $1.0843 \times 10^{11}$ | $1.0724 \times 10^3$ | 0.26700050 | 0.0000 | 1.0000 | 0.7149 | $4.2503 \times 10^2$ | $2.0197 \times 10^2$ |
| 0.00030 | 3.4715 | $-0.9120$ | $9.2769 \times 10^{10}$ | $1.4651 \times 10^3$ | 0.26700050 | 0.0000 | 1.0000 | 0.7149 | $4.1879 \times 10^2$ | $1.9987 \times 10^2$ |
| 0.00050 | 3.4777 | $-1.0561$ | $7.6375 \times 10^{10}$ | $2.1616 \times 10^3$ | 0.26700050 | 0.0000 | 1.0000 | 0.7149 | $4.1226 \times 10^2$ | $1.9760 \times 10^2$ |
| 0.00070 | 3.4813 | $-1.1520$ | $6.7283 \times 10^{10}$ | $2.7852 \times 10^3$ | 0.26700050 | 0.0000 | 1.0000 | 0.7149 | $4.0855 \times 10^2$ | $1.9626 \times 10^2$ |
| 0.00100 | 3.4844 | $-1.2549$ | $5.8909 \times 10^{10}$ | $3.6333 \times 10^3$ | 0.26700050 | 0.0000 | 1.0000 | 0.7149 | $4.0543 \times 10^2$ | $1.9508 \times 10^2$ |
| 0.00200 | 3.4893 | $-1.4558$ | $4.5703 \times 10^{10}$ | $6.0365 \times 10^3$ | 0.26700050 | 0.0000 | 1.0000 | 0.7149 | $4.0061 \times 10^2$ | $1.9356 \times 10^2$ |
| 0.00300 | 3.4913 | $-1.5748$ | $3.9488 \times 10^{10}$ | $8.0862 \times 10^3$ | 0.26700050 | 0.0000 | 1.0000 | 0.7149 | $3.9847 \times 10^2$ | $1.9260 \times 10^2$ |
| 0.00500 | 3.4933 | $-1.7247$ | $3.2927 \times 10^{10}$ | $1.1630 \times 10^4$ | 0.26700050 | 0.0000 | 1.0000 | 0.7149 | $3.9627 \times 10^2$ | $1.9190 \times 10^2$ |
| 0.00700 | 3.4941 | $-1.8242$ | $2.9256 \times 10^{10}$ | $1.4732 \times 10^4$ | 0.26700050 | 0.0000 | 1.0000 | 0.7149 | $3.9552 \times 10^2$ | $1.9167 \times 10^2$ |
| 0.01000 | 3.4945 | $-1.9307$ | $2.5830 \times 10^{10}$ | $1.8898 \times 10^4$ | 0.26700050 | 0.0000 | 1.0000 | 0.7149 | $3.9591 \times 10^2$ | $1.9194 \times 10^2$ |
| 0.02000 | 3.4942 | $-2.1391$ | $2.0350 \times 10^{10}$ | $3.0446 \times 10^4$ | 0.26700050 | 0.0000 | 1.0000 | 0.7149 | $3.9680 \times 10^2$ | $1.9334 \times 10^2$ |
| 0.03000 | 3.4930 | $-2.2634$ | $1.7727 \times 10^{10}$ | $4.0124 \times 10^4$ | 0.26700050 | 0.0000 | 1.0000 | 0.7149 | $3.9786 \times 10^2$ | $1.9402 \times 10^2$ |
| 0.05000 | 3.4903 | $-2.4221$ | $1.4955 \times 10^{10}$ | $5.6376 \times 10^4$ | 0.26700050 | 0.0000 | 1.0000 | 0.7149 | $4.0199 \times 10^2$ | $1.9583 \times 10^2$ |
| 0.07000 | 3.4880 | $-2.5264$ | $1.3404 \times 10^{10}$ | $7.0178 \times 10^4$ | 0.26700050 | 0.0013 | 0.9987 | 0.7149 | $4.0409 \times 10^2$ | $1.9770 \times 10^2$ |
| 0.10000 | 3.4853 | $-2.6362$ | $1.1960 \times 10^{10}$ | $8.8153 \times 10^4$ | 0.26700050 | 0.0119 | 0.9881 | 0.7149 | $4.0516 \times 10^2$ | $1.9919 \times 10^2$ |
| 0.20000 | 3.4749 | $-2.8598$ | $9.6981 \times 10^9$ | $1.3406 \times 10^5$ | 0.26700050 | 0.1389 | 0.8611 | 0.7149 | $3.9071 \times 10^2$ | $1.9491 \times 10^2$ |

Table 1. Continued

| Age(Gyr) | logT$_{\rm eff}$ | logL/L$_\odot$ | R(cm) | g(cm/s$^2$) | Y$_c$ | L(PP1) | L$_g$ | X$_{\rm env}$ | $\tau_{gc}$ (day) | $\tau_{lc}$ (day) |
|---|---|---|---|---|---|---|---|---|---|---|
| 0.24607 | 3.4718 | −2.9233 | 9.1436 × 10$^9$ | 1.5081 × 10$^5$ | 0.26700050 | 0.2386 | 0.7614 | 0.7149 | 3.8028 × 10$^2$ | 1.8974 × 10$^2$ |

$0.096 M_\odot$

| Age(Gyr) | logT$_{\rm eff}$ | logL/L$_\odot$ | R(cm) | g(cm/s$^2$) | Y$_c$ | L(PP1) | L$_g$ | X$_{\rm env}$ | $\tau_{gc}$ (day) | $\tau_{lc}$ (day) |
|---|---|---|---|---|---|---|---|---|---|---|
| 0.00002 | 3.4238 | −0.1964 | 2.6335 × 10$^{11}$ | 1.8372 × 10$^2$ | 0.26700050 | 0.0000 | 1.0000 | 0.7149 | 4.7546 × 10$^2$ | 2.1693 × 10$^2$ |
| 0.00010 | 3.4539 | −0.6098 | 1.4245 × 10$^{11}$ | 6.2786 × 10$^2$ | 0.26700050 | 0.0000 | 1.0000 | 0.7149 | 4.3958 × 10$^2$ | 2.0689 × 10$^2$ |
| 0.00020 | 3.4661 | −0.7937 | 1.0897 × 10$^{11}$ | 1.0729 × 10$^3$ | 0.26700050 | 0.0000 | 1.0000 | 0.7149 | 4.2585 × 10$^2$ | 2.0235 × 10$^2$ |
| 0.00030 | 3.4719 | −0.9057 | 9.3256 × 10$^{10}$ | 1.4651 × 10$^3$ | 0.26700050 | 0.0000 | 1.0000 | 0.7149 | 4.1961 × 10$^2$ | 2.0028 × 10$^2$ |
| 0.00050 | 3.4780 | −1.0501 | 7.6788 × 10$^{10}$ | 2.1609 × 10$^3$ | 0.26700050 | 0.0000 | 1.0000 | 0.7149 | 4.1323 × 10$^2$ | 1.9804 × 10$^2$ |
| 0.00070 | 3.4816 | −1.1458 | 6.7655 × 10$^{10}$ | 2.7837 × 10$^3$ | 0.26700050 | 0.0000 | 1.0000 | 0.7149 | 4.0947 × 10$^2$ | 1.9673 × 10$^2$ |
| 0.00100 | 3.4848 | −1.2485 | 5.9243 × 10$^{10}$ | 3.6303 × 10$^3$ | 0.26700050 | 0.0000 | 1.0000 | 0.7149 | 4.0630 × 10$^2$ | 1.9550 × 10$^2$ |
| 0.00200 | 3.4896 | −1.4494 | 4.5969 × 10$^{10}$ | 6.0297 × 10$^3$ | 0.26700050 | 0.0000 | 1.0000 | 0.7149 | 4.0153 × 10$^2$ | 1.9400 × 10$^2$ |
| 0.00300 | 3.4916 | −1.5683 | 3.9715 × 10$^{10}$ | 8.0782 × 10$^3$ | 0.26700050 | 0.0000 | 1.0000 | 0.7149 | 3.9937 × 10$^2$ | 1.9301 × 10$^2$ |
| 0.00500 | 3.4937 | −1.7180 | 3.3120 × 10$^{10}$ | 1.1616 × 10$^4$ | 0.26700050 | 0.0000 | 1.0000 | 0.7149 | 3.9712 × 10$^2$ | 1.9229 × 10$^2$ |
| 0.00700 | 3.4944 | −1.8177 | 2.9424 × 10$^{10}$ | 1.4716 × 10$^4$ | 0.26700050 | 0.0000 | 1.0000 | 0.7149 | 3.9642 × 10$^2$ | 1.9209 × 10$^2$ |
| 0.01000 | 3.4949 | −1.9240 | 2.5980 × 10$^{10}$ | 1.8878 × 10$^4$ | 0.26700050 | 0.0000 | 1.0000 | 0.7149 | 3.9676 × 10$^2$ | 1.9234 × 10$^2$ |
| 0.02000 | 3.4946 | −2.1324 | 2.0465 × 10$^{10}$ | 3.0421 × 10$^4$ | 0.26700050 | 0.0000 | 1.0000 | 0.7149 | 3.9762 × 10$^2$ | 1.9369 × 10$^2$ |
| 0.03000 | 3.4935 | −2.2566 | 1.7827 × 10$^{10}$ | 4.0092 × 10$^4$ | 0.26700050 | 0.0000 | 1.0000 | 0.7149 | 3.9864 × 10$^2$ | 1.9437 × 10$^2$ |
| 0.05000 | 3.4908 | −2.4151 | 1.5038 × 10$^{10}$ | 5.6341 × 10$^4$ | 0.26700050 | 0.0000 | 1.0000 | 0.7149 | 4.0271 × 10$^2$ | 1.9615 × 10$^2$ |
| 0.07000 | 3.4885 | −2.5196 | 1.3477 × 10$^{10}$ | 7.0146 × 10$^4$ | 0.26700050 | 0.0016 | 0.9984 | 0.7149 | 4.0471 × 10$^2$ | 1.9803 × 10$^2$ |
| 0.10000 | 3.4859 | −2.6289 | 1.2026 × 10$^{10}$ | 8.8096 × 10$^4$ | 0.26700050 | 0.0130 | 0.9870 | 0.7149 | 4.0546 × 10$^2$ | 1.9935 × 10$^2$ |
| 0.20000 | 3.4762 | −2.8493 | 9.7571 × 10$^9$ | 1.3384 × 10$^5$ | 0.26700050 | 0.1453 | 0.8547 | 0.7149 | 3.8954 × 10$^2$ | 1.9414 × 10$^2$ |
| 0.23065 | 3.4732 | −2.8956 | 9.3798 × 10$^9$ | 1.4482 × 10$^5$ | 0.26700050 | 0.2156 | 0.7844 | 0.7149 | 3.8327 × 10$^2$ | 1.9092 × 10$^2$ |

$0.097 M_\odot$

| Age(Gyr) | logT$_{\rm eff}$ | logL/L$_\odot$ | R(cm) | g(cm/s$^2$) | Y$_c$ | L(PP1) | L$_g$ | X$_{\rm env}$ | $\tau_{gc}$ (day) | $\tau_{lc}$ (day) |
|---|---|---|---|---|---|---|---|---|---|---|
| 0.00002 | 3.4247 | −0.1911 | 2.6384 × 10$^{11}$ | 1.8494 × 10$^2$ | 0.26700050 | 0.0000 | 1.0000 | 0.7149 | 4.7580 × 10$^2$ | 2.1715 × 10$^2$ |
| 0.00010 | 3.4545 | −0.6038 | 1.4308 × 10$^{11}$ | 6.2884 × 10$^2$ | 0.26700050 | 0.0000 | 1.0000 | 0.7149 | 4.4032 × 10$^2$ | 2.0726 × 10$^2$ |
| 0.00020 | 3.4666 | −0.7876 | 1.0951 × 10$^{11}$ | 1.0735 × 10$^3$ | 0.26700050 | 0.0000 | 1.0000 | 0.7149 | 4.2665 × 10$^2$ | 2.0274 × 10$^2$ |
| 0.00030 | 3.4724 | −0.8996 | 9.3736 × 10$^{10}$ | 1.4652 × 10$^3$ | 0.26700050 | 0.0000 | 1.0000 | 0.7149 | 4.2045 × 10$^2$ | 2.0065 × 10$^2$ |
| 0.00050 | 3.4784 | −1.0441 | 7.7197 × 10$^{10}$ | 2.1603 × 10$^3$ | 0.26700050 | 0.0000 | 1.0000 | 0.7149 | 4.1420 × 10$^2$ | 1.9852 × 10$^2$ |
| 0.00070 | 3.4819 | −1.1398 | 6.8021 × 10$^{10}$ | 2.7825 × 10$^3$ | 0.26700050 | 0.0000 | 1.0000 | 0.7149 | 4.1039 × 10$^2$ | 1.9719 × 10$^2$ |
| 0.00100 | 3.4851 | −1.2421 | 5.9578 × 10$^{10}$ | 3.6270 × 10$^3$ | 0.26700050 | 0.0000 | 1.0000 | 0.7149 | 4.0718 × 10$^2$ | 1.9592 × 10$^2$ |
| 0.00200 | 3.4900 | −1.4431 | 4.6229 × 10$^{10}$ | 6.0240 × 10$^3$ | 0.26700050 | 0.0000 | 1.0000 | 0.7149 | 4.0247 × 10$^2$ | 1.9442 × 10$^2$ |
| 0.00300 | 3.4920 | −1.5618 | 3.9942 × 10$^{10}$ | 8.0698 × 10$^3$ | 0.26700050 | 0.0000 | 1.0000 | 0.7149 | 4.0017 × 10$^2$ | 1.9341 × 10$^2$ |
| 0.00500 | 3.4940 | −1.7115 | 3.3308 × 10$^{10}$ | 1.1605 × 10$^4$ | 0.26700050 | 0.0000 | 1.0000 | 0.7149 | 3.9794 × 10$^2$ | 1.9272 × 10$^2$ |
| 0.00700 | 3.4948 | −1.8113 | 2.9592 × 10$^{10}$ | 1.4702 × 10$^4$ | 0.26700050 | 0.0000 | 1.0000 | 0.7149 | 3.9730 × 10$^2$ | 1.9252 × 10$^2$ |
| 0.01000 | 3.4953 | −1.9175 | 2.6127 × 10$^{10}$ | 1.8860 × 10$^4$ | 0.26700050 | 0.0000 | 1.0000 | 0.7149 | 3.9760 × 10$^2$ | 1.9273 × 10$^2$ |
| 0.02000 | 3.4950 | −2.1259 | 2.0580 × 10$^{10}$ | 3.0398 × 10$^4$ | 0.26700050 | 0.0000 | 1.0000 | 0.7149 | 3.9839 × 10$^2$ | 1.9402 × 10$^2$ |
| 0.03000 | 3.4940 | −2.2500 | 1.7925 × 10$^{10}$ | 4.0068 × 10$^4$ | 0.26700050 | 0.0000 | 1.0000 | 0.7149 | 3.9941 × 10$^2$ | 1.9476 × 10$^2$ |
| 0.05000 | 3.4914 | −2.4082 | 1.5120 × 10$^{10}$ | 5.6310 × 10$^4$ | 0.26700050 | 0.0000 | 1.0000 | 0.7149 | 4.0343 × 10$^2$ | 1.9647 × 10$^2$ |
| 0.07000 | 3.4890 | −2.5128 | 1.3550 × 10$^{10}$ | 7.0116 × 10$^4$ | 0.26700050 | 0.0020 | 0.9980 | 0.7149 | 4.0532 × 10$^2$ | 1.9835 × 10$^2$ |
| 0.10000 | 3.4865 | −2.6218 | 1.2092 × 10$^{10}$ | 8.8052 × 10$^4$ | 0.26700050 | 0.0141 | 0.9859 | 0.7149 | 4.0584 × 10$^2$ | 1.9953 × 10$^2$ |
| 0.20000 | 3.4775 | −2.8390 | 9.8146 × 10$^9$ | 1.3365 × 10$^5$ | 0.26700050 | 0.1524 | 0.8476 | 0.7149 | 3.8835 × 10$^2$ | 1.9338 × 10$^2$ |
| 0.25243 | 3.4733 | −2.9106 | 9.2168 × 10$^9$ | 1.5155 × 10$^5$ | 0.26700050 | 0.2787 | 0.7213 | 0.7149 | 3.7686 × 10$^2$ | 1.8703 × 10$^2$ |

$0.098 M_\odot$

| Age(Gyr) | logT$_{\rm eff}$ | logL/L$_\odot$ | R(cm) | g(cm/s$^2$) | Y$_c$ | L(PP1) | L$_g$ | X$_{\rm env}$ | $\tau_{gc}$ (day) | $\tau_{lc}$ (day) |
|---|---|---|---|---|---|---|---|---|---|---|
| 0.00002 | 3.4258 | −0.1857 | 2.6428 × 10$^{11}$ | 1.8623 × 10$^2$ | 0.26700050 | 0.0000 | 1.0000 | 0.7149 | 4.7577 × 10$^2$ | 2.1721 × 10$^2$ |
| 0.00010 | 3.4550 | −0.5981 | 1.4367 × 10$^{11}$ | 6.3015 × 10$^2$ | 0.26700050 | 0.0000 | 1.0000 | 0.7149 | 4.4106 × 10$^2$ | 2.0762 × 10$^2$ |
| 0.00020 | 3.4670 | −0.7817 | 1.1004 × 10$^{11}$ | 1.0743 × 10$^3$ | 0.26700050 | 0.0000 | 1.0000 | 0.7149 | 4.2746 × 10$^2$ | 2.0313 × 10$^2$ |
| 0.00030 | 3.4728 | −0.8936 | 9.4203 × 10$^{10}$ | 1.4657 × 10$^3$ | 0.26700050 | 0.0000 | 1.0000 | 0.7149 | 4.2127 × 10$^2$ | 2.0106 × 10$^2$ |
| 0.00050 | 3.4788 | −1.0380 | 7.7601 × 10$^{10}$ | 2.1599 × 10$^3$ | 0.26700050 | 0.0000 | 1.0000 | 0.7149 | 4.1505 × 10$^2$ | 1.9889 × 10$^2$ |
| 0.00070 | 3.4823 | −1.1338 | 6.8389 × 10$^{10}$ | 2.7810 × 10$^3$ | 0.26700050 | 0.0000 | 1.0000 | 0.7149 | 4.1133 × 10$^2$ | 1.9764 × 10$^2$ |
| 0.00100 | 3.4855 | −1.2359 | 5.9902 × 10$^{10}$ | 3.6249 × 10$^3$ | 0.26700050 | 0.0000 | 1.0000 | 0.7149 | 4.0804 × 10$^2$ | 1.9636 × 10$^2$ |
| 0.00200 | 3.4903 | −1.4369 | 4.6488 × 10$^{10}$ | 6.0186 × 10$^3$ | 0.26700050 | 0.0000 | 1.0000 | 0.7149 | 4.0341 × 10$^2$ | 1.9483 × 10$^2$ |
| 0.00300 | 3.4924 | −1.5553 | 4.0166 × 10$^{10}$ | 8.0622 × 10$^3$ | 0.26700050 | 0.0000 | 1.0000 | 0.7149 | 4.0101 × 10$^2$ | 1.9382 × 10$^2$ |
| 0.00500 | 3.4944 | −1.7052 | 3.3494 × 10$^{10}$ | 1.1594 × 10$^4$ | 0.26700050 | 0.0000 | 1.0000 | 0.7149 | 3.9880 × 10$^2$ | 1.9314 × 10$^2$ |
| 0.00700 | 3.4952 | −1.8050 | 2.9754 × 10$^{10}$ | 1.4693 × 10$^4$ | 0.26700050 | 0.0000 | 1.0000 | 0.7149 | 3.9818 × 10$^2$ | 1.9293 × 10$^2$ |
| 0.01000 | 3.4957 | −1.9111 | 2.6273 × 10$^{10}$ | 1.8843 × 10$^4$ | 0.26700050 | 0.0000 | 1.0000 | 0.7149 | 3.9843 × 10$^2$ | 1.9312 × 10$^2$ |
| 0.02000 | 3.4954 | −2.1192 | 2.0699 × 10$^{10}$ | 3.0357 × 10$^4$ | 0.26700050 | 0.0000 | 1.0000 | 0.7149 | 3.9915 × 10$^2$ | 1.9440 × 10$^2$ |
| 0.03000 | 3.4944 | −2.2432 | 1.8028 × 10$^{10}$ | 4.0021 × 10$^4$ | 0.26700050 | 0.0000 | 1.0000 | 0.7149 | 4.0023 × 10$^2$ | 1.9514 × 10$^2$ |
| 0.05000 | 3.4919 | −2.4012 | 1.5206 × 10$^{10}$ | 5.6255 × 10$^4$ | 0.26700050 | 0.0000 | 1.0000 | 0.7149 | 4.0411 × 10$^2$ | 1.9678 × 10$^2$ |
| 0.07000 | 3.4896 | −2.5060 | 1.3623 × 10$^{10}$ | 7.0082 × 10$^4$ | 0.26700050 | 0.0022 | 0.9978 | 0.7149 | 4.0597 × 10$^2$ | 1.9864 × 10$^2$ |
| 0.10000 | 3.4871 | −2.6151 | 1.2155 × 10$^{10}$ | 8.8039 × 10$^4$ | 0.26700050 | 0.0151 | 0.9849 | 0.7149 | 4.0633 × 10$^2$ | 1.9973 × 10$^2$ |
| 0.20000 | 3.4788 | −2.8289 | 9.8728 × 10$^9$ | 1.3344 × 10$^5$ | 0.26700050 | 0.1610 | 0.8390 | 0.7149 | 3.8733 × 10$^2$ | 1.9269 × 10$^2$ |
| 0.30000 | 3.4726 | −2.9436 | 8.9015 × 10$^9$ | 1.6415 × 10$^5$ | 0.26700050 | 0.4109 | 0.5891 | 0.7149 | 3.6569 × 10$^2$ | 1.8056 × 10$^2$ |
| 0.32453 | 3.4721 | −2.9609 | 8.7443 × 10$^9$ | 1.7011 × 10$^5$ | 0.26700050 | 0.4703 | 0.5297 | 0.7149 | 3.6096 × 10$^2$ | 1.7797 × 10$^2$ |

Table 1. Continued

| Age(Gyr) | log$T_{\rm eff}$ | log$L/L_\odot$ | R(cm) | g(cm/s$^2$) | $Y_c$ | L(PP1) | $L_g$ | $X_{\rm env}$ | $\tau_{gc}$ (day) | $\tau_{lc}$ (day) |
|---|---|---|---|---|---|---|---|---|---|---|
| | | | | $0.099 M_\odot$ | | | | | | |
| 0.00002 | 3.4270 | −0.1791 | 2.6474 × 10$^{11}$ | 1.8747 × 10$^2$ | 0.26700050 | 0.0000 | 1.0000 | 0.7149 | 4.7555 × 10$^2$ | 2.1717 × 10$^2$ |
| 0.00010 | 3.4555 | −0.5925 | 1.4424 × 10$^{11}$ | 6.3152 × 10$^2$ | 0.26700050 | 0.0000 | 1.0000 | 0.7149 | 4.4179 × 10$^2$ | 2.0799 × 10$^2$ |
| 0.00020 | 3.4675 | −0.7759 | 1.1055 × 10$^{11}$ | 1.0751 × 10$^3$ | 0.26700050 | 0.0000 | 1.0000 | 0.7149 | 4.2827 × 10$^2$ | 2.0352 × 10$^2$ |
| 0.00030 | 3.4732 | −0.8876 | 9.4674 × 10$^{10}$ | 1.4660 × 10$^3$ | 0.26700050 | 0.0000 | 1.0000 | 0.7149 | 4.2211 × 10$^2$ | 2.0147 × 10$^2$ |
| 0.00050 | 3.4792 | −1.0319 | 7.8007 × 10$^{10}$ | 2.1593 × 10$^3$ | 0.26700050 | 0.0000 | 1.0000 | 0.7149 | 4.1586 × 10$^2$ | 1.9930 × 10$^2$ |
| 0.00070 | 3.4826 | −1.1279 | 6.8761 × 10$^{10}$ | 2.7791 × 10$^3$ | 0.26700050 | 0.0000 | 1.0000 | 0.7149 | 4.1228 × 10$^2$ | 1.9809 × 10$^2$ |
| 0.00100 | 3.4859 | −1.2298 | 6.0229 × 10$^{10}$ | 3.6221 × 10$^3$ | 0.26700050 | 0.0000 | 1.0000 | 0.7149 | 4.0890 × 10$^2$ | 1.9680 × 10$^2$ |
| 0.00200 | 3.4906 | −1.4309 | 4.6740 × 10$^{10}$ | 6.0147 × 10$^3$ | 0.26700050 | 0.0000 | 1.0000 | 0.7149 | 4.0429 × 10$^2$ | 1.9527 × 10$^2$ |
| 0.00300 | 3.4928 | −1.5491 | 4.0384 × 10$^{10}$ | 8.0567 × 10$^3$ | 0.26700050 | 0.0000 | 1.0000 | 0.7149 | 4.0184 × 10$^2$ | 1.9425 × 10$^2$ |
| 0.00500 | 3.4948 | −1.6990 | 3.3678 × 10$^{10}$ | 1.1585 × 10$^4$ | 0.26700050 | 0.0000 | 1.0000 | 0.7149 | 3.9966 × 10$^2$ | 1.9357 × 10$^2$ |
| 0.00700 | 3.4955 | −1.7988 | 2.9918 × 10$^{10}$ | 1.4679 × 10$^4$ | 0.26700050 | 0.0000 | 1.0000 | 0.7149 | 3.9901 × 10$^2$ | 1.9335 × 10$^2$ |
| 0.01000 | 3.4961 | −1.9047 | 2.6419 × 10$^{10}$ | 1.8825 × 10$^4$ | 0.26700050 | 0.0000 | 1.0000 | 0.7149 | 3.9925 × 10$^2$ | 1.9351 × 10$^2$ |
| 0.02000 | 3.4959 | −2.1126 | 2.0814 × 10$^{10}$ | 3.0329 × 10$^4$ | 0.26700050 | 0.0000 | 1.0000 | 0.7149 | 3.9993 × 10$^2$ | 1.9473 × 10$^2$ |
| 0.03000 | 3.4949 | −2.2366 | 1.8129 × 10$^{10}$ | 3.9978 × 10$^4$ | 0.26700050 | 0.0000 | 1.0000 | 0.7149 | 4.0105 × 10$^2$ | 1.9550 × 10$^2$ |
| 0.05000 | 3.4924 | −2.3944 | 1.5288 × 10$^{10}$ | 5.6221 × 10$^4$ | 0.26700050 | 0.0000 | 1.0000 | 0.7149 | 4.0480 × 10$^2$ | 1.9710 × 10$^2$ |
| 0.07000 | 3.4901 | −2.4993 | 1.3695 × 10$^{10}$ | 7.0058 × 10$^4$ | 0.26700050 | 0.0025 | 0.9975 | 0.7149 | 4.0666 × 10$^2$ | 1.9893 × 10$^2$ |
| 0.10000 | 3.4876 | −2.6085 | 1.2218 × 10$^{10}$ | 8.8016 × 10$^4$ | 0.26700050 | 0.0162 | 0.9838 | 0.7149 | 4.0685 × 10$^2$ | 1.9996 × 10$^2$ |
| 0.20000 | 3.4798 | −2.8202 | 9.9258 × 10$^9$ | 1.3337 × 10$^5$ | 0.26700050 | 0.1681 | 0.8319 | 0.7149 | 3.8648 × 10$^2$ | 1.9215 × 10$^2$ |
| 0.30000 | 3.4735 | −2.9332 | 8.9721 × 10$^9$ | 1.6323 × 10$^5$ | 0.26700050 | 0.4269 | 0.5731 | 0.7149 | 3.6474 × 10$^2$ | 1.7985 × 10$^2$ |
| 0.32549 | 3.4730 | −2.9505 | 8.8141 × 10$^9$ | 1.6913 × 10$^5$ | 0.26700050 | 0.4906 | 0.5094 | 0.7149 | 3.5988 × 10$^2$ | 1.7717 × 10$^2$ |
| | | | | $0.100 M_\odot$ | | | | | | |
| 0.00002 | 3.4283 | −0.1735 | 2.6495 × 10$^{11}$ | 1.8907 × 10$^2$ | 0.26700050 | 0.0000 | 1.0000 | 0.7149 | 4.7534 × 10$^2$ | 2.1715 × 10$^2$ |
| 0.00010 | 3.4560 | −0.5871 | 1.4480 × 10$^{11}$ | 6.3298 × 10$^2$ | 0.26700050 | 0.0000 | 1.0000 | 0.7149 | 4.4249 × 10$^2$ | 2.0835 × 10$^2$ |
| 0.00020 | 3.4679 | −0.7702 | 1.1106 × 10$^{11}$ | 1.0761 × 10$^3$ | 0.26700050 | 0.0000 | 1.0000 | 0.7149 | 4.2909 × 10$^2$ | 2.0392 × 10$^2$ |
| 0.00030 | 3.4736 | −0.8819 | 9.5130 × 10$^{10}$ | 1.4666 × 10$^3$ | 0.26700050 | 0.0000 | 1.0000 | 0.7149 | 4.2295 × 10$^2$ | 2.0187 × 10$^2$ |
| 0.00050 | 3.4796 | −1.0259 | 7.8404 × 10$^{10}$ | 2.1591 × 10$^3$ | 0.26700050 | 0.0000 | 1.0000 | 0.7149 | 4.1669 × 10$^2$ | 1.9969 × 10$^2$ |
| 0.00070 | 3.4829 | −1.1221 | 6.9123 × 10$^{10}$ | 2.7778 × 10$^3$ | 0.26700050 | 0.0000 | 1.0000 | 0.7149 | 4.1329 × 10$^2$ | 1.9848 × 10$^2$ |
| 0.00100 | 3.4862 | −1.2238 | 6.0553 × 10$^{10}$ | 3.6198 × 10$^3$ | 0.26700050 | 0.0000 | 1.0000 | 0.7149 | 4.0980 × 10$^2$ | 1.9723 × 10$^2$ |
| 0.00200 | 3.4910 | −1.4249 | 4.6994 × 10$^{10}$ | 6.0098 × 10$^3$ | 0.26700050 | 0.0000 | 1.0000 | 0.7149 | 4.0527 × 10$^2$ | 1.9565 × 10$^2$ |
| 0.00300 | 3.4932 | −1.5429 | 4.0606 × 10$^{10}$ | 8.0496 × 10$^3$ | 0.26700050 | 0.0000 | 1.0000 | 0.7149 | 4.0270 × 10$^2$ | 1.9467 × 10$^2$ |
| 0.00500 | 3.4951 | −1.6928 | 3.3863 × 10$^{10}$ | 1.1574 × 10$^4$ | 0.26700050 | 0.0000 | 1.0000 | 0.7149 | 4.0054 × 10$^2$ | 1.9398 × 10$^2$ |
| 0.00700 | 3.4959 | −1.7924 | 3.0084 × 10$^{10}$ | 1.4665 × 10$^4$ | 0.26700050 | 0.0000 | 1.0000 | 0.7149 | 3.9983 × 10$^2$ | 1.9373 × 10$^2$ |
| 0.01000 | 3.4964 | −1.8983 | 2.6569 × 10$^{10}$ | 1.8801 × 10$^4$ | 0.26700050 | 0.0000 | 1.0000 | 0.7149 | 4.0006 × 10$^2$ | 1.9389 × 10$^2$ |
| 0.02000 | 3.4963 | −2.1061 | 2.0930 × 10$^{10}$ | 3.0297 × 10$^4$ | 0.26700050 | 0.0000 | 1.0000 | 0.7149 | 4.0072 × 10$^2$ | 1.9507 × 10$^2$ |
| 0.03000 | 3.4953 | −2.2300 | 1.8231 × 10$^{10}$ | 3.9934 × 10$^4$ | 0.26700050 | 0.0000 | 1.0000 | 0.7149 | 4.0199 × 10$^2$ | 1.9582 × 10$^2$ |
| 0.05000 | 3.4929 | −2.3879 | 1.5367 × 10$^{10}$ | 5.6206 × 10$^4$ | 0.26700050 | 0.0000 | 1.0000 | 0.7149 | 4.0560 × 10$^2$ | 1.9742 × 10$^2$ |
| 0.07000 | 3.4906 | −2.4926 | 1.3767 × 10$^{10}$ | 7.0025 × 10$^4$ | 0.26700050 | 0.0029 | 0.9971 | 0.7149 | 4.0731 × 10$^2$ | 1.9923 × 10$^2$ |
| 0.10000 | 3.4881 | −2.6019 | 1.2283 × 10$^{10}$ | 8.7970 × 10$^4$ | 0.26700050 | 0.0173 | 0.9827 | 0.7149 | 4.0731 × 10$^2$ | 2.0018 × 10$^2$ |
| 0.20000 | 3.4807 | −2.8115 | 9.9841 × 10$^9$ | 1.3315 × 10$^5$ | 0.26700050 | 0.1754 | 0.8246 | 0.7149 | 3.8581 × 10$^2$ | 1.9166 × 10$^2$ |
| 0.26568 | 3.4745 | −2.8974 | 9.3039 × 10$^9$ | 1.5333 × 10$^5$ | 0.26700050 | 0.3512 | 0.6488 | 0.7149 | 3.7199 × 10$^2$ | 1.8355 × 10$^2$ |
| | | | | $0.110 M_\odot$ | | | | | | |
| 0.00002 | 3.4401 | −0.1130 | 2.6898 × 10$^{11}$ | 2.0179 × 10$^2$ | 0.26700050 | 0.0000 | 1.0000 | 0.7149 | 4.7283 × 10$^2$ | 2.1668 × 10$^2$ |
| 0.00010 | 3.4616 | −0.5327 | 1.5030 × 10$^{11}$ | 6.4632 × 10$^2$ | 0.26700050 | 0.0000 | 1.0000 | 0.7149 | 4.4856 × 10$^2$ | 2.1138 × 10$^2$ |
| 0.00020 | 3.4722 | −0.7153 | 1.1595 × 10$^{11}$ | 1.0859 × 10$^3$ | 0.26700050 | 0.0000 | 1.0000 | 0.7149 | 4.3647 × 10$^2$ | 2.0753 × 10$^2$ |
| 0.00030 | 3.4774 | −0.8268 | 9.9587 × 10$^{10}$ | 1.4721 × 10$^3$ | 0.26700050 | 0.0000 | 1.0000 | 0.7149 | 4.3098 × 10$^2$ | 2.0573 × 10$^2$ |
| 0.00050 | 3.4830 | −0.9697 | 8.2309 × 10$^{10}$ | 2.1550 × 10$^3$ | 0.26700050 | 0.0000 | 1.0000 | 0.7149 | 4.2500 × 10$^2$ | 2.0369 × 10$^2$ |
| 0.00070 | 3.4865 | −1.0644 | 7.2642 × 10$^{10}$ | 2.7667 × 10$^3$ | 0.26700050 | 0.0000 | 1.0000 | 0.7149 | 4.2135 × 10$^2$ | 2.0231 × 10$^2$ |
| 0.00100 | 3.4896 | −1.1663 | 6.3688 × 10$^{10}$ | 3.5994 × 10$^3$ | 0.26700050 | 0.0000 | 1.0000 | 0.7149 | 4.1812 × 10$^2$ | 2.0112 × 10$^2$ |
| 0.00200 | 3.4945 | −1.3661 | 4.9477 × 10$^{10}$ | 5.9638 × 10$^3$ | 0.26700050 | 0.0000 | 1.0000 | 0.7149 | 4.1329 × 10$^2$ | 1.9946 × 10$^2$ |
| 0.00300 | 3.4965 | −1.4847 | 4.2763 × 10$^{10}$ | 7.9838 × 10$^3$ | 0.26700050 | 0.0000 | 1.0000 | 0.7149 | 4.1118 × 10$^2$ | 1.9865 × 10$^2$ |
| 0.00500 | 3.4986 | −1.6336 | 3.5671 × 10$^{10}$ | 1.1474 × 10$^4$ | 0.26700050 | 0.0000 | 1.0000 | 0.7149 | 4.0873 × 10$^2$ | 1.9777 × 10$^2$ |
| 0.00700 | 3.4995 | −1.7328 | 3.1695 × 10$^{10}$ | 1.4533 × 10$^4$ | 0.26700050 | 0.0000 | 1.0000 | 0.7149 | 4.0798 × 10$^2$ | 1.9754 × 10$^2$ |
| 0.01000 | 3.5001 | −1.8387 | 2.7983 × 10$^{10}$ | 1.8644 × 10$^4$ | 0.26700050 | 0.0000 | 1.0000 | 0.7149 | 4.0807 × 10$^2$ | 1.9763 × 10$^2$ |
| 0.02000 | 3.5001 | −2.0459 | 2.2039 × 10$^{10}$ | 3.0057 × 10$^4$ | 0.26700050 | 0.0000 | 1.0000 | 0.7149 | 4.0864 × 10$^2$ | 1.9850 × 10$^2$ |
| 0.03000 | 3.4994 | −2.1687 | 1.9196 × 10$^{10}$ | 3.9620 × 10$^4$ | 0.26700050 | 0.0000 | 1.0000 | 0.7149 | 4.0992 × 10$^2$ | 1.9938 × 10$^2$ |
| 0.05000 | 3.4975 | −2.3251 | 1.6175 × 10$^{10}$ | 5.5803 × 10$^4$ | 0.26700050 | 0.0007 | 0.9993 | 0.7149 | 4.1233 × 10$^2$ | 2.0073 × 10$^2$ |
| 0.07000 | 3.4955 | −2.4289 | 1.4483 × 10$^{10}$ | 6.9605 × 10$^4$ | 0.26700050 | 0.0065 | 0.9935 | 0.7149 | 4.1373 × 10$^2$ | 2.0215 × 10$^2$ |
| 0.10000 | 3.4931 | −2.5380 | 1.2920 × 10$^{10}$ | 8.7455 × 10$^4$ | 0.26700050 | 0.0290 | 0.9710 | 0.7149 | 4.1170 × 10$^2$ | 2.0217 × 10$^2$ |
| 0.20000 | 3.4877 | −2.7347 | 1.0561 × 10$^{10}$ | 1.3091 × 10$^5$ | 0.26700050 | 0.2487 | 0.7513 | 0.7149 | 3.8132 × 10$^2$ | 1.8893 × 10$^2$ |
| 0.30000 | 3.4835 | −2.8258 | 9.6925 × 10$^9$ | 1.5541 × 10$^5$ | 0.26700050 | 0.5660 | 0.4340 | 0.7148 | 3.5591 × 10$^2$ | 1.7387 × 10$^2$ |
| 0.50000 | 3.4794 | −2.8851 | 9.2258 × 10$^9$ | 1.7153 × 10$^5$ | 0.26700050 | 0.9035 | 0.0965 | 0.7148 | 3.4013 × 10$^2$ | 1.6302 × 10$^2$ |
| 0.70000 | 3.4788 | −2.8943 | 9.1531 × 10$^9$ | 1.7426 × 10$^5$ | 0.26700050 | 0.9776 | 0.0224 | 0.7147 | 3.3740 × 10$^2$ | 1.6114 × 10$^2$ |
| 1.26797 | 3.4788 | −2.8965 | 9.1331 × 10$^9$ | 1.7502 × 10$^5$ | 0.26700050 | 0.9998 | 0.0002 | 0.7145 | 3.3657 × 10$^2$ | 1.6069 × 10$^2$ |

Table 1. Continued

| Age(Gyr) | log$T_{\text{eff}}$ | log$L/L_\odot$ | R(cm) | g(cm/s$^2$) | $Y_c$ | L(PP1) | $L_g$ | $X_{\text{env}}$ | $\tau_{gc}$ (day) | $\tau_{lc}$ (day) |
|---|---|---|---|---|---|---|---|---|---|---|
| | | | | 0.120$M_\odot$ | | | | | | |
| 0.00002 | 3.4523 | −0.0535 | 2.7232 × 10$^{11}$ | 2.1477 × 10$^2$ | 0.26700050 | 0.0000 | 1.0000 | 0.7149 | 4.6875 × 10$^2$ | 2.1540 × 10$^2$ |
| 0.00010 | 3.4672 | −0.4835 | 1.5501 × 10$^{11}$ | 6.6286 × 10$^2$ | 0.26700050 | 0.0000 | 1.0000 | 0.7149 | 4.5341 × 10$^2$ | 2.1390 × 10$^2$ |
| 0.00020 | 3.4763 | −0.6667 | 1.2033 × 10$^{11}$ | 1.0999 × 10$^3$ | 0.26700050 | 0.0000 | 1.0000 | 0.7149 | 4.4326 × 10$^2$ | 2.1084 × 10$^2$ |
| 0.00030 | 3.4811 | −0.7771 | 1.0367 × 10$^{11}$ | 1.4819 × 10$^3$ | 0.26700050 | 0.0000 | 1.0000 | 0.7149 | 4.3817 × 10$^2$ | 2.0919 × 10$^2$ |
| 0.00050 | 3.4868 | −0.9177 | 8.5906 × 10$^{10}$ | 2.1581 × 10$^3$ | 0.26700050 | 0.0000 | 1.0000 | 0.7149 | 4.3205 × 10$^2$ | 2.0708 × 10$^2$ |
| 0.00070 | 3.4899 | −1.0126 | 7.5907 × 10$^{10}$ | 2.7642 × 10$^3$ | 0.26700050 | 0.0000 | 1.0000 | 0.7149 | 4.2871 × 10$^2$ | 2.0587 × 10$^2$ |
| 0.00100 | 3.4928 | −1.1145 | 6.6618 × 10$^{10}$ | 3.5887 × 10$^3$ | 0.26700050 | 0.0000 | 1.0000 | 0.7149 | 4.2581 × 10$^2$ | 2.0480 × 10$^2$ |
| 0.00200 | 3.4978 | −1.3130 | 5.1807 × 10$^{10}$ | 5.9339 × 10$^3$ | 0.26700050 | 0.0000 | 1.0000 | 0.7149 | 4.2071 × 10$^2$ | 2.0293 × 10$^2$ |
| 0.00300 | 3.4999 | −1.4308 | 4.4799 × 10$^{10}$ | 7.9358 × 10$^3$ | 0.26700050 | 0.0000 | 1.0000 | 0.7149 | 4.1852 × 10$^2$ | 2.0214 × 10$^2$ |
| 0.00500 | 3.5018 | −1.5805 | 3.7373 × 10$^{10}$ | 1.1403 × 10$^4$ | 0.26700050 | 0.0000 | 1.0000 | 0.7149 | 4.1657 × 10$^2$ | 2.0147 × 10$^2$ |
| 0.00700 | 3.5029 | −1.6786 | 3.3214 × 10$^{10}$ | 1.4438 × 10$^4$ | 0.26700050 | 0.0000 | 1.0000 | 0.7149 | 4.1539 × 10$^2$ | 2.0101 × 10$^2$ |
| 0.01000 | 3.5035 | −1.7841 | 2.9331 × 10$^{10}$ | 1.8513 × 10$^4$ | 0.26700050 | 0.0000 | 1.0000 | 0.7149 | 4.1536 × 10$^2$ | 2.0107 × 10$^2$ |
| 0.02000 | 3.5037 | −1.9905 | 2.3109 × 10$^{10}$ | 2.9825 × 10$^4$ | 0.26700050 | 0.0000 | 1.0000 | 0.7149 | 4.1611 × 10$^2$ | 2.0182 × 10$^2$ |
| 0.03000 | 3.5031 | −2.1124 | 2.0133 × 10$^{10}$ | 3.9294 × 10$^4$ | 0.26700050 | 0.0000 | 1.0000 | 0.7149 | 4.1762 × 10$^2$ | 2.0275 × 10$^2$ |
| 0.05000 | 3.5016 | −2.2679 | 1.6953 × 10$^{10}$ | 5.5414 × 10$^4$ | 0.26700050 | 0.0020 | 0.9980 | 0.7149 | 4.1954 × 10$^2$ | 2.0413 × 10$^2$ |
| 0.07000 | 3.4999 | −2.3711 | 1.5173 × 10$^{10}$ | 6.9179 × 10$^4$ | 0.26700050 | 0.0103 | 0.9897 | 0.7149 | 4.2036 × 10$^2$ | 2.0504 × 10$^2$ |
| 0.10000 | 3.4975 | −2.4795 | 1.3538 × 10$^{10}$ | 8.6905 × 10$^4$ | 0.26700050 | 0.0412 | 0.9588 | 0.7149 | 4.1654 × 10$^2$ | 2.0422 × 10$^2$ |
| 0.20000 | 3.4926 | −2.6694 | 1.1129 × 10$^{10}$ | 1.2860 × 10$^5$ | 0.26700050 | 0.3220 | 0.6780 | 0.7149 | 3.7930 × 10$^2$ | 1.8746 × 10$^2$ |
| 0.30000 | 3.4904 | −2.7433 | 1.0324 × 10$^{10}$ | 1.4943 × 10$^5$ | 0.26700050 | 0.6652 | 0.3348 | 0.7148 | 3.5301 × 10$^2$ | 1.7161 × 10$^2$ |
| 0.50000 | 3.4890 | −2.7809 | 9.9534 × 10$^9$ | 1.6076 × 10$^5$ | 0.26700050 | 0.9435 | 0.0565 | 0.7148 | 3.3908 × 10$^2$ | 1.6234 × 10$^2$ |
| 0.70000 | 3.4888 | −2.7853 | 9.9099 × 10$^9$ | 1.6218 × 10$^5$ | 0.26700050 | 0.9888 | 0.0112 | 0.7147 | 3.3733 × 10$^2$ | 1.6114 × 10$^2$ |
| 1.15835 | 3.4888 | −2.7862 | 9.9004 × 10$^9$ | 1.6249 × 10$^5$ | 0.26700050 | 0.9996 | 0.0004 | 0.7145 | 3.3692 × 10$^2$ | 1.6096 × 10$^2$ |
| | | | | 0.130$M_\odot$ | | | | | | |
| 0.00002 | 3.4635 | 0.0014 | 2.7548 × 10$^{11}$ | 2.2736 × 10$^2$ | 0.26700050 | 0.0000 | 1.0000 | 0.7149 | 4.6502 × 10$^2$ | 2.1423 × 10$^2$ |
| 0.00010 | 3.4727 | −0.4390 | 1.5906 × 10$^{11}$ | 6.8196 × 10$^2$ | 0.26700050 | 0.0000 | 1.0000 | 0.7149 | 4.5763 × 10$^2$ | 2.1611 × 10$^2$ |
| 0.00020 | 3.4807 | −0.6212 | 1.2429 × 10$^{11}$ | 1.1168 × 10$^3$ | 0.26700050 | 0.0000 | 1.0000 | 0.7149 | 4.4888 × 10$^2$ | 2.1362 × 10$^2$ |
| 0.00030 | 3.4852 | −0.7303 | 1.0735 × 10$^{11}$ | 1.4972 × 10$^3$ | 0.26700050 | 0.0000 | 1.0000 | 0.7149 | 4.4394 × 10$^2$ | 2.1199 × 10$^2$ |
| 0.00050 | 3.4903 | −0.8709 | 8.9189 × 10$^{10}$ | 2.1691 × 10$^3$ | 0.26700050 | 0.0000 | 1.0000 | 0.7149 | 4.3849 × 10$^2$ | 2.1016 × 10$^2$ |
| 0.00070 | 3.4932 | −0.9654 | 7.8929 × 10$^{10}$ | 2.7696 × 10$^3$ | 0.26700050 | 0.0000 | 1.0000 | 0.7149 | 4.3549 × 10$^2$ | 2.0910 × 10$^2$ |
| 0.00100 | 3.4964 | −1.0652 | 6.9354 × 10$^{10}$ | 3.5872 × 10$^3$ | 0.26700050 | 0.0000 | 1.0000 | 0.7149 | 4.3207 × 10$^2$ | 2.0782 × 10$^2$ |
| 0.00200 | 3.5009 | −1.2645 | 5.3994 × 10$^{10}$ | 5.9183 × 10$^3$ | 0.26700050 | 0.0000 | 1.0000 | 0.7149 | 4.2755 × 10$^2$ | 2.0620 × 10$^2$ |
| 0.00300 | 3.5032 | −1.3814 | 4.6707 × 10$^{10}$ | 7.9090 × 10$^3$ | 0.26700050 | 0.0000 | 1.0000 | 0.7149 | 4.2516 × 10$^2$ | 2.0527 × 10$^2$ |
| 0.00500 | 3.5052 | −1.5304 | 3.8984 × 10$^{10}$ | 1.1353 × 10$^4$ | 0.26700050 | 0.0000 | 1.0000 | 0.7149 | 4.2320 × 10$^2$ | 2.0465 × 10$^2$ |
| 0.00700 | 3.5060 | −1.6292 | 3.4650 × 10$^{10}$ | 1.4371 × 10$^4$ | 0.26700050 | 0.0000 | 1.0000 | 0.7149 | 4.2235 × 10$^2$ | 2.0431 × 10$^2$ |
| 0.01000 | 3.5068 | −1.7339 | 3.0611 × 10$^{10}$ | 1.8414 × 10$^4$ | 0.26700050 | 0.0000 | 1.0000 | 0.7149 | 4.2205 × 10$^2$ | 2.0422 × 10$^2$ |
| 0.02000 | 3.5071 | −1.9399 | 2.4117 × 10$^{10}$ | 2.9664 × 10$^4$ | 0.26700050 | 0.0000 | 1.0000 | 0.7149 | 4.2310 × 10$^2$ | 2.0500 × 10$^2$ |
| 0.03000 | 3.5066 | −2.0615 | 2.1014 × 10$^{10}$ | 3.9072 × 10$^4$ | 0.26700050 | 0.0000 | 1.0000 | 0.7149 | 4.2499 × 10$^2$ | 2.0605 × 10$^2$ |
| 0.05000 | 3.5053 | −2.2159 | 1.7699 × 10$^{10}$ | 5.5083 × 10$^4$ | 0.26700050 | 0.0032 | 0.9968 | 0.7149 | 4.2701 × 10$^2$ | 2.0752 × 10$^2$ |
| 0.07000 | 3.5038 | −2.3183 | 1.5838 × 10$^{10}$ | 6.8786 × 10$^4$ | 0.26700050 | 0.0145 | 0.9855 | 0.7149 | 4.2704 × 10$^2$ | 2.0807 × 10$^2$ |
| 0.10000 | 3.5017 | −2.4257 | 1.4133 × 10$^{10}$ | 8.6384 × 10$^4$ | 0.26700050 | 0.0543 | 0.9457 | 0.7149 | 4.2140 × 10$^2$ | 2.0652 × 10$^2$ |
| 0.20000 | 3.4971 | −2.6085 | 1.1692 × 10$^{10}$ | 1.2621 × 10$^5$ | 0.26700050 | 0.3864 | 0.6136 | 0.7149 | 3.7850 × 10$^2$ | 1.8657 × 10$^2$ |
| 0.30000 | 3.4954 | −2.6721 | 1.0952 × 10$^{10}$ | 1.4386 × 10$^5$ | 0.26700050 | 0.7447 | 0.2553 | 0.7148 | 3.5327 × 10$^2$ | 1.7108 × 10$^2$ |
| 0.50000 | 3.4946 | −2.6989 | 1.0662 × 10$^{10}$ | 1.5179 × 10$^5$ | 0.26700050 | 0.9665 | 0.0335 | 0.7147 | 3.4227 × 10$^2$ | 1.6374 × 10$^2$ |
| 0.70000 | 3.4945 | −2.7016 | 1.0633 × 10$^{10}$ | 1.5261 × 10$^5$ | 0.26700050 | 0.9935 | 0.0065 | 0.7146 | 3.4113 × 10$^2$ | 1.6291 × 10$^2$ |
| 0.95343 | 3.4945 | −2.7017 | 1.0631 × 10$^{10}$ | 1.5267 × 10$^5$ | 0.26700050 | 0.9999 | 0.0001 | 0.7145 | 3.4099 × 10$^2$ | 1.6291 × 10$^2$ |
| | | | | 0.140$M_\odot$ | | | | | | |
| 0.00002 | 3.4740 | 0.0530 | 2.7853 × 10$^{11}$ | 2.3951 × 10$^2$ | 0.26700050 | 0.0000 | 1.0000 | 0.7149 | 4.6150 × 10$^2$ | 2.1312 × 10$^2$ |
| 0.00010 | 3.4787 | −0.3958 | 1.6264 × 10$^{11}$ | 7.0248 × 10$^2$ | 0.26700050 | 0.0000 | 1.0000 | 0.7149 | 4.6028 × 10$^2$ | 2.1756 × 10$^2$ |
| 0.00020 | 3.4854 | −0.5783 | 1.2777 × 10$^{11}$ | 1.1381 × 10$^3$ | 0.26700050 | 0.0000 | 1.0000 | 0.7149 | 4.4903 × 10$^2$ | 2.1578 × 10$^2$ |
| 0.00030 | 3.4893 | −0.6876 | 1.1066 × 10$^{11}$ | 1.5173 × 10$^3$ | 0.26700050 | 0.0000 | 1.0000 | 0.7149 | 4.4903 × 10$^2$ | 2.1449 × 10$^2$ |
| 0.00050 | 3.4940 | −0.8272 | 9.2209 × 10$^{10}$ | 2.1854 × 10$^3$ | 0.26700050 | 0.0000 | 1.0000 | 0.7149 | 4.4404 × 10$^2$ | 2.1285 × 10$^2$ |
| 0.00070 | 3.4970 | −0.9202 | 8.1717 × 10$^{10}$ | 2.7826 × 10$^3$ | 0.26700050 | 0.0000 | 1.0000 | 0.7149 | 4.4087 × 10$^2$ | 2.1169 × 10$^2$ |
| 0.00100 | 3.4997 | −1.0209 | 7.1878 × 10$^{10}$ | 3.5965 × 10$^3$ | 0.26700050 | 0.0000 | 1.0000 | 0.7149 | 4.3807 × 10$^2$ | 2.1066 × 10$^2$ |
| 0.00200 | 3.5043 | −1.2185 | 5.6054 × 10$^{10}$ | 5.9138 × 10$^3$ | 0.26700050 | 0.0000 | 1.0000 | 0.7149 | 4.3336 × 10$^2$ | 2.0899 × 10$^2$ |
| 0.00300 | 3.5062 | −1.3362 | 4.8525 × 10$^{10}$ | 7.8913 × 10$^3$ | 0.26700050 | 0.0000 | 1.0000 | 0.7149 | 4.3152 × 10$^2$ | 2.0831 × 10$^2$ |
| 0.00500 | 3.5083 | −1.4845 | 4.0519 × 10$^{10}$ | 1.1318 × 10$^4$ | 0.26700050 | 0.0000 | 1.0000 | 0.7149 | 4.2954 × 10$^2$ | 2.0754 × 10$^2$ |
| 0.00700 | 3.5092 | −1.5829 | 3.6022 × 10$^{10}$ | 1.4320 × 10$^4$ | 0.26700050 | 0.0000 | 1.0000 | 0.7149 | 4.2878 × 10$^2$ | 2.0732 × 10$^2$ |
| 0.01000 | 3.5098 | −1.6880 | 3.1819 × 10$^{10}$ | 1.8352 × 10$^4$ | 0.26700050 | 0.0000 | 1.0000 | 0.7149 | 4.2836 × 10$^2$ | 2.0730 × 10$^2$ |
| 0.02000 | 3.5102 | −1.8935 | 2.5076 × 10$^{10}$ | 2.9551 × 10$^4$ | 0.26700050 | 0.0000 | 1.0000 | 0.7149 | 4.2986 × 10$^2$ | 2.0813 × 10$^2$ |
| 0.03000 | 3.5098 | −2.0150 | 2.1846 × 10$^{10}$ | 3.8934 × 10$^4$ | 0.26700050 | 0.0002 | 0.9998 | 0.7149 | 4.3164 × 10$^2$ | 2.0929 × 10$^2$ |
| 0.05000 | 3.5086 | −2.1687 | 1.8405 × 10$^{10}$ | 5.4856 × 10$^4$ | 0.26700050 | 0.0047 | 0.9953 | 0.7149 | 4.3433 × 10$^2$ | 2.1095 × 10$^2$ |
| 0.07000 | 3.5073 | −2.2699 | 1.6474 × 10$^{10}$ | 6.8470 × 10$^4$ | 0.26700050 | 0.0191 | 0.9809 | 0.7149 | 4.3383 × 10$^2$ | 2.1118 × 10$^2$ |
| 0.10000 | 3.5055 | −2.3758 | 1.4707 × 10$^{10}$ | 8.5912 × 10$^4$ | 0.26700050 | 0.0683 | 0.9317 | 0.7149 | 4.2614 × 10$^2$ | 2.0903 × 10$^2$ |
| 0.20000 | 3.5013 | −2.5517 | 1.2245 × 10$^{10}$ | 1.2392 × 10$^5$ | 0.26700050 | 0.4515 | 0.5485 | 0.7149 | 3.7859 × 10$^2$ | 1.8617 × 10$^2$ |

Table 1. Continued

| Age(Gyr) | $\log T_{eff}$ | $\log L/L_\odot$ | R(cm) | g(cm/s$^2$) | $Y_c$ | L(PP1) | $L_g$ | $X_{env}$ | $\tau_{gc}$ (day) | $\tau_{lc}$ (day) |
|---|---|---|---|---|---|---|---|---|---|---|
| 0.30000 | 3.4999 | −2.6069 | $1.1567 \times 10^{10}$ | $1.3888 \times 10^5$ | 0.26700050 | 0.8007 | 0.1993 | 0.7148 | $3.5500 \times 10^2$ | $1.7150 \times 10^2$ |
| 0.50000 | 3.4993 | −2.6263 | $1.1341 \times 10^{10}$ | $1.4448 \times 10^5$ | 0.26700050 | 0.9802 | 0.0198 | 0.7147 | $3.4633 \times 10^2$ | $1.6567 \times 10^2$ |
| 0.87756 | 3.4992 | −2.6284 | $1.1317 \times 10^{10}$ | $1.4509 \times 10^5$ | 0.26700050 | 0.9995 | 0.0005 | 0.7145 | $3.4539 \times 10^2$ | $1.6514 \times 10^2$ |
| | | | | $0.150 M_\odot$ | | | | | | |
| 0.00002 | 3.4835 | 0.1002 | $2.8145 \times 10^{11}$ | $2.5132 \times 10^2$ | 0.26700050 | 0.0000 | 1.0000 | 0.7149 | $4.5864 \times 10^2$ | $2.1228 \times 10^2$ |
| 0.00010 | 3.4851 | −0.3530 | $1.6582 \times 10^{11}$ | $7.2403 \times 10^2$ | 0.26700050 | 0.0000 | 1.0000 | 0.7149 | $4.6148 \times 10^2$ | $2.1836 \times 10^2$ |
| 0.00020 | 3.4903 | −0.5378 | $1.3089 \times 10^{11}$ | $1.1620 \times 10^3$ | 0.26700050 | 0.0000 | 1.0000 | 0.7149 | $4.5656 \times 10^2$ | $2.1750 \times 10^2$ |
| 0.00030 | 3.4935 | −0.6475 | $1.1366 \times 10^{11}$ | $1.5410 \times 10^3$ | 0.26700050 | 0.0000 | 1.0000 | 0.7149 | $4.5332 \times 10^2$ | $2.1662 \times 10^2$ |
| 0.00050 | 3.4981 | −0.7857 | $9.4945 \times 10^{10}$ | $2.2085 \times 10^3$ | 0.26700050 | 0.0000 | 1.0000 | 0.7149 | $4.4845 \times 10^2$ | $2.1500 \times 10^2$ |
| 0.00070 | 3.5006 | −0.8792 | $8.4258 \times 10^{10}$ | $2.8042 \times 10^3$ | 0.26700050 | 0.0000 | 1.0000 | 0.7149 | $4.4586 \times 10^2$ | $2.1412 \times 10^2$ |
| 0.00100 | 3.5031 | −0.9791 | $7.4224 \times 10^{10}$ | $3.6136 \times 10^3$ | 0.26700050 | 0.0000 | 1.0000 | 0.7149 | $4.4321 \times 10^2$ | $2.1315 \times 10^2$ |
| 0.00200 | 3.5073 | −1.1768 | $5.7993 \times 10^{10}$ | $5.9195 \times 10^3$ | 0.26700050 | 0.0000 | 1.0000 | 0.7149 | $4.3907 \times 10^2$ | $2.1165 \times 10^2$ |
| 0.00300 | 3.5095 | −1.2928 | $5.0228 \times 10^{10}$ | $7.8912 \times 10^3$ | 0.26700050 | 0.0000 | 1.0000 | 0.7149 | $4.3670 \times 10^2$ | $2.1084 \times 10^2$ |
| 0.00500 | 3.5114 | −1.4416 | $4.1964 \times 10^{10}$ | $1.1306 \times 10^4$ | 0.26700050 | 0.0000 | 1.0000 | 0.7149 | $4.3509 \times 10^2$ | $2.1025 \times 10^2$ |
| 0.00700 | 3.5123 | −1.5398 | $3.7311 \times 10^{10}$ | $1.4301 \times 10^4$ | 0.26700050 | 0.0000 | 1.0000 | 0.7149 | $4.3447 \times 10^2$ | $2.1005 \times 10^2$ |
| 0.01000 | 3.5130 | −1.6444 | $3.2981 \times 10^{10}$ | $1.8303 \times 10^4$ | 0.26700050 | 0.0000 | 1.0000 | 0.7149 | $4.3423 \times 10^2$ | $2.0996 \times 10^2$ |
| 0.02000 | 3.5134 | −1.8493 | $2.6000 \times 10^{10}$ | $2.9449 \times 10^4$ | 0.26700050 | 0.0000 | 1.0000 | 0.7149 | $4.3601 \times 10^2$ | $2.1099 \times 10^2$ |
| 0.03000 | 3.5129 | −1.9706 | $2.2659 \times 10^{10}$ | $3.8774 \times 10^4$ | 0.26700050 | 0.0004 | 0.9996 | 0.7149 | $4.3823 \times 10^2$ | $2.1222 \times 10^2$ |
| 0.05000 | 3.5118 | −2.1237 | $1.9093 \times 10^{10}$ | $5.4612 \times 10^4$ | 0.26700050 | 0.0062 | 0.9938 | 0.7149 | $4.4146 \times 10^2$ | $2.1413 \times 10^2$ |
| 0.07000 | 3.5106 | −2.2248 | $1.7092 \times 10^{10}$ | $6.8149 \times 10^4$ | 0.26700050 | 0.0234 | 0.9766 | 0.7149 | $4.4098 \times 10^2$ | $2.1440 \times 10^2$ |
| 0.10000 | 3.5089 | −2.3299 | $1.5263 \times 10^{10}$ | $8.5454 \times 10^4$ | 0.26700050 | 0.0827 | 0.9173 | 0.7149 | $4.3129 \times 10^2$ | $2.1165 \times 10^2$ |
| 0.20000 | 3.5052 | −2.4982 | $1.2790 \times 10^{10}$ | $1.2170 \times 10^5$ | 0.26700050 | 0.5085 | 0.4915 | 0.7148 | $3.7947 \times 10^2$ | $1.8625 \times 10^2$ |
| 0.30000 | 3.5039 | −2.5466 | $1.2171 \times 10^{10}$ | $1.3441 \times 10^5$ | 0.26700050 | 0.8447 | 0.1553 | 0.7148 | $3.5777 \times 10^2$ | $1.7267 \times 10^2$ |
| 0.60467 | 3.5034 | −2.5625 | $1.1977 \times 10^{10}$ | $1.3880 \times 10^5$ | 0.26700050 | 0.9944 | 0.0056 | 0.7147 | $3.5045 \times 10^2$ | $1.6782 \times 10^2$ |
| | | | | $0.160 M_\odot$ | | | | | | |
| 0.00002 | 3.4921 | 0.1429 | $2.8422 \times 10^{11}$ | $2.6287 \times 10^2$ | 0.26700050 | 0.0000 | 1.0000 | 0.7149 | $4.5652 \times 10^2$ | $2.1172 \times 10^2$ |
| 0.00010 | 3.4917 | −0.3117 | $1.6868 \times 10^{11}$ | $7.4633 \times 10^2$ | 0.26700050 | 0.0000 | 1.0000 | 0.7149 | $4.6196 \times 10^2$ | $2.1878 \times 10^2$ |
| 0.00020 | 3.4955 | −0.4988 | $1.3369 \times 10^{11}$ | $1.1882 \times 10^3$ | 0.26700050 | 0.0000 | 1.0000 | 0.7149 | $4.5891 \times 10^2$ | $2.1875 \times 10^2$ |
| 0.00030 | 3.4984 | −0.6077 | $1.1635 \times 10^{11}$ | $1.5687 \times 10^3$ | 0.26700050 | 0.0000 | 1.0000 | 0.7149 | $4.5662 \times 10^2$ | $2.1800 \times 10^2$ |
| 0.00050 | 3.5020 | −0.7477 | $9.7413 \times 10^{10}$ | $2.2378 \times 10^3$ | 0.26700050 | 0.0000 | 1.0000 | 0.7149 | $4.5250 \times 10^2$ | $2.1697 \times 10^2$ |
| 0.00070 | 3.5044 | −0.8403 | $8.6603 \times 10^{10}$ | $2.8314 \times 10^3$ | 0.26700050 | 0.0000 | 1.0000 | 0.7149 | $4.5000 \times 10^2$ | $2.1614 \times 10^2$ |
| 0.00100 | 3.5068 | −0.9396 | $7.6379 \times 10^{10}$ | $3.6402 \times 10^3$ | 0.26700050 | 0.0000 | 1.0000 | 0.7149 | $4.4750 \times 10^2$ | $2.1521 \times 10^2$ |
| 0.00200 | 3.5108 | −1.1364 | $5.9782 \times 10^{10}$ | $5.9420 \times 10^3$ | 0.26700050 | 0.0000 | 1.0000 | 0.7149 | $4.4352 \times 10^2$ | $2.1382 \times 10^2$ |
| 0.00300 | 3.5126 | −1.2536 | $5.1818 \times 10^{10}$ | $7.9087 \times 10^3$ | 0.26700050 | 0.0000 | 1.0000 | 0.7149 | $4.4190 \times 10^2$ | $2.1326 \times 10^2$ |
| 0.00500 | 3.5146 | −1.4010 | $4.3328 \times 10^{10}$ | $1.1312 \times 10^4$ | 0.26700050 | 0.0000 | 1.0000 | 0.7149 | $4.4010 \times 10^2$ | $2.1252 \times 10^2$ |
| 0.00700 | 3.5153 | −1.4998 | $3.8531 \times 10^{10}$ | $1.4304 \times 10^4$ | 0.26700050 | 0.0000 | 1.0000 | 0.7149 | $4.3979 \times 10^2$ | $2.1258 \times 10^2$ |
| 0.01000 | 3.5160 | −1.6045 | $3.4053 \times 10^{10}$ | $1.8313 \times 10^4$ | 0.26700050 | 0.0000 | 1.0000 | 0.7149 | $4.3971 \times 10^2$ | $2.1264 \times 10^2$ |
| 0.02000 | 3.5163 | −1.8090 | $2.6871 \times 10^{10}$ | $2.9411 \times 10^4$ | 0.26700050 | 0.0000 | 1.0000 | 0.7149 | $4.4209 \times 10^2$ | $2.1380 \times 10^2$ |
| 0.03000 | 3.5159 | −1.9300 | $2.3417 \times 10^{10}$ | $3.8727 \times 10^4$ | 0.26700050 | 0.0006 | 0.9994 | 0.7149 | $4.4460 \times 10^2$ | $2.1521 \times 10^2$ |
| 0.05000 | 3.5148 | −2.0829 | $1.9737 \times 10^{10}$ | $5.4512 \times 10^4$ | 0.26700050 | 0.0078 | 0.9922 | 0.7149 | $4.4867 \times 10^2$ | $2.1742 \times 10^2$ |
| 0.07000 | 3.5138 | −2.1827 | $1.7677 \times 10^{10}$ | $6.7958 \times 10^4$ | 0.26700050 | 0.0287 | 0.9713 | 0.7149 | $4.4787 \times 10^2$ | $2.1772 \times 10^2$ |
| 0.10000 | 3.5122 | −2.2866 | $1.5802 \times 10^{10}$ | $8.5041 \times 10^4$ | 0.26700050 | 0.0972 | 0.9028 | 0.7149 | $4.3654 \times 10^2$ | $2.1435 \times 10^2$ |
| 0.20000 | 3.5088 | −2.4481 | $1.3326 \times 10^{10}$ | $1.1957 \times 10^5$ | 0.26700050 | 0.5619 | 0.4381 | 0.7148 | $3.8120 \times 10^2$ | $1.8689 \times 10^2$ |
| 0.30000 | 3.5077 | −2.4899 | $1.2763 \times 10^{10}$ | $1.3036 \times 10^5$ | 0.26700050 | 0.8768 | 0.1232 | 0.7148 | $3.6121 \times 10^2$ | $1.7434 \times 10^2$ |
| 0.50000 | 3.5075 | −2.5008 | $1.2618 \times 10^{10}$ | $1.3337 \times 10^5$ | 0.26700050 | 0.9929 | 0.0071 | 0.7147 | $3.5565 \times 10^2$ | $1.7069 \times 10^2$ |
| 0.63888 | 3.5075 | −2.5010 | $1.2614 \times 10^{10}$ | $1.3346 \times 10^5$ | 0.26700051 | 0.9991 | 0.0009 | 0.7146 | $3.5544 \times 10^2$ | $1.7055 \times 10^2$ |
| | | | | $0.170 M_\odot$ | | | | | | |
| 0.00002 | 3.5000 | 0.1823 | $2.8682 \times 10^{11}$ | $2.7427 \times 10^2$ | 0.26700050 | 0.0000 | 1.0000 | 0.7149 | $4.5477 \times 10^2$ | $2.1131 \times 10^2$ |
| 0.00010 | 3.4986 | −0.2706 | $1.7134 \times 10^{11}$ | $7.6858 \times 10^2$ | 0.26700050 | 0.0000 | 1.0000 | 0.7149 | $4.6140 \times 10^2$ | $2.1872 \times 10^2$ |
| 0.00020 | 3.5008 | −0.4612 | $1.3621 \times 10^{11}$ | $1.2162 \times 10^3$ | 0.26700050 | 0.0000 | 1.0000 | 0.7149 | $4.6054 \times 10^2$ | $2.1962 \times 10^2$ |
| 0.00030 | 3.5031 | −0.5710 | $1.1876 \times 10^{11}$ | $1.5998 \times 10^3$ | 0.26700050 | 0.0000 | 1.0000 | 0.7149 | $4.5845 \times 10^2$ | $2.1923 \times 10^2$ |
| 0.00050 | 3.5061 | −0.7110 | $9.9687 \times 10^{10}$ | $2.2705 \times 10^3$ | 0.26700050 | 0.0000 | 1.0000 | 0.7149 | $4.5568 \times 10^2$ | $2.1853 \times 10^2$ |
| 0.00070 | 3.5084 | −0.8033 | $8.8718 \times 10^{10}$ | $2.8666 \times 10^3$ | 0.26700050 | 0.0000 | 1.0000 | 0.7149 | $4.5340 \times 10^2$ | $2.1778 \times 10^2$ |
| 0.00100 | 3.5104 | −0.9030 | $7.8359 \times 10^{10}$ | $3.6746 \times 10^3$ | 0.26700050 | 0.0000 | 1.0000 | 0.7149 | $4.5139 \times 10^2$ | $2.1711 \times 10^2$ |
| 0.00200 | 3.5141 | −1.0992 | $6.1463 \times 10^{10}$ | $5.9726 \times 10^3$ | 0.26700050 | 0.0000 | 1.0000 | 0.7149 | $4.4787 \times 10^2$ | $2.1584 \times 10^2$ |
| 0.00300 | 3.5160 | −1.2154 | $5.3310 \times 10^{10}$ | $7.9392 \times 10^3$ | 0.26700050 | 0.0000 | 1.0000 | 0.7149 | $4.4602 \times 10^2$ | $2.1526 \times 10^2$ |
| 0.00500 | 3.5176 | −1.3637 | $4.4605 \times 10^{10}$ | $1.1340 \times 10^4$ | 0.26700050 | 0.0000 | 1.0000 | 0.7149 | $4.4492 \times 10^2$ | $2.1480 \times 10^2$ |
| 0.00700 | 3.5186 | −1.4612 | $3.9690 \times 10^{10}$ | $1.4323 \times 10^4$ | 0.26700050 | 0.0000 | 1.0000 | 0.7149 | $4.4431 \times 10^2$ | $2.1472 \times 10^2$ |
| 0.01000 | 3.5190 | −1.5660 | $3.5104 \times 10^{10}$ | $1.8310 \times 10^4$ | 0.26700050 | 0.0000 | 1.0000 | 0.7149 | $4.4477 \times 10^2$ | $2.1495 \times 10^2$ |
| 0.02000 | 3.5195 | −1.7700 | $2.7700 \times 10^{10}$ | $2.9407 \times 10^4$ | 0.26700050 | 0.0000 | 1.0000 | 0.7149 | $4.4740 \times 10^2$ | $2.1633 \times 10^2$ |
| 0.03000 | 3.5189 | −1.8911 | $2.4152 \times 10^{10}$ | $3.8681 \times 10^4$ | 0.26700050 | 0.0010 | 0.9990 | 0.7149 | $4.5092 \times 10^2$ | $2.1802 \times 10^2$ |
| 0.05000 | 3.5179 | −2.0436 | $2.0364 \times 10^{10}$ | $5.4410 \times 10^4$ | 0.26700050 | 0.0096 | 0.9904 | 0.7149 | $4.4591 \times 10^2$ | $2.2048 \times 10^2$ |
| 0.07000 | 3.5167 | −2.1437 | $1.8243 \times 10^{10}$ | $6.7795 \times 10^4$ | 0.26700050 | 0.0339 | 0.9661 | 0.7149 | $4.5550 \times 10^2$ | $2.2121 \times 10^2$ |
| 0.10000 | 3.5153 | −2.2464 | $1.6319 \times 10^{10}$ | $8.4729 \times 10^4$ | 0.26700050 | 0.1135 | 0.8865 | 0.7149 | $4.4202 \times 10^2$ | $2.1726 \times 10^2$ |
| 0.20000 | 3.5122 | −2.4008 | $1.3854 \times 10^{10}$ | $1.1755 \times 10^5$ | 0.26700050 | 0.6113 | 0.3887 | 0.7148 | $3.8357 \times 10^2$ | $1.8796 \times 10^2$ |

Table 1. Continued

| Age(Gyr) | logT$_{eff}$ | logL/L$_\odot$ | R(cm) | g(cm/s$^2$) | Y$_c$ | L(PP1) | L$_g$ | X$_{env}$ | $\tau_{gc}$ (day) | $\tau_{lc}$ (day) |
|---|---|---|---|---|---|---|---|---|---|---|
| 0.30000 | 3.5114 | −2.4368 | 1.3343 × 10$^{10}$ | 1.2673 × 10$^5$ | 0.26700050 | 0.9009 | 0.0991 | 0.7148 | 3.6522 × 10$^2$ | 1.7641 × 10$^2$ |
| 0.56614 | 3.5111 | −2.4467 | 1.3209 × 10$^{10}$ | 1.2931 × 10$^5$ | 0.26700051 | 0.9967 | 0.0033 | 0.7146 | 3.6020 × 10$^2$ | 1.7309 × 10$^2$ |

0.180$M_\odot$

| Age(Gyr) | logT$_{eff}$ | logL/L$_\odot$ | R(cm) | g(cm/s$^2$) | Y$_c$ | L(PP1) | L$_g$ | X$_{env}$ | $\tau_{gc}$ (day) | $\tau_{lc}$ (day) |
|---|---|---|---|---|---|---|---|---|---|---|
| 0.00002 | 3.5072 | 0.2056 | 2.8495 × 10$^{11}$ | 2.9423 × 10$^2$ | 0.26700050 | 0.0000 | 1.0000 | 0.7149 | 4.5347 × 10$^2$ | 2.1125 × 10$^2$ |
| 0.00010 | 3.5052 | −0.2348 | 1.7326 × 10$^{11}$ | 7.9581 × 10$^2$ | 0.26700050 | 0.0000 | 1.0000 | 0.7149 | 4.6093 × 10$^2$ | 2.1868 × 10$^2$ |
| 0.00020 | 3.5065 | −0.4254 | 1.3829 × 10$^{11}$ | 1.2493 × 10$^3$ | 0.26700050 | 0.0000 | 1.0000 | 0.7149 | 4.6119 × 10$^2$ | 2.2006 × 10$^2$ |
| 0.00030 | 3.5079 | −0.5368 | 1.2084 × 10$^{11}$ | 1.6361 × 10$^3$ | 0.26700050 | 0.0000 | 1.0000 | 0.7149 | 4.6033 × 10$^2$ | 2.2021 × 10$^2$ |
| 0.00050 | 3.5106 | −0.6758 | 1.0168 × 10$^{11}$ | 2.3106 × 10$^3$ | 0.26700050 | 0.0000 | 1.0000 | 0.7149 | 4.5788 × 10$^2$ | 2.1965 × 10$^2$ |
| 0.00070 | 3.5122 | −0.7694 | 9.0624 × 10$^{10}$ | 2.9089 × 10$^3$ | 0.26700050 | 0.0000 | 1.0000 | 0.7149 | 4.5648 × 10$^2$ | 2.1928 × 10$^2$ |
| 0.00100 | 3.5143 | −0.8678 | 8.0158 × 10$^{10}$ | 3.7182 × 10$^3$ | 0.26700050 | 0.0000 | 1.0000 | 0.7149 | 4.5444 × 10$^2$ | 2.1858 × 10$^2$ |
| 0.00200 | 3.5176 | −1.0634 | 6.3016 × 10$^{10}$ | 6.0161 × 10$^3$ | 0.26700050 | 0.0000 | 1.0000 | 0.7149 | 4.5139 × 10$^2$ | 2.1751 × 10$^2$ |
| 0.00300 | 3.5192 | −1.1800 | 5.4704 × 10$^{10}$ | 7.9833 × 10$^3$ | 0.26700050 | 0.0000 | 1.0000 | 0.7149 | 4.5000 × 10$^2$ | 2.1715 × 10$^2$ |
| 0.00500 | 3.5209 | −1.3274 | 4.5805 × 10$^{10}$ | 1.1386 × 10$^4$ | 0.26700050 | 0.0000 | 1.0000 | 0.7149 | 4.4882 × 10$^2$ | 2.1662 × 10$^2$ |
| 0.00700 | 3.5215 | −1.4258 | 4.0779 × 10$^{10}$ | 1.4367 × 10$^4$ | 0.26700050 | 0.0000 | 1.0000 | 0.7149 | 4.4892 × 10$^2$ | 2.1678 × 10$^2$ |
| 0.01000 | 3.5221 | −1.5299 | 3.6071 × 10$^{10}$ | 1.8362 × 10$^4$ | 0.26700050 | 0.0000 | 1.0000 | 0.7149 | 4.4934 × 10$^2$ | 2.1704 × 10$^2$ |
| 0.02000 | 3.5223 | −1.7342 | 2.8486 × 10$^{10}$ | 2.9441 × 10$^4$ | 0.26700050 | 0.0001 | 0.9999 | 0.7149 | 4.5300 × 10$^2$ | 2.1879 × 10$^2$ |
| 0.03000 | 3.5220 | −1.8545 | 2.4843 × 10$^{10}$ | 3.8710 × 10$^4$ | 0.26700050 | 0.0013 | 0.9987 | 0.7149 | 4.5712 × 10$^2$ | 2.2073 × 10$^2$ |
| 0.05000 | 3.5208 | −2.0073 | 2.0952 × 10$^{10}$ | 5.4421 × 10$^4$ | 0.26700050 | 0.0115 | 0.9885 | 0.7149 | 4.6341 × 10$^2$ | 2.2379 × 10$^2$ |
| 0.07000 | 3.5198 | −2.1061 | 1.8781 × 10$^{10}$ | 6.7728 × 10$^4$ | 0.26700050 | 0.0397 | 0.9603 | 0.7149 | 4.6292 × 10$^2$ | 2.2455 × 10$^2$ |
| 0.10000 | 3.5183 | −2.2079 | 1.6824 × 10$^{10}$ | 8.4404 × 10$^4$ | 0.26700050 | 0.1293 | 0.8707 | 0.7149 | 4.4793 × 10$^2$ | 2.2024 × 10$^2$ |
| 0.20000 | 3.5155 | −2.3554 | 1.4376 × 10$^{10}$ | 1.1559 × 10$^5$ | 0.26700050 | 0.6599 | 0.3401 | 0.7148 | 3.8656 × 10$^2$ | 1.8951 × 10$^2$ |
| 0.30000 | 3.5147 | −2.3874 | 1.3907 × 10$^{10}$ | 1.2352 × 10$^5$ | 0.26700050 | 0.9200 | 0.0800 | 0.7148 | 3.6964 × 10$^2$ | 1.7877 × 10$^2$ |
| 0.53224 | 3.5145 | −2.3950 | 1.3799 × 10$^{10}$ | 1.2546 × 10$^5$ | 0.26700051 | 0.9966 | 0.0034 | 0.7146 | 3.6562 × 10$^2$ | 1.7612 × 10$^2$ |

0.190$M_\odot$

| Age(Gyr) | logT$_{eff}$ | logL/L$_\odot$ | R(cm) | g(cm/s$^2$) | Y$_c$ | L(PP1) | L$_g$ | X$_{env}$ | $\tau_{gc}$ (day) | $\tau_{lc}$ (day) |
|---|---|---|---|---|---|---|---|---|---|---|
| 0.00002 | 3.5138 | 0.2429 | 2.8853 × 10$^{11}$ | 3.0292 × 10$^2$ | 0.26700050 | 0.0000 | 1.0000 | 0.7149 | 4.5251 × 10$^2$ | 2.1109 × 10$^2$ |
| 0.00010 | 3.5116 | −0.1961 | 1.7586 × 10$^{11}$ | 8.1541 × 10$^2$ | 0.26700050 | 0.0000 | 1.0000 | 0.7149 | 4.6006 × 10$^2$ | 2.1843 × 10$^2$ |
| 0.00020 | 3.5122 | −0.3886 | 1.4052 × 10$^{11}$ | 1.2770 × 10$^3$ | 0.26700050 | 0.0000 | 1.0000 | 0.7149 | 4.6133 × 10$^2$ | 2.2021 × 10$^2$ |
| 0.00030 | 3.5131 | −0.5010 | 1.2293 × 10$^{11}$ | 1.6687 × 10$^3$ | 0.26700050 | 0.0000 | 1.0000 | 0.7149 | 4.6113 × 10$^2$ | 2.2067 × 10$^2$ |
| 0.00050 | 3.5151 | −0.6419 | 1.0360 × 10$^{11}$ | 2.3497 × 10$^3$ | 0.26700050 | 0.0000 | 1.0000 | 0.7149 | 4.5977 × 10$^2$ | 2.2058 × 10$^2$ |
| 0.00070 | 3.5166 | −0.7349 | 9.2434 × 10$^{10}$ | 2.9515 × 10$^3$ | 0.26700050 | 0.0000 | 1.0000 | 0.7149 | 4.5851 × 10$^2$ | 2.2028 × 10$^2$ |
| 0.00100 | 3.5182 | −0.8342 | 8.1839 × 10$^{10}$ | 3.7651 × 10$^3$ | 0.26700050 | 0.0000 | 1.0000 | 0.7149 | 4.5706 × 10$^2$ | 2.1986 × 10$^2$ |
| 0.00200 | 3.5212 | −1.0296 | 6.4461 × 10$^{10}$ | 6.0689 × 10$^3$ | 0.26700050 | 0.0000 | 1.0000 | 0.7149 | 4.5449 × 10$^2$ | 2.1900 × 10$^2$ |
| 0.00300 | 3.5226 | −1.1456 | 5.6016 × 10$^{10}$ | 8.0367 × 10$^3$ | 0.26700050 | 0.0000 | 1.0000 | 0.7149 | 4.5324 × 10$^2$ | 2.1868 × 10$^2$ |
| 0.00500 | 3.5240 | −1.2938 | 4.6941 × 10$^{10}$ | 1.1445 × 10$^4$ | 0.26700050 | 0.0000 | 1.0000 | 0.7149 | 4.5268 × 10$^2$ | 2.1850 × 10$^2$ |
| 0.00700 | 3.5249 | −1.3908 | 4.1810 × 10$^{10}$ | 1.4426 × 10$^4$ | 0.26700050 | 0.0000 | 1.0000 | 0.7149 | 4.5244 × 10$^2$ | 2.1845 × 10$^2$ |
| 0.01000 | 3.5252 | −1.4957 | 3.6996 × 10$^{10}$ | 1.8424 × 10$^4$ | 0.26700050 | 0.0000 | 1.0000 | 0.7149 | 4.5360 × 10$^2$ | 2.1901 × 10$^2$ |
| 0.02000 | 3.5255 | −1.6994 | 2.9224 × 10$^{10}$ | 2.9526 × 10$^4$ | 0.26700050 | 0.0002 | 0.9998 | 0.7149 | 4.5779 × 10$^2$ | 2.2106 × 10$^2$ |
| 0.03000 | 3.5249 | −1.8202 | 2.5494 × 10$^{10}$ | 3.8800 × 10$^4$ | 0.26700050 | 0.0017 | 0.9983 | 0.7149 | 4.6329 × 10$^2$ | 2.2343 × 10$^2$ |
| 0.05000 | 3.5237 | −1.9725 | 2.1516 × 10$^{10}$ | 5.4473 × 10$^4$ | 0.26700050 | 0.0136 | 0.9864 | 0.7149 | 4.7129 × 10$^2$ | 2.2706 × 10$^2$ |
| 0.07000 | 3.5226 | −2.0717 | 1.9293 × 10$^{10}$ | 6.7750 × 10$^4$ | 0.26700050 | 0.0461 | 0.9539 | 0.7149 | 4.7144 × 10$^2$ | 2.2832 × 10$^2$ |
| 0.10000 | 3.5212 | −2.1719 | 1.7303 × 10$^{10}$ | 8.4228 × 10$^4$ | 0.26700050 | 0.1477 | 0.8523 | 0.7149 | 4.5412 × 10$^2$ | 2.2342 × 10$^2$ |
| 0.20000 | 3.5186 | −2.3128 | 1.4886 × 10$^{10}$ | 1.1380 × 10$^5$ | 0.26700050 | 0.6992 | 0.3008 | 0.7148 | 3.9008 × 10$^2$ | 1.9141 × 10$^2$ |
| 0.30000 | 3.5180 | −2.3403 | 1.4462 × 10$^{10}$ | 1.2057 × 10$^5$ | 0.26700050 | 0.9335 | 0.0665 | 0.7148 | 3.7458 × 10$^2$ | 1.8154 × 10$^2$ |
| 0.45870 | 3.5179 | −2.3454 | 1.4384 × 10$^{10}$ | 1.2188 × 10$^5$ | 0.26700051 | 0.9968 | 0.0032 | 0.7147 | 3.7159 × 10$^2$ | 1.7959 × 10$^2$ |

0.200$M_\odot$

| Age(Gyr) | logT$_{eff}$ | logL/L$_\odot$ | R(cm) | g(cm/s$^2$) | Y$_c$ | L(PP1) | L$_g$ | X$_{env}$ | $\tau_{gc}$ (day) | $\tau_{lc}$ (day) |
|---|---|---|---|---|---|---|---|---|---|---|
| 0.00002 | 3.5200 | 0.2782 | 2.9213 × 10$^{11}$ | 3.1105 × 10$^2$ | 0.26700050 | 0.0000 | 1.0000 | 0.7149 | 4.5171 × 10$^2$ | 2.1102 × 10$^2$ |
| 0.00010 | 3.5178 | −0.1591 | 1.7837 × 10$^{11}$ | 8.3436 × 10$^2$ | 0.26700050 | 0.0000 | 1.0000 | 0.7149 | 4.5916 × 10$^2$ | 2.1815 × 10$^2$ |
| 0.00020 | 3.5176 | −0.3540 | 1.4266 × 10$^{11}$ | 1.3043 × 10$^3$ | 0.26700050 | 0.0000 | 1.0000 | 0.7149 | 4.6149 × 10$^2$ | 2.2041 × 10$^2$ |
| 0.00030 | 3.5183 | −0.4663 | 1.2491 × 10$^{11}$ | 1.7013 × 10$^3$ | 0.26700050 | 0.0000 | 1.0000 | 0.7149 | 4.6150 × 10$^2$ | 2.2092 × 10$^2$ |
| 0.00050 | 3.5197 | −0.6084 | 1.0540 × 10$^{11}$ | 2.3893 × 10$^3$ | 0.26700050 | 0.0000 | 1.0000 | 0.7149 | 4.6095 × 10$^2$ | 2.2119 × 10$^2$ |
| 0.00070 | 3.5209 | −0.7020 | 9.4120 × 10$^{10}$ | 2.9965 × 10$^3$ | 0.26700050 | 0.0000 | 1.0000 | 0.7149 | 4.6015 × 10$^2$ | 2.2109 × 10$^2$ |
| 0.00100 | 3.5221 | −0.8020 | 8.3416 × 10$^{10}$ | 3.8149 × 10$^3$ | 0.26700050 | 0.0000 | 1.0000 | 0.7149 | 4.5927 × 10$^2$ | 2.2093 × 10$^2$ |
| 0.00200 | 3.5247 | −0.9971 | 6.5830 × 10$^{10}$ | 6.1253 × 10$^3$ | 0.26700050 | 0.0000 | 1.0000 | 0.7149 | 4.5716 × 10$^2$ | 2.2028 × 10$^2$ |
| 0.00300 | 3.5260 | −1.1131 | 5.7252 × 10$^{10}$ | 8.0983 × 10$^3$ | 0.26700050 | 0.0000 | 1.0000 | 0.7149 | 4.5632 × 10$^2$ | 2.2003 × 10$^2$ |
| 0.00500 | 3.5274 | −1.2605 | 4.8021 × 10$^{10}$ | 1.1511 × 10$^4$ | 0.26700050 | 0.0000 | 1.0000 | 0.7149 | 4.5570 × 10$^2$ | 2.1997 × 10$^2$ |
| 0.00700 | 3.5279 | −1.3588 | 4.2783 × 10$^{10}$ | 1.4502 × 10$^4$ | 0.26700050 | 0.0000 | 1.0000 | 0.7149 | 4.5625 × 10$^2$ | 2.2025 × 10$^2$ |
| 0.01000 | 3.5284 | −1.4624 | 3.7873 × 10$^{10}$ | 1.8506 × 10$^4$ | 0.26700050 | 0.0000 | 1.0000 | 0.7149 | 4.5714 × 10$^2$ | 2.2074 × 10$^2$ |
| 0.02000 | 3.5283 | −1.6670 | 2.9939 × 10$^{10}$ | 2.9613 × 10$^4$ | 0.26700050 | 0.0002 | 0.9998 | 0.7149 | 4.6310 × 10$^2$ | 2.2335 × 10$^2$ |
| 0.03000 | 3.5280 | −1.7865 | 2.6130 × 10$^{10}$ | 3.8876 × 10$^4$ | 0.26700050 | 0.0021 | 0.9979 | 0.7149 | 4.6920 × 10$^2$ | 2.2595 × 10$^2$ |
| 0.05000 | 3.5267 | −1.9391 | 2.2057 × 10$^{10}$ | 5.4562 × 10$^4$ | 0.26700050 | 0.0156 | 0.9844 | 0.7149 | 4.7973 × 10$^2$ | 2.3045 × 10$^2$ |
| 0.07000 | 3.5256 | −2.0373 | 1.9792 × 10$^{10}$ | 6.7762 × 10$^4$ | 0.26700050 | 0.0525 | 0.9475 | 0.7149 | 4.8019 × 10$^2$ | 2.3195 × 10$^2$ |
| 0.10000 | 3.5241 | −2.1368 | 1.7775 × 10$^{10}$ | 8.4018 × 10$^4$ | 0.26700050 | 0.1650 | 0.8350 | 0.7149 | 4.6074 × 10$^2$ | 2.2687 × 10$^2$ |
| 0.20000 | 3.5217 | −2.2714 | 1.5391 × 10$^{10}$ | 1.1206 × 10$^5$ | 0.26700050 | 0.7273 | 0.2727 | 0.7148 | 3.9421 × 10$^2$ | 1.9364 × 10$^2$ |
| 0.30000 | 3.5212 | −2.2957 | 1.5004 × 10$^{10}$ | 1.1792 × 10$^5$ | 0.26700050 | 0.9445 | 0.0555 | 0.7148 | 3.7984 × 10$^2$ | 1.8461 × 10$^2$ |
| 0.49957 | 3.5210 | −2.3007 | 1.4927 × 10$^{10}$ | 1.1913 × 10$^5$ | 0.26700053 | 0.9972 | 0.0028 | 0.7146 | 3.7693 × 10$^2$ | 1.8271 × 10$^2$ |

Table 1. Continued

| Age(Gyr) | log$T_{eff}$ | log$L/L_\odot$ | R(cm) | g(cm/s$^2$) | $Y_c$ | L(PP1) | $L_g$ | $X_{env}$ | $\tau_{gc}$ (day) | $\tau_{lc}$ (day) |
|---|---|---|---|---|---|---|---|---|---|---|
| | | | | 0.210$M_\odot$ | | | | | | |
| 0.00002 | 3.5258 | 0.3111 | 2.9537 × 10$^{11}$ | 3.1948 × 10$^2$ | 0.26700050 | 0.0000 | 1.0000 | 0.7149 | 4.5104 × 10$^2$ | 2.1097 × 10$^2$ |
| 0.00010 | 3.5235 | −0.1246 | 1.8080 × 10$^{11}$ | 8.5266 × 10$^2$ | 0.26700050 | 0.0000 | 1.0000 | 0.7149 | 4.5860 × 10$^2$ | 2.1802 × 10$^2$ |
| 0.00020 | 3.5229 | −0.3203 | 1.4471 × 10$^{11}$ | 1.3309 × 10$^3$ | 0.26700050 | 0.0000 | 1.0000 | 0.7149 | 4.6142 × 10$^2$ | 2.2045 × 10$^2$ |
| 0.00030 | 3.5233 | −0.4336 | 1.2678 × 10$^{11}$ | 1.7341 × 10$^3$ | 0.26700050 | 0.0000 | 1.0000 | 0.7149 | 4.6192 × 10$^2$ | 2.2120 × 10$^2$ |
| 0.00050 | 3.5243 | −0.5763 | 1.0708 × 10$^{11}$ | 2.4306 × 10$^3$ | 0.26700050 | 0.0000 | 1.0000 | 0.7149 | 4.6189 × 10$^2$ | 2.2168 × 10$^2$ |
| 0.00070 | 3.5252 | −0.6702 | 9.5708 × 10$^{10}$ | 3.0428 × 10$^3$ | 0.26700050 | 0.0000 | 1.0000 | 0.7149 | 4.6139 × 10$^2$ | 2.2173 × 10$^2$ |
| 0.00100 | 3.5263 | −0.7701 | 8.4880 × 10$^{10}$ | 3.8686 × 10$^3$ | 0.26700050 | 0.0000 | 1.0000 | 0.7149 | 4.6071 × 10$^2$ | 2.2165 × 10$^2$ |
| 0.00200 | 3.5284 | −0.9659 | 6.7088 × 10$^{10}$ | 6.1926 × 10$^3$ | 0.26700050 | 0.0000 | 1.0000 | 0.7149 | 4.5935 × 10$^2$ | 2.2133 × 10$^2$ |
| 0.00300 | 3.5295 | −1.0818 | 5.8417 × 10$^{10}$ | 8.1675 × 10$^3$ | 0.26700050 | 0.0000 | 1.0000 | 0.7149 | 4.5879 × 10$^2$ | 2.2125 × 10$^2$ |
| 0.00500 | 3.5306 | −1.2296 | 4.9026 × 10$^{10}$ | 1.1596 × 10$^4$ | 0.26700050 | 0.0000 | 1.0000 | 0.7149 | 4.5872 × 10$^2$ | 2.2134 × 10$^2$ |
| 0.00700 | 3.5312 | −1.3270 | 4.3699 × 10$^{10}$ | 1.4596 × 10$^4$ | 0.26700050 | 0.0000 | 1.0000 | 0.7149 | 4.5923 × 10$^2$ | 2.2158 × 10$^2$ |
| 0.01000 | 3.5315 | −1.4315 | 3.8701 × 10$^{10}$ | 1.8609 × 10$^4$ | 0.26700050 | 0.0000 | 1.0000 | 0.7149 | 4.6095 × 10$^2$ | 2.2238 × 10$^2$ |
| 0.02000 | 3.5315 | −1.6350 | 3.0615 × 10$^{10}$ | 2.9738 × 10$^4$ | 0.26700050 | 0.0004 | 0.9996 | 0.7149 | 4.6772 × 10$^2$ | 2.2535 × 10$^2$ |
| 0.03000 | 3.5309 | −1.7552 | 2.6726 × 10$^{10}$ | 3.9022 × 10$^4$ | 0.26700050 | 0.0024 | 0.9976 | 0.7149 | 4.7570 × 10$^2$ | 2.2858 × 10$^2$ |
| 0.05000 | 3.5296 | −1.9073 | 2.2569 × 10$^{10}$ | 5.4720 × 10$^4$ | 0.26700050 | 0.0182 | 0.9818 | 0.7149 | 4.8875 × 10$^2$ | 2.3407 × 10$^2$ |
| 0.07000 | 3.5284 | −2.0061 | 2.0258 × 10$^{10}$ | 6.7916 × 10$^4$ | 0.26700050 | 0.0599 | 0.9401 | 0.7149 | 4.9063 × 10$^2$ | 2.3622 × 10$^2$ |
| 0.10000 | 3.5270 | −2.1037 | 1.8217 × 10$^{10}$ | 8.3987 × 10$^4$ | 0.26700050 | 0.1846 | 0.8154 | 0.7149 | 4.6822 × 10$^2$ | 2.3060 × 10$^2$ |
| 0.20000 | 3.5247 | −2.2322 | 1.5882 × 10$^{10}$ | 1.1050 × 10$^5$ | 0.26700050 | 0.7631 | 0.2369 | 0.7148 | 3.9876 × 10$^2$ | 1.9641 × 10$^2$ |
| 0.30000 | 3.5242 | −2.2532 | 1.5540 × 10$^{10}$ | 1.1541 × 10$^5$ | 0.26700050 | 0.9590 | 0.0410 | 0.7148 | 3.8586 × 10$^2$ | 1.8822 × 10$^2$ |
| 0.58319 | 3.5241 | −2.2573 | 1.5473 × 10$^{10}$ | 1.1641 × 10$^5$ | 0.26700058 | 0.9998 | 0.0002 | 0.7145 | 3.8320 × 10$^2$ | 1.8653 × 10$^2$ |
| | | | | 0.220$M_\odot$ | | | | | | |
| 0.00002 | 3.5309 | 0.3428 | 2.9921 × 10$^{11}$ | 3.2616 × 10$^2$ | 0.26700050 | 0.0000 | 1.0000 | 0.7149 | 4.5099 × 10$^2$ | 2.1118 × 10$^2$ |
| 0.00010 | 3.5290 | −0.0906 | 1.8330 × 10$^{11}$ | 8.6909 × 10$^2$ | 0.26700050 | 0.0000 | 1.0000 | 0.7149 | 4.5792 × 10$^2$ | 2.1781 × 10$^2$ |
| 0.00020 | 3.5283 | −0.2863 | 1.4675 × 10$^{11}$ | 1.3558 × 10$^3$ | 0.26700050 | 0.0000 | 1.0000 | 0.7149 | 4.6079 × 10$^2$ | 2.2022 × 10$^2$ |
| 0.00030 | 3.5281 | −0.4017 | 1.2860 × 10$^{11}$ | 1.7655 × 10$^3$ | 0.26700050 | 0.0000 | 1.0000 | 0.7149 | 4.6213 × 10$^2$ | 2.2137 × 10$^2$ |
| 0.00050 | 3.5287 | −0.5454 | 1.0870 × 10$^{11}$ | 2.4711 × 10$^3$ | 0.26700050 | 0.0000 | 1.0000 | 0.7149 | 4.6263 × 10$^2$ | 2.2207 × 10$^2$ |
| 0.00070 | 3.5295 | −0.6394 | 9.7206 × 10$^{10}$ | 3.0902 × 10$^3$ | 0.26700050 | 0.0000 | 1.0000 | 0.7149 | 4.6238 × 10$^2$ | 2.2222 × 10$^2$ |
| 0.00100 | 3.5304 | −0.7396 | 8.6275 × 10$^{10}$ | 3.9228 × 10$^3$ | 0.26700050 | 0.0000 | 1.0000 | 0.7149 | 4.6198 × 10$^2$ | 2.2229 × 10$^2$ |
| 0.00200 | 3.5319 | −0.9365 | 6.8284 × 10$^{10}$ | 6.2623 × 10$^3$ | 0.26700050 | 0.0000 | 1.0000 | 0.7149 | 4.6145 × 10$^2$ | 2.2236 × 10$^2$ |
| 0.00300 | 3.5330 | −1.0519 | 5.9490 × 10$^{10}$ | 8.2505 × 10$^3$ | 0.26700050 | 0.0000 | 1.0000 | 0.7149 | 4.6098 × 10$^2$ | 2.2227 × 10$^2$ |
| 0.00500 | 3.5340 | −1.1992 | 4.9986 × 10$^{10}$ | 1.1686 × 10$^4$ | 0.26700050 | 0.0000 | 1.0000 | 0.7149 | 4.6131 × 10$^2$ | 2.2250 × 10$^2$ |
| 0.00700 | 3.5343 | −1.2975 | 4.4566 × 10$^{10}$ | 1.4702 × 10$^4$ | 0.26700050 | 0.0000 | 1.0000 | 0.7149 | 4.6226 × 10$^2$ | 2.2305 × 10$^2$ |
| 0.01000 | 3.5347 | −1.4011 | 3.9489 × 10$^{10}$ | 1.8725 × 10$^4$ | 0.26700050 | 0.0000 | 1.0000 | 0.7149 | 4.6408 × 10$^2$ | 2.2380 × 10$^2$ |
| 0.02000 | 3.5344 | −1.6057 | 3.1242 × 10$^{10}$ | 2.9915 × 10$^4$ | 0.26700050 | 0.0005 | 0.9995 | 0.7149 | 4.7275 × 10$^2$ | 2.2760 × 10$^2$ |
| 0.03000 | 3.5339 | −1.7250 | 2.7290 × 10$^{10}$ | 3.9207 × 10$^4$ | 0.26700050 | 0.0029 | 0.9971 | 0.7149 | 4.8216 × 10$^2$ | 2.3116 × 10$^2$ |
| 0.05000 | 3.5326 | −1.8766 | 2.3061 × 10$^{10}$ | 5.4906 × 10$^4$ | 0.26700050 | 0.0208 | 0.9792 | 0.7149 | 4.9916 × 10$^2$ | 2.3776 × 10$^2$ |
| 0.07000 | 3.5314 | −1.9747 | 2.0714 × 10$^{10}$ | 6.8050 × 10$^4$ | 0.26700050 | 0.0675 | 0.9325 | 0.7149 | 5.0216 × 10$^2$ | 2.4061 × 10$^2$ |
| 0.10000 | 3.5299 | −2.0716 | 1.8652 × 10$^{10}$ | 8.3929 × 10$^4$ | 0.26700050 | 0.2048 | 0.7952 | 0.7149 | 4.7661 × 10$^2$ | 2.3471 × 10$^2$ |
| 0.20000 | 3.5277 | −2.1938 | 1.6371 × 10$^{10}$ | 1.0895 × 10$^5$ | 0.26700050 | 0.7914 | 0.2086 | 0.7148 | 4.0408 × 10$^2$ | 1.9964 × 10$^2$ |
| 0.30000 | 3.5273 | −2.2125 | 1.6052 × 10$^{10}$ | 1.1333 × 10$^5$ | 0.26700051 | 0.9599 | 0.0401 | 0.7147 | 3.9167 × 10$^2$ | 1.9182 × 10$^2$ |
| 0.44256 | 3.5272 | −2.2157 | 1.5997 × 10$^{10}$ | 1.1409 × 10$^5$ | 0.26700054 | 0.9959 | 0.0040 | 0.7146 | 3.8949 × 10$^2$ | 1.9044 × 10$^2$ |
| | | | | 0.230$M_\odot$ | | | | | | |
| 0.00002 | 3.5359 | 0.3788 | 3.0477 × 10$^{11}$ | 3.2865 × 10$^2$ | 0.26700050 | 0.0000 | 1.0000 | 0.7149 | 4.5074 × 10$^2$ | 2.1125 × 10$^2$ |
| 0.00010 | 3.5343 | −0.0528 | 1.8685 × 10$^{11}$ | 8.7439 × 10$^2$ | 0.26700050 | 0.0000 | 1.0000 | 0.7149 | 4.5719 × 10$^2$ | 2.1754 × 10$^2$ |
| 0.00020 | 3.5333 | −0.2501 | 1.4956 × 10$^{11}$ | 1.3647 × 10$^3$ | 0.26700050 | 0.0000 | 1.0000 | 0.7149 | 4.6057 × 10$^2$ | 2.2017 × 10$^2$ |
| 0.00030 | 3.5332 | −0.3649 | 1.3111 × 10$^{11}$ | 1.7758 × 10$^3$ | 0.26700050 | 0.0000 | 1.0000 | 0.7149 | 4.6177 × 10$^2$ | 2.2120 × 10$^2$ |
| 0.00050 | 3.5334 | −0.5097 | 1.1088 × 10$^{11}$ | 2.4830 × 10$^3$ | 0.26700050 | 0.0000 | 1.0000 | 0.7149 | 4.6281 × 10$^2$ | 2.2216 × 10$^2$ |
| 0.00070 | 3.5337 | −0.6052 | 9.9184 × 10$^{10}$ | 3.1031 × 10$^3$ | 0.26700050 | 0.0000 | 1.0000 | 0.7149 | 4.6318 × 10$^2$ | 2.2260 × 10$^2$ |
| 0.00100 | 3.5344 | −0.7056 | 8.8081 × 10$^{10}$ | 3.9347 × 10$^3$ | 0.26700050 | 0.0000 | 1.0000 | 0.7149 | 4.6305 × 10$^2$ | 2.2278 × 10$^2$ |
| 0.00200 | 3.5357 | −0.9024 | 6.9797 × 10$^{10}$ | 6.2661 × 10$^3$ | 0.26700050 | 0.0000 | 1.0000 | 0.7149 | 4.6290 × 10$^2$ | 2.2302 × 10$^2$ |
| 0.00300 | 3.5364 | −1.0185 | 6.0880 × 10$^{10}$ | 8.2360 × 10$^3$ | 0.26700050 | 0.0000 | 1.0000 | 0.7149 | 4.6306 × 10$^2$ | 2.2326 × 10$^2$ |
| 0.00500 | 3.5372 | −1.1652 | 5.1226 × 10$^{10}$ | 1.1633 × 10$^4$ | 0.26700050 | 0.0000 | 1.0000 | 0.7149 | 4.6366 × 10$^2$ | 2.2363 × 10$^2$ |
| 0.00700 | 3.5376 | −1.2627 | 4.5688 × 10$^{10}$ | 1.4624 × 10$^4$ | 0.26700050 | 0.0000 | 1.0000 | 0.7149 | 4.6473 × 10$^2$ | 2.2414 × 10$^2$ |
| 0.01000 | 3.5378 | −1.3675 | 4.0469 × 10$^{10}$ | 1.8639 × 10$^4$ | 0.26700050 | 0.0000 | 1.0000 | 0.7149 | 4.6713 × 10$^2$ | 2.2528 × 10$^2$ |
| 0.02000 | 3.5375 | −1.5724 | 3.2002 × 10$^{10}$ | 2.9807 × 10$^4$ | 0.26700050 | 0.0006 | 0.9994 | 0.7149 | 4.7723 × 10$^2$ | 2.2945 × 10$^2$ |
| 0.03000 | 3.5370 | −1.6926 | 2.7938 × 10$^{10}$ | 3.9109 × 10$^4$ | 0.26700050 | 0.0032 | 0.9968 | 0.7149 | 4.8876 × 10$^2$ | 2.3381 × 10$^2$ |
| 0.05000 | 3.5356 | −1.8442 | 2.3608 × 10$^{10}$ | 5.4771 × 10$^4$ | 0.26700050 | 0.0227 | 0.9773 | 0.7149 | 5.1085 × 10$^2$ | 2.4178 × 10$^2$ |
| 0.07000 | 3.5343 | −1.9428 | 2.1206 × 10$^{10}$ | 6.7880 × 10$^4$ | 0.26700050 | 0.0737 | 0.9263 | 0.7149 | 5.1688 × 10$^2$ | 2.4562 × 10$^2$ |
| 0.10000 | 3.5329 | −2.0378 | 1.9130 × 10$^{10}$ | 8.3415 × 10$^4$ | 0.26700050 | 0.2184 | 0.7816 | 0.7149 | 4.8797 × 10$^2$ | 2.3983 × 10$^2$ |
| 0.20000 | 3.5307 | −2.1570 | 1.6849 × 10$^{10}$ | 1.0753 × 10$^5$ | 0.26700050 | 0.7962 | 0.2038 | 0.7148 | 4.1028 × 10$^2$ | 2.0352 × 10$^2$ |
| 0.30000 | 3.5302 | −2.1751 | 1.6537 × 10$^{10}$ | 1.1163 × 10$^5$ | 0.26700051 | 0.9582 | 0.0418 | 0.7147 | 3.9788 × 10$^2$ | 1.9574 × 10$^2$ |
| 0.47450 | 3.5301 | −2.1780 | 1.6488 × 10$^{10}$ | 1.1230 × 10$^5$ | 0.26700059 | 0.9988 | 0.0012 | 0.7146 | 3.9574 × 10$^2$ | 1.9440 × 10$^2$ |

Table 1. Continued

| Age(Gyr) | logT$_{\text{eff}}$ | logL/L$_\odot$ | R(cm) | g(cm/s$^2$) | Y$_c$ | L(PP1) | L$_g$ | X$_{\text{env}}$ | $\tau_{gc}$ (day) | $\tau_{lc}$ (day) |
|---|---|---|---|---|---|---|---|---|---|---|
| \multicolumn{11}{c}{0.240$M_\odot$} |
| 0.00002 | 3.5406 | 0.4024 | 3.0646 × 10$^{11}$ | 3.3917 × 10$^2$ | 0.26700050 | 0.0000 | 1.0000 | 0.7149 | 4.5074 × 10$^2$ | 2.1152 × 10$^2$ |
| 0.00010 | 3.5391 | −0.0235 | 1.8902 × 10$^{11}$ | 8.9155 × 10$^2$ | 0.26700050 | 0.0000 | 1.0000 | 0.7149 | 4.5694 × 10$^2$ | 2.1756 × 10$^2$ |
| 0.00020 | 3.5379 | −0.2207 | 1.5142 × 10$^{11}$ | 1.3893 × 10$^3$ | 0.26700050 | 0.0000 | 1.0000 | 0.7149 | 4.6048 × 10$^2$ | 2.2019 × 10$^2$ |
| 0.00030 | 3.5377 | −0.3355 | 1.3279 × 10$^{11}$ | 1.8063 × 10$^3$ | 0.26700050 | 0.0000 | 1.0000 | 0.7149 | 4.6180 × 10$^2$ | 2.2128 × 10$^2$ |
| 0.00050 | 3.5378 | −0.4805 | 1.1237 × 10$^{11}$ | 2.5226 × 10$^3$ | 0.26700050 | 0.0000 | 1.0000 | 0.7149 | 4.6302 × 10$^2$ | 2.2230 × 10$^2$ |
| 0.00070 | 3.5379 | −0.5762 | 1.0057 × 10$^{11}$ | 3.1496 × 10$^3$ | 0.26700050 | 0.0000 | 1.0000 | 0.7149 | 4.6368 × 10$^2$ | 2.2287 × 10$^2$ |
| 0.00100 | 3.5383 | −0.6775 | 8.9345 × 10$^{10}$ | 3.9904 × 10$^3$ | 0.26700050 | 0.0000 | 1.0000 | 0.7149 | 4.6400 × 10$^2$ | 2.2325 × 10$^2$ |
| 0.00200 | 3.5392 | −0.8748 | 7.0877 × 10$^{10}$ | 6.3408 × 10$^3$ | 0.26700050 | 0.0000 | 1.0000 | 0.7149 | 4.6440 × 10$^2$ | 2.2374 × 10$^2$ |
| 0.00300 | 3.5399 | −0.9905 | 6.1866 × 10$^{10}$ | 8.3226 × 10$^3$ | 0.26700050 | 0.0000 | 1.0000 | 0.7149 | 4.6468 × 10$^2$ | 2.2402 × 10$^2$ |
| 0.00500 | 3.5405 | −1.1373 | 5.2093 × 10$^{10}$ | 1.1738 × 10$^4$ | 0.26700050 | 0.0000 | 1.0000 | 0.7149 | 4.6575 × 10$^2$ | 2.2460 × 10$^2$ |
| 0.00700 | 3.5407 | −1.2352 | 4.6491 × 10$^{10}$ | 1.4738 × 10$^4$ | 0.26700050 | 0.0000 | 1.0000 | 0.7149 | 4.6733 × 10$^2$ | 2.2534 × 10$^2$ |
| 0.01000 | 3.5409 | −1.3393 | 4.1209 × 10$^{10}$ | 1.8758 × 10$^4$ | 0.26700050 | 0.0000 | 1.0000 | 0.7149 | 4.7009 × 10$^2$ | 2.2654 × 10$^2$ |
| 0.02000 | 3.5405 | −1.5449 | 3.2582 × 10$^{10}$ | 3.0006 × 10$^4$ | 0.26700050 | 0.0007 | 0.9993 | 0.7149 | 4.8225 × 10$^2$ | 2.3148 × 10$^2$ |
| 0.03000 | 3.5399 | −1.6652 | 2.8450 × 10$^{10}$ | 3.9355 × 10$^4$ | 0.26700050 | 0.0038 | 0.9962 | 0.7149 | 4.9644 × 10$^2$ | 2.3661 × 10$^2$ |
| 0.05000 | 3.5386 | −1.8165 | 2.4048 × 10$^{10}$ | 5.5080 × 10$^4$ | 0.26700050 | 0.0257 | 0.9743 | 0.7149 | 5.2547 × 10$^2$ | 2.4620 × 10$^2$ |
| 0.07000 | 3.5373 | −1.9137 | 2.1627 × 10$^{10}$ | 6.8102 × 10$^4$ | 0.26700050 | 0.0822 | 0.9178 | 0.7149 | 5.3498 × 10$^2$ | 2.5112 × 10$^2$ |
| 0.10000 | 3.5358 | −2.0082 | 1.9527 × 10$^{10}$ | 8.3540 × 10$^4$ | 0.26700050 | 0.2364 | 0.7636 | 0.7149 | 5.0038 × 10$^2$ | 2.4538 × 10$^2$ |
| 0.20000 | 3.5337 | −2.1208 | 1.7323 × 10$^{10}$ | 1.0615 × 10$^5$ | 0.26700050 | 0.8182 | 0.1818 | 0.7148 | 4.1713 × 10$^2$ | 2.0789 × 10$^2$ |
| 0.30000 | 3.5334 | −2.1356 | 1.7054 × 10$^{10}$ | 1.0952 × 10$^5$ | 0.26700052 | 0.9630 | 0.0370 | 0.7147 | 4.0573 × 10$^2$ | 2.0084 × 10$^2$ |
| 0.43111 | 3.5333 | −2.1386 | 1.6999 × 10$^{10}$ | 1.1024 × 10$^5$ | 0.26700059 | 0.9947 | 0.0053 | 0.7146 | 4.0334 × 10$^2$ | 1.9936 × 10$^2$ |
| \multicolumn{11}{c}{0.250$M_\odot$} |
| 0.00002 | 3.5448 | 0.4693 | 3.2475 × 10$^{11}$ | 3.1462 × 10$^2$ | 0.26700050 | 0.0000 | 1.0000 | 0.7149 | 4.5080 × 10$^2$ | 2.1135 × 10$^2$ |
| 0.00010 | 3.5437 | 0.0146 | 1.9337 × 10$^{11}$ | 8.8734 × 10$^2$ | 0.26700050 | 0.0000 | 1.0000 | 0.7149 | 4.5664 × 10$^2$ | 2.1748 × 10$^2$ |
| 0.00020 | 3.5426 | −0.1863 | 1.5417 × 10$^{11}$ | 1.3961 × 10$^3$ | 0.26700050 | 0.0000 | 1.0000 | 0.7149 | 4.6004 × 10$^2$ | 2.2003 × 10$^2$ |
| 0.00030 | 3.5421 | −0.3040 | 1.3499 × 10$^{11}$ | 1.8208 × 10$^3$ | 0.26700050 | 0.0000 | 1.0000 | 0.7149 | 4.6192 × 10$^2$ | 2.2138 × 10$^2$ |
| 0.00050 | 3.5418 | −0.4509 | 1.1410 × 10$^{11}$ | 2.5485 × 10$^3$ | 0.26700050 | 0.0000 | 1.0000 | 0.7149 | 4.6349 × 10$^2$ | 2.2256 × 10$^2$ |
| 0.00070 | 3.5421 | −0.5465 | 1.0209 × 10$^{11}$ | 3.1838 × 10$^3$ | 0.26700050 | 0.0000 | 1.0000 | 0.7149 | 4.6396 × 10$^2$ | 2.2303 × 10$^2$ |
| 0.00100 | 3.5422 | −0.6490 | 9.0699 × 10$^{10}$ | 4.0335 × 10$^3$ | 0.26700050 | 0.0000 | 1.0000 | 0.7149 | 4.6476 × 10$^2$ | 2.2363 × 10$^2$ |
| 0.00200 | 3.5429 | −0.8468 | 7.1981 × 10$^{10}$ | 6.4041 × 10$^3$ | 0.26700050 | 0.0000 | 1.0000 | 0.7149 | 4.6549 × 10$^2$ | 2.2426 × 10$^2$ |
| 0.00300 | 3.5432 | −0.9637 | 6.2836 × 10$^{10}$ | 8.4036 × 10$^3$ | 0.26700050 | 0.0000 | 1.0000 | 0.7149 | 4.6638 × 10$^2$ | 2.2482 × 10$^2$ |
| 0.00500 | 3.5436 | −1.1108 | 5.2937 × 10$^{10}$ | 1.1840 × 10$^4$ | 0.26700050 | 0.0000 | 1.0000 | 0.7149 | 4.6789 × 10$^2$ | 2.2559 × 10$^2$ |
| 0.00700 | 3.5439 | −1.2083 | 4.7266 × 10$^{10}$ | 1.4852 × 10$^4$ | 0.26700050 | 0.0000 | 1.0000 | 0.7149 | 4.6973 × 10$^2$ | 2.2641 × 10$^2$ |
| 0.01000 | 3.5439 | −1.3125 | 4.1908 × 10$^{10}$ | 1.8892 × 10$^4$ | 0.26700050 | 0.0000 | 1.0000 | 0.7149 | 4.7301 × 10$^2$ | 2.2787 × 10$^2$ |
| 0.02000 | 3.5435 | −1.5173 | 3.3174 × 10$^{10}$ | 3.0151 × 10$^4$ | 0.26700050 | 0.0008 | 0.9992 | 0.7149 | 4.8737 × 10$^2$ | 2.3348 × 10$^2$ |
| 0.03000 | 3.5428 | −1.6375 | 2.8973 × 10$^{10}$ | 3.9528 × 10$^4$ | 0.26700050 | 0.0044 | 0.9956 | 0.7149 | 5.0478 × 10$^2$ | 2.3949 × 10$^2$ |
| 0.05000 | 3.5416 | −1.7881 | 2.4506 × 10$^{10}$ | 5.5252 × 10$^4$ | 0.26700050 | 0.0289 | 0.9711 | 0.7149 | 5.4368 × 10$^2$ | 2.5091 × 10$^2$ |
| 0.07000 | 3.5402 | −1.8855 | 2.2043 × 10$^{10}$ | 6.8286 × 10$^4$ | 0.26700050 | 0.0909 | 0.9091 | 0.7149 | 5.6022 × 10$^2$ | 2.5746 × 10$^2$ |
| 0.10000 | 3.5387 | −1.9793 | 1.9921 × 10$^{10}$ | 8.3607 × 10$^4$ | 0.26700050 | 0.2593 | 0.7407 | 0.7149 | 5.1503 × 10$^2$ | 2.5155 × 10$^2$ |
| 0.20000 | 3.5367 | −2.0861 | 1.7781 × 10$^{10}$ | 1.0495 × 10$^5$ | 0.26700050 | 0.8402 | 0.1597 | 0.7148 | 4.2461 × 10$^2$ | 2.1275 × 10$^2$ |
| 0.30000 | 3.5364 | −2.1003 | 1.7516 × 10$^{10}$ | 1.0814 × 10$^5$ | 0.26700053 | 0.9664 | 0.0336 | 0.7147 | 4.1309 × 10$^2$ | 2.0573 × 10$^2$ |
| 0.44279 | 3.5364 | −2.1025 | 1.7476 × 10$^{10}$ | 1.0864 × 10$^5$ | 0.26700066 | 0.9982 | 0.0018 | 0.7146 | 4.1106 × 10$^2$ | 2.0452 × 10$^2$ |
| \multicolumn{11}{c}{0.260$M_\odot$} |
| 0.00002 | 3.5488 | 0.4921 | 3.2731 × 10$^{11}$ | 3.2212 × 10$^2$ | 0.26700050 | 0.0000 | 1.0000 | 0.7149 | 4.5122 × 10$^2$ | 2.1175 × 10$^2$ |
| 0.00010 | 3.5481 | 0.0431 | 1.9573 × 10$^{11}$ | 9.0073 × 10$^2$ | 0.26700050 | 0.0000 | 1.0000 | 0.7149 | 4.5633 × 10$^2$ | 2.1740 × 10$^2$ |
| 0.00020 | 3.5470 | −0.1580 | 1.5610 × 10$^{11}$ | 1.4161 × 10$^3$ | 0.26700050 | 0.0000 | 1.0000 | 0.7149 | 4.5999 × 10$^2$ | 2.1999 × 10$^2$ |
| 0.00030 | 3.5464 | −0.2759 | 1.3667 × 10$^{11}$ | 1.8473 × 10$^3$ | 0.26700050 | 0.0000 | 1.0000 | 0.7149 | 4.6175 × 10$^2$ | 2.2134 × 10$^2$ |
| 0.00050 | 3.5460 | −0.4234 | 1.1554 × 10$^{11}$ | 2.5850 × 10$^3$ | 0.26700050 | 0.0000 | 1.0000 | 0.7149 | 4.6365 × 10$^2$ | 2.2267 × 10$^2$ |
| 0.00070 | 3.5459 | −0.5200 | 1.0342 × 10$^{11}$ | 3.2262 × 10$^3$ | 0.26700050 | 0.0000 | 1.0000 | 0.7149 | 4.6458 × 10$^2$ | 2.2335 × 10$^2$ |
| 0.00100 | 3.5460 | −0.6223 | 9.1903 × 10$^{10}$ | 4.0857 × 10$^3$ | 0.26700050 | 0.0000 | 1.0000 | 0.7149 | 4.6537 × 10$^2$ | 2.2393 × 10$^2$ |
| 0.00200 | 3.5463 | −0.8210 | 7.3002 × 10$^{10}$ | 6.4751 × 10$^3$ | 0.26700050 | 0.0000 | 1.0000 | 0.7149 | 4.6670 × 10$^2$ | 2.2483 × 10$^2$ |
| 0.00300 | 3.5465 | −0.9376 | 6.3766 × 10$^{10}$ | 8.4868 × 10$^3$ | 0.26700050 | 0.0000 | 1.0000 | 0.7149 | 4.6777 × 10$^2$ | 2.2547 × 10$^2$ |
| 0.00500 | 3.5468 | −1.0849 | 5.3756 × 10$^{10}$ | 1.1942 × 10$^4$ | 0.26700050 | 0.0000 | 1.0000 | 0.7149 | 4.6980 × 10$^2$ | 2.2646 × 10$^2$ |
| 0.00700 | 3.5468 | −1.1826 | 4.8026 × 10$^{10}$ | 1.4961 × 10$^4$ | 0.26700050 | 0.0000 | 1.0000 | 0.7149 | 4.7225 × 10$^2$ | 2.2750 × 10$^2$ |
| 0.01000 | 3.5469 | −1.2866 | 4.2597 × 10$^{10}$ | 1.9018 × 10$^4$ | 0.26700050 | 0.0000 | 1.0000 | 0.7149 | 4.7603 × 10$^2$ | 2.2912 × 10$^2$ |
| 0.02000 | 3.5463 | −1.4912 | 3.3740 × 10$^{10}$ | 3.0313 × 10$^4$ | 0.26700050 | 0.0010 | 0.9990 | 0.7149 | 4.9293 × 10$^2$ | 2.3557 × 10$^2$ |
| 0.03000 | 3.5457 | −1.6120 | 2.9451 × 10$^{10}$ | 3.9786 × 10$^4$ | 0.26700050 | 0.0049 | 0.9951 | 0.7149 | 5.1480 × 10$^2$ | 2.4272 × 10$^2$ |
| 0.05000 | 3.5444 | −1.7625 | 2.4915 × 10$^{10}$ | 5.5591 × 10$^4$ | 0.26700050 | 0.0323 | 0.9677 | 0.7149 | 5.5709 × 10$^2$ | 2.5658 × 10$^2$ |
| 0.07000 | 3.5432 | −1.8586 | 2.2429 × 10$^{10}$ | 6.8598 × 10$^4$ | 0.26700050 | 0.0989 | 0.9011 | 0.7149 | 6.0290 × 10$^2$ | 2.6507 × 10$^2$ |
| 0.10000 | 3.5417 | −1.9501 | 2.0319 × 10$^{10}$ | 8.3580 × 10$^4$ | 0.26700050 | 0.2837 | 0.7163 | 0.7149 | 5.3493 × 10$^2$ | 2.5894 × 10$^2$ |
| 0.20000 | 3.5397 | −2.0523 | 1.8228 × 10$^{10}$ | 1.0385 × 10$^5$ | 0.26700050 | 0.8515 | 0.1485 | 0.7148 | 4.3340 × 10$^2$ | 2.1855 × 10$^2$ |
| 0.30000 | 3.5395 | −2.0649 | 1.7986 × 10$^{10}$ | 1.0667 × 10$^5$ | 0.26700055 | 0.9710 | 0.0290 | 0.7147 | 4.2185 × 10$^2$ | 2.1164 × 10$^2$ |
| 0.43200 | 3.5395 | −2.0665 | 1.7954 × 10$^{10}$ | 1.0705 × 10$^5$ | 0.26700072 | 0.9988 | 0.0012 | 0.7146 | 4.1997 × 10$^2$ | 2.1058 × 10$^2$ |

Table 1. Continued

| Age(Gyr) | log$T_{\text{eff}}$ | log$L/L_\odot$ | R(cm) | g(cm/s$^2$) | $Y_c$ | L(PP1) | $L_g$ | $X_{\text{env}}$ | $\tau_{gc}$ (day) | $\tau_{lc}$ (day) |
|---|---|---|---|---|---|---|---|---|---|---|
| | | | | 0.270$M_\odot$ | | | | | | |
| 0.00002 | 3.5525 | 0.5136 | 3.2978 × 10$^{11}$ | 3.2950 × 10$^2$ | 0.26700050 | 0.0000 | 1.0000 | 0.7149 | 4.5181 × 10$^2$ | 2.1224 × 10$^2$ |
| 0.00010 | 3.5522 | 0.0691 | 1.9796 × 10$^{11}$ | 9.1441 × 10$^2$ | 0.26700050 | 0.0000 | 1.0000 | 0.7149 | 4.5641 × 10$^2$ | 2.1753 × 10$^2$ |
| 0.00020 | 3.5509 | −0.1321 | 1.5795 × 10$^{11}$ | 1.4365 × 10$^3$ | 0.26700050 | 0.0000 | 1.0000 | 0.7149 | 4.6006 × 10$^2$ | 2.2017 × 10$^2$ |
| 0.00030 | 3.5505 | −0.2489 | 1.3836 × 10$^{11}$ | 1.8720 × 10$^3$ | 0.26700050 | 0.0000 | 1.0000 | 0.7149 | 4.6176 × 10$^2$ | 2.2139 × 10$^2$ |
| 0.00050 | 3.5499 | −0.3969 | 1.1701 × 10$^{11}$ | 2.6173 × 10$^3$ | 0.26700050 | 0.0000 | 1.0000 | 0.7149 | 4.6392 × 10$^2$ | 2.2283 × 10$^2$ |
| 0.00070 | 3.5496 | −0.4943 | 1.0472 × 10$^{11}$ | 3.2675 × 10$^3$ | 0.26700050 | 0.0000 | 1.0000 | 0.7149 | 4.6512 × 10$^2$ | 2.2364 × 10$^2$ |
| 0.00100 | 3.5496 | −0.5963 | 9.3121 × 10$^{10}$ | 4.1325 × 10$^3$ | 0.26700050 | 0.0000 | 1.0000 | 0.7149 | 4.6595 × 10$^2$ | 2.2422 × 10$^2$ |
| 0.00200 | 3.5495 | −0.7961 | 7.4011 × 10$^{10}$ | 6.5422 × 10$^3$ | 0.26700050 | 0.0000 | 1.0000 | 0.7149 | 4.6802 × 10$^2$ | 2.2547 × 10$^2$ |
| 0.00300 | 3.5497 | −0.9127 | 6.4663 × 10$^{10}$ | 8.5705 × 10$^3$ | 0.26700050 | 0.0000 | 1.0000 | 0.7149 | 4.6923 × 10$^2$ | 2.2615 × 10$^2$ |
| 0.00500 | 3.5498 | −1.0602 | 5.4546 × 10$^{10}$ | 1.2045 × 10$^4$ | 0.26700050 | 0.0000 | 1.0000 | 0.7149 | 4.7178 × 10$^2$ | 2.2736 × 10$^2$ |
| 0.00700 | 3.5497 | −1.1578 | 4.8763 × 10$^{10}$ | 1.5071 × 10$^4$ | 0.26700050 | 0.0000 | 1.0000 | 0.7149 | 4.7461 × 10$^2$ | 2.2859 × 10$^2$ |
| 0.01000 | 3.5497 | −1.2619 | 4.3258 × 10$^{10}$ | 1.9150 × 10$^4$ | 0.26700050 | 0.0000 | 1.0000 | 0.7149 | 4.7912 × 10$^2$ | 2.3048 × 10$^2$ |
| 0.02000 | 3.5491 | −1.4663 | 3.4279 × 10$^{10}$ | 3.0496 × 10$^4$ | 0.26700050 | 0.0011 | 0.9989 | 0.7149 | 4.9913 × 10$^2$ | 2.3782 × 10$^2$ |
| 0.03000 | 3.5484 | −1.5861 | 2.9957 × 10$^{10}$ | 3.9932 × 10$^4$ | 0.26700050 | 0.0055 | 0.9945 | 0.7149 | 5.2623 × 10$^2$ | 2.4607 × 10$^2$ |
| 0.05000 | 3.5472 | −1.7364 | 2.5342 × 10$^{10}$ | 5.5799 × 10$^4$ | 0.26700050 | 0.0354 | 0.9646 | 0.7149 | 6.1646 × 10$^2$ | 2.6286 × 10$^2$ |
| 0.07000 | 3.5460 | −1.8317 | 2.2833 × 10$^{10}$ | 6.8733 × 10$^4$ | 0.26700976 | 0.1082 | 0.8918 | 0.7149 | 7.4429 × 10$^2$ | 2.7280 × 10$^2$ |
| 0.10000 | 3.5445 | −1.9232 | 2.0686 × 10$^{10}$ | 8.3742 × 10$^4$ | 0.26700050 | 0.3012 | 0.6988 | 0.7149 | 5.6470 × 10$^2$ | 2.6843 × 10$^2$ |
| 0.20000 | 3.5427 | −2.0201 | 1.8658 × 10$^{10}$ | 1.0294 × 10$^5$ | 0.26700051 | 0.8698 | 0.1302 | 0.7148 | 4.4309 × 10$^2$ | 2.2503 × 10$^2$ |
| 0.30000 | 3.5426 | −2.0298 | 1.8467 × 10$^{10}$ | 1.0507 × 10$^5$ | 0.26700057 | 0.9759 | 0.0240 | 0.7147 | 4.3287 × 10$^2$ | 2.1906 × 10$^2$ |
| 0.39674 | 3.5425 | −2.0314 | 1.8435 × 10$^{10}$ | 1.0544 × 10$^5$ | 0.26700072 | 0.9956 | 0.0044 | 0.7146 | 4.3093 × 10$^2$ | 2.1800 × 10$^2$ |
| | | | | 0.280$M_\odot$ | | | | | | |
| 0.00002 | 3.5561 | 0.5353 | 3.3255 × 10$^{11}$ | 3.3604 × 10$^2$ | 0.26700050 | 0.0000 | 1.0000 | 0.7149 | 4.5232 × 10$^2$ | 2.1268 × 10$^2$ |
| 0.00010 | 3.5559 | 0.0942 | 2.0031 × 10$^{11}$ | 9.2623 × 10$^2$ | 0.26700050 | 0.0000 | 1.0000 | 0.7149 | 4.5672 × 10$^2$ | 2.1779 × 10$^2$ |
| 0.00020 | 3.5548 | −0.1062 | 1.5987 × 10$^{11}$ | 1.4541 × 10$^3$ | 0.26700050 | 0.0000 | 1.0000 | 0.7149 | 4.6019 × 10$^2$ | 2.2029 × 10$^2$ |
| 0.00030 | 3.5542 | −0.2235 | 1.4004 × 10$^{11}$ | 1.8950 × 10$^3$ | 0.26700050 | 0.0000 | 1.0000 | 0.7149 | 4.6204 × 10$^2$ | 2.2157 × 10$^2$ |
| 0.00050 | 3.5536 | −0.3711 | 1.1847 × 10$^{11}$ | 2.6480 × 10$^3$ | 0.26700050 | 0.0000 | 1.0000 | 0.7149 | 4.6408 × 10$^2$ | 2.2295 × 10$^2$ |
| 0.00070 | 3.5531 | −0.4693 | 1.0605 × 10$^{11}$ | 3.3041 × 10$^3$ | 0.26700050 | 0.0000 | 1.0000 | 0.7149 | 4.6572 × 10$^2$ | 2.2395 × 10$^2$ |
| 0.00100 | 3.5530 | −0.5718 | 9.4308 × 10$^{10}$ | 4.1784 × 10$^3$ | 0.26700050 | 0.0000 | 1.0000 | 0.7149 | 4.6681 × 10$^2$ | 2.2466 × 10$^2$ |
| 0.00200 | 3.5526 | −0.7721 | 7.5004 × 10$^{10}$ | 6.6059 × 10$^3$ | 0.26700050 | 0.0000 | 1.0000 | 0.7149 | 4.6931 × 10$^2$ | 2.2609 × 10$^2$ |
| 0.00300 | 3.5527 | −0.8888 | 6.5567 × 10$^{10}$ | 8.6445 × 10$^3$ | 0.26700050 | 0.0000 | 1.0000 | 0.7149 | 4.7079 × 10$^2$ | 2.2688 × 10$^2$ |
| 0.00500 | 3.5526 | −1.0363 | 5.5330 × 10$^{10}$ | 1.2139 × 10$^4$ | 0.26700050 | 0.0000 | 1.0000 | 0.7149 | 4.7373 × 10$^2$ | 2.2824 × 10$^2$ |
| 0.00700 | 3.5525 | −1.1339 | 4.9479 × 10$^{10}$ | 1.5180 × 10$^4$ | 0.26700050 | 0.0000 | 1.0000 | 0.7149 | 4.7708 × 10$^2$ | 2.2967 × 10$^2$ |
| 0.01000 | 3.5524 | −1.2382 | 4.3909 × 10$^{10}$ | 1.9275 × 10$^4$ | 0.26700050 | 0.0000 | 1.0000 | 0.7149 | 4.8246 × 10$^2$ | 2.3181 × 10$^2$ |
| 0.02000 | 3.5518 | −1.4420 | 3.4823 × 10$^{10}$ | 3.0646 × 10$^4$ | 0.26700050 | 0.0013 | 0.9987 | 0.7149 | 5.0590 × 10$^2$ | 2.4018 × 10$^2$ |
| 0.03000 | 3.5511 | −1.5617 | 3.0436 × 10$^{10}$ | 4.0116 × 10$^4$ | 0.26700050 | 0.0062 | 0.9938 | 0.7149 | 5.4056 × 10$^2$ | 2.4981 × 10$^2$ |
| 0.05000 | 3.5499 | −1.7119 | 2.5744 × 10$^{10}$ | 5.6073 × 10$^4$ | 0.26700420 | 0.0395 | 0.9605 | 0.7149 | 6.8319 × 10$^2$ | 2.6083 × 10$^2$ |
| 0.07000 | 3.5487 | −1.8072 | 2.3193 × 10$^{10}$ | 6.9088 × 10$^4$ | 0.26703303 | 0.1188 | 0.8812 | 0.7149 | 5.4447 × 10$^2$ | 2.3746 × 10$^2$ |
| 0.10000 | 3.5474 | −1.8960 | 2.1067 × 10$^{10}$ | 8.3730 × 10$^4$ | 0.26700050 | 0.3256 | 0.6744 | 0.7149 | 6.1164 × 10$^2$ | 2.7972 × 10$^2$ |
| 0.20000 | 3.5457 | −1.9877 | 1.9105 × 10$^{10}$ | 1.0182 × 10$^5$ | 0.26700051 | 0.8793 | 0.1207 | 0.7148 | 4.5630 × 10$^2$ | 2.3368 × 10$^2$ |
| 0.33655 | 3.5455 | −1.9986 | 1.8884 × 10$^{10}$ | 1.0422 × 10$^5$ | 0.26700066 | 0.9903 | 0.0097 | 0.7146 | 4.4311 × 10$^2$ | 2.2630 × 10$^2$ |
| | | | | 0.290$M_\odot$ | | | | | | |
| 0.00002 | 3.5595 | 0.5548 | 3.3479 × 10$^{11}$ | 3.4339 × 10$^2$ | 0.26700050 | 0.0000 | 1.0000 | 0.7149 | 4.5295 × 10$^2$ | 2.1319 × 10$^2$ |
| 0.00010 | 3.5596 | 0.1186 | 2.0260 × 10$^{11}$ | 9.3772 × 10$^2$ | 0.26700050 | 0.0000 | 1.0000 | 0.7149 | 4.5699 × 10$^2$ | 2.1797 × 10$^2$ |
| 0.00020 | 3.5584 | −0.0815 | 1.6176 × 10$^{11}$ | 1.4710 × 10$^3$ | 0.26700050 | 0.0000 | 1.0000 | 0.7149 | 4.6046 × 10$^2$ | 2.2047 × 10$^2$ |
| 0.00030 | 3.5577 | −0.1994 | 1.4168 × 10$^{11}$ | 1.9176 × 10$^3$ | 0.26700050 | 0.0000 | 1.0000 | 0.7149 | 4.6244 × 10$^2$ | 2.2181 × 10$^2$ |
| 0.00050 | 3.5570 | −0.3473 | 1.1987 × 10$^{11}$ | 2.6785 × 10$^3$ | 0.26700050 | 0.0000 | 1.0000 | 0.7149 | 4.6471 × 10$^2$ | 2.2328 × 10$^2$ |
| 0.00070 | 3.5566 | −0.4447 | 1.0737 × 10$^{11}$ | 3.3389 × 10$^3$ | 0.26700050 | 0.0000 | 1.0000 | 0.7149 | 4.6617 × 10$^2$ | 2.2419 × 10$^2$ |
| 0.00100 | 3.5562 | −0.5484 | 9.5478 × 10$^{10}$ | 4.2222 × 10$^3$ | 0.26700050 | 0.0000 | 1.0000 | 0.7149 | 4.6774 × 10$^2$ | 2.2511 × 10$^2$ |
| 0.00200 | 3.5557 | −0.7487 | 7.5982 × 10$^{10}$ | 6.6668 × 10$^3$ | 0.26700050 | 0.0000 | 1.0000 | 0.7149 | 4.7051 × 10$^2$ | 2.2666 × 10$^2$ |
| 0.00300 | 3.5555 | −0.8658 | 6.6457 × 10$^{10}$ | 8.7150 × 10$^3$ | 0.26700050 | 0.0000 | 1.0000 | 0.7149 | 4.7243 × 10$^2$ | 2.2765 × 10$^2$ |
| 0.00500 | 3.5553 | −1.0134 | 5.6110 × 10$^{10}$ | 1.2225 × 10$^4$ | 0.26700050 | 0.0000 | 1.0000 | 0.7149 | 4.7591 × 10$^2$ | 2.2922 × 10$^2$ |
| 0.00700 | 3.5552 | −1.1110 | 5.0181 × 10$^{10}$ | 1.5285 × 10$^4$ | 0.26700050 | 0.0000 | 1.0000 | 0.7149 | 4.7962 × 10$^2$ | 2.3077 × 10$^2$ |
| 0.01000 | 3.5550 | −1.2152 | 4.4557 × 10$^{10}$ | 1.9388 × 10$^4$ | 0.26700050 | 0.0000 | 1.0000 | 0.7149 | 4.8585 × 10$^2$ | 2.3328 × 10$^2$ |
| 0.02000 | 3.5543 | −1.4188 | 3.5348 × 10$^{10}$ | 3.0804 × 10$^4$ | 0.26700050 | 0.0014 | 0.9986 | 0.7149 | 5.1366 × 10$^2$ | 2.4275 × 10$^2$ |
| 0.03000 | 3.5536 | −1.5385 | 3.0900 × 10$^{10}$ | 4.0311 × 10$^4$ | 0.26700050 | 0.0068 | 0.9932 | 0.7149 | 5.5969 × 10$^2$ | 2.5408 × 10$^2$ |
| 0.05000 | 3.5524 | −1.6876 | 2.6167 × 10$^{10}$ | 5.6211 × 10$^4$ | 0.26701365 | 0.0429 | 0.9571 | 0.7149 | 5.5190 × 10$^2$ | 2.3801 × 10$^2$ |
| 0.07000 | 3.5511 | −1.7813 | 2.3602 × 10$^{10}$ | 6.9095 × 10$^4$ | 0.26704250 | 0.1288 | 0.8712 | 0.7149 | 6.6650 × 10$^2$ | 2.1515 × 10$^2$ |
| 0.10000 | 3.5501 | −1.8696 | 2.1445 × 10$^{10}$ | 8.3692 × 10$^4$ | 0.26709240 | 0.3469 | 0.6531 | 0.7149 | 5.1795 × 10$^2$ | 2.3511 × 10$^2$ |
| 0.20000 | 3.5486 | −1.9569 | 1.9536 × 10$^{10}$ | 1.0085 × 10$^5$ | 0.26700049 | 0.8943 | 0.1057 | 0.7148 | 4.7218 × 10$^2$ | 2.4400 × 10$^2$ |
| 0.33982 | 3.5484 | −1.9666 | 1.9334 × 10$^{10}$ | 1.0296 × 10$^5$ | 0.26700071 | 0.9919 | 0.0081 | 0.7146 | 4.5804 × 10$^2$ | 2.3654 × 10$^2$ |
| | | | | 0.300$M_\odot$ | | | | | | |
| 0.00002 | 3.5628 | 0.5681 | 3.3488 × 10$^{11}$ | 3.5505 × 10$^2$ | 0.26700050 | 0.0000 | 1.0000 | 0.7149 | 4.5354 × 10$^2$ | 2.1370 × 10$^2$ |

Table 1. Continued

| Age(Gyr) | log$T_{eff}$ | log$L/L_\odot$ | R(cm) | g(cm/s$^2$) | $Y_c$ | L(PP1) | $L_g$ | $X_{env}$ | $\tau_{gc}$ (day) | $\tau_{lc}$ (day) |
|---|---|---|---|---|---|---|---|---|---|---|
| 0.00010 | 3.5630 | 0.1369 | 2.0362 × 10$^{11}$ | 9.6031 × 10$^2$ | 0.26700050 | 0.0000 | 1.0000 | 0.7149 | 4.5731 × 10$^2$ | 2.1825 × 10$^2$ |
| 0.00020 | 3.5619 | −0.0624 | 1.6269 × 10$^{11}$ | 1.5043 × 10$^3$ | 0.26700050 | 0.0000 | 1.0000 | 0.7149 | 4.6068 × 10$^2$ | 2.2065 × 10$^2$ |
| 0.00030 | 3.5611 | −0.1805 | 1.4256 × 10$^{11}$ | 1.9592 × 10$^3$ | 0.26700050 | 0.0000 | 1.0000 | 0.7149 | 4.6292 × 10$^2$ | 2.2210 × 10$^2$ |
| 0.00050 | 3.5602 | −0.3289 | 1.2066 × 10$^{11}$ | 2.7351 × 10$^3$ | 0.26700050 | 0.0000 | 1.0000 | 0.7149 | 4.6543 × 10$^2$ | 2.2367 × 10$^2$ |
| 0.00070 | 3.5599 | −0.4257 | 1.0807 × 10$^{11}$ | 3.4093 × 10$^3$ | 0.26700050 | 0.0000 | 1.0000 | 0.7149 | 4.6669 × 10$^2$ | 2.2450 × 10$^2$ |
| 0.00100 | 3.5592 | −0.5301 | 9.6141 × 10$^{10}$ | 4.3078 × 10$^3$ | 0.26700050 | 0.0000 | 1.0000 | 0.7149 | 4.6864 × 10$^2$ | 2.2558 × 10$^2$ |
| 0.00200 | 3.5585 | −0.7307 | 7.6552 × 10$^{10}$ | 6.7945 × 10$^3$ | 0.26700050 | 0.0000 | 1.0000 | 0.7149 | 4.7179 × 10$^2$ | 2.2731 × 10$^2$ |
| 0.00300 | 3.5582 | −0.8482 | 6.6978 × 10$^{10}$ | 8.8757 × 10$^3$ | 0.26700050 | 0.0000 | 1.0000 | 0.7149 | 4.7415 × 10$^2$ | 2.2847 × 10$^2$ |
| 0.00500 | 3.5579 | −0.9963 | 5.6559 × 10$^{10}$ | 1.2447 × 10$^4$ | 0.26700050 | 0.0000 | 1.0000 | 0.7149 | 4.7827 × 10$^2$ | 2.3030 × 10$^2$ |
| 0.00700 | 3.5577 | −1.0938 | 5.0584 × 10$^{10}$ | 1.5561 × 10$^4$ | 0.26700050 | 0.0000 | 1.0000 | 0.7149 | 4.8257 × 10$^2$ | 2.3200 × 10$^2$ |
| 0.01000 | 3.5574 | −1.1981 | 4.4929 × 10$^{10}$ | 1.9725 × 10$^4$ | 0.26700050 | 0.0001 | 0.9999 | 0.7149 | 4.8991 × 10$^2$ | 2.3486 × 10$^2$ |
| 0.02000 | 3.5567 | −1.4008 | 3.5695 × 10$^{10}$ | 3.1250 × 10$^4$ | 0.26700050 | 0.0017 | 0.9983 | 0.7149 | 5.2396 × 10$^2$ | 2.4583 × 10$^2$ |
| 0.03000 | 3.5560 | −1.5198 | 3.1224 × 10$^{10}$ | 4.0841 × 10$^4$ | 0.26700050 | 0.0079 | 0.9921 | 0.7149 | 5.9063 × 10$^2$ | 2.5926 × 10$^2$ |
| 0.05000 | 3.5548 | −1.6681 | 2.6466 × 10$^{10}$ | 5.6845 × 10$^4$ | 0.26701933 | 0.0486 | 0.9514 | 0.7149 | 4.9236 × 10$^2$ | 2.2195 × 10$^2$ |
| 0.07000 | 3.5539 | −1.7606 | 2.3890 × 10$^{10}$ | 6.9765 × 10$^4$ | 0.26704824 | 0.1433 | 0.8567 | 0.7149 | 4.1715 × 10$^2$ | 1.9847 × 10$^2$ |
| 0.10000 | 3.5527 | −1.8460 | 2.1773 × 10$^{10}$ | 8.3992 × 10$^4$ | 0.26709809 | 0.3759 | 0.6241 | 0.7149 | 4.1320 × 10$^2$ | 1.9842 × 10$^2$ |
| 0.20000 | 3.5513 | −1.9271 | 1.9960 × 10$^{10}$ | 9.9939 × 10$^4$ | 0.26700044 | 0.9177 | 0.0822 | 0.7148 | 4.9276 × 10$^2$ | 2.5698 × 10$^2$ |
| 0.32246 | 3.5512 | −1.9342 | 1.9810 × 10$^{10}$ | 1.0146 × 10$^5$ | 0.26700065 | 0.9937 | 0.0063 | 0.7146 | 4.7908 × 10$^2$ | 2.5039 × 10$^2$ |

$0.400 M_\odot$

| Age(Gyr) | log$T_{eff}$ | log$L/L_\odot$ | R(cm) | g(cm/s$^2$) | $Y_c$ | L(PP1) | $L_g$ | $X_{env}$ | $\tau_{gc}$ (day) | $\tau_{lc}$ (day) |
|---|---|---|---|---|---|---|---|---|---|---|
| 0.00002 | 3.5868 | 0.7457 | 3.6782 × 10$^{11}$ | 3.9240 × 10$^2$ | 0.26700050 | 0.0000 | 1.0000 | 0.7149 | 4.6422 × 10$^2$ | 2.1992 × 10$^2$ |
| 0.00010 | 3.5891 | 0.3353 | 2.2690 × 10$^{11}$ | 1.0312 × 10$^3$ | 0.26700050 | 0.0000 | 1.0000 | 0.7149 | 4.6474 × 10$^2$ | 2.2233 × 10$^2$ |
| 0.00020 | 3.5883 | 0.1377 | 1.8143 × 10$^{11}$ | 1.6129 × 10$^3$ | 0.26700050 | 0.0000 | 1.0000 | 0.7149 | 4.6765 × 10$^2$ | 2.2435 × 10$^2$ |
| 0.00030 | 3.5872 | 0.0191 | 1.5902 × 10$^{11}$ | 2.0995 × 10$^3$ | 0.26700050 | 0.0000 | 1.0000 | 0.7149 | 4.7023 × 10$^2$ | 2.2588 × 10$^2$ |
| 0.00050 | 3.5859 | −0.1307 | 1.3466 × 10$^{11}$ | 2.9279 × 10$^3$ | 0.26700050 | 0.0000 | 1.0000 | 0.7149 | 4.7356 × 10$^2$ | 2.2777 × 10$^2$ |
| 0.00070 | 3.5850 | −0.2295 | 1.2069 × 10$^{11}$ | 3.6446 × 10$^3$ | 0.26700050 | 0.0000 | 1.0000 | 0.7149 | 4.7590 × 10$^2$ | 2.2903 × 10$^2$ |
| 0.00100 | 3.5837 | −0.3353 | 1.0748 × 10$^{11}$ | 4.5956 × 10$^3$ | 0.26700050 | 0.0000 | 1.0000 | 0.7149 | 4.7901 × 10$^2$ | 2.3064 × 10$^2$ |
| 0.00200 | 3.5814 | −0.5398 | 8.5856 × 10$^{10}$ | 7.2022 × 10$^3$ | 0.26700050 | 0.0000 | 1.0000 | 0.7149 | 4.8633 × 10$^2$ | 2.3407 × 10$^2$ |
| 0.00300 | 3.5801 | −0.6588 | 7.5290 × 10$^{10}$ | 9.3655 × 10$^3$ | 0.26700050 | 0.0000 | 1.0000 | 0.7149 | 4.9226 × 10$^2$ | 2.3664 × 10$^2$ |
| 0.00500 | 3.5786 | −0.8086 | 6.3807 × 10$^{10}$ | 1.3040 × 10$^4$ | 0.26700050 | 0.0000 | 1.0000 | 0.7149 | 5.0441 × 10$^2$ | 2.4140 × 10$^2$ |
| 0.00700 | 3.5777 | −0.9070 | 5.7207 × 10$^{10}$ | 1.6222 × 10$^4$ | 0.26700050 | 0.0000 | 1.0000 | 0.7149 | 5.1871 × 10$^2$ | 2.4632 × 10$^2$ |
| 0.01000 | 3.5770 | −1.0108 | 5.0939 × 10$^{10}$ | 2.0460 × 10$^4$ | 0.26700050 | 0.0003 | 0.9997 | 0.7149 | 5.4765 × 10$^2$ | 2.5453 × 10$^2$ |
| 0.02000 | 3.5755 | −1.2131 | 4.0624 × 10$^{10}$ | 3.2170 × 10$^4$ | 0.26700504 | 0.0039 | 0.9961 | 0.7149 | 5.0105 × 10$^2$ | 2.2555 × 10$^2$ |
| 0.03000 | 3.5749 | −1.3294 | 3.5640 × 10$^{10}$ | 4.1796 × 10$^4$ | 0.26701553 | 0.0163 | 0.9837 | 0.7149 | 3.7869 × 10$^2$ | 1.8276 × 10$^2$ |
| 0.05000 | 3.5747 | −1.4648 | 3.0521 × 10$^{10}$ | 5.6991 × 10$^4$ | 0.26704064 | 0.0866 | 0.9134 | 0.7149 | 2.8692 × 10$^2$ | 1.4271 × 10$^2$ |
| 0.07000 | 3.5753 | −1.5405 | 2.7901 × 10$^{10}$ | 6.8200 × 10$^4$ | 0.26707057 | 0.2310 | 0.7690 | 0.7149 | 2.4582 × 10$^2$ | 1.2270 × 10$^2$ |
| 0.10000 | 3.5758 | −1.6031 | 2.5892 × 10$^{10}$ | 7.9189 × 10$^4$ | 0.26712414 | 0.5240 | 0.4760 | 0.7149 | 2.1429 × 10$^2$ | 1.0787 × 10$^2$ |
| 0.17132 | 3.5751 | −1.6547 | 2.4485 × 10$^{10}$ | 8.8552 × 10$^4$ | 0.26729845 | 0.9900 | 0.0095 | 0.7150 | 2.0528 × 10$^2$ | 1.0370 × 10$^2$ |

$0.500 M_\odot$

| Age(Gyr) | log$T_{eff}$ | log$L/L_\odot$ | R(cm) | g(cm/s$^2$) | $Y_c$ | L(PP1) | $L_g$ | $X_{env}$ | $\tau_{gc}$ (day) | $\tau_{lc}$ (day) |
|---|---|---|---|---|---|---|---|---|---|---|
| 0.00002 | 3.6010 | 0.9248 | 4.2352 × 10$^{11}$ | 3.6997 × 10$^2$ | 0.26700050 | 0.0000 | 1.0000 | 0.7149 | 4.7876 × 10$^2$ | 2.2729 × 10$^2$ |
| 0.00010 | 3.6062 | 0.4973 | 2.5271 × 10$^{11}$ | 1.0391 × 10$^3$ | 0.26700050 | 0.0000 | 1.0000 | 0.7149 | 4.7516 × 10$^2$ | 2.2763 × 10$^2$ |
| 0.00020 | 3.6058 | 0.2954 | 2.0071 × 10$^{11}$ | 1.6474 × 10$^3$ | 0.26700050 | 0.0000 | 1.0000 | 0.7149 | 4.7773 × 10$^2$ | 2.2940 × 10$^2$ |
| 0.00030 | 3.6050 | 0.1756 | 1.7550 × 10$^{11}$ | 2.1547 × 10$^3$ | 0.26700050 | 0.0000 | 1.0000 | 0.7149 | 4.8009 × 10$^2$ | 2.3078 × 10$^2$ |
| 0.00050 | 3.6037 | 0.0243 | 1.4833 × 10$^{11}$ | 3.0160 × 10$^3$ | 0.26700050 | 0.0000 | 1.0000 | 0.7149 | 4.8365 × 10$^2$ | 2.3272 × 10$^2$ |
| 0.00070 | 3.6024 | −0.0764 | 1.3284 × 10$^{11}$ | 3.7607 × 10$^3$ | 0.26700050 | 0.0000 | 1.0000 | 0.7149 | 4.8688 × 10$^2$ | 2.3436 × 10$^2$ |
| 0.00100 | 3.6012 | −0.1825 | 1.1825 × 10$^{11}$ | 4.7459 × 10$^3$ | 0.26700050 | 0.0000 | 1.0000 | 0.7149 | 4.9072 × 10$^2$ | 2.3618 × 10$^2$ |
| 0.00200 | 3.5982 | −0.3896 | 9.4452 × 10$^{10}$ | 7.4386 × 10$^3$ | 0.26700050 | 0.0000 | 1.0000 | 0.7149 | 5.0249 × 10$^2$ | 2.4124 × 10$^2$ |
| 0.00300 | 3.5963 | −0.5107 | 8.2882 × 10$^{10}$ | 9.6605 × 10$^3$ | 0.26700050 | 0.0000 | 1.0000 | 0.7149 | 5.1455 × 10$^2$ | 2.4593 × 10$^2$ |
| 0.00500 | 3.5940 | −0.6625 | 7.0332 × 10$^{10}$ | 1.3416 × 10$^4$ | 0.26700050 | 0.0000 | 1.0000 | 0.7149 | 5.4510 × 10$^2$ | 2.5578 × 10$^2$ |
| 0.00700 | 3.5926 | −0.7618 | 6.3138 × 10$^{10}$ | 1.6647 × 10$^4$ | 0.26700050 | 0.0001 | 0.9999 | 0.7149 | 5.9911 × 10$^2$ | 2.6749 × 10$^2$ |
| 0.01000 | 3.5912 | −0.8669 | 5.6312 × 10$^{10}$ | 2.0927 × 10$^4$ | 0.26700170 | 0.0005 | 0.9995 | 0.7149 | 5.6711 × 10$^2$ | 2.4456 × 10$^2$ |
| 0.02000 | 3.5893 | −1.0651 | 4.5206 × 10$^{10}$ | 3.2473 × 10$^4$ | 0.26701050 | 0.0064 | 0.9936 | 0.7149 | 3.4656 × 10$^2$ | 1.6906 × 10$^2$ |
| 0.03000 | 3.5894 | −1.1704 | 4.0027 × 10$^{10}$ | 4.1420 × 10$^4$ | 0.26702119 | 0.0247 | 0.9753 | 0.7149 | 2.7517 × 10$^2$ | 1.3699 × 10$^2$ |
| 0.05000 | 3.5919 | −1.2785 | 3.4937 × 10$^{10}$ | 5.4367 × 10$^4$ | 0.26704761 | 0.1246 | 0.8754 | 0.7149 | 2.0891 × 10$^2$ | 1.0339 × 10$^2$ |
| 0.07000 | 3.5954 | −1.3244 | 3.2607 × 10$^{10}$ | 6.2414 × 10$^4$ | 0.26708058 | 0.3214 | 0.6786 | 0.7149 | 1.6790 × 10$^2$ | 8.5190 × 10$^1$ |
| 0.10000 | 3.5995 | −1.3512 | 3.1028 × 10$^{10}$ | 6.8930 × 10$^4$ | 0.26717083 | 0.6994 | 0.3002 | 0.7150 | 1.3903 × 10$^2$ | 7.1783 × 10$^1$ |
| 0.12793 | 3.5993 | −1.3721 | 3.0319 × 10$^{10}$ | 7.2190 × 10$^4$ | 0.26725855 | 0.9904 | 0.0083 | 0.7150 | 1.3415 × 10$^2$ | 6.9417 × 10$^1$ |

$0.600 M_\odot$

| Age(Gyr) | log$T_{eff}$ | log$L/L_\odot$ | R(cm) | g(cm/s$^2$) | $Y_c$ | L(PP1) | $L_g$ | $X_{env}$ | $\tau_{gc}$ (day) | $\tau_{lc}$ (day) |
|---|---|---|---|---|---|---|---|---|---|---|
| 0.00002 | 3.6106 | 1.0567 | 4.7170 × 10$^{11}$ | 3.5790 × 10$^2$ | 0.26700050 | 0.0000 | 1.0000 | 0.7149 | 4.9423 × 10$^2$ | 2.3506 × 10$^2$ |
| 0.00010 | 3.6187 | 0.6259 | 2.7663 × 10$^{11}$ | 1.0406 × 10$^3$ | 0.26700050 | 0.0000 | 1.0000 | 0.7149 | 4.8625 × 10$^2$ | 2.3318 × 10$^2$ |
| 0.00020 | 3.6192 | 0.4237 | 2.1869 × 10$^{11}$ | 1.6652 × 10$^3$ | 0.26700050 | 0.0000 | 1.0000 | 0.7149 | 4.8753 × 10$^2$ | 2.3429 × 10$^2$ |
| 0.00030 | 3.6185 | 0.3023 | 1.9081 × 10$^{11}$ | 2.1872 × 10$^3$ | 0.26700050 | 0.0000 | 1.0000 | 0.7149 | 4.9012 × 10$^2$ | 2.3573 × 10$^2$ |
| 0.00050 | 3.6170 | 0.1488 | 1.6099 × 10$^{11}$ | 3.0726 × 10$^3$ | 0.26700050 | 0.0000 | 1.0000 | 0.7149 | 4.9458 × 10$^2$ | 2.3799 × 10$^2$ |
| 0.00070 | 3.6159 | 0.0481 | 1.4410 × 10$^{11}$ | 3.8351 × 10$^3$ | 0.26700050 | 0.0000 | 1.0000 | 0.7149 | 4.9829 × 10$^2$ | 2.3975 × 10$^2$ |
| 0.00100 | 3.6144 | −0.0592 | 1.2822 × 10$^{11}$ | 4.8442 × 10$^3$ | 0.26700050 | 0.0000 | 1.0000 | 0.7149 | 5.0379 × 10$^2$ | 2.4217 × 10$^2$ |
| 0.00200 | 3.6113 | −0.2671 | 1.0238 × 10$^{11}$ | 7.5974 × 10$^3$ | 0.26700050 | 0.0000 | 1.0000 | 0.7149 | 5.2237 × 10$^2$ | 2.4941 × 10$^2$ |

Table 1. Continued

| Age(Gyr) | log$T_{eff}$ | log$L/L_\odot$ | R(cm) | g(cm/s$^2$) | $Y_c$ | L(PP1) | $L_g$ | $X_{env}$ | $\tau_{gc}$ (day) | $\tau_{lc}$ (day) |
|---|---|---|---|---|---|---|---|---|---|---|
| 0.00300 | 3.6093 | $-0.3887$ | $8.9841 \times 10^{10}$ | $9.8661 \times 10^3$ | 0.26700050 | 0.0000 | 1.0000 | 0.7149 | $5.4616 \times 10^2$ | $2.5721 \times 10^2$ |
| 0.00500 | 3.6066 | $-0.5418$ | $7.6275 \times 10^{10}$ | $1.3688 \times 10^4$ | 0.26700050 | 0.0001 | 0.9999 | 0.7149 | $6.5855 \times 10^2$ | $2.7689 \times 10^2$ |
| 0.00700 | 3.6048 | $-0.6421$ | $6.8512 \times 10^{10}$ | $1.6965 \times 10^4$ | 0.26700159 | 0.0002 | 0.9998 | 0.7149 | $5.2008 \times 10^2$ | $2.3317 \times 10^2$ |
| 0.01000 | 3.6031 | $-0.7466$ | $6.1218 \times 10^{10}$ | $2.1249 \times 10^4$ | 0.26700379 | 0.0008 | 0.9992 | 0.7149 | $4.0151 \times 10^2$ | $1.9271 \times 10^2$ |
| 0.02000 | 3.6020 | $-0.9319$ | $4.9704 \times 10^{10}$ | $3.2234 \times 10^4$ | 0.26701277 | 0.0092 | 0.9908 | 0.7149 | $2.7126 \times 10^2$ | $1.3425 \times 10^2$ |
| 0.03000 | 3.6041 | $-1.0176$ | $4.4604 \times 10^{10}$ | $4.0027 \times 10^4$ | 0.26702402 | 0.0347 | 0.9653 | 0.7149 | $2.1067 \times 10^2$ | $1.0614 \times 10^2$ |
| 0.05000 | 3.6131 | $-1.0707$ | $4.0257 \times 10^{10}$ | $4.9138 \times 10^4$ | 0.26705317 | 0.1792 | 0.8208 | 0.7149 | $1.4605 \times 10^2$ | $7.4386 \times 10^1$ |
| 0.07000 | 3.6243 | $-1.0581$ | $3.8799 \times 10^{10}$ | $5.2899 \times 10^4$ | 0.26710486 | 0.4749 | 0.5247 | 0.7149 | $1.1039 \times 10^2$ | $5.7095 \times 10^1$ |
| 0.09431 | 3.6278 | $-1.0764$ | $3.7373 \times 10^{10}$ | $5.7014 \times 10^4$ | 0.26723634 | 0.9828 | 0.0139 | 0.7150 | $9.5916 \times 10^1$ | $5.0670 \times 10^1$ |
| | | | | $0.700 M_\odot$ | | | | | | |
| 0.00002 | 3.6181 | 1.1540 | $5.0967 \times 10^{11}$ | $3.5766 \times 10^2$ | 0.26700050 | 0.0000 | 1.0000 | 0.7149 | $5.0888 \times 10^2$ | $2.4244 \times 10^2$ |
| 0.00010 | 3.6282 | 0.7299 | $2.9845 \times 10^{11}$ | $1.0431 \times 10^3$ | 0.26700050 | 0.0000 | 1.0000 | 0.7149 | $4.9769 \times 10^2$ | $2.3884 \times 10^2$ |
| 0.00020 | 3.6298 | 0.5296 | $2.3532 \times 10^{11}$ | $1.6777 \times 10^3$ | 0.26700050 | 0.0000 | 1.0000 | 0.7149 | $4.9774 \times 10^2$ | $2.3929 \times 10^2$ |
| 0.00030 | 3.6296 | 0.4093 | $2.0501 \times 10^{11}$ | $2.2104 \times 10^3$ | 0.26700050 | 0.0000 | 1.0000 | 0.7149 | $4.9962 \times 10^2$ | $2.4036 \times 10^2$ |
| 0.00050 | 3.6285 | 0.2555 | $1.7267 \times 10^{11}$ | $3.1161 \times 10^3$ | 0.26700050 | 0.0000 | 1.0000 | 0.7149 | $5.0453 \times 10^2$ | $2.4271 \times 10^2$ |
| 0.00070 | 3.6272 | 0.1534 | $1.5444 \times 10^{11}$ | $3.8952 \times 10^3$ | 0.26700050 | 0.0000 | 1.0000 | 0.7149 | $5.0971 \times 10^2$ | $2.4500 \times 10^2$ |
| 0.00100 | 3.6255 | 0.0451 | $1.3736 \times 10^{11}$ | $4.9242 \times 10^3$ | 0.26700050 | 0.0000 | 1.0000 | 0.7149 | $5.1764 \times 10^2$ | $2.4823 \times 10^2$ |
| 0.00200 | 3.6221 | $-0.1642$ | $1.0968 \times 10^{11}$ | $7.7236 \times 10^3$ | 0.26700050 | 0.0000 | 1.0000 | 0.7149 | $5.4958 \times 10^2$ | $2.5926 \times 10^2$ |
| 0.00300 | 3.6200 | $-0.2861$ | $9.6257 \times 10^{10}$ | $1.0027 \times 10^4$ | 0.26700050 | 0.0000 | 1.0000 | 0.7149 | $6.0766 \times 10^2$ | $2.7270 \times 10^2$ |
| 0.00500 | 3.6171 | $-0.4394$ | $8.1758 \times 10^{10}$ | $1.3899 \times 10^4$ | 0.26700132 | 0.0001 | 0.9999 | 0.7149 | $5.0512 \times 10^2$ | $2.3005 \times 10^2$ |
| 0.00700 | 3.6154 | $-0.5382$ | $7.3539 \times 10^{10}$ | $1.7179 \times 10^4$ | 0.26700263 | 0.0003 | 0.9997 | 0.7149 | $4.0073 \times 10^2$ | $1.9240 \times 10^2$ |
| 0.01000 | 3.6140 | $-0.6386$ | $6.5944 \times 10^{10}$ | $2.1365 \times 10^4$ | 0.26700486 | 0.0012 | 0.9988 | 0.7149 | $3.2558 \times 10^2$ | $1.6004 \times 10^2$ |
| 0.02000 | 3.6156 | $-0.7992$ | $5.4405 \times 10^{10}$ | $3.1388 \times 10^4$ | 0.26701422 | 0.0124 | 0.9876 | 0.7149 | $2.1634 \times 10^2$ | $1.0840 \times 10^2$ |
| 0.03000 | 3.6226 | $-0.8467$ | $4.9880 \times 10^{10}$ | $3.7342 \times 10^4$ | 0.26702643 | 0.0489 | 0.9511 | 0.7149 | $1.6140 \times 10^2$ | $8.1443 \times 10^1$ |
| 0.05000 | 3.6485 | $-0.7881$ | $4.7347 \times 10^{10}$ | $4.1445 \times 10^4$ | 0.26706425 | 0.2834 | 0.7164 | 0.7149 | $9.4789 \times 10^1$ | $4.9620 \times 10^1$ |
| 0.07052 | 3.6619 | $-0.7881$ | $4.4524 \times 10^{10}$ | $4.6866 \times 10^4$ | 0.26724304 | 0.9770 | 0.0154 | 0.7150 | $7.2282 \times 10^1$ | $3.8930 \times 10^1$ |
| | | | | $0.800 M_\odot$ | | | | | | |
| 0.00002 | 3.6246 | 1.2224 | $5.3495 \times 10^{11}$ | $3.7104 \times 10^2$ | 0.26700050 | 0.0000 | 1.0000 | 0.7149 | $5.2202 \times 10^2$ | $2.4913 \times 10^2$ |
| 0.00010 | 3.6360 | 0.8153 | $3.1780 \times 10^{11}$ | $1.0513 \times 10^3$ | 0.26700050 | 0.0000 | 1.0000 | 0.7149 | $5.0891 \times 10^2$ | $2.4437 \times 10^2$ |
| 0.00020 | 3.6385 | 0.6195 | $2.5066 \times 10^{11}$ | $1.6900 \times 10^3$ | 0.26700050 | 0.0000 | 1.0000 | 0.7149 | $5.0757 \times 10^2$ | $2.4412 \times 10^2$ |
| 0.00030 | 3.6390 | 0.5011 | $2.1820 \times 10^{11}$ | $2.2301 \times 10^3$ | 0.26700050 | 0.0000 | 1.0000 | 0.7149 | $5.0903 \times 10^2$ | $2.4489 \times 10^2$ |
| 0.00050 | 3.6384 | 0.3484 | $1.8353 \times 10^{11}$ | $3.1523 \times 10^3$ | 0.26700050 | 0.0000 | 1.0000 | 0.7149 | $5.1432 \times 10^2$ | $2.4724 \times 10^2$ |
| 0.00070 | 3.6373 | 0.2462 | $1.6399 \times 10^{11}$ | $3.9480 \times 10^3$ | 0.26700050 | 0.0000 | 1.0000 | 0.7149 | $5.2092 \times 10^2$ | $2.4993 \times 10^2$ |
| 0.00100 | 3.6356 | 0.1369 | $1.4574 \times 10^{11}$ | $4.9993 \times 10^3$ | 0.26700050 | 0.0000 | 1.0000 | 0.7149 | $5.3262 \times 10^2$ | $2.5434 \times 10^2$ |
| 0.00200 | 3.6317 | $-0.0747$ | $1.1633 \times 10^{11}$ | $7.8456 \times 10^3$ | 0.26700050 | 0.0000 | 1.0000 | 0.7149 | $5.9440 \times 10^2$ | $2.7166 \times 10^2$ |
| 0.00300 | 3.6292 | $-0.1976$ | $1.0213 \times 10^{11}$ | $1.0179 \times 10^4$ | 0.26700070 | 0.0000 | 1.0000 | 0.7149 | $6.1431 \times 10^2$ | $2.5921 \times 10^2$ |
| 0.00500 | 3.6264 | $-0.3493$ | $8.6890 \times 10^{10}$ | $1.4064 \times 10^4$ | 0.26700186 | 0.0001 | 0.9999 | 0.7149 | $4.0064 \times 10^2$ | $1.9511 \times 10^2$ |
| 0.00700 | 3.6250 | $-0.4444$ | $7.8390 \times 10^{10}$ | $1.7279 \times 10^4$ | 0.26700320 | 0.0005 | 0.9995 | 0.7149 | $3.3391 \times 10^2$ | $1.6398 \times 10^2$ |
| 0.01000 | 3.6246 | $-0.5362$ | $7.0660 \times 10^{10}$ | $2.1266 \times 10^4$ | 0.26700549 | 0.0016 | 0.9984 | 0.7149 | $2.7391 \times 10^2$ | $1.3547 \times 10^2$ |
| 0.02000 | 3.6318 | $-0.6538$ | $5.9682 \times 10^{10}$ | $2.9809 \times 10^4$ | 0.26701543 | 0.0169 | 0.9831 | 0.7149 | $1.7252 \times 10^2$ | $8.6824 \times 10^1$ |
| 0.03000 | 3.6508 | $-0.6241$ | $5.6579 \times 10^{10}$ | $3.3169 \times 10^4$ | 0.26702917 | 0.0732 | 0.9268 | 0.7149 | $1.1564 \times 10^2$ | $5.9332 \times 10^1$ |
| 0.05000 | 3.7049 | $-0.4783$ | $5.2179 \times 10^{10}$ | $3.8999 \times 10^4$ | 0.26717137 | 0.7737 | 0.2147 | 0.7150 | $5.7873 \times 10^1$ | $3.1555 \times 10^1$ |
| 0.05722 | 3.6967 | $-0.5594$ | $4.9346 \times 10^{10}$ | $4.3604 \times 10^4$ | 0.26737379 | 0.9780 | 0.0045 | 0.7150 | $5.8011 \times 10^1$ | $3.1660 \times 10^1$ |
| | | | | $0.900 M_\odot$ | | | | | | |
| 0.00002 | 3.6294 | 1.2958 | $5.6945 \times 10^{11}$ | $3.6836 \times 10^2$ | 0.26700050 | 0.0000 | 1.0000 | 0.7149 | $5.3555 \times 10^2$ | $2.5584 \times 10^2$ |
| 0.00010 | 3.6424 | 0.8941 | $3.3779 \times 10^{11}$ | $1.0469 \times 10^3$ | 0.26700050 | 0.0000 | 1.0000 | 0.7149 | $5.1970 \times 10^2$ | $2.4968 \times 10^2$ |
| 0.00020 | 3.6458 | 0.6997 | $2.6584 \times 10^{11}$ | $1.6902 \times 10^3$ | 0.26700050 | 0.0000 | 1.0000 | 0.7149 | $5.1767 \times 10^2$ | $2.4901 \times 10^2$ |
| 0.00030 | 3.6469 | 0.5825 | $2.3112 \times 10^{11}$ | $2.2362 \times 10^3$ | 0.26700050 | 0.0000 | 1.0000 | 0.7149 | $5.1882 \times 10^2$ | $2.4954 \times 10^2$ |
| 0.00050 | 3.6471 | 0.4312 | $1.9402 \times 10^{11}$ | $3.1731 \times 10^3$ | 0.26700050 | 0.0000 | 1.0000 | 0.7149 | $5.2472 \times 10^2$ | $2.5190 \times 10^2$ |
| 0.00070 | 3.6463 | 0.3293 | $1.7313 \times 10^{11}$ | $3.9851 \times 10^3$ | 0.26700050 | 0.0000 | 1.0000 | 0.7149 | $5.3341 \times 10^2$ | $2.5513 \times 10^2$ |
| 0.00100 | 3.6450 | 0.2205 | $1.5368 \times 10^{11}$ | $5.0578 \times 10^3$ | 0.26700050 | 0.0000 | 1.0000 | 0.7149 | $5.5026 \times 10^2$ | $2.6083 \times 10^2$ |
| 0.00200 | 3.6407 | 0.0062 | $1.2247 \times 10^{11}$ | $7.9640 \times 10^3$ | 0.26700051 | 0.0000 | 1.0000 | 0.7149 | $7.5105 \times 10^2$ | $2.8618 \times 10^2$ |
| 0.00300 | 3.6379 | $-0.1179$ | $1.0755 \times 10^{11}$ | $1.0327 \times 10^4$ | 0.26700101 | 0.0000 | 1.0000 | 0.7149 | $4.7661 \times 10^2$ | $2.2104 \times 10^2$ |
| 0.00500 | 3.6352 | $-0.2666$ | $9.1791 \times 10^{10}$ | $1.4177 \times 10^4$ | 0.26700220 | 0.0002 | 0.9998 | 0.7149 | $3.4656 \times 10^2$ | $1.6937 \times 10^2$ |
| 0.00700 | 3.6345 | $-0.3551$ | $8.3165 \times 10^{10}$ | $1.7271 \times 10^4$ | 0.26700356 | 0.0006 | 0.9994 | 0.7149 | $2.8626 \times 10^2$ | $1.4165 \times 10^2$ |
| 0.01000 | 3.6357 | $-0.4338$ | $7.5511 \times 10^{10}$ | $2.0949 \times 10^4$ | 0.26700595 | 0.0020 | 0.9980 | 0.7149 | $2.3070 \times 10^2$ | $1.1531 \times 10^2$ |
| 0.02000 | 3.6558 | $-0.4709$ | $6.5955 \times 10^{10}$ | $2.7460 \times 10^4$ | 0.26701675 | 0.0238 | 0.9762 | 0.7149 | $1.3179 \times 10^2$ | $6.7130 \times 10^1$ |
| 0.03000 | 3.7049 | $-0.3067$ | $6.3574 \times 10^{10}$ | $2.9555 \times 10^4$ | 0.26703479 | 0.1318 | 0.8680 | 0.7149 | $7.3294 \times 10^1$ | $3.8845 \times 10^1$ |
| 0.04670 | 3.7298 | $-0.3380$ | $5.4671 \times 10^{10}$ | $3.9965 \times 10^4$ | 0.26751727 | 0.9556 | 0.0020 | 0.7150 | $4.5129 \times 10^1$ | $2.5391 \times 10^1$ |
| | | | | $1.000 M_\odot$ | | | | | | |
| 0.00002 | 3.6343 | 1.3422 | $5.8725 \times 10^{11}$ | $3.8486 \times 10^2$ | 0.26700050 | 0.0000 | 1.0000 | 0.7149 | $5.4719 \times 10^2$ | $2.6169 \times 10^2$ |
| 0.00010 | 3.6479 | 0.9587 | $3.5470 \times 10^{11}$ | $1.0549 \times 10^3$ | 0.26700050 | 0.0000 | 1.0000 | 0.7149 | $5.3042 \times 10^2$ | $2.5490 \times 10^2$ |
| 0.00020 | 3.6521 | 0.7685 | $2.7955 \times 10^{11}$ | $1.6983 \times 10^3$ | 0.26700050 | 0.0000 | 1.0000 | 0.7149 | $5.2782 \times 10^2$ | $2.5386 \times 10^2$ |

Table 1. Continued

| Age(Gyr) | log$T_{\rm eff}$ | log$L/L_\odot$ | R(cm) | g(cm/s$^2$) | $Y_c$ | L(PP1) | $L_g$ | $X_{\rm env}$ | $\tau_{gc}$ (day) | $\tau_{lc}$ (day) |
|---|---|---|---|---|---|---|---|---|---|---|
| 0.00030 | 3.6537 | 0.6532 | 2.4299 × 10$^{11}$ | 2.2478 × 10$^3$ | 0.26700050 | 0.0000 | 1.0000 | 0.7149 | 5.2910 × 10$^2$ | 2.5432 × 10$^2$ |
| 0.00050 | 3.6545 | 0.5037 | 2.0381 × 10$^{11}$ | 3.1951 × 10$^3$ | 0.26700050 | 0.0000 | 1.0000 | 0.7149 | 5.3658 × 10$^2$ | 2.5702 × 10$^2$ |
| 0.00070 | 3.6543 | 0.4032 | 1.8173 × 10$^{11}$ | 4.0189 × 10$^3$ | 0.26700050 | 0.0000 | 1.0000 | 0.7149 | 5.4837 × 10$^2$ | 2.6098 × 10$^2$ |
| 0.00100 | 3.6534 | 0.2950 | 1.6113 × 10$^{11}$ | 5.1118 × 10$^3$ | 0.26700050 | 0.0000 | 1.0000 | 0.7149 | 5.7489 × 10$^2$ | 2.6866 × 10$^2$ |
| 0.00200 | 3.6495 | 0.0807 | 1.2816 × 10$^{11}$ | 8.0800 × 10$^3$ | 0.26700071 | 0.0000 | 1.0000 | 0.7149 | 5.4009 × 10$^2$ | 2.4194 × 10$^2$ |
| 0.00300 | 3.6466 | −0.0434 | 1.1261 × 10$^{11}$ | 1.0467 × 10$^4$ | 0.26700123 | 0.0000 | 1.0000 | 0.7149 | 4.0270 × 10$^2$ | 1.9420 × 10$^2$ |
| 0.00500 | 3.6443 | −0.1864 | 9.6514 × 10$^{10}$ | 1.4248 × 10$^4$ | 0.26700243 | 0.0003 | 0.9997 | 0.7149 | 3.0090 × 10$^2$ | 1.4852 × 10$^2$ |
| 0.00700 | 3.6449 | −0.2650 | 8.7904 × 10$^{10}$ | 1.7176 × 10$^4$ | 0.26700385 | 0.0008 | 0.9992 | 0.7149 | 2.4691 × 10$^2$ | 1.2283 × 10$^2$ |
| 0.01000 | 3.6495 | −0.3226 | 8.0575 × 10$^{10}$ | 2.0443 × 10$^4$ | 0.26700636 | 0.0026 | 0.9974 | 0.7149 | 1.9425 × 10$^2$ | 9.7528 × 10$^1$ |
| 0.02000 | 3.6974 | −0.2192 | 7.2775 × 10$^{10}$ | 2.5060 × 10$^4$ | 0.26701858 | 0.0384 | 0.9616 | 0.7149 | 9.2436 × 10$^1$ | 4.8160 × 10$^1$ |
| 0.03000 | 3.7597 | 0.0011 | 7.0382 × 10$^{10}$ | 2.6793 × 10$^4$ | 0.26714656 | 0.5267 | 0.4496 | 0.7149 | 3.6291 × 10$^1$ | 2.1060 × 10$^1$ |
| 0.03628 | 3.7586 | −0.1162 | 6.1806 × 10$^{10}$ | 3.4745 × 10$^4$ | 0.26753638 | 0.8959 | 0.0017 | 0.7150 | 3.2984 × 10$^1$ | 1.9503 × 10$^1$ |
| 1.100$M_\odot$ | | | | | | | | | | |
| 0.00002 | 3.6389 | 1.3853 | 6.0446 × 10$^{11}$ | 3.9958 × 10$^2$ | 0.26700050 | 0.0000 | 1.0000 | 0.7149 | 5.5796 × 10$^2$ | 2.6708 × 10$^2$ |
| 0.00010 | 3.6528 | 1.0170 | 3.7085 × 10$^{11}$ | 1.0615 × 10$^3$ | 0.26700050 | 0.0000 | 1.0000 | 0.7149 | 5.4081 × 10$^2$ | 2.5996 × 10$^2$ |
| 0.00020 | 3.6576 | 0.8306 | 2.9271 × 10$^{11}$ | 1.7040 × 10$^3$ | 0.26700050 | 0.0000 | 1.0000 | 0.7149 | 5.3810 × 10$^2$ | 2.5872 × 10$^2$ |
| 0.00030 | 3.6596 | 0.7166 | 2.5440 × 10$^{11}$ | 2.2558 × 10$^3$ | 0.26700050 | 0.0000 | 1.0000 | 0.7149 | 5.4004 × 10$^2$ | 2.5934 × 10$^2$ |
| 0.00050 | 3.6611 | 0.5691 | 2.1323 × 10$^{11}$ | 3.2110 × 10$^3$ | 0.26700050 | 0.0000 | 1.0000 | 0.7149 | 5.5022 × 10$^2$ | 2.6265 × 10$^2$ |
| 0.00070 | 3.6613 | 0.4696 | 1.8998 × 10$^{11}$ | 4.0451 × 10$^3$ | 0.26700050 | 0.0000 | 1.0000 | 0.7149 | 5.6764 × 10$^2$ | 2.6786 × 10$^2$ |
| 0.00100 | 3.6608 | 0.3624 | 1.6829 × 10$^{11}$ | 5.1548 × 10$^3$ | 0.26700050 | 0.0000 | 1.0000 | 0.7149 | 6.1534 × 10$^2$ | 2.7852 × 10$^2$ |
| 0.00200 | 3.6578 | 0.1502 | 1.3363 × 10$^{11}$ | 8.1759 × 10$^3$ | 0.26700085 | 0.0000 | 1.0000 | 0.7149 | 4.5435 × 10$^2$ | 2.1422 × 10$^2$ |
| 0.00300 | 3.6555 | 0.0293 | 1.1750 × 10$^{11}$ | 1.0574 × 10$^4$ | 0.26700137 | 0.0001 | 0.9999 | 0.7149 | 3.5436 × 10$^2$ | 1.7283 × 10$^2$ |
| 0.00500 | 3.6547 | −0.1040 | 1.0115 × 10$^{11}$ | 1.4270 × 10$^4$ | 0.26700261 | 0.0003 | 0.9997 | 0.7149 | 2.6244 × 10$^2$ | 1.3074 × 10$^2$ |
| 0.00700 | 3.6579 | −0.1674 | 9.2681 × 10$^{10}$ | 1.6997 × 10$^4$ | 0.26700429 | 0.0010 | 0.9990 | 0.7149 | 2.1229 × 10$^2$ | 1.0617 × 10$^2$ |
| 0.01000 | 3.6689 | −0.1896 | 8.5879 × 10$^{10}$ | 1.9795 × 10$^4$ | 0.26700678 | 0.0035 | 0.9965 | 0.7149 | 1.5978 × 10$^2$ | 8.0916 × 10$^1$ |
| 0.02000 | 3.7453 | 0.0798 | 8.2345 × 10$^{10}$ | 2.1531 × 10$^4$ | 0.26702316 | 0.0830 | 0.9167 | 0.7149 | 5.4151 × 10$^1$ | 2.9686 × 10$^1$ |
| 0.02839 | 3.7807 | 0.0903 | 7.0824 × 10$^{10}$ | 2.9105 × 10$^4$ | 0.26746793 | 0.7854 | 0.0131 | 0.7150 | 2.0896 × 10$^1$ | 1.3579 × 10$^1$ |
| 1.200$M_\odot$ | | | | | | | | | | |
| 0.00002 | 3.6422 | 1.4368 | 6.3143 × 10$^{11}$ | 3.9946 × 10$^2$ | 0.26700050 | 0.0000 | 1.0000 | 0.7149 | 5.6934 × 10$^2$ | 2.7265 × 10$^2$ |
| 0.00010 | 3.6570 | 1.0732 | 3.8809 × 10$^{11}$ | 1.0575 × 10$^3$ | 0.26700050 | 0.0000 | 1.0000 | 0.7149 | 5.5134 × 10$^2$ | 2.6504 × 10$^2$ |
| 0.00020 | 3.6624 | 0.8886 | 3.0608 × 10$^{11}$ | 1.7001 × 10$^3$ | 0.26700050 | 0.0000 | 1.0000 | 0.7149 | 5.4883 × 10$^2$ | 2.6375 × 10$^2$ |
| 0.00030 | 3.6648 | 0.7755 | 2.6582 × 10$^{11}$ | 2.2539 × 10$^3$ | 0.26700050 | 0.0000 | 1.0000 | 0.7149 | 5.5184 × 10$^2$ | 2.6463 × 10$^2$ |
| 0.00050 | 3.6668 | 0.6293 | 2.2253 × 10$^{11}$ | 3.2162 × 10$^3$ | 0.26700050 | 0.0000 | 1.0000 | 0.7149 | 5.6648 × 10$^2$ | 2.6900 × 10$^2$ |
| 0.00070 | 3.6674 | 0.5304 | 1.9808 × 10$^{11}$ | 4.0591 × 10$^3$ | 0.26700050 | 0.0000 | 1.0000 | 0.7149 | 5.9427 × 10$^2$ | 2.7608 × 10$^2$ |
| 0.00100 | 3.6674 | 0.4242 | 1.7528 × 10$^{11}$ | 5.1839 × 10$^3$ | 0.26700050 | 0.0000 | 1.0000 | 0.7149 | 7.2785 × 10$^2$ | 2.9128 × 10$^2$ |
| 0.00200 | 3.6655 | 0.2152 | 1.3902 × 10$^{11}$ | 8.2405 × 10$^3$ | 0.26700094 | 0.0000 | 1.0000 | 0.7149 | 4.0061 × 10$^2$ | 1.9293 × 10$^2$ |
| 0.00300 | 3.6643 | 0.0999 | 1.2239 × 10$^{11}$ | 1.0633 × 10$^4$ | 0.26700146 | 0.0001 | 0.9999 | 0.7149 | 3.1461 × 10$^2$ | 1.5517 × 10$^2$ |
| 0.00500 | 3.6663 | −0.0180 | 1.0586 × 10$^{11}$ | 1.4211 × 10$^4$ | 0.26700275 | 0.0004 | 0.9996 | 0.7149 | 2.3068 × 10$^2$ | 1.1524 × 10$^2$ |
| 0.00700 | 3.6736 | −0.0590 | 9.7678 × 10$^{10}$ | 1.6693 × 10$^4$ | 0.26700431 | 0.0012 | 0.9988 | 0.7149 | 1.8118 × 10$^2$ | 9.1033 × 10$^1$ |
| 0.01000 | 3.6928 | −0.0354 | 9.1857 × 10$^{10}$ | 1.8876 × 10$^4$ | 0.26700726 | 0.0049 | 0.9951 | 0.7149 | 1.2845 × 10$^2$ | 6.5403 × 10$^1$ |
| 0.02000 | 3.7926 | 0.3928 | 9.4968 × 10$^{10}$ | 1.7659 × 10$^4$ | 0.26717428 | 0.3801 | 0.5657 | 0.7149 | 1.0868 × 10$^1$ | 8.7656 × 10$^0$ |
| 0.02281 | 3.7976 | 0.2748 | 8.1035 × 10$^{10}$ | 2.4254 × 10$^4$ | 0.26738981 | 0.6633 | 0.0098 | 0.7150 | 8.7224 × 10$^0$ | 7.7002 × 10$^0$ |
| 1.300$M_\odot$ | | | | | | | | | | |
| 0.00002 | 3.6456 | 1.4780 | 6.5189 × 10$^{11}$ | 4.0601 × 10$^2$ | 0.26700050 | 0.0000 | 1.0000 | 0.7149 | 5.7970 × 10$^2$ | 2.7774 × 10$^2$ |
| 0.00010 | 3.6607 | 1.1258 | 4.0547 × 10$^{11}$ | 1.0495 × 10$^3$ | 0.26700050 | 0.0000 | 1.0000 | 0.7149 | 5.6201 × 10$^2$ | 2.7010 × 10$^2$ |
| 0.00020 | 3.6666 | 0.9443 | 3.2018 × 10$^{11}$ | 1.6831 × 10$^3$ | 0.26700050 | 0.0000 | 1.0000 | 0.7149 | 5.6014 × 10$^2$ | 2.6897 × 10$^2$ |
| 0.00030 | 3.6693 | 0.8330 | 2.7811 × 10$^{11}$ | 2.2307 × 10$^3$ | 0.26700050 | 0.0000 | 1.0000 | 0.7149 | 5.6457 × 10$^2$ | 2.7021 × 10$^2$ |
| 0.00050 | 3.6719 | 0.6883 | 2.3272 × 10$^{11}$ | 3.1857 × 10$^3$ | 0.26700050 | 0.0000 | 1.0000 | 0.7149 | 5.8596 × 10$^2$ | 2.7605 × 10$^2$ |
| 0.00070 | 3.6728 | 0.5906 | 2.0703 × 10$^{11}$ | 4.0254 × 10$^3$ | 0.26700050 | 0.0000 | 1.0000 | 0.7149 | 6.3285 × 10$^2$ | 2.8545 × 10$^2$ |
| 0.00100 | 3.6732 | 0.4853 | 1.8307 × 10$^{11}$ | 5.1480 × 10$^3$ | 0.26700056 | 0.0000 | 1.0000 | 0.7149 | 6.0654 × 10$^2$ | 2.6178 × 10$^2$ |
| 0.00200 | 3.6726 | 0.2815 | 1.4517 × 10$^{11}$ | 8.1867 × 10$^3$ | 0.26700100 | 0.0000 | 1.0000 | 0.7149 | 3.6069 × 10$^2$ | 1.7610 × 10$^2$ |
| 0.00300 | 3.6729 | 0.1730 | 1.2797 × 10$^{11}$ | 1.0535 × 10$^4$ | 0.26700154 | 0.0001 | 0.9999 | 0.7149 | 2.8340 × 10$^2$ | 1.4076 × 10$^2$ |
| 0.00500 | 3.6784 | 0.0734 | 1.1125 × 10$^{11}$ | 1.3941 × 10$^4$ | 0.26700286 | 0.0005 | 0.9995 | 0.7149 | 2.0299 × 10$^2$ | 1.0210 × 10$^2$ |
| 0.00700 | 3.6903 | 0.0583 | 1.0352 × 10$^{11}$ | 1.6099 × 10$^4$ | 0.26700453 | 0.0016 | 0.9984 | 0.7149 | 1.5348 × 10$^2$ | 7.7830 × 10$^1$ |
| 0.01000 | 3.7182 | 0.1382 | 9.9778 × 10$^{10}$ | 1.7331 × 10$^4$ | 0.26700783 | 0.0072 | 0.9928 | 0.7149 | 9.8456 × 10$^1$ | 5.1204 × 10$^1$ |
| 0.01870 | 3.8121 | 0.4501 | 9.2723 × 10$^{10}$ | 2.0068 × 10$^4$ | 0.26732463 | 0.5409 | 0.0253 | 0.7149 | | |
| 1.400$M_\odot$ | | | | | | | | | | |
| 0.00002 | 3.6483 | 1.5220 | 6.7752 × 10$^{11}$ | 4.0480 × 10$^2$ | 0.26700050 | 0.0000 | 1.0000 | 0.7149 | 5.9055 × 10$^2$ | 2.8303 × 10$^2$ |
| 0.00010 | 3.6640 | 1.1737 | 4.2187 × 10$^{11}$ | 1.0440 × 10$^3$ | 0.26700050 | 0.0000 | 1.0000 | 0.7149 | 5.7268 × 10$^2$ | 2.7516 × 10$^2$ |
| 0.00020 | 3.6703 | 0.9932 | 3.3292 × 10$^{11}$ | 1.6765 × 10$^3$ | 0.26700050 | 0.0000 | 1.0000 | 0.7149 | 5.7224 × 10$^2$ | 2.7448 × 10$^2$ |
| 0.00030 | 3.6734 | 0.8826 | 2.8896 × 10$^{11}$ | 2.2254 × 10$^3$ | 0.26700050 | 0.0000 | 1.0000 | 0.7149 | 5.7916 × 10$^2$ | 2.7636 × 10$^2$ |
| 0.00050 | 3.6764 | 0.7388 | 2.4157 × 10$^{11}$ | 3.1842 × 10$^3$ | 0.26700050 | 0.0000 | 1.0000 | 0.7149 | 6.1286 × 10$^2$ | 2.8449 × 10$^2$ |
| 0.00070 | 3.6777 | 0.6419 | 2.1474 × 10$^{11}$ | 4.0294 × 10$^3$ | 0.26700050 | 0.0000 | 1.0000 | 0.7149 | 7.3501 × 10$^2$ | 2.9776 × 10$^2$ |

Table 1. Continued

| Age(Gyr) | log$T_{\rm eff}$ | log$L/L_\odot$ | R(cm) | g(cm/s$^2$) | $Y_c$ | L(PP1) | $L_g$ | $X_{\rm env}$ | $\tau_{gc}$ (day) | $\tau_{lc}$ (day) |
|---|---|---|---|---|---|---|---|---|---|---|
| 0.00100 | 3.6785 | 0.5376 | 1.8976 × 10$^{11}$ | 5.1601 × 10$^3$ | 0.26700061 | 0.0000 | 1.0000 | 0.7149 | 5.1678 × 10$^2$ | 2.3724 × 10$^2$ |
| 0.00200 | 3.6792 | 0.3395 | 1.5056 × 10$^{11}$ | 8.1970 × 10$^3$ | 0.26700105 | 0.0000 | 1.0000 | 0.7149 | 3.2822 × 10$^2$ | 1.6116 × 10$^2$ |
| 0.00300 | 3.6812 | 0.2400 | 1.3306 × 10$^{11}$ | 1.0495 × 10$^4$ | 0.26700160 | 0.0001 | 0.9999 | 0.7149 | 2.5452 × 10$^2$ | 1.2749 × 10$^2$ |
| 0.00500 | 3.6909 | 0.1650 | 1.1669 × 10$^{11}$ | 1.3646 × 10$^4$ | 0.26700299 | 0.0006 | 0.9994 | 0.7149 | 1.7751 × 10$^2$ | 8.9526 × 10$^1$ |
| 0.00700 | 3.7084 | 0.1863 | 1.1033 × 10$^{11}$ | 1.5264 × 10$^4$ | 0.26700480 | 0.0021 | 0.9979 | 0.7149 | 1.2612 × 10$^2$ | 6.4785 × 10$^1$ |
| 0.01000 | 3.7465 | 0.3549 | 1.1241 × 10$^{11}$ | 1.4705 × 10$^4$ | 0.26700858 | 0.0120 | 0.9880 | 0.7149 | 6.5809 × 10$^1$ | 3.5659 × 10$^1$ |
| 0.01552 | 3.8260 | 0.6112 | 1.0472 × 10$^{11}$ | 1.6944 × 10$^4$ | 0.26727591 | 0.4478 | 0.0238 | 0.7149 | | |
| | | | | 1.500$M_\odot$ | | | | | | |
| 0.00002 | 3.6515 | 1.5468 | 6.8679 × 10$^{11}$ | 4.2208 × 10$^2$ | 0.26700050 | 0.0000 | 1.0000 | 0.7149 | 5.9997 × 10$^2$ | 2.8768 × 10$^2$ |
| 0.00010 | 3.6673 | 1.2132 | 4.3494 × 10$^{11}$ | 1.0524 × 10$^3$ | 0.26700050 | 0.0000 | 1.0000 | 0.7149 | 5.8356 × 10$^2$ | 2.8028 × 10$^2$ |
| 0.00020 | 3.6738 | 1.0359 | 3.4410 × 10$^{11}$ | 1.6814 × 10$^3$ | 0.26700050 | 0.0000 | 1.0000 | 0.7149 | 5.8545 × 10$^2$ | 2.8037 × 10$^2$ |
| 0.00030 | 3.6771 | 0.9263 | 2.9880 × 10$^{11}$ | 2.2298 × 10$^3$ | 0.26700050 | 0.0000 | 1.0000 | 0.7149 | 5.9665 × 10$^2$ | 2.8335 × 10$^2$ |
| 0.00050 | 3.6805 | 0.7844 | 2.4983 × 10$^{11}$ | 3.1898 × 10$^3$ | 0.26700050 | 0.0000 | 1.0000 | 0.7149 | 6.5465 × 10$^2$ | 2.9464 × 10$^2$ |
| 0.00070 | 3.6820 | 0.6882 | 2.2204 × 10$^{11}$ | 4.0382 × 10$^3$ | 0.26700054 | 0.0000 | 1.0000 | 0.7149 | 6.3449 × 10$^2$ | 2.7153 × 10$^2$ |
| 0.00100 | 3.6833 | 0.5857 | 1.9619 × 10$^{11}$ | 5.1724 × 10$^3$ | 0.26700065 | 0.0000 | 1.0000 | 0.7149 | 4.5922 × 10$^2$ | 2.1809 × 10$^2$ |
| 0.00200 | 3.6856 | 0.3956 | 1.5599 × 10$^{11}$ | 8.1821 × 10$^3$ | 0.26700109 | 0.0000 | 1.0000 | 0.7149 | 2.9762 × 10$^2$ | 1.4806 × 10$^2$ |
| 0.00300 | 3.6892 | 0.3062 | 1.3836 × 10$^{11}$ | 1.0399 × 10$^4$ | 0.26700166 | 0.0001 | 0.9999 | 0.7149 | 2.3009 × 10$^2$ | 1.1561 × 10$^2$ |
| 0.00500 | 3.7039 | 0.2623 | 1.2295 × 10$^{11}$ | 1.3169 × 10$^4$ | 0.26700312 | 0.0007 | 0.9993 | 0.7149 | 1.5354 × 10$^2$ | 7.7787 × 10$^1$ |
| 0.00700 | 3.7278 | 0.3340 | 1.1960 × 10$^{11}$ | 1.3919 × 10$^4$ | 0.26700510 | 0.0030 | 0.9970 | 0.7149 | 9.9550 × 10$^1$ | 5.1881 × 10$^1$ |
| 0.01000 | 3.7791 | 0.6513 | 1.3607 × 10$^{11}$ | 1.0752 × 10$^4$ | 0.26700999 | 0.0233 | 0.9765 | 0.7149 | 2.0771 × 10$^1$ | 1.3851 × 10$^1$ |
| 0.01314 | 3.8411 | 0.7519 | 1.1488 × 10$^{11}$ | 1.5085 × 10$^4$ | 0.26725628 | 0.3647 | 0.0199 | 0.7149 | | |
| | | | | 1.600$M_\odot$ | | | | | | |
| 0.00002 | 3.6561 | 1.5394 | 6.6655 × 10$^{11}$ | 4.7797 × 10$^2$ | 0.26700050 | 0.0000 | 1.0000 | 0.7149 | 6.0724 × 10$^2$ | 2.9132 × 10$^2$ |
| 0.00010 | 3.6707 | 1.2397 | 4.4144 × 10$^{11}$ | 1.0898 × 10$^3$ | 0.26700050 | 0.0000 | 1.0000 | 0.7149 | 5.9483 × 10$^2$ | 2.8553 × 10$^2$ |
| 0.00020 | 3.6773 | 1.0700 | 3.5230 × 10$^{11}$ | 1.7110 × 10$^3$ | 0.26700050 | 0.0000 | 1.0000 | 0.7149 | 6.0069 × 10$^2$ | 2.8681 × 10$^2$ |
| 0.00030 | 3.6806 | 0.9636 | 3.0690 × 10$^{11}$ | 2.2546 × 10$^3$ | 0.26700050 | 0.0000 | 1.0000 | 0.7149 | 6.1782 × 10$^2$ | 2.9131 × 10$^2$ |
| 0.00050 | 3.6843 | 0.8244 | 2.5710 × 10$^{11}$ | 3.2126 × 10$^3$ | 0.26700050 | 0.0000 | 1.0000 | 0.7149 | 7.7149 × 10$^2$ | 3.0753 × 10$^2$ |
| 0.00070 | 3.6861 | 0.7299 | 2.2865 × 10$^{11}$ | 4.0618 × 10$^3$ | 0.26700056 | 0.0000 | 1.0000 | 0.7149 | 5.3972 × 10$^2$ | 2.4720 × 10$^2$ |
| 0.00100 | 3.6877 | 0.6297 | 2.0224 × 10$^{11}$ | 5.1918 × 10$^3$ | 0.26700069 | 0.0000 | 1.0000 | 0.7149 | 4.1826 × 10$^2$ | 2.0172 × 10$^2$ |
| 0.00200 | 3.6915 | 0.4491 | 1.6143 × 10$^{11}$ | 8.1493 × 10$^3$ | 0.26700112 | 0.0000 | 1.0000 | 0.7149 | 2.7398 × 10$^2$ | 1.3625 × 10$^2$ |
| 0.00300 | 3.6972 | 0.3727 | 1.4398 × 10$^{11}$ | 1.0244 × 10$^4$ | 0.26700171 | 0.0001 | 0.9999 | 0.7149 | 2.0824 × 10$^2$ | 1.0467 × 10$^2$ |
| 0.00500 | 3.7174 | 0.3688 | 1.3060 × 10$^{11}$ | 1.2451 × 10$^4$ | 0.26700325 | 0.0009 | 0.9991 | 0.7149 | 1.2912 × 10$^2$ | 6.6411 × 10$^1$ |
| 0.00700 | 3.7490 | 0.5192 | 1.3425 × 10$^{11}$ | 1.1782 × 10$^4$ | 0.26700548 | 0.0045 | 0.9955 | 0.7149 | 6.8499 × 10$^1$ | 3.7163 × 10$^1$ |
| 0.01000 | 3.8276 | 0.9487 | 1.5331 × 10$^{11}$ | 9.0349 × 10$^3$ | 0.26707017 | 0.1087 | 0.8709 | 0.7149 | | |
| 0.01128 | 3.8598 | 0.8755 | 1.2149 × 10$^{11}$ | 1.4388 × 10$^4$ | 0.26724475 | 0.3110 | 0.0029 | 0.7149 | | |
| | | | | 1.700$M_\odot$ | | | | | | |
| 0.00002 | 3.6591 | 1.5650 | 6.7723 × 10$^{11}$ | 4.9195 × 10$^2$ | 0.26700050 | 0.0000 | 1.0000 | 0.7149 | 6.1643 × 10$^2$ | 2.9572 × 10$^2$ |
| 0.00010 | 3.6733 | 1.2738 | 4.5358 × 10$^{11}$ | 1.0967 × 10$^3$ | 0.26700050 | 0.0000 | 1.0000 | 0.7149 | 6.0736 × 10$^2$ | 2.9125 × 10$^2$ |
| 0.00020 | 3.6801 | 1.1070 | 3.6277 × 10$^{11}$ | 1.7144 × 10$^3$ | 0.26700050 | 0.0000 | 1.0000 | 0.7149 | 6.1802 × 10$^2$ | 2.9392 × 10$^2$ |
| 0.00030 | 3.6838 | 1.0022 | 3.1619 × 10$^{11}$ | 2.2568 × 10$^3$ | 0.26700050 | 0.0000 | 1.0000 | 0.7149 | 6.4651 × 10$^2$ | 3.0023 × 10$^2$ |
| 0.00050 | 3.6876 | 0.8638 | 2.6495 × 10$^{11}$ | 3.2141 × 10$^3$ | 0.26700052 | 0.0000 | 1.0000 | 0.7149 | 6.6411 × 10$^2$ | 2.8260 × 10$^2$ |
| 0.00070 | 3.6898 | 0.7709 | 2.3568 × 10$^{11}$ | 4.0621 × 10$^3$ | 0.26700059 | 0.0000 | 1.0000 | 0.7149 | 4.8627 × 10$^2$ | 2.2944 × 10$^2$ |
| 0.00100 | 3.6918 | 0.6728 | 2.0858 × 10$^{11}$ | 5.1863 × 10$^3$ | 0.26700070 | 0.0000 | 1.0000 | 0.7149 | 3.8506 × 10$^2$ | 1.8805 × 10$^2$ |
| 0.00200 | 3.6973 | 0.5027 | 1.6720 × 10$^{11}$ | 8.0714 × 10$^3$ | 0.26700116 | 0.0000 | 1.0000 | 0.7149 | 2.5041 × 10$^2$ | 1.2571 × 10$^2$ |
| 0.00300 | 3.7052 | 0.4413 | 1.5018 × 10$^{11}$ | 1.0004 × 10$^4$ | 0.26700176 | 0.0001 | 0.9999 | 0.7149 | 1.8681 × 10$^2$ | 9.4579 × 10$^1$ |
| 0.00500 | 3.7317 | 0.4912 | 1.4077 × 10$^{11}$ | 1.1386 × 10$^4$ | 0.26700342 | 0.0012 | 0.9988 | 0.7149 | 1.0522 × 10$^2$ | 5.4843 × 10$^1$ |
| 0.00700 | 3.7744 | 0.7771 | 1.6074 × 10$^{11}$ | 8.7332 × 10$^3$ | 0.26700599 | 0.0073 | 0.9926 | 0.7149 | 2.4926 × 10$^1$ | 1.6104 × 10$^1$ |
| 0.00973 | 3.8815 | 1.0048 | 1.2760 × 10$^{11}$ | 1.3857 × 10$^4$ | 0.26722327 | 0.2589 | 0.0187 | 0.7149 | | |
| | | | | 1.800$M_\odot$ | | | | | | |
| 0.00002 | 3.6615 | 1.5876 | 6.8731 × 10$^{11}$ | 5.0573 × 10$^2$ | 0.26700050 | 0.0000 | 1.0000 | 0.7149 | 6.2655 × 10$^2$ | 3.0050 × 10$^2$ |
| 0.00010 | 3.6759 | 1.3064 | 4.6533 × 10$^{11}$ | 1.1033 × 10$^3$ | 0.26700050 | 0.0000 | 1.0000 | 0.7149 | 6.2054 × 10$^2$ | 2.9714 × 10$^2$ |
| 0.00020 | 3.6828 | 1.1419 | 3.7296 × 10$^{11}$ | 1.7175 × 10$^3$ | 0.26700050 | 0.0000 | 1.0000 | 0.7149 | 6.3842 × 10$^2$ | 3.0175 × 10$^2$ |
| 0.00030 | 3.6865 | 1.0379 | 3.2529 × 10$^{11}$ | 2.2578 × 10$^3$ | 0.26700050 | 0.0000 | 1.0000 | 0.7149 | 6.8751 × 10$^2$ | 3.1086 × 10$^2$ |
| 0.00050 | 3.6908 | 0.9015 | 2.7265 × 10$^{11}$ | 3.2136 × 10$^3$ | 0.26700054 | 0.0000 | 1.0000 | 0.7149 | 5.7217 × 10$^2$ | 2.6032 × 10$^2$ |
| 0.00070 | 3.6931 | 0.8095 | 2.4264 × 10$^{11}$ | 4.0579 × 10$^3$ | 0.26700061 | 0.0000 | 1.0000 | 0.7149 | 4.4553 × 10$^2$ | 2.1454 × 10$^2$ |
| 0.00100 | 3.6955 | 0.7140 | 2.1494 × 10$^{11}$ | 5.1712 × 10$^3$ | 0.26700072 | 0.0000 | 1.0000 | 0.7149 | 3.5717 × 10$^2$ | 1.7604 × 10$^2$ |
| 0.00200 | 3.7029 | 0.5560 | 1.7323 × 10$^{11}$ | 7.9607 × 10$^3$ | 0.26700119 | 0.0000 | 1.0000 | 0.7149 | 2.3157 × 10$^2$ | 1.1600 × 10$^2$ |
| 0.00300 | 3.7133 | 0.5131 | 1.5714 × 10$^{11}$ | 9.6751 × 10$^3$ | 0.26700181 | 0.0002 | 0.9998 | 0.7149 | 1.6653 × 10$^2$ | 8.4924 × 10$^1$ |
| 0.00500 | 3.7476 | 0.6438 | 1.5602 × 10$^{11}$ | 9.8142 × 10$^3$ | 0.26700361 | 0.0016 | 0.9984 | 0.7149 | 7.6114 × 10$^1$ | 4.1083 × 10$^1$ |
| 0.00700 | 3.8106 | 1.0541 | 1.8719 × 10$^{11}$ | 6.8183 × 10$^3$ | 0.26700720 | 0.0179 | 0.9816 | 0.7149 | | |
| 0.00851 | 3.9011 | 1.1137 | 1.3217 × 10$^{11}$ | 1.3676 × 10$^4$ | 0.26721787 | 0.2170 | 0.0168 | 0.7149 | | |

Table 1. Continued

| Age(Gyr) | $\log T_{\text{eff}}$ | $\log L/L_\odot$ | R(cm) | g(cm/s$^2$) | $Y_c$ | L(PP1) | $L_g$ | $X_{\text{env}}$ | $\tau_{gc}$ (day) | $\tau_{lc}$ (day) |
|---|---|---|---|---|---|---|---|---|---|---|
| \multicolumn{11}{c}{1.900$M_\odot$} ||||||||||
| 0.00002 | 3.6652 | 1.5876 | 6.7586 × 10$^{11}$ | 5.5207 × 10$^2$ | 0.26700050 | 0.0000 | 1.0000 | 0.7149 | 6.3573 × 10$^2$ | 3.0479 × 10$^2$ |
| 0.00010 | 3.6787 | 1.3257 | 4.6964 × 10$^{11}$ | 1.1433 × 10$^3$ | 0.26700050 | 0.0000 | 1.0000 | 0.7149 | 6.3624 × 10$^2$ | 3.0386 × 10$^2$ |
| 0.00020 | 3.6856 | 1.1664 | 3.7876 × 10$^{11}$ | 1.7578 × 10$^3$ | 0.26700050 | 0.0000 | 1.0000 | 0.7149 | 6.6739 × 10$^2$ | 3.1145 × 10$^2$ |
| 0.00030 | 3.6895 | 1.0648 | 3.3103 × 10$^{11}$ | 2.3013 × 10$^3$ | 0.26700050 | 0.0000 | 1.0000 | 0.7149 | 7.8870 × 10$^2$ | 3.2467 × 10$^2$ |
| 0.00050 | 3.6938 | 0.9305 | 2.7803 × 10$^{11}$ | 3.2623 × 10$^3$ | 0.26700055 | 0.0000 | 1.0000 | 0.7149 | 5.1042 × 10$^2$ | 2.4102 × 10$^2$ |
| 0.00070 | 3.6964 | 0.8411 | 2.4778 × 10$^{11}$ | 4.1075 × 10$^3$ | 0.26700062 | 0.0000 | 1.0000 | 0.7149 | 4.0844 × 10$^2$ | 2.0011 × 10$^2$ |
| 0.00100 | 3.6994 | 0.7494 | 2.1996 × 10$^{11}$ | 5.2123 × 10$^3$ | 0.26700073 | 0.0000 | 1.0000 | 0.7149 | 3.3048 × 10$^2$ | 1.6410 × 10$^2$ |
| 0.00200 | 3.7087 | 0.6077 | 1.7897 × 10$^{11}$ | 7.8734 × 10$^3$ | 0.26700121 | 0.0000 | 1.0000 | 0.7149 | 2.1028 × 10$^2$ | 1.0605 × 10$^2$ |
| 0.00300 | 3.7222 | 0.5909 | 1.6501 × 10$^{11}$ | 9.2612 × 10$^3$ | 0.26700186 | 0.0002 | 0.9998 | 0.7149 | 1.4607 × 10$^2$ | 7.4677 × 10$^1$ |
| 0.00500 | 3.7679 | 0.8677 | 1.8381 × 10$^{11}$ | 7.4638 × 10$^3$ | 0.26700384 | 0.0024 | 0.9976 | 0.7149 | 3.3769 × 10$^1$ | 2.0393 × 10$^1$ |
| 0.00700 | 3.9040 | 1.2987 | 1.6136 × 10$^{11}$ | 9.6851 × 10$^3$ | 0.26710331 | 0.1165 | 0.7511 | 0.7149 | | |
| 0.00747 | 3.9195 | 1.2143 | 1.3633 × 10$^{11}$ | 1.3567 × 10$^4$ | 0.26721075 | 0.1875 | 0.0012 | 0.7149 | | |
| \multicolumn{11}{c}{2.000$M_\odot$} ||||||||||
| 0.00002 | 3.6677 | 1.6073 | 6.8347 × 10$^{11}$ | 5.6825 × 10$^2$ | 0.26700050 | 0.0000 | 1.0000 | 0.7149 | 6.4646 × 10$^2$ | 3.0969 × 10$^2$ |
| 0.00010 | 3.6810 | 1.3533 | 4.7983 × 10$^{11}$ | 1.1529 × 10$^3$ | 0.26700050 | 0.0000 | 1.0000 | 0.7149 | 6.5368 × 10$^2$ | 3.1113 × 10$^2$ |
| 0.00020 | 3.6879 | 1.1965 | 3.8801 × 10$^{11}$ | 1.7632 × 10$^3$ | 0.26700050 | 0.0000 | 1.0000 | 0.7149 | 7.0508 × 10$^2$ | 3.2221 × 10$^2$ |
| 0.00030 | 3.6918 | 1.0961 | 3.3942 × 10$^{11}$ | 2.3041 × 10$^3$ | 0.26700051 | 0.0000 | 1.0000 | 0.7149 | 7.3404 × 10$^2$ | 3.0544 × 10$^2$ |
| 0.00050 | 3.6965 | 0.9641 | 2.8536 × 10$^{11}$ | 3.2597 × 10$^3$ | 0.26700057 | 0.0000 | 1.0000 | 0.7149 | 4.7250 × 10$^2$ | 2.2643 × 10$^2$ |
| 0.00070 | 3.6994 | 0.8763 | 2.5458 × 10$^{11}$ | 4.0958 × 10$^3$ | 0.26700063 | 0.0000 | 1.0000 | 0.7149 | 3.8244 × 10$^2$ | 1.8847 × 10$^2$ |
| 0.00100 | 3.7027 | 0.7880 | 2.2644 × 10$^{11}$ | 5.1771 × 10$^3$ | 0.26700075 | 0.0000 | 1.0000 | 0.7149 | 3.0962 × 10$^2$ | 1.5414 × 10$^2$ |
| 0.00200 | 3.7144 | 0.6638 | 1.8603 × 10$^{11}$ | 7.6700 × 10$^3$ | 0.26700124 | 0.0000 | 1.0000 | 0.7149 | 1.9211 × 10$^2$ | 9.7295 × 10$^1$ |
| 0.00300 | 3.7310 | 0.6784 | 1.7522 × 10$^{11}$ | 8.6461 × 10$^3$ | 0.26700192 | 0.0002 | 0.9998 | 0.7149 | 1.2498 × 10$^2$ | 6.4760 × 10$^1$ |
| 0.00500 | 3.7940 | 1.1113 | 2.1580 × 10$^{11}$ | 5.6998 × 10$^3$ | 0.26700414 | 0.0043 | 0.9957 | 0.7149 | | |
| 0.00663 | 3.9366 | 1.3107 | 1.4077 × 10$^{11}$ | 1.3395 × 10$^4$ | 0.26720409 | 0.1643 | 0.0040 | 0.7149 | | |
| \multicolumn{11}{c}{2.100$M_\odot$} ||||||||||
| 0.00002 | 3.6699 | 1.6252 | 6.9064 × 10$^{11}$ | 5.8434 × 10$^2$ | 0.26700050 | 0.0000 | 1.0000 | 0.7149 | 6.5837 × 10$^2$ | 3.1508 × 10$^2$ |
| 0.00010 | 3.6831 | 1.3797 | 4.8980 × 10$^{11}$ | 1.1618 × 10$^3$ | 0.26700050 | 0.0000 | 1.0000 | 0.7149 | 6.7406 × 10$^2$ | 3.1919 × 10$^2$ |
| 0.00020 | 3.6902 | 1.2256 | 3.9704 × 10$^{11}$ | 1.7681 × 10$^3$ | 0.26700050 | 0.0000 | 1.0000 | 0.7149 | 7.6725 × 10$^2$ | 3.3470 × 10$^2$ |
| 0.00030 | 3.6942 | 1.1262 | 3.4766 × 10$^{11}$ | 2.3060 × 10$^3$ | 0.26700050 | 0.0000 | 1.0000 | 0.7149 | 6.2889 × 10$^2$ | 2.8272 × 10$^2$ |
| 0.00050 | 3.6990 | 0.9959 | 2.9265 × 10$^{11}$ | 3.2545 × 10$^3$ | 0.26700057 | 0.0000 | 1.0000 | 0.7149 | 4.3801 × 10$^2$ | 2.1353 × 10$^2$ |
| 0.00070 | 3.7022 | 0.9109 | 2.6144 × 10$^{11}$ | 4.0777 × 10$^3$ | 0.26700064 | 0.0000 | 1.0000 | 0.7149 | 3.5780 × 10$^2$ | 1.7783 × 10$^2$ |
| 0.00100 | 3.7061 | 0.8268 | 2.3308 × 10$^{11}$ | 5.1306 × 10$^3$ | 0.26700075 | 0.0000 | 1.0000 | 0.7149 | 2.9051 × 10$^2$ | 1.4491 × 10$^2$ |
| 0.00200 | 3.7200 | 0.7224 | 1.9388 × 10$^{11}$ | 7.4148 × 10$^3$ | 0.26700126 | 0.0000 | 1.0000 | 0.7149 | 1.7410 × 10$^2$ | 8.8710 × 10$^1$ |
| 0.00300 | 3.7406 | 0.7820 | 1.8885 × 10$^{11}$ | 7.8149 × 10$^3$ | 0.26700198 | 0.0003 | 0.9997 | 0.7149 | 1.0034 × 10$^2$ | 5.3150 × 10$^1$ |
| 0.00500 | 3.8512 | 1.3513 | 2.1855 × 10$^{11}$ | 5.8352 × 10$^3$ | 0.26700483 | 0.0121 | 0.9872 | 0.7149 | | |
| 0.00591 | 3.9526 | 1.3994 | 1.4485 × 10$^{11}$ | 1.3284 × 10$^4$ | 0.26720477 | 0.1405 | 0.0036 | 0.7149 | | |
| \multicolumn{11}{c}{2.200$M_\odot$} ||||||||||
| 0.00002 | 3.6727 | 1.6312 | 6.8632 × 10$^{11}$ | 6.1990 × 10$^2$ | 0.26700055 | 0.0000 | 1.0000 | 0.7149 | 6.7149 × 10$^2$ | 3.2080 × 10$^2$ |
| 0.00010 | 3.6853 | 1.3992 | 4.9600 × 10$^{11}$ | 1.1869 × 10$^3$ | 0.26700055 | 0.0000 | 1.0000 | 0.7149 | 7.0125 × 10$^2$ | 3.2894 × 10$^2$ |
| 0.00020 | 3.6923 | 1.2497 | 4.0429 × 10$^{11}$ | 1.7864 × 10$^3$ | 0.26700055 | 0.0000 | 1.0000 | 0.7149 | 8.7036 × 10$^2$ | 3.3583 × 10$^2$ |
| 0.00030 | 3.6964 | 1.1528 | 3.5479 × 10$^{11}$ | 2.3196 × 10$^3$ | 0.26700057 | 0.0000 | 1.0000 | 0.7149 | 5.6382 × 10$^2$ | 2.6355 × 10$^2$ |
| 0.00050 | 3.7015 | 1.0255 | 2.9941 × 10$^{11}$ | 3.2572 × 10$^3$ | 0.26700063 | 0.0000 | 1.0000 | 0.7149 | 4.0861 × 10$^2$ | 2.0131 × 10$^2$ |
| 0.00070 | 3.7049 | 0.9435 | 2.6808 × 10$^{11}$ | 4.0630 × 10$^3$ | 0.26700069 | 0.0000 | 1.0000 | 0.7149 | 3.3653 × 10$^2$ | 1.6770 × 10$^2$ |
| 0.00100 | 3.7093 | 0.8641 | 2.3977 × 10$^{11}$ | 5.0790 × 10$^3$ | 0.26700081 | 0.0000 | 1.0000 | 0.7149 | 2.7062 × 10$^2$ | 1.3605 × 10$^2$ |
| 0.00200 | 3.7259 | 0.7854 | 2.0288 × 10$^{11}$ | 7.0943 × 10$^3$ | 0.26700133 | 0.0001 | 0.9999 | 0.7149 | 1.5516 × 10$^2$ | 7.9867 × 10$^1$ |
| 0.00300 | 3.7521 | 0.9183 | 2.0953 × 10$^{11}$ | 6.6507 × 10$^3$ | 0.26700210 | 0.0004 | 0.9996 | 0.7149 | 6.8366 × 10$^1$ | 3.7423 × 10$^1$ |
| 0.00500 | 3.9525 | 1.5656 | 1.7550 × 10$^{11}$ | 9.4799 × 10$^3$ | 0.26709444 | 0.0785 | 0.7752 | 0.7149 | | |
| 0.00530 | 3.9675 | 1.4809 | 1.4855 × 10$^{11}$ | 1.3232 × 10$^4$ | 0.26720411 | 0.1239 | 0.0002 | 0.7149 | | |
| \multicolumn{11}{c}{2.300$M_\odot$} ||||||||||
| 0.00002 | 3.6811 | 1.5365 | 5.9228 × 10$^{11}$ | 8.7019 × 10$^2$ | 0.26700070 | 0.0000 | 1.0000 | 0.7149 | 6.9841 × 10$^2$ | 3.3077 × 10$^2$ |
| 0.00010 | 3.6899 | 1.3674 | 4.6811 × 10$^{11}$ | 1.3931 × 10$^3$ | 0.26700070 | 0.0000 | 1.0000 | 0.7149 | 7.8557 × 10$^2$ | 3.4749 × 10$^2$ |
| 0.00020 | 3.6957 | 1.2417 | 3.9426 × 10$^{11}$ | 1.9639 × 10$^3$ | 0.26700071 | 0.0000 | 1.0000 | 0.7149 | 6.2126 × 10$^2$ | 2.8389 × 10$^2$ |
| 0.00030 | 3.6995 | 1.1559 | 3.5106 × 10$^{11}$ | 2.4769 × 10$^3$ | 0.26700074 | 0.0000 | 1.0000 | 0.7149 | 4.8881 × 10$^2$ | 2.3586 × 10$^2$ |
| 0.00050 | 3.7044 | 1.0409 | 3.0060 × 10$^{11}$ | 3.3783 × 10$^3$ | 0.26700080 | 0.0000 | 1.0000 | 0.7149 | 3.7046 × 10$^2$ | 1.8450 × 10$^2$ |
| 0.00070 | 3.7081 | 0.9665 | 2.7130 × 10$^{11}$ | 4.1474 × 10$^3$ | 0.26700087 | 0.0000 | 1.0000 | 0.7149 | 3.0912 × 10$^2$ | 1.5455 × 10$^2$ |
| 0.00100 | 3.7131 | 0.8967 | 2.4468 × 10$^{11}$ | 5.0990 × 10$^3$ | 0.26700099 | 0.0000 | 1.0000 | 0.7149 | 2.4767 × 10$^2$ | 1.2519 × 10$^2$ |
| 0.00200 | 3.7329 | 0.8578 | 2.1354 × 10$^{11}$ | 6.6944 × 10$^3$ | 0.26700153 | 0.0001 | 0.9999 | 0.7149 | 1.3232 × 10$^2$ | 6.8890 × 10$^1$ |
| 0.00300 | 3.7689 | 1.1188 | 2.4431 × 10$^{11}$ | 5.1142 × 10$^3$ | 0.26700238 | 0.0005 | 0.9995 | 0.7149 | 2.6877 × 10$^1$ | 1.7314 × 10$^1$ |
| 0.00475 | 3.9818 | 1.5602 | 1.5241 × 10$^{11}$ | 1.3142 × 10$^4$ | 0.26720060 | 0.1120 | 0.0010 | 0.7149 | | |

Table 1. Continued

| Age(Gyr) | $\log T_{\rm eff}$ | $\log L/L_\odot$ | R(cm) | g(cm/s$^2$) | $Y_c$ | L(PP1) | $L_g$ | $X_{\rm env}$ | $\tau_{gc}$ (day) | $\tau_{lc}$ (day) |
|---|---|---|---|---|---|---|---|---|---|---|
| | | | | $2.400 M_\odot$ | | | | | | |
| 0.00002 | 3.6835 | 1.5466 | $5.9252 \times 10^{11}$ | $9.0730 \times 10^2$ | 0.26700050 | 0.0000 | 1.0000 | 0.7149 | $7.2573 \times 10^2$ | $3.4054 \times 10^2$ |
| 0.00010 | 3.6919 | 1.3871 | $4.7449 \times 10^{11}$ | $1.4148 \times 10^3$ | 0.26700050 | 0.0000 | 1.0000 | 0.7149 | $9.2712 \times 10^2$ | $3.5697 \times 10^2$ |
| 0.00020 | 3.6977 | 1.2662 | $4.0197 \times 10^{11}$ | $1.9714 \times 10^3$ | 0.26700052 | 0.0000 | 1.0000 | 0.7149 | $5.6076 \times 10^2$ | $2.6488 \times 10^2$ |
| 0.00030 | 3.7014 | 1.1831 | $3.5902 \times 10^{11}$ | $2.4713 \times 10^3$ | 0.26700055 | 0.0000 | 1.0000 | 0.7149 | $4.5297 \times 10^2$ | $2.2236 \times 10^2$ |
| 0.00050 | 3.7065 | 1.0722 | $3.0860 \times 10^{11}$ | $3.3448 \times 10^3$ | 0.26700060 | 0.0000 | 1.0000 | 0.7149 | $3.4989 \times 10^2$ | $1.7473 \times 10^2$ |
| 0.00070 | 3.7106 | 1.0015 | $2.7928 \times 10^{11}$ | $4.0839 \times 10^3$ | 0.26700068 | 0.0000 | 1.0000 | 0.7149 | $2.9081 \times 10^2$ | $1.4614 \times 10^2$ |
| 0.00100 | 3.7162 | 0.9379 | $2.5291 \times 10^{11}$ | $4.9798 \times 10^3$ | 0.26700080 | 0.0000 | 1.0000 | 0.7149 | $2.3272 \times 10^2$ | $1.1750 \times 10^2$ |
| 0.00200 | 3.7396 | 0.9401 | $2.2768 \times 10^{11}$ | $6.1448 \times 10^3$ | 0.26700136 | 0.0001 | 0.9999 | 0.7149 | $1.0990 \times 10^2$ | $5.7873 \times 10^1$ |
| 0.00300 | 3.7886 | 1.3084 | $2.7762 \times 10^{11}$ | $4.1329 \times 10^3$ | 0.26700228 | 0.0009 | 0.9991 | 0.7149 | | |
| 0.00431 | 3.9945 | 1.6271 | $1.5524 \times 10^{11}$ | $1.3217 \times 10^4$ | 0.26721080 | 0.0981 | 0.0034 | 0.7149 | | |
| | | | | $2.500 M_\odot$ | | | | | | |
| 0.00002 | 3.6857 | 1.5484 | $5.8784 \times 10^{11}$ | $9.6022 \times 10^2$ | 0.26700050 | 0.0000 | 1.0000 | 0.7149 | $7.6959 \times 10^2$ | $3.5381 \times 10^2$ |
| 0.00010 | 3.6941 | 1.3984 | $4.7582 \times 10^{11}$ | $1.4655 \times 10^3$ | 0.26700051 | 0.0000 | 1.0000 | 0.7149 | $7.2986 \times 10^2$ | $3.1864 \times 10^2$ |
| 0.00020 | 3.6998 | 1.2818 | $4.0526 \times 10^{11}$ | $2.0204 \times 10^3$ | 0.26700053 | 0.0000 | 1.0000 | 0.7149 | $5.1058 \times 10^2$ | $2.4654 \times 10^2$ |
| 0.00030 | 3.7036 | 1.2015 | $3.6311 \times 10^{11}$ | $2.5165 \times 10^3$ | 0.26700055 | 0.0000 | 1.0000 | 0.7149 | $4.2203 \times 10^2$ | $2.0846 \times 10^2$ |
| 0.00050 | 3.7090 | 1.0958 | $3.1355 \times 10^{11}$ | $3.3751 \times 10^3$ | 0.26700061 | 0.0000 | 1.0000 | 0.7149 | $3.2760 \times 10^2$ | $1.6399 \times 10^2$ |
| 0.00070 | 3.7134 | 1.0302 | $2.8490 \times 10^{11}$ | $4.0879 \times 10^3$ | 0.26700068 | 0.0000 | 1.0000 | 0.7149 | $2.6977 \times 10^2$ | $1.3657 \times 10^2$ |
| 0.00100 | 3.7197 | 0.9752 | $2.5973 \times 10^{11}$ | $4.9186 \times 10^3$ | 0.26700081 | 0.0000 | 1.0000 | 0.7149 | $2.1251 \times 10^2$ | $1.0843 \times 10^2$ |
| 0.00200 | 3.7490 | 1.0548 | $2.4880 \times 10^{11}$ | $5.3604 \times 10^3$ | 0.26700140 | 0.0001 | 0.9999 | 0.7149 | $7.6716 \times 10^1$ | $4.1765 \times 10^1$ |
| 0.00300 | 3.8309 | 1.5156 | $2.9007 \times 10^{11}$ | $3.9435 \times 10^3$ | 0.26700202 | 0.0018 | 0.9982 | 0.7149 | | |
| 0.00388 | 4.0082 | 1.7075 | $1.5987 \times 10^{11}$ | $1.2982 \times 10^4$ | 0.26720180 | 0.0873 | 0.0049 | 0.7149 | | |
| | | | | $2.600 M_\odot$ | | | | | | |
| 0.00002 | 3.6877 | 1.5707 | $5.9764 \times 10^{11}$ | $9.6614 \times 10^2$ | 0.26700050 | 0.0000 | 1.0000 | 0.7149 | $8.2161 \times 10^2$ | $3.6630 \times 10^2$ |
| 0.00010 | 3.6956 | 1.4230 | $4.8618 \times 10^{11}$ | $1.4599 \times 10^3$ | 0.26700051 | 0.0000 | 1.0000 | 0.7149 | $6.4756 \times 10^2$ | $2.9709 \times 10^2$ |
| 0.00020 | 3.7013 | 1.3088 | $4.1517 \times 10^{11}$ | $2.0020 \times 10^3$ | 0.26700053 | 0.0000 | 1.0000 | 0.7149 | $4.7727 \times 10^2$ | $2.3368 \times 10^2$ |
| 0.00030 | 3.7052 | 1.2304 | $3.7262 \times 10^{11}$ | $2.4854 \times 10^3$ | 0.26700056 | 0.0000 | 1.0000 | 0.7149 | $3.9957 \times 10^2$ | $1.9829 \times 10^2$ |
| 0.00050 | 3.7110 | 1.1284 | $3.2262 \times 10^{11}$ | $3.3155 \times 10^3$ | 0.26700061 | 0.0000 | 1.0000 | 0.7149 | $3.1030 \times 10^2$ | $1.5587 \times 10^2$ |
| 0.00070 | 3.7158 | 1.0667 | $2.9393 \times 10^{11}$ | $3.9942 \times 10^3$ | 0.26700069 | 0.0000 | 1.0000 | 0.7149 | $2.5356 \times 10^2$ | $1.2912 \times 10^2$ |
| 0.00100 | 3.7229 | 1.0193 | $2.6934 \times 10^{11}$ | $4.7568 \times 10^3$ | 0.26700081 | 0.0000 | 1.0000 | 0.7149 | $1.9728 \times 10^2$ | $1.0091 \times 10^2$ |
| 0.00200 | 3.7594 | 1.1965 | $2.7917 \times 10^{11}$ | $4.4279 \times 10^3$ | 0.26700144 | 0.0001 | 0.9999 | 0.7149 | $4.3318 \times 10^1$ | $2.5631 \times 10^1$ |
| 0.00300 | 3.9086 | 1.7184 | $2.5611 \times 10^{11}$ | $5.2609 \times 10^3$ | 0.26700252 | 0.0049 | 0.9947 | 0.7149 | | |
| 0.00354 | 4.0199 | 1.7706 | $1.6287 \times 10^{11}$ | $1.3008 \times 10^4$ | 0.26720860 | 0.0777 | 0.0096 | 0.7149 | | |
| | | | | $2.700 M_\odot$ | | | | | | |
| 0.00002 | 3.6896 | 1.5815 | $5.9974 \times 10^{11}$ | $9.9627 \times 10^2$ | 0.26700050 | 0.0000 | 1.0000 | 0.7149 | $9.7713 \times 10^2$ | $3.8421 \times 10^2$ |
| 0.00010 | 3.6974 | 1.4381 | $4.9050 \times 10^{11}$ | $1.4895 \times 10^3$ | 0.26700051 | 0.0000 | 1.0000 | 0.7149 | $5.8382 \times 10^2$ | $2.7618 \times 10^2$ |
| 0.00020 | 3.7032 | 1.3267 | $4.2009 \times 10^{11}$ | $2.0306 \times 10^3$ | 0.26700054 | 0.0000 | 1.0000 | 0.7149 | $4.4484 \times 10^2$ | $2.1986 \times 10^2$ |
| 0.00030 | 3.7072 | 1.2509 | $3.7791 \times 10^{11}$ | $2.5092 \times 10^3$ | 0.26700056 | 0.0000 | 1.0000 | 0.7149 | $3.7408 \times 10^2$ | $1.8688 \times 10^2$ |
| 0.00050 | 3.7134 | 1.1539 | $3.2855 \times 10^{11}$ | $3.3198 \times 10^3$ | 0.26700062 | 0.0000 | 1.0000 | 0.7149 | $2.9056 \times 10^2$ | $1.4632 \times 10^2$ |
| 0.00070 | 3.7186 | 1.0978 | $3.0068 \times 10^{11}$ | $3.9636 \times 10^3$ | 0.26700069 | 0.0000 | 1.0000 | 0.7149 | $2.3657 \times 10^2$ | $1.2014 \times 10^2$ |
| 0.00100 | 3.7266 | 1.0626 | $2.7825 \times 10^{11}$ | $4.6286 \times 10^3$ | 0.26700082 | 0.0000 | 1.0000 | 0.7149 | $1.7757 \times 10^2$ | $9.1357 \times 10^1$ |
| 0.00200 | 3.7749 | 1.3740 | $3.1884 \times 10^{11}$ | $3.5251 \times 10^3$ | 0.26700150 | 0.0002 | 0.9998 | 0.7149 | $1.1532 \times 10^1$ | $1.0240 \times 10^1$ |
| 0.00300 | 4.0053 | 1.9162 | $2.0600 \times 10^{11}$ | $8.4442 \times 10^3$ | 0.26707308 | 0.0353 | 0.9060 | 0.7149 | | |
| 0.00322 | 4.0321 | 1.8392 | $1.6667 \times 10^{11}$ | $1.2900 \times 10^4$ | 0.26720193 | 0.0719 | 0.0055 | 0.7149 | | |
| | | | | $2.800 M_\odot$ | | | | | | |
| 0.00002 | 3.6920 | 1.5829 | $5.9393 \times 10^{11}$ | $1.0535 \times 10^3$ | 0.26700150 | 0.0000 | 1.0000 | 0.7149 | $7.8383 \times 10^2$ | $3.3818 \times 10^2$ |
| 0.00010 | 3.6993 | 1.4483 | $4.9190 \times 10^{11}$ | $1.5359 \times 10^3$ | 0.26700152 | 0.0000 | 1.0000 | 0.7149 | $5.2828 \times 10^2$ | $2.5457 \times 10^2$ |
| 0.00020 | 3.7050 | 1.3428 | $4.2439 \times 10^{11}$ | $2.0633 \times 10^3$ | 0.26700154 | 0.0000 | 1.0000 | 0.7149 | $4.1345 \times 10^2$ | $2.0543 \times 10^2$ |
| 0.00030 | 3.7091 | 1.2712 | $3.8345 \times 10^{11}$ | $2.5275 \times 10^3$ | 0.26700156 | 0.0000 | 1.0000 | 0.7149 | $3.5028 \times 10^2$ | $1.7521 \times 10^2$ |
| 0.00050 | 3.7156 | 1.1808 | $3.3545 \times 10^{11}$ | $3.3026 \times 10^3$ | 0.26700163 | 0.0000 | 1.0000 | 0.7149 | $2.7001 \times 10^2$ | $1.3680 \times 10^2$ |
| 0.00070 | 3.7214 | 1.1322 | $3.0883 \times 10^{11}$ | $3.8963 \times 10^3$ | 0.26700170 | 0.0000 | 1.0000 | 0.7149 | $2.1812 \times 10^2$ | $1.1100 \times 10^2$ |
| 0.00100 | 3.7305 | 1.1121 | $2.8932 \times 10^{11}$ | $4.4397 \times 10^3$ | 0.26700184 | 0.0000 | 1.0000 | 0.7149 | $1.5660 \times 10^2$ | $8.1122 \times 10^1$ |
| 0.00200 | 3.7963 | 1.5402 | $3.4978 \times 10^{11}$ | $3.0375 \times 10^3$ | 0.26700245 | 0.0003 | 0.9997 | 0.7149 | | |
| 0.00294 | 4.0429 | 1.8971 | $1.6946 \times 10^{11}$ | $1.2940 \times 10^4$ | 0.26720465 | 0.0667 | 0.0019 | 0.7149 | | |
| | | | | $2.900 M_\odot$ | | | | | | |
| 0.00002 | 3.6938 | 1.5939 | $5.9663 \times 10^{11}$ | $1.0813 \times 10^3$ | 0.26700150 | 0.0000 | 1.0000 | 0.7149 | $6.7771 \times 10^2$ | $3.0972 \times 10^2$ |
| 0.00010 | 3.7010 | 1.4645 | $4.9744 \times 10^{11}$ | $1.5555 \times 10^3$ | 0.26700152 | 0.0000 | 1.0000 | 0.7149 | $4.8616 \times 10^2$ | $2.3813 \times 10^2$ |
| 0.00020 | 3.7067 | 1.3630 | $4.3102 \times 10^{11}$ | $2.0718 \times 10^3$ | 0.26700154 | 0.0000 | 1.0000 | 0.7149 | $3.8660 \times 10^2$ | $1.9333 \times 10^2$ |
| 0.00030 | 3.7109 | 1.2944 | $3.9059 \times 10^{11}$ | $2.5229 \times 10^3$ | 0.26700157 | 0.0000 | 1.0000 | 0.7149 | $3.2611 \times 10^2$ | $1.6493 \times 10^2$ |

Table 1. Continued

| Age(Gyr) | log$T_{eff}$ | log$L/L_\odot$ | R(cm) | g(cm/s$^2$) | $Y_c$ | L(PP1) | $L_g$ | $X_{env}$ | $\tau_{gc}$ (day) | $\tau_{lc}$ (day) |
|---|---|---|---|---|---|---|---|---|---|---|
| 0.00050 | 3.7178 | 1.2103 | $3.4350 \times 10^{11}$ | $3.2621 \times 10^3$ | 0.26700163 | 0.0000 | 1.0000 | 0.7149 | $2.5194 \times 10^2$ | $1.2781 \times 10^2$ |
| 0.00070 | 3.7242 | 1.1694 | $3.1821 \times 10^{11}$ | $3.8012 \times 10^3$ | 0.26700170 | 0.0000 | 1.0000 | 0.7149 | $1.9909 \times 10^2$ | $1.0190 \times 10^2$ |
| 0.00100 | 3.7348 | 1.1692 | $3.0300 \times 10^{11}$ | $4.1924 \times 10^3$ | 0.26700185 | 0.0000 | 1.0000 | 0.7149 | $1.3347 \times 10^2$ | $6.9859 \times 10^1$ |
| 0.00200 | 3.8433 | 1.7161 | $3.4500 \times 10^{11}$ | $3.2337 \times 10^3$ | 0.26700232 | 0.0006 | 0.9994 | 0.7149 | | |
| 0.00269 | 4.0536 | 1.9556 | $1.7261 \times 10^{11}$ | $1.2919 \times 10^4$ | 0.26720771 | 0.0600 | 0.0018 | 0.7149 | | |
| | | | | | 3.000$M_\odot$ | | | | | |
| 0.00002 | 3.6963 | 1.5894 | $5.8678 \times 10^{11}$ | $1.1564 \times 10^3$ | 0.26700050 | 0.0000 | 1.0000 | 0.7149 | $6.0080 \times 10^2$ | $2.8192 \times 10^2$ |
| 0.00010 | 3.7031 | 1.4715 | $4.9651 \times 10^{11}$ | $1.6151 \times 10^3$ | 0.26700052 | 0.0000 | 1.0000 | 0.7149 | $4.4480 \times 10^2$ | $2.1937 \times 10^2$ |
| 0.00020 | 3.7088 | 1.3773 | $4.3406 \times 10^{11}$ | $2.1133 \times 10^3$ | 0.26700054 | 0.0000 | 1.0000 | 0.7149 | $3.5980 \times 10^2$ | $1.7944 \times 10^2$ |
| 0.00030 | 3.7131 | 1.3140 | $3.9555 \times 10^{11}$ | $2.5449 \times 10^3$ | 0.26700057 | 0.0000 | 1.0000 | 0.7149 | $3.0486 \times 10^2$ | $1.5318 \times 10^2$ |
| 0.00050 | 3.7204 | 1.2389 | $3.5084 \times 10^{11}$ | $3.2349 \times 10^3$ | 0.26700063 | 0.0000 | 1.0000 | 0.7149 | $2.3038 \times 10^2$ | $1.1751 \times 10^2$ |
| 0.00070 | 3.7274 | 1.2091 | $3.2819 \times 10^{11}$ | $3.6968 \times 10^3$ | 0.26700071 | 0.0000 | 1.0000 | 0.7149 | $1.7678 \times 10^2$ | $9.1216 \times 10^1$ |
| 0.00100 | 3.7401 | 1.2404 | $3.2092 \times 10^{11}$ | $3.8661 \times 10^3$ | 0.26700086 | 0.0000 | 1.0000 | 0.7149 | $1.0461 \times 10^2$ | $5.6009 \times 10^1$ |
| 0.00200 | 3.9179 | 1.9100 | $3.0599 \times 10^{11}$ | $4.2525 \times 10^3$ | 0.26700128 | 0.0016 | 0.9984 | 0.7149 | | |
| 0.00246 | 4.0641 | 2.0152 | $1.7613 \times 10^{11}$ | $1.2835 \times 10^4$ | 0.26720588 | 0.0554 | 0.0063 | 0.7149 | | |
| | | | | | 3.200$M_\odot$ | | | | | |
| 0.00002 | 3.7001 | 1.6000 | $5.8366 \times 10^{11}$ | $1.2468 \times 10^3$ | 0.26700050 | 0.0000 | 1.0000 | 0.7149 | $5.1541 \times 10^2$ | $2.4877 \times 10^2$ |
| 0.00010 | 3.7066 | 1.4968 | $5.0314 \times 10^{11}$ | $1.6777 \times 10^3$ | 0.26700052 | 0.0000 | 1.0000 | 0.7149 | $3.8636 \times 10^2$ | $1.9150 \times 10^2$ |
| 0.00020 | 3.7123 | 1.4140 | $4.4547 \times 10^{11}$ | $2.1402 \times 10^3$ | 0.26700054 | 0.0000 | 1.0000 | 0.7149 | $3.0973 \times 10^2$ | $1.5624 \times 10^2$ |
| 0.00030 | 3.7170 | 1.3600 | $4.0971 \times 10^{11}$ | $2.5301 \times 10^3$ | 0.26700057 | 0.0000 | 1.0000 | 0.7149 | $2.6061 \times 10^2$ | $1.3219 \times 10^2$ |
| 0.00050 | 3.7255 | 1.3057 | $3.7005 \times 10^{11}$ | $3.1015 \times 10^3$ | 0.26700064 | 0.0000 | 1.0000 | 0.7149 | $1.8934 \times 10^2$ | $9.6995 \times 10^1$ |
| 0.00070 | 3.7347 | 1.3080 | $3.5572 \times 10^{11}$ | $3.3564 \times 10^3$ | 0.26700072 | 0.0000 | 1.0000 | 0.7149 | $1.2948 \times 10^2$ | $6.7911 \times 10^1$ |
| 0.00100 | 3.7557 | 1.4584 | $3.8391 \times 10^{11}$ | $2.8816 \times 10^3$ | 0.26700088 | 0.0000 | 1.0000 | 0.7149 | $4.1937 \times 10^1$ | $2.4943 \times 10^1$ |
| 0.00200 | 4.0839 | 2.2063 | $2.0037 \times 10^{11}$ | $1.0579 \times 10^4$ | 0.26711840 | 0.0474 | 0.4094 | 0.7149 | | |
| 0.00207 | 4.0832 | 2.1200 | $1.8193 \times 10^{11}$ | $1.2831 \times 10^4$ | 0.26720880 | 0.0480 | 0.0056 | 0.7149 | | |
| | | | | | 3.400$M_\odot$ | | | | | |
| 0.00002 | 3.6958 | 1.7406 | $7.0001 \times 10^{11}$ | $9.2091 \times 10^2$ | 0.26700050 | 0.0000 | 1.0000 | 0.7149 | $5.4359 \times 10^2$ | $2.5991 \times 10^2$ |
| 0.00010 | 3.7049 | 1.5997 | $5.7079 \times 10^{11}$ | $1.3851 \times 10^3$ | 0.26700052 | 0.0000 | 1.0000 | 0.7149 | $3.8785 \times 10^2$ | $1.9219 \times 10^2$ |
| 0.00020 | 3.7121 | 1.4989 | $4.9162 \times 10^{11}$ | $1.8671 \times 10^3$ | 0.26700054 | 0.0000 | 1.0000 | 0.7149 | $3.0008 \times 10^2$ | $1.5170 \times 10^2$ |
| 0.00030 | 3.7178 | 1.4387 | $4.4686 \times 10^{11}$ | $2.2599 \times 10^3$ | 0.26700057 | 0.0000 | 1.0000 | 0.7149 | $2.4515 \times 10^2$ | $1.2485 \times 10^2$ |
| 0.00050 | 3.7280 | 1.3890 | $4.0260 \times 10^{11}$ | $2.7840 \times 10^3$ | 0.26700063 | 0.0000 | 1.0000 | 0.7149 | $1.6482 \times 10^2$ | $8.5365 \times 10^1$ |
| 0.00070 | 3.7401 | 1.4249 | $3.9682 \times 10^{11}$ | $2.8658 \times 10^3$ | 0.26700071 | 0.0000 | 1.0000 | 0.7149 | $9.4627 \times 10^1$ | $5.1246 \times 10^1$ |
| 0.00100 | 3.7717 | 1.6609 | $4.5013 \times 10^{11}$ | $2.2271 \times 10^3$ | 0.26700089 | 0.0000 | 1.0000 | 0.7149 | $7.5434 \times 10^0$ | $3.2537 \times 10^1$ |
| 0.00182 | 4.1002 | 2.2089 | $1.8636 \times 10^{11}$ | $1.2993 \times 10^4$ | 0.26722195 | 0.0411 | 0.0002 | 0.7149 | | |
| | | | | | 3.600$M_\odot$ | | | | | |
| 0.00002 | 3.6969 | 1.7849 | $7.3280 \times 10^{11}$ | $8.8977 \times 10^2$ | 0.26700050 | 0.0000 | 1.0000 | 0.7149 | $4.9456 \times 10^2$ | $2.4001 \times 10^2$ |
| 0.00010 | 3.7065 | 1.6468 | $5.9821 \times 10^{11}$ | $1.3352 \times 10^3$ | 0.26700052 | 0.0000 | 1.0000 | 0.7149 | $3.4837 \times 10^2$ | $1.7389 \times 10^2$ |
| 0.00020 | 3.7143 | 1.5519 | $5.1722 \times 10^{11}$ | $1.7861 \times 10^3$ | 0.26700054 | 0.0000 | 1.0000 | 0.7149 | $2.6690 \times 10^2$ | $1.3458 \times 10^2$ |
| 0.00030 | 3.7208 | 1.5000 | $4.7289 \times 10^{11}$ | $2.1367 \times 10^3$ | 0.26700057 | 0.0000 | 1.0000 | 0.7149 | $2.1054 \times 10^2$ | $1.0764 \times 10^2$ |
| 0.00050 | 3.7334 | 1.4822 | $4.3722 \times 10^{11}$ | $2.4995 \times 10^3$ | 0.26700064 | 0.0000 | 1.0000 | 0.7149 | $1.2369 \times 10^2$ | $6.5402 \times 10^1$ |
| 0.00070 | 3.7519 | 1.6052 | $4.6253 \times 10^{11}$ | $2.2334 \times 10^3$ | 0.26700072 | 0.0000 | 1.0000 | 0.7149 | $4.2959 \times 10^1$ | $2.5738 \times 10^1$ |
| 0.00100 | 3.8134 | 1.8969 | $4.8759 \times 10^{11}$ | $2.0097 \times 10^3$ | 0.26700082 | 0.0000 | 1.0000 | 0.7149 | | |
| 0.00159 | 4.1140 | 2.2739 | $1.8848 \times 10^{11}$ | $1.3449 \times 10^4$ | 0.26725586 | 0.0341 | 0.0074 | 0.7149 | | |
| | | | | | 3.800$M_\odot$ | | | | | |
| 0.00002 | 3.6975 | 1.8341 | $7.7338 \times 10^{11}$ | $8.4322 \times 10^2$ | 0.26700050 | 0.0000 | 1.0000 | 0.7149 | $4.6498 \times 10^2$ | $2.2685 \times 10^2$ |
| 0.00010 | 3.7080 | 1.6942 | $6.2744 \times 10^{11}$ | $1.2811 \times 10^3$ | 0.26700052 | 0.0000 | 1.0000 | 0.7149 | $3.1507 \times 10^2$ | $1.5839 \times 10^2$ |
| 0.00020 | 3.7167 | 1.6046 | $5.4361 \times 10^{11}$ | $1.7067 \times 10^3$ | 0.26700054 | 0.0000 | 1.0000 | 0.7149 | $2.3197 \times 10^2$ | $1.1834 \times 10^2$ |
| 0.00030 | 3.7242 | 1.5645 | $5.0148 \times 10^{11}$ | $2.0055 \times 10^3$ | 0.26700057 | 0.0000 | 1.0000 | 0.7149 | $1.7333 \times 10^2$ | $9.0158 \times 10^1$ |
| 0.00050 | 3.7412 | 1.6106 | $4.8891 \times 10^{11}$ | $2.1099 \times 10^3$ | 0.26700064 | 0.0000 | 1.0000 | 0.7149 | $7.6777 \times 10^1$ | $4.2800 \times 10^1$ |
| 0.00070 | 3.7709 | 1.8092 | $5.3605 \times 10^{11}$ | $1.7551 \times 10^3$ | 0.26700074 | 0.0000 | 1.0000 | 0.7149 | $4.4412 \times 10^0$ | $2.6528 \times 10^1$ |
| 0.00100 | 3.9107 | 2.1781 | $4.3065 \times 10^{11}$ | $2.7195 \times 10^3$ | 0.26700075 | 0.0002 | 0.9998 | 0.7149 | | |
| 0.00137 | 4.1311 | 2.3746 | $1.9563 \times 10^{11}$ | $1.3178 \times 10^4$ | 0.26723873 | 0.0320 | 0.0001 | 0.7149 | | |
| | | | | | 4.000$M_\odot$ | | | | | |
| 0.00002 | 3.7002 | 1.8471 | $7.7526 \times 10^{11}$ | $8.8331 \times 10^2$ | 0.26700050 | 0.0000 | 1.0000 | 0.7149 | $4.1814 \times 10^2$ | $2.0553 \times 10^2$ |
| 0.00010 | 3.7107 | 1.7267 | $6.4329 \times 10^{11}$ | $1.2829 \times 10^3$ | 0.26700052 | 0.0000 | 1.0000 | 0.7149 | $2.7578 \times 10^2$ | $1.3879 \times 10^2$ |
| 0.00020 | 3.7202 | 1.6561 | $5.6754 \times 10^{11}$ | $1.6482 \times 10^3$ | 0.26700054 | 0.0000 | 1.0000 | 0.7149 | $1.9237 \times 10^2$ | $9.9052 \times 10^1$ |
| 0.00030 | 3.7292 | 1.6422 | $5.3591 \times 10^{11}$ | $1.8485 \times 10^3$ | 0.26700057 | 0.0000 | 1.0000 | 0.7149 | $1.3145 \times 10^2$ | $6.9507 \times 10^1$ |
| 0.00050 | 3.7552 | 1.7972 | $5.6822 \times 10^{11}$ | $1.6443 \times 10^3$ | 0.26700065 | 0.0000 | 1.0000 | 0.7149 | $2.6024 \times 10^1$ | $1.7991 \times 10^1$ |
| 0.00070 | 3.8144 | 2.0257 | $5.6305 \times 10^{11}$ | $1.6746 \times 10^3$ | 0.26700069 | 0.0000 | 1.0000 | 0.7149 | | |

Table 1. Continued

| Age(Gyr) | log$T_{\rm eff}$ | log$L/L_\odot$ | R(cm) | g(cm/s$^2$) | $Y_c$ | L(PP1) | $L_g$ | $X_{\rm env}$ | $\tau_{gc}$ (day) | $\tau_{lc}$ (day) |
|---|---|---|---|---|---|---|---|---|---|---|
| 0.00100 | 4.0460 | 2.4855 | 3.2901 × 10$^{11}$ | 4.9044 × 10$^3$ | 0.26700140 | 0.0015 | 0.9981 | 0.7149 | | |
| 0.00118 | 4.1457 | 2.4556 | 2.0084 × 10$^{11}$ | 1.3161 × 10$^4$ | 0.26724459 | 0.0281 | 0.0047 | 0.7149 | | |

4.500$M_\odot$

| Age(Gyr) | log$T_{\rm eff}$ | log$L/L_\odot$ | R(cm) | g(cm/s$^2$) | $Y_c$ | L(PP1) | $L_g$ | $X_{\rm env}$ | $\tau_{gc}$ (day) | $\tau_{lc}$ (day) |
|---|---|---|---|---|---|---|---|---|---|---|
| 0.00002 | 3.6991 | 1.9901 | 9.1895 × 10$^{11}$ | 7.0725 × 10$^2$ | 0.26700050 | 0.0000 | 1.0000 | 0.7149 | 3.7413 × 10$^2$ | 1.8670 × 10$^2$ |
| 0.00010 | 3.7131 | 1.8679 | 7.4826 × 10$^{11}$ | 1.0667 × 10$^3$ | 0.26700052 | 0.0000 | 1.0000 | 0.7149 | 2.1660 × 10$^2$ | 1.1093 × 10$^2$ |
| 0.00020 | 3.7271 | 1.8415 | 6.8063 × 10$^{11}$ | 1.2892 × 10$^3$ | 0.26700054 | 0.0000 | 1.0000 | 0.7149 | 1.1830 × 10$^2$ | 6.3101 × 10$^1$ |
| 0.00030 | 3.7457 | 1.9378 | 6.9815 × 10$^{11}$ | 1.2253 × 10$^3$ | 0.26700057 | 0.0000 | 1.0000 | 0.7149 | 3.8014 × 10$^1$ | 2.4146 × 10$^1$ |
| 0.00050 | 3.8339 | 2.2129 | 6.3836 × 10$^{11}$ | 1.4656 × 10$^3$ | 0.26700061 | 0.0000 | 1.0000 | 0.7149 | | |
| 0.00070 | 4.0423 | 2.6285 | 3.9454 × 10$^{11}$ | 3.8368 × 10$^3$ | 0.26700067 | 0.0004 | 0.9995 | 0.7149 | | |
| 0.00087 | 4.1791 | 2.6454 | 2.1424 × 10$^{11}$ | 1.3012 × 10$^4$ | 0.26724505 | 0.0231 | 0.0146 | 0.7149 | | |

5.000$M_\odot$

| Age(Gyr) | log$T_{\rm eff}$ | log$L/L_\odot$ | R(cm) | g(cm/s$^2$) | $Y_c$ | L(PP1) | $L_g$ | $X_{\rm env}$ | $\tau_{gc}$ (day) | $\tau_{lc}$ (day) |
|---|---|---|---|---|---|---|---|---|---|---|
| 0.00002 | 3.7007 | 2.0825 | 1.0146 × 10$^{12}$ | 6.4471 × 10$^2$ | 0.26700050 | 0.0000 | 1.0000 | 0.7149 | 3.2367 × 10$^2$ | 1.6185 × 10$^2$ |
| 0.00010 | 3.7182 | 2.0128 | 8.6391 × 10$^{11}$ | 8.8915 × 10$^2$ | 0.26700052 | 0.0000 | 1.0000 | 0.7149 | 1.4662 × 10$^2$ | 7.6777 × 10$^1$ |
| 0.00020 | 3.7452 | 2.1250 | 8.6804 × 10$^{11}$ | 8.8071 × 10$^2$ | 0.26700054 | 0.0000 | 1.0000 | 0.7149 | 2.7883 × 10$^1$ | 1.9601 × 10$^1$ |
| 0.00030 | 3.7934 | 2.2610 | 8.1296 × 10$^{11}$ | 1.0041 × 10$^3$ | 0.26700056 | 0.0000 | 1.0000 | 0.7149 | | |
| 0.00050 | 4.0469 | 2.7682 | 4.5365 × 10$^{11}$ | 3.2246 × 10$^3$ | 0.26700057 | 0.0002 | 0.9998 | 0.7149 | | |
| 0.00065 | 4.2075 | 2.8045 | 2.2580 × 10$^{11}$ | 1.3016 × 10$^4$ | 0.26724996 | 0.0207 | 0.0253 | 0.7149 | | |